\documentclass[%
 reprint,
preprintnumbers,
superscriptaddress,
prd,
nofootinbib,
 amsmath,amssymb, aps,
floatfix,
onecolumn
]{revtex4-2}

\makeatletter
\def\p@subsection{}
\makeatother

\usepackage{graphicx,amsmath,amssymb,lipsum,bm}

\usepackage[dvipsnames]{xcolor}
\definecolor{xlinkcolor}{rgb}{0.7752941176470588, 0.22078431372549023, 0.2262745098039215}
\definecolor{BrickRed}{rgb}{0.7752941176470588, 0.22078431372549023, 0.2262745098039215}
\definecolor{xlinkcolor}{HTML}{1c1e94}
\usepackage{booktabs} % Include this package in the preamble
\usepackage[normalem]{ulem}
\usepackage[utf8]{inputenc} 
\usepackage{mathtools}
\usepackage{aux_files/aas_macro}
\usepackage{xspace}

\usepackage{ulem}
\usepackage{diagbox}

\usepackage[dvipsnames]{xcolor}
\usepackage{mathrsfs}
\usepackage{tabularx} 

\usepackage[colorlinks=true,citecolor=xlinkcolor,linkcolor=xlinkcolor,urlcolor=xlinkcolor, backref=false,pdfborder={0 0 0}]{hyperref}
\newcommand{\beqa}{\begin{eqnarray}}
\newcommand{\eeqa}{\end{eqnarray}}

\newcommand{\be}{\begin{equation}}
\newcommand{\ee}{\end{equation}}
\newcommand{\beq}{\begin{equation}}
\newcommand{\eeq}{\end{equation}}

\newcommand\p{{\bm p}}
\renewcommand\k{{\bm k}}
\newcommand\q{\bm{q}}

\newcommand\G{\mathcal{G}_2}

\renewcommand\b{\beta}

\newcommand{\bseq}{\begin{subequations}}
\newcommand{\eseq}{\end{subequations}}

\renewcommand{\ln}{\mathop{\rm ln}\nolimits}

\def\ltsima{$\; \buildrel < \over \sim \;$\xspace}
\def\gtsima{$\; \buildrel > \over \sim \;$\xspace}
\def\simlt{\lower.5ex\hbox{\ltsima}}
\def\simgt{\lower.5ex\hbox{\gtsima}}

\newcommand{\hinvMpc}{\,h^{-1}\, {\rm Mpc}\xspace}

\newcommand{\hMpcinv}{\,h\, {\rm Mpc}^{-1}\xspace}
\newcommand{\hinvGpc}{\,h^{-1}\, {\rm Gpc}\xspace}

\newcommand{\Mpch}{\, h^{-1}\mathrm{Mpc}\, }
\newcommand{\hMpc}{\, h\mathrm{Mpc}^{-1}\, }

\newcommand{\kmax}{\, k_{\rm max}\, }
\newcommand{\knl}{\, k_{\rm NL}\, }
\newcommand{\kp}{\, {k_\parallel}\, }

\newcommand{\zhat}{\hat{\bm z}}

\newcommand{\Lya}{Ly-$\alpha$\xspace}
\newcommand{\td}{\delta}

\newcommand{\qvec}{\mathbf{q}}

\newcommand{\kvec}{\mathbf{k}}

\newcommand{\kpar}{k_{\parallel}}

\newcommand{\kvperp}{\mathbf{k}_{\perp}}
\newcommand{\apar}{\alpha_{\parallel}}
\newcommand{\aperp}{\alpha_{\perp}}

\newcommand{\vk}{\mathbf{k}}

\newcommand{\vpsi}{\boldsymbol{\psi}}

\def\gsim{\raise0.3ex\hbox{$\;>$\kern-0.75em\raise-1.1ex\hbox{$\sim\;$}}}
\def\lsim{\raise0.3ex\hbox{$\;<$\kern-0.75em\raise-1.1ex\hbox{$\sim\;$}}}

\def\beqn#1{\begin{equation}\label{#1}}
\def\eeqn{\end{equation}}

\def\beqa#1{\begin{eqnarray}\label{#1}}
\def\eeqa{\end{eqnarray}}

\newcommand{\abacus}{\textsc{AbacusSummit}\xspace}

\def\kmax{{k_\text{max}}}
\def\hMpc{h{\text{Mpc}}^{-1}}
\def\Mpch{h^{-1}{\text{Mpc}}}

\def\Z2{$\mathcal{Z_2}$}

\def\vpsi{{\boldsymbol{\psi}}}

\newcommand {\ignore}[1]{}

\renewcommand{\arraystretch}{1.3} 

\DeclareRobustCommand{\ion}[2]{%
\relax\ifmmode
\ifx\testbx\f@series
{\mathbf{#1\,\mathsc{#2}}}\else
{\mathrm{#1\,\mathsc{#2}}}\fi
\else\textup{#1\,{\mdseries\textsc{#2}}}%
\fi}
\newcommand{\MIT}{Center for Theoretical Physics -- a Leinweber Institute, Massachusetts Institute of Technology, Cambridge, MA 02139, USA}
\newcommand{\IAIFI}{The NSF AI Institute for Artificial Intelligence and Fundamental Interactions, Cambridge, MA 02139, USA}

\usepackage{slashbox}

\begin{document}

\preprint{MIT-CTP/5998}
\preprint{KEK-Cosmo-0408}

\title{Lyman-$\alpha$ Forest and  its 
Cross-Correlation with High-Redshift Galaxies \\
in Effective Field Theory at the Field Level}

\author{Roger de Belsunce}
\email{belsunce@mit.edu}
\affiliation{\MIT}
\affiliation{\IAIFI}
\author{Mikhail M. Ivanov}
\email{ivanov99@mit.edu}
\affiliation{\MIT}
\affiliation{\IAIFI}
\author{James M.~Sullivan}
\email{jms3@mit.edu}\thanks{Brinson Prize Fellow}
\affiliation{\MIT}
\affiliation{\IAIFI}
\author{Shi-Fan Chen}
\affiliation{Department of Physics, Columbia University, New York, NY 10027, USA}
\affiliation{NASA Hubble Fellowship Program, Einstein Fellow}
\author{Kazuyuki Akitsu}
\affiliation{Theory Center, Institute of Particle and Nuclear Studies, High Energy Accelerator Research Organization (KEK), Tsukuba, Ibaraki 305-0801, Japan}

%              %  but any date may be explicitly specified

\begin{abstract}
We present a field-level perturbative forward model for the Lyman-$\alpha$ (\Lya) forest flux decrement.
We validate it on two simulation suites: large-volume \abacus
$N$-body simulations with the \Lya forest painted onto the dark matter field,
and the Sherwood hydrodynamic simulations. 
Across the redshift range of the simulations ($z=2.0$--$3.2$), the 3D and 1D power spectra of the model match the simulated \Lya fields at the 1\% (5\%) level up to $k \approx 0.3\, (1.0)\hMpcinv$, with similar performance for the cross-correlation with massive dark matter halos.
The counts-in-cells statistic shows
excellent agreement down to cell radii of $2\hinvMpc$.
Leveraging cosmic variance cancellation, the model
enables precision measurements of \Lya bias parameters and robustly detects the
full set of quadratic line-of-sight bias operators, consistent with the notion of 
naturalness 
in effective field theory (EFT). 
We quantify the stochasticity of the \Lya forest 
(the analog to 
the one-halo term), and find it to be white (scale- and 
orientation-independent)
on large scales, matching EFT predictions.
We further find that phenomenological flux power spectrum models, based on modulations
of the linear-theory power spectrum, fail at the field level even on
quasi-linear scales. 
For the currently observing Dark Energy Spectroscopic
Instrument (DESI), we generate 
large-scale clustering
mocks of the \Lya forest to validate cosmological parameter inference
pipelines. Looking ahead to its successor, DESI-II, we produce large-volume 
mocks
of representative samples of 
Lyman-break galaxies (LBGs) and 
\Lya emitters (LAEs), calibrated on 
Astrid hydrodynamic simulations and matched to observations at \(z=3\), enabling joint analyses of \Lya forest and high-redshift galaxy data. 
\end{abstract}

\maketitle
\makeatletter
\renewcommand{\l@subsubsection}[2]{}
\makeatother
\tableofcontents

\section{\label{sec:intro}Introduction and key results}
The Lyman-$\alpha$ (\Lya) forest consists of a series of absorption features in the spectra of distant quasars, produced by intervening neutral hydrogen along the line of sight. Since the 1990s, high-resolution observations with instruments such as the High Resolution Echelle Spectrometer \citep[HIRES;][]{Vogt:1994, OMeara:2021} and the Ultraviolet and Visual Echelle Spectrograph \citep[UVES;][]{Dekker:2000, Murphy:2019} have enabled precise measurements of small-scale fluctuations in the neutral hydrogen density. These observations established that the absorbing gas resides in the low-density, highly ionized intergalactic medium (IGM), providing a transparent link between the neutral hydrogen distribution and the underlying dark matter field. This connection, in turn, enables precision simulations of the \Lya forest. Because the neutral gas is in photoionization equilibrium with an approximately uniform ultraviolet background, the \Lya forest probes density fluctuations from cosmological down to Mpc scales and below over a wide redshift range $(2 \leq z \leq 5)$ using ground-based observations.

High-resolution spectra have enabled analyses deep into the small-scale regime ($\kmax \simlt 10\hMpcinv$) through measurements of the line-of-sight (or one-dimensional) power spectrum \cite{Seljak:2005, Viel:2005, McDonald06, PYB13, Chabanier:2019, Pedersen:2020, 2023MNRAS.526.5118R, 2024MNRAS.tmp..176K}. These measurements are sensitive to a broad range of fundamental physics, including neutrino properties \cite{Seljak:2005, Viel:2010, PYB13, Palanque2020, Ivanov:2024jtl, He:2023oke, He:2025jwp}, primordial black holes \cite{Afshordi:2003, Murgia:2019, Ivanov:2025pbu}, dark matter models \cite{Viel:2013, Baur:2016, Irsic17, Kobayashi:2017, Armengaud:2017, Murgia:2018, Garzilli:2019, Irsic:2020, Rogers:2022, Villasenor:2023, Irsic:2023}, the thermal history of the ionized IGM \cite{Zaldarriaga:2002, Meiksin:2009, McQuinn:2016, Viel:2006, Walther:2019, Bolton:2008, Garzilli:2012, Gaikwad:2019, Boera:2019, Gaikwad:2021, Wilson:2022, Villasenor:2022}, non-minimal cosmological models \cite{Goldstein:2023gnw, Fuss:2022zyt, Garny:2018byk}, and the running of the spectral index \cite{Seljak:2006bg, Ivanov:2024jtl}.

Over the past two decades, \Lya forest surveys have expanded dramatically in both spectral resolution and the total cosmological volume probed. Large samples of medium-resolution quasar spectra from the extended Baryon Oscillation Spectroscopic Survey \citep[eBOSS;][]{Dawson:2016} and, in particular, from the ongoing Dark Energy Spectroscopic Instrument \citep[DESI;][]{DESI:2016, DESI_BAO_2024, DESI_lya_2024,DESI_lya_dr2} now enable cross-correlation analyses across many independent lines of sight. These multi-skewer maps trace the large-scale structure of our Universe, containing information similar to that provided by spectroscopic galaxy and quasar surveys at high redshift, but with a significantly wider dynamic range \cite{DESI:2016, Cuceu:2021, Gerardi:2022ncj}. 

Observations of the \Lya forest constrain the expansion history of the Universe through measurements of the baryon acoustic oscillation (BAO) feature \cite{McDonald:2007, Slosar2013, Busca:2013, dMdB:2020, DESI_lya_2024} and through the broadband shape of the three-dimensional correlation function \cite{Slosar2013, Cuceu:2021, Cuceu:2023, Gordon:2023, Cuceu:2025nvl}. The mapping between neutral hydrogen and underlying dark matter has motivated the development of high-fidelity numerical simulations of the \Lya forest \cite{1994ApJ...437L...9C,1996ApJ...471..582M,2001MNRAS.327..296M, Sexton2021, Bolton17,2017MNRAS.465.3291W, Bird:2023evb, Pillepich:2017jle,CAMELS_presentation, 2021JCAP...04..059W,Chabanier:2024knr}. These simulations are commonly used to calibrate linear-theory-based models augmented by phenomenological fitting functions that account for nonlinear growth, pressure smoothing, and line-of-sight velocity broadening \cite{McDonald:2001fe, Arinyo-i-Prats:2015vqa, Givans:2022qgb}. 

At the current precision set by the finite sampling of quasar sightlines (i.e.,~the shot-noise limited regime for galaxies), such modeling approaches have yielded robust cosmological constraints \cite{dMdB:2018, DESI_BAO_2024, DESI_lya_2024, Cuceu:2025nvl}. However, DESI is expected to observe up to one million quasar spectra over its lifetime, with forecasts indicating a cumulative precision below the 0.2\% level when combining all tracers and redshift bins \cite{DESI:2016}. Achieving this level of precision requires exquisite control over theoretical systematics. Indeed, recent studies have demonstrated that phenomenological \Lya forest models  bias the inferred BAO scaling parameters -- the primary observables for constraining the cosmic expansion history with DESI -- at the 0.3\% level \cite{deBelsunce:2024rvv, Hadzhiyska:2025cvk}. These findings indicate that existing modeling frameworks are approaching their limits, motivating the development of more accurate and robust theoretical descriptions for current DESI data and forthcoming surveys such as DESI-II, the WEAVE-QSO survey \citep{2016sf2a.conf..259P}, the Prime Focus Spectrograph \citep[PFS;][]{2022PFSGE} and 4MOST \citep{2019Msngr.175....3D}.

\begin{figure*}
    \centering
    \hfill
    \includegraphics[width=0.40\linewidth]{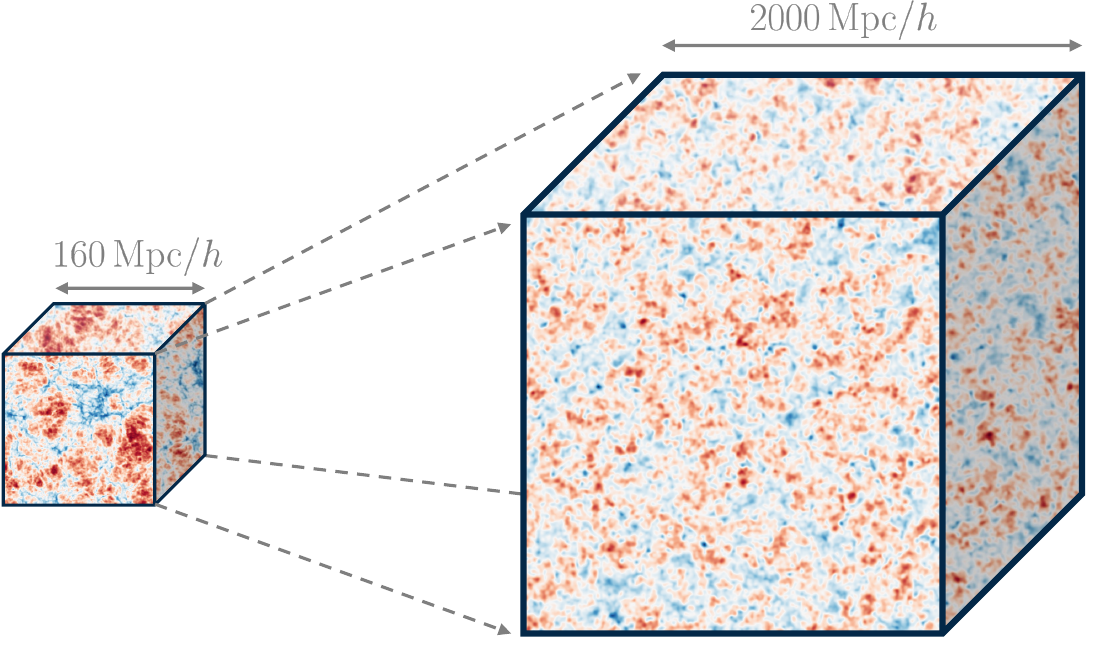}\hfill
    \includegraphics[width=0.40\linewidth]{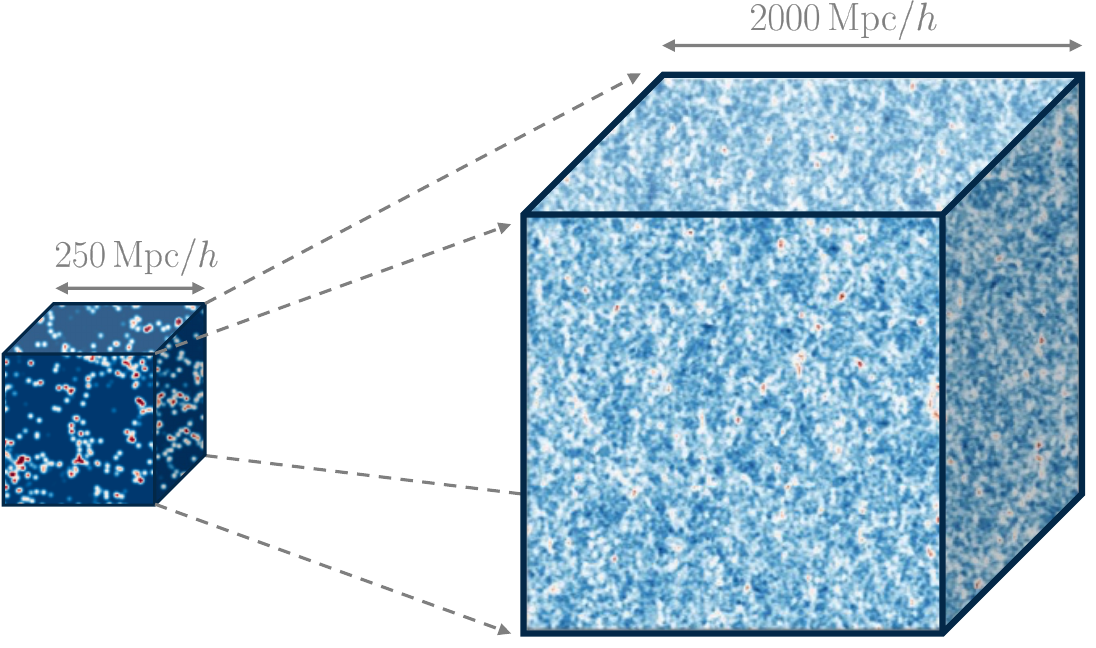}
    \hfill
    \vspace{-0.1in}
    \caption{
    \textbf{Summary of Results:} We fit an analytic, perturbative forward model at the field level to high-resolution hydrodynamic simulations, and use it to generate large-volume realizations. In the left panel, we fit small-volume ($V=(160\, h^{-1}\mathrm{Mpc})^3$) Sherwood hydrodynamic simulations of the \Lya forest \cite{Bolton:2016bfs}. In the right panel, a similar framework is applied to high-redshift galaxy simulations (such as Lyman-$\alpha$ emitters and Lyman-break galaxies) from Astrid in a $(250\,h^{-1}\mathrm{Mpc})^3$ volume \cite{2022MNRAS.512.3703B,2022MNRAS.513..670N}. 
    These calibrated models enable the generation of large-volume simulations (here: $V=(2000\,h^{-1}\mathrm{Mpc})^3$) here depicted as larger 3D boxes. Red (blue) regions correspond to over- (under-)dense regions. The field-level fits benefit from cosmic variance cancellation when using the same set of initial conditions (ICs) for the forward model as for the input hydrodynamic simulation. Coincidentally, two simulations calibrated on different simulations using the same ICs can be used for cross-correlation analyses of the \Lya forest and high-redshift galaxies -- a key science driver for DESI-II. 
    }
    \label{fig:EFT_sims}
    \vspace{-0.2in}
\end{figure*}

One can remove the bias on the BAO inference within the framework of the effective field theory (EFT) of large-scale structure, which has recently been extended to the \Lya forest \cite{Ivanov:2023yla,deBelsunce:2024rvv,Belsunce_Sullivan_skewspectrum,deBelsunce:2025bqc}. 
The EFT formalism provides a perturbative description of large-scale dynamics by incorporating only the symmetries relevant to the tracer~\cite{McDonald:2009dh,Baumann:2010tm,Carrasco:2013mua,Ivanov:2022mrd}. 
In the case of the \Lya\ forest, these include the equivalence principle and rotational invariance around the line-of-sight direction $\hat{z}$, corresponding to the $SO(2)$ group~\cite{McDonald:1999dt,Givans:2020sez,Desjacques:2018pfv,Chen:2021rnb,Ivanov:2023yla,Ivanov:2024jtl, Belsunce_Sullivan_skewspectrum}. 
Whilst this paves the way to directly constrain cosmological parameters through the one-loop power spectrum (in the context of low-redshift galaxy surveys, see, e.g.,~\cite{Ivanov:2019pdj,DAmico:2019fhj,Chen:2021wdi,Philcox:2021kcw,Chen:2022jzq,Chen:2024vuf}) a key challenge for cosmological analyses is that the large range of scales involved require large-volume, high-resolution simulations to validate inference pipelines.

High-resolution hydrodynamic simulations accurately capture the involved physics on small to intermediate scales but running a set of simulations covering a wide range of cosmological and astrophysical parameters, whilst capturing cosmological volumes is prohibitively expensive. To mitigate this, several approaches have been developed to connect the observed flux to the underlying matter field for large-scale clustering simulations~\cite{Hadzhiyska:2023, Hadzhiyska:2025cvk, Croft98,2022ApJ...930..109Q, Peirani:2014,Peirani:2022, Sorini:2016, 2022ApJ...927..230S, 2024A&A...682A..21S}. In particular, deep learning-based reconstruction methods \cite{Horowitz:2021olb, Harrington:2021srm, Jacobus:2024yev, Horowitz:2025rke, Hafezianzadeh:2025ifw} yield promising results at intermediate to small scales but do not fully capture the long-wavelength quasi-linear modes. 
An alternative avenue are emulators that directly predict a summary statistic such as the power spectrum~\cite{Bird:2018efe, Pedersen:2020kaw,Ho:2023alj,Chaves-Montero:2024gnz}. Whilst these offer per cent level accuracy for a wide range of scales they depend on (i) fits to power spectra of small-volume hydrodynamic simulations suffering from cosmic variance; and 
(ii) are restricted to the power spectrum.
Whilst paired-fixed simulations reduce cosmic variance limitations~\cite{Pontzen:2015eoh}, the resulting emulators cannot directly be generalized to other summary statistics. 
A forward model at the field level, however, needs to match \emph{all} the amplitudes and phases of \emph{all} Fourier modes -- a more stringent test of the theoretical framework than comparing summary statistics -- all  whilst yielding higher-order moments of the field.

To enable cosmological analyses of the \Lya forest from DESI and DESI-II that incorporate two- and, eventually, three-point statistics, we present a computationally efficient framework for generating large-volume simulations calibrated on high-fidelity hydrodynamic simulations. This work provides the theoretical background for Ref.~\cite{deBelsunce:2025bqc}, where this technique was first introduced for the \Lya forest, and extends earlier developments in perturbative, field-level modeling \cite{Seljak:2012tp, Cieplak:2015kra, Schmittfull:2018yuk, Schmittfull:2020trd, Obuljen:2022cjo, Ivanov:2024xgb}. The same framework  generalizes to simulations of high-redshift galaxies, enabling validation of cross-correlation measurements between the \Lya forest and galaxy positions. Such joint analyses help break the degeneracy between the growth rate \(f\) and the otherwise poorly constrained velocity-gradient bias \cite{Font-Ribera:2013fha, Chudaykin:2025gsh}. More broadly, the simulations developed here provide a controlled environment for validating end-to-end full-shape inference pipelines and for quantifying key physical systematics, including shifts of the BAO feature \cite{Chen:2024tfp, deBelsunce:2024rvv}.

We validate our perturbative forward model using two complementary sets of simulations. First, we employ the  large-volume \(N\)-body simulations from the \abacus suite, onto which the \Lya forest is painted, spanning a volume of \(V = 2^3\,(\hinvGpc)^3\) \cite{Hadzhiyska:2023}. Second, we use two sets of the Sherwood hydrodynamic simulations, which cover a volume of \(V = 160^3\,(\hinvMpc)^3\) and \(V = 80^3\,(\hinvMpc)^3\) \cite{Bolton:2016bfs}. A key advantage of performing field-level fits is the resulting cancellation of cosmic variance, which enables tight constraints on both cosmological and nuisance parameters even for small-volume simulations.

In this work, we present two main results. First, we fit and \emph{predict} the shapes of the bias transfer functions using a perturbative bias expansion directly at the field level. Second, using the resulting field-level fits, we generate large-scale clustering simulations. Examples of these simulations are shown in Fig.~\ref{fig:EFT_sims}, with the \Lya forest displayed in the left panel and high-redshift galaxies in the right panel. The methodology developed here is critical for forthcoming cosmological analyses, as existing large-scale clustering mocks \cite{Farr20, Ramirez-Perez:2021cpq} are approaching their limits when targeting scales an order of magnitude smaller than the current baseline for analyses of the \Lya broadband shape of \(r\approx 25\,\hinvMpc\) \cite{Cuceu:2025nvl}. Especially, since the \Lya forest is one of the only ways of probing the high-redshift Universe (\(2 \lesssim z \lesssim 5\)) ahead of next-generation surveys such as DESI-II and Spec-S5 \cite{schlegel_megamapper_concept}. 

% Details of structure
This paper provides the theoretical foundation of the \Lya forward model presented in a \textit{Letter} in Ref.~\cite{deBelsunce:2025bqc} and is organized as follows: 
We review the perturbative forward model of the \Lya forest and halos (as proxies for high-redshift galaxies) in redshift space in Sec.~\ref{sec:model_setup}. 
In Sec.~\ref{sec:simulations} we present the used synthetic data. 
We assess the performance of our perturbative forward model in Sec.~\ref{sec:results} and investigate the obtained transfer functions from our field-level fits in Sec.~\ref{sec:tranfer_func}. 
In Sec.~\ref{sec:theory_transfer_func} we compare the measured transfer functions to theoretically expected ones and use these transfer functions to create large-scale clustering mocks encompassing cosmological volumes in Sec.~\ref{sec:EFT_mocks}.
We conclude and discuss future work in Sec.~\ref{sec:conclusions}.

\section{Building the field-level model} \label{sec:model_setup}

The description of the 
cosmological 
\Lya
forest correlations begins 
with the linear theory model \cite{McDonald:1999dt,McDonald:2001fe},
\be 
\label{eq:lin_model}
\delta^{\rm lin}_F(\k,z) = (b_1-b_\eta f\mu^2) \delta_{1}(\k,z) ~\,,
\ee 
where $b_1$, $b_\eta$
are linear bias parameters, 
$f=d\ln D_+/d\ln a$ is the logarithmic growth 
factor, 
$\mu\equiv k_\parallel/k = \hat{\mathbf{k}}\cdot\zhat $ is the cosine of the angle to the line-of-sight, $\zhat$,
and 
$\delta_{1}$ the linear
density field, which can be rewritten as 
\be 
\delta_{1}(\k,z) = \delta_{1,0}(\k)D_+(z)\,,
\ee 
where $\delta_{1,0}(\k)$
is the initial 
condition field extrapolated to 
redshift zero and $D_+(z)$
is the linear growth factor.
In simulations, $\delta_{1,0}(\k)$
is the random scalar field
generated from the linear matter
power spectrum of initial conditions. 
For brevity, we will 
suppress the time dependence
of $\delta_1$, implicitly assuming
that this quantity is always evaluated at the simulation
redshift $z$.
The model in Eq.~\eqref{eq:lin_model}
reproduces the well-known 
linear theory model for the flux
power spectrum, 
\be 
\begin{split}
\langle\delta_F(\k) \delta_F(\k')\rangle 
& ~= (2\pi)^3 \delta_D^{(3)}(\k+\k')
\langle\delta_F(\k) \delta_F(\k')\rangle'=
(2\pi)^3 \delta_D^{(3)}(\k+\k')P(k,\mu)~\,,\\
P_{11}(k) & \underbrace{=}_{\rm linear} (b_1-f b_\eta \mu^2)^2 P_{\rm lin}(k)\,,
\end{split}
\ee 
where $P_{\rm lin}(k)$ is the linear matter power spectrum (evaluated at redshift $z$) where we use $\int_{\k} = \int \mathrm{d}^3 k/(2\pi)^3$. The above coefficient
$(b_1-fb_\eta\mu^2)^2$
is the well-known generalization of the Kaiser factor for galaxies~\cite{Kaiser:1987,McDonald:1999dt,McDonald:2001fe},
which can be recovered by setting 
$b_\eta=-1$.

From this discussion it is clear
that all phases of the \Lya
field are captured by 
the field $\delta_{1}(\k)$
in the linear approximation. 
It is thus convenient to
split the theory model from Eq.~\eqref{eq:lin_model}
into parts that make the amplitude
and phase dependence manifest. 
Absorbing the 
Kaiser factor into Eq.~\eqref{eq:lin_model}
into a momentum-dependent 
transfer function yields
\be 
\label{eq:lin_model_2}
\delta^{\rm lin}_F(\k,z) = \beta_1(k,\mu)\delta_{1}(\k,z) \,.
\ee
% As we shall see, 
The advantage of this approach will become evident 
at the non-linear level, where the transfer function 
$\beta_1$ will also account for the 
one-loop corrections. 

The success of the linear theory 
model in describing the 
simulated \Lya field $\delta_F^{\rm truth}$
can be estimated using the 
error power spectrum, 
\be \label{eq:Perr}
P_{\rm err}(\k) =\langle |\delta_F^{\rm truth}(\k) - 
\delta^{\rm lin}_F(\k)
|^2\rangle' ~\,.
\ee 
On very large scales (i.e. in the limit $k\to 0$), the non-linear corrections to our model
are expected to be small, 
so the
only expected source
of error in the model should be the stochastic 
field $\epsilon$, 
\be 
\delta_F(\k)
=\delta_F^{\rm lin}(\k)+\epsilon(\k)~\,,
\ee 
which by definition does not 
correlate with $\delta_F^{\rm lin}$ generated by cosmological
fluctuations. In general, 
$\epsilon(\k)$
is the \Lya flux decrement 
component produced by 
small-scale processes
unrelated to large-scale 
initial conditions.
In the context of galaxies and halos, 
the field $\epsilon(\k)$
captures the shot noise which arises 
due to the discreteness 
of tracers. Using this analogy, the stochastic field of the \Lya forest could be thought to 
originate from the discreteness
of absorption lines. Due to their
large numbers it is expected to be extremely small. This logic, 
however, is not fully correct because it assumes that the discreteness is generated at the level of absorption lines. 
One can consider that \Lya
absorption is produced by 
neutral hydrogen clouds whose 
distribution also has a stochastic
component. This discreteness of the \Lya clouds can be used as a first proxy to understand the 
stochasticity of the forest. 

Within the EFT for LSS framework~\cite{Baumann:2010tm,Carrasco:2013mua,Mirbabayi:2014zca,Pajer:2013jj,Desjacques:2016bnm,Desjacques:2018pfv,Ivanov:2022mrd}, 
one uses perturbative Taylor 
expansions to parameterize 
unknown functions. 
In this approach, 
the error power spectrum assumes
the following expansion
valid in the $k\to 0$ limit:
\be 
\label{eq:EFTPerr}
P_{\rm err}(k,\mu)= n_0 (1+ \alpha_1 k^2 + \alpha_2 k^2\mu^2 + ...)\,,
\ee 
where $n_0,\alpha_{1,2}$ are dimensional constants whose values are set by the \Lya physics. For galaxies, $n_0$ can be estimated at leading order as $1/\bar n$, where $\bar n=N/V$
is the galaxy number density ($N$ is the number of galaxies in the comoving volume $V$).
The relevant distance scale in this case is the mean
separation $R=(V/N)^{1/3}$ between the
individual galaxies. 
Applying the same 
argument to the \Lya
forest and assuming 
for simplicity that the individual tracer 
is a neutral hydrogen 
cloud whose separation
to its neighboring cloud is about 
$0.5~\Mpch$, we get an estimate 
\be \label{eq:EFT_Perr_estimate}
n_0\sim R^3\sim 0.1~[\Mpch]^3\,\quad 
\alpha_{1,2}\sim R^2\sim 0.3~[\Mpch]^2~\,.
\ee 
As we shall see, this 
naive 
estimate will turn
out to be quite accurate 
for the actual \Lya
simulations. 
Importantly, the stochasticity expansion
features scales
that are in general
different from 
those that appear
in the perturbative 
bias
expansion. The latter can be estimated as a
non-linear scale, where matter density fluctuations 
become of order one~\cite{Ivanov:2024lya}, 
\be 
\frac{k_{\rm NL}^3}{2\pi^2}P_{\rm lin}(k_{\rm NL},z)=1\,,\quad 
\Rightarrow \quad 
k_{\rm NL}\approx 5~\hMpc\quad \text{at}\quad z=2.8\,.
\ee

If the theory model is accurate, 
we expect to recover the scale-dependence suggested by Eq.~\eqref{eq:EFTPerr}
on large scales. If there is a significant correction to the naive linear model~\eqref{eq:lin_model_2},
this will generate a noticeable 
scale and orientation
dependence not captured 
by Eq.~\eqref{eq:EFTPerr}.
What corrections do we expect? 

First, the displacements of 
dark matter particles 
in our simulations are large,
and have to be treated non-perturbatively. If unaccounted for, they produce large distortions 
of baryon acoustic 
oscillations, which show up 
as a mismatch between
the linear model and actual phases
of the density field.
This effect is well understood, 
and can be corrected for by 
using the linear density field $\tilde\delta_1$ shifted by the  Zel'dovich
displacement 
in lieu of $\delta_1$~\cite{1970A&A.....5...84Z}.
In what follows $\tilde\delta_1$ 
will be referred to as the shifted linear field. 
\be 
\tilde{\delta}_1(\k)
=\int d^3 \q~\delta_1(\q)
e^{-i\k\cdot(\q+\vpsi_1(\q)+f\zhat(\vpsi_1(\q)\cdot \zhat))}~\,,
\ee 
where $\q$ denotes Lagrangian space (initial) coordinates, and $\vpsi_1$ is the Zel’dovich displacement  
\be 
\vpsi_1(\q) = \int d^3k\, e^{i\q\cdot \k}\, \frac{i\k}{k^2} \delta_1(\k).
\label{eq:zel_displacement}
\ee
Note that $\tilde{\delta}_1$ above has an infinite Taylor expansion
in the linear field $\delta_1$. In
perturbation theory one can write this as: 
\be 
\label{eq:d1_spt}
\tilde \delta_1 = \sum_{n=1}^3\left(
\prod_{i=1}^n
\int_{\k_i}\delta_1(\k_i)
\right)(2\pi)^3\delta_D^{(3)}(\k-\k_{1...n})\tilde K_n
(\k_1,...,\k_n)
\,,
\ee 
where $\tilde K_1 (\k)  = 1$ and 
\be 
\label{eq:delta1_kern_rsd}
\begin{split}
\tilde K_2(\k_1,\k_2) &  = 
\frac{\k\cdot \k_1}{2k_1^2} 
+ \frac{\k\cdot \k_2}{2k_2^2} 
+
\frac{(f\mu k )}{2}\left(\frac{k_{1z}}{k_1^2}+\frac{k_{2z}}{k_2^2}
\right)\,,\quad \text{etc.}
\end{split}
\ee 
where $k_{iz}=(\k_i \cdot\hat{\bm z})$,
and $\k\equiv \k_1+...+\k_n$ for the n'th kernel.
This expansion 
is very similar to
that of the Zel'dovich matter density field~\cite{1970A&A.....5...84Z,Scoccimarro:1995if,Scoccimarro:1996se,Scoccimarro:1996jy} 
\be 
\delta_{\rm Z}(\k)
=\int d^3 \q~
e^{-i\k\cdot(\q+\vpsi_1(\q)+f\zhat(\vpsi_1(\q)\cdot \zhat))}=\sum_{n=1}\left(
\prod_{i=1}^n
\int_{\k_i}
\delta_1(\k_i)
\right)(2\pi)^3\delta_D^{(3)}(\k-\k_{1...n})F^{\rm ZA}_n
(\k_1,...,\k_n)
\,,
\ee 
with $F_{1}^{\rm ZA}=1+f (\hat{\k}\cdot \zhat)^2$, and 
\be 
F_2^{\rm ZA}= \frac{1}{2}\frac{(\k\cdot \k_1)(\k\cdot \k_2)}{k_1^2k_2^2}+\frac{f}{2}\frac{(\k_1\cdot\k)k_{2z}(k\mu)}{k_1^2k_2^2}
+\frac{f}{2}\frac{(\k_2\cdot\k)k_{1z}(k\mu)}{k_1^2k_2^2}
+\frac{f^2(\mu k )^2}{2k_1^2k_2^2}k_{1z}k_{2z}
\,, \quad \text{etc.}
\ee 
These building blocks
will be useful 
in our future discussion. 
The shifts implemented by the Zel'dovich displacement
introduce higher
order non-linear effects. In the following, we will discuss these effects
more systematically. 

Nonlinearities in the bias expansion
are a second 
important source of 
corrections.
In the perturbative Eulerian bias 
formulation these are
given by
\be
\begin{split}
\delta_F(\k)=\sum_{n=1}\delta_F^{(n)}= &\sum_{n=1} \Big[ \prod_{j=1}^n\int_{}\frac{d^3{\bf k}_j}{(2\pi)^3} \delta_1(\k_j)\Big]
 K_n(\k_1,...,\k_n) (2\pi)^3\delta^{(3)}_D(\k-\k_1-...-\k_n)\,,
 \end{split}
\ee
where $K_n$ are non-linear kernels. 
General perturbative corrections to the power spectrum from individual powers in this series have the form 
\be 
P_{nm}=s_{nm}\langle \delta_F^{(n)}(\k)\delta_F^{(m)}(-\k)\rangle'\,,
\ee 
where $s_{nm}$
is the combinatorial factor and $n=m=1$ at the linear level, $n+m=4,6,8$ etc. at the one, two, three-loop orders, respectively.
At the quadratic order one has
the following general set of 
operators 
consistent with the line-of-sight rotations~\cite{Desjacques:2018pfv,Ivanov:2023yla}: 
\be
\label{eq:K2full}
\begin{split}
% & K_1(\k) = b_1-b_\eta f\mu^2\,,\\
 K_2
 (\k_1,\k_2)
\equiv b_1 F_\delta
% (\k_1,\k_2) 
+ b_2 F_{\delta^2}+b_{\G}F_{\G}+b_\eta F_{\eta} +b_{\delta \eta}F_{\delta \eta}+
b_{\eta^2} F_{\eta^2}
+b_{\Pi^{[2]}_\parallel}
F_{\Pi^{[2]}_\parallel}
+b_{(KK)_\parallel}
F_{(KK)_\parallel}~\,,
\end{split}
\ee
where the momentum-dependent quadratic kernels are given by 
\be 
\begin{split}
& F_{\delta^2}=\frac{1}{2}\,,\quad F_{\G}=\left(\frac{(\k_1\cdot \k_2)^2}{k_1^2 k_2^2}-1\right)\,,\quad F_{\delta}=1+
\frac{(\k_1\cdot \k_2)}{2k^2_1}
+\frac{(\k_1\cdot \k_2)}{2k^2_2}+\frac{2}{7}F_{\G}
+f\frac{\mu_1\mu_2}{2}\left(\frac{k_2}{k_1} + \frac{k_1}{k_2}\right)
\,,\\
& F_{\eta}=- f\mu^2 \left(
1+
\frac{(\k_1\cdot \k_2)}{2k^2_1}
+\frac{(\k_1\cdot \k_2)}{2k^2_2}+\frac{4}{7}F_{\G}
\right)-f^2\frac{\mu_1\mu_2}{2}\left(\frac{k_2}{k_1}\mu_2^2 + \frac{k_1}{k_2}\mu_1^2\right)\,,\quad 
F_{\delta\eta}=- f\frac{\mu_2^2+\mu_1^2}{2}\,, 
\\
& 
F_{\eta^2}=f^2\mu_1^2\mu_2^2\,,\quad 
F_{(KK)_\parallel}=\mu_1\mu_2 \frac{(\k_1\cdot \k_2)}{k_1k_2}
-\frac{\mu_1^2+\mu_2^2}{3}+\frac{1}{9}\,,
\quad F_{\Pi^{[2]}_\parallel}=\mu_1\mu_2 \frac{(\k_1\cdot \k_2)}{k_1k_2}-\frac{5}{7}\mu^2 F_{\G}\,,\\
\end{split}
\ee 
where $\mu_i \equiv \frac{\hat{\mathbf z}\cdot \mathbf k_i}{k_i}$ are the cosines between the line-of-sight and momentum vectors. 
One can see that the kernels above can be related to the 
shifted density 
and Zel'dovich kernels that 
we have introduced before. For instance, it is easy to see that  
\be 
F_\delta=\tilde{K}_2 +F_{\delta \eta} + \frac{2}{7}F_{\mathcal{G}_2}~\,.
\ee 
Likewise,
the kernel $F_\eta$ can be rewritten as 
\be 
F_\eta = F_{\delta}-\left(F_2^{\rm ZA}-\frac{3}{7}f\mu^2F_{\G}\right)+\frac{3}{14}F_{\G}+F_{\eta^2}-F_{\delta \eta}=\tilde{K}_2-\left(F_2^{\rm ZA}-\frac{3}{7}f\mu^2F_{\G}\right)+\frac{1}{2}F_{\G}+F_{\eta^2}~\,.
\ee 
Finally, we have 
\be 
F_{(KK)_\parallel}=F_{\Pi^{[2]}_\parallel}+\frac{5}{7}\mu^2 F_{\G}-\frac{\mu_1^2+\mu_2^2}{3}+\frac{1}{9}
=F_{\Pi^{[2]}_\parallel}+\frac{5}{7}\mu^2 F_{\G}
+\frac{2}{3f}F_{\delta \eta}+\frac{2}{9}F_{\delta^2}~\,.
\ee 
Therefore, using the new
line-of-sight velocity 
divergence field
\be \label{eq:eta_new}
\eta_{\rm new}\equiv -\left(\delta_{Z}-\frac{3}{7}f\mu^2\G
\right)\,,
\ee 
the quadratic kernels 
can be rewritten as 
\be
\label{eq:K2full_2}
\begin{split}
 K_2(\k_1,\k_2)
\equiv & (b_1+b_\eta) \tilde{K}_2
+b_\eta F_{\eta_{\rm new}}
+ b_2F_{\delta^2}
+\left(\frac{2}{7}b_1 
+\frac{1}{2}b_\eta 
+b_{\G}\right)F_{\G} \\
&+(b_1+b_{\delta \eta})F_{\delta \eta}+
(b_{\eta^2}+b_\eta) F_{\eta^2}
+b_{\Pi^{[2]}_\parallel}
\left(
F_{(KK)_\parallel}
% F_{(KK)_\parallel}-\frac{2}{3}f
% F_{\delta\eta}-\frac{1}{9}
% F_{\Pi^{(2)}}
-\frac{5}{7}\mu^2 F_{\G}
-\frac{2}{3f}F_{\delta \eta}-\frac{2}{9}F_{\delta^2}
\right)
+b_{(KK)_\parallel}
F_{(KK)_\parallel}\\
=& (b_1+b_\eta ) \tilde{K}_2
-b_\eta \left(
F_2^{\rm ZA}-\frac{3}{7}f\mu^2F_{\G}
\right)
+ \left(b_2-\frac{2}{9}b_{\Pi^{[2]}_\parallel}
\right) F_{\delta^2}
+\left(\frac{2}{7}b_1 
+\frac{1}{2}b_\eta 
+b_{\G}
-\frac{5}{7}\mu^2
b_{\Pi^{(2)}}
\right)F_{\G} \\
&+\left(b_1+b_{\delta \eta}-\frac{2}{3f}b_{\Pi^{[2]}_\parallel}\right)F_{\delta \eta}+
(b_{\eta^2}+b_\eta) F_{\eta^2}
% \left(
% F_{(KK)_\parallel}
% % F_{(KK)_\parallel}-\frac{2}{3}f
% % F_{\delta\eta}-\frac{1}{9}
% % F_{\Pi^{(2)}}
%  % F_{\G}
% % F_{\delta \eta}-\frac{2}{9}F_{\delta^2}
% \right)
+(b_{(KK)_\parallel}+b_{\Pi^{[2]}_\parallel})
F_{(KK)_\parallel}
\end{split}
\ee
If we use
the $k$ and $\mu$
dependent transfer functions, there are 
only seven 
independent quadratic operators: 
the effect of $\Pi_\parallel^{(2)}$
is fully absorbed into
transfer functions.
In addition, the 
velocity bias 
terms can be fully 
captured by the
Zel'dovich field. 
This implies that at the quadratic level the full EFT can be described by the following forward model:
\begin{align}
\label{eqn:lya_model_0}
   & \td^{\rm model}_F(\vk) = 
   \beta^F_1(k,\mu)\tilde \delta_1(\vk)
   +\beta^F_{\eta}(k,\mu)\left(
   \delta_Z(\vk)-\frac{3}{7} f \mu^2\tilde{\mathcal{G}_2} \right)
   %^\perp 
   +\beta^F_2(k,\mu)
\tilde{(\delta_1^2)}
%^\perp (\vk)  \nonumber 
\\
&+\beta^F_{\mathcal{G}_2}(k,\mu)
\tilde{\mathcal{G}_2}
% ^{\perp}
(\vk) +\beta^F_{\delta \eta}(k,\mu)
\tilde{[\delta \eta]}
% ^\perp 
(\vk) +\beta^F_{\eta^2}(k,\mu)
\tilde{(\eta^{2})} (\vk)  
+\beta^F_{KK_\parallel}(k,\mu)
\tilde{(KK)}_\parallel
% ^\perp 
\,, \nonumber 
\end{align}
where we promoted the 
bias parameters 
in Eq.~\eqref{eq:K2full_2}
to momentum-dependent 
transfer functions,
and also shifted each operator with the 
Zel'dovich displacement,
\be \label{eq:shifted_operators}
\tilde{\mathcal{O}}(\k)
=\int d^3 \q~\mathcal{O}(\q)
e^{-i\k\cdot(\q+\vpsi(\q)+f\zhat(\vpsi(\q)\cdot \zhat))}~\,,
\ee 
which accounts for IR-resummation
of bulk flows that affect
the phases from these 
operators at higher 
orders~\cite{Crocce:2007dt,Baldauf:2015xfa,Senatore:2014via,Blas:2015qsi,Blas:2016sfa,Ivanov:2018gjr,Vasudevan:2019ewf}. It is easy to see that the model~\eqref{eqn:lya_model_0} reproduces the linear \Lya forest bias on large scales, 
\be 
\delta_F^{\rm model}(\k)|_{k\to 0}= ((\beta^F_1 + \beta^F_{\eta})+
\beta^F_{\eta} f\mu^2 )\delta_1(\k)\,,
\ee 
so that the transfer functions $\beta_1$
and $\beta_\eta$
are constant in this limit, and the relationship
\be 
b_1=\beta^F_1 + \beta^F_{\eta}\,,\quad b_\eta = -\beta^F_{\eta}\,,
\ee 
consistently
holds through the quadratic 
order in our forward model. 
Beyond the $k\to 0$ limit the linear transfer functions 
$\beta_1$
and $\beta_\eta$
absorb any physical corrections 
correlated
with the linear density field. 
For instance, they automatically 
account for all cubic operators 
contributing 
to the \Lya
power spectrum at the one-loop order and all the relevant counterterms. Indeed, 
the cross-correlation between these terms and the linear matter field 
in EFT 
can be cast as a non-linear correction $\Delta\beta_1$
to the $\beta_1$ transfer function,
\be 
\begin{split}
& \langle \delta_1 \delta_F^{(3)}\rangle'+\langle \delta_1 \delta_F^{\rm ctr}\rangle'=3
% (b_1-f\mu^2 b_\eta)
P_{\rm lin}(k)\int_\p K_3(\p,-\p,\k)P_{11}(p)
+
% (b_1-f\mu^2 b_\eta)
k^2P_{11}(k)\sum_{m=0}^2 c_m \mu^{2m}\\
&\equiv 
\Delta 
\beta_1(k,\mu)
% (\Delta 
% \beta_1(k,\mu)-f\mu^2 \Delta \beta_\eta(k,\mu))
P_{\rm lin}(k)=
\langle 
\Delta 
\beta_1(k,\mu)
% (\Delta \beta_1(k,\mu)-f\mu^2 \Delta \beta_\eta(k,\mu))
\delta_1\delta_1\rangle'~\,.
\end{split}
\ee 
Thus, by construction, our
forward model
absorbs
small-scale effects such as
baryonic feedback 
and gas smoothing
into the 
transfer functions. 
In particular, the linear forward model with the transfer function $\beta_1$ 
produces the power spectrum 
\be 
P_{\rm F}(k,\kp)=F_{\rm NL}(k,\kp)P_{11}(k)\,,
\ee 
% where $F_{\rm NL}(k,\kp)$ is an arbitrary function,
equivalent to 
expressions 
which appear 
in many popular phenomenological models~\cite{McDonald:2001fe, Arinyo-i-Prats:2015vqa, mcquinn2011}. 
Likewise, the transfer functions
of the quadratic operators 
automatically encapsulate
all the bispectrum higher order corrections 
that correlate with the linear and quadratic fields, including the deterministic 
bispectrum 
counterterms. 

It is important to stress 
that fitting the density field
is a much more difficult 
task than fitting the power 
spectrum. For instance, 
a forward model
\be 
\delta^{\rm model}_F(\k)\bigg|_{\rm pheno}=\sqrt{\frac{P_{\rm F,NL}(\k)}{P_{\rm lin}(k)}}\delta_1(\k)\,,
\ee 
where $P_{\rm F,NL}$
is the simulated non-linear flux power
spectrum, 
by construction 
reproduces the \Lya power spectrum perfectly. 
However, this model is expected 
to fail at the field 
level because it misses
higher order \Lya bias
operators and proper
IR resummation. We will
demonstrate the breakdown
of this model below. 

Our full forward model is equivalent to the full EFT at the quadratic order, and includes some 
additional terms beyond that order. 
Thanks to the transfer functions, at the power spectrum level our model is 
equivalent to the full 1-loop
EFT including the cubic operators. To reflect this fact, with some abuse of
terminology, we will refer to our model as ``effectively cubic'' in what follows. 
However, we stress that the 
model does not currently 
contain all the 
necessary cubic terms, like $\Gamma_3$ or $\eta^3$. The cubic terms that cannot be absorbed into the transfer functions will then contribute 
to the model error, and in particular to the noise power spectrum. The effect of these terms, however, 
vanishes in the 
$k\to 0$ limit, except those 
proportional to
the 
$\delta^3$
operator, entering through $P_{33}$ in the EFT nomenclature. This is the only independent cubic operator whose power 
spectrum is constant 
in the $k\to 0$
limit.
Hence, in principle, it could contribute to  $n_0$. 
In order to reduce the noise power spectrum
in the $k\to 0$
limit, we explicitly include
the $\delta^3$
operator in our model. 
At fourth order, 
there are two power spectrum contributions
constant on large scales, $P_{24}$ (two-loop power spectrum order)
and $P_{44}$ (three-loop power spectrum order).
Our quadratic transfer functions, however, account for all higher order corrections 
that correlate with the quadratic fields. 
Hence, the contributions from  $P_{24}$ terms are already included in our model, and the first missing perturbative contribution to the noise 
on large scales is the constant piece
from $P_{44}$ at the three-loop
order.
This can be estimated to be negligibly 
small on the redshifts 
of interest, 
and we will ignore this term
in what follows. 

Finally, in order to 
reduce the numerical noise due to degeneracies in our bias expansion, we 
orthogonalize all the relevant operators 
using the Gram-Schmidt algorithm~\cite{Schmittfull:2018yuk} for each $(k,\mu)$ bin, 
\be 
\tilde{\mathcal{O}}^\perp_a = \sum_{b}M_{ab}\tilde{\mathcal{O}}_b~\,,
\ee 
where 
$\tilde{\mathcal{O}}_b=\{\tilde\delta_1,
\tilde\delta_1^2,
\tilde\G,
\tilde\delta_1^3,
\tilde{(KK)_\parallel},
\tilde\eta_{\rm new},
\tilde{\eta^2},
\tilde{[\delta\eta]}\}$, and $M_{ab}$ is the rotation matrix~\cite{Schmittfull:2018yuk} constructed from 
\[
O_{ab}\equiv \langle \tilde{\mathcal{O}}_a(\k)  \tilde{\mathcal{O}}_b^*(\k)\rangle'\,,
\]
using the Cholesky 
decomposition procedure. 
One can check that the operator $\Pi^{[2]}_\parallel$
does not contribute to 
our model because 
$\langle 
|\tilde{\mathcal{O}}^{\perp}_{\Pi^{[2]}_\parallel}|^2\rangle\approx 0$,
which is equivalent to the statement that its contribution
is absorbed by the transfer
functions. 
This yields the final forward model
\begin{align}
\label{eqn:lya_model}
   & \td^{\rm model}_F(\vk) = 
   \beta^F_1(k,\mu)\tilde \delta_1(\vk)
   +\beta^F_{\eta}(k,\mu)\eta_{\rm new}^\perp +\beta^F_2(k,\mu)
\tilde{(\delta_1^2)}^\perp (\vk) 
% +
+\beta^F_3(k,\mu)
\tilde{(\delta_1^3)}^\perp (\vk)
\nonumber \\
&+\beta^F_{\mathcal{G}_2}(k,\mu)
\tilde{\mathcal{G}_2}^{\perp}(\vk) +\beta^F_{\delta \eta}(k,\mu)
\tilde{[\delta \eta]}^\perp (\vk)+\beta^F_{\eta^2}(k,\mu)
\tilde{\eta}^{2,\, \perp} (\vk)  
+\beta^F_{KK_\parallel}(k,\mu)
\tilde{(KK)}_\parallel^\perp (\vk)\,. 
\end{align}
Note that the above model
appears as a simple generalization of the 
forward model 
for redshift-space galaxies, 
\begin{align}
\label{eqn:g_model}
   & \td^{\rm model}_g(\vk) = 
   \beta_1(k,\mu)\tilde \delta_1(\k)
   +\eta_{\rm new}(\k)  +\beta_2(k,\mu)
\tilde{(\delta_1^2)}^\perp (\vk) 
+\beta_3(k,\mu)
\tilde{(\delta_1^3)}^\perp (\k)+\beta_{\mathcal{G}_2}(k,\mu)
\tilde{\mathcal{G}_2}^{\perp}(\k) \,.
\end{align}
Note that the above model \emph{does not have} a transfer function in front 
of $\eta_{\rm new}$, defined in Eq.~\eqref{eq:eta_new}, which reflects that halos conserve denisty when switching 
from real space to redshift space.

In what follows we will explore the implications of our \Lya forward model.

\section{Simulation data} \label{sec:simulations}
We fit our field-level perturbative model to two synthetic \Lya forest data sets: First, large \Lya forest mocks constructed from the $N$-body simulation suite \textsc{AbacusSummit} (hereinafter Abacus) \cite{Hadzhiyska:2023bcx}. Second, hydrodynamic simulations of the intergalactic medium (IGM) from the Sherwood suite \cite{Bolton:2016bfs,Givans:2022qgb}. The Abacus simulation will be the primary data set of this work. We now introduce both simulations, briefly summarizing \cite{Bolton:2016bfs, Givans:2022qgb, Hadzhiyska:2023}, to which the reader is referred for a fuller presentation.

The Abacus simulations, centered at redshift $z=2.5$, are large $N$-body simulations with the \Lya forest painted on top of them using a simplistic fluctuating Gunn-Peterson approximation (FGPA) \cite{Hadzhiyska:2023}. Note that Ref.~\cite{Hadzhiyska:2023} provides four different FGPA implementations resulting in four different realizations of the Abacus \Lya simulations. The main difference arises for the value of the bias parameter associated with the gradient of the peculiar velocity, $b_\eta$, which is lower for models I and II and larger for models III and IV than current constraints obtained from DESI data \cite{DESI_lya_2024}. The simulations encompass a comoving volume of $V=2000^3(\hinvMpc)^3$ with $6912^3$ particles, each of mass $M_{\rm part} = 2.1 \times 10^9$  \cite{2021MNRAS.508.4017M,Hadzhiyska:2023,Hadzhiyska:2025cvk}. The simulation is based on a fiducial \textit{Planck} 2018 cosmology with $\Omega_b h^2 = 0.02237$, $\Omega_c h^2 = 0.12$, $h = 0.6736$, $A_s = 2.0830 \times 10^{-9}$, $n_s = 0.9649$, $w_0 = -1$, $w_a = 0$. The quasi-stellar objects (QSO) use a simplistic halo occupation distribution (HOD) model given in equations (1) and (2) in \cite{Hadzhiyska:2025cvk} with as mass range approximately yielding a linear bias that matches observations $b_q\approx 3.3$ and a number density of $\sim 1.75\times 10^{-4} \ (\hinvMpc)^{-3}$. The parameters describing the HOD are $\log_{10}{(M_{\rm cut})} = 13.2$ which characterizes the minimum halo mass to host a central
galaxy, $\log_{10}{(M_1)} = 13.8$ the typical halo mass that hosts one satellite galaxy, $\sigma = 0.65$ the steepness of the transition from 0 to 1 in the number of central
galaxies, $\alpha=0.8$ the power law index on the number of satellite galaxies,
ic the incompleteness parameter, and $\kappa=1.11$ multiplied by $ M_{\rm cut}$ gives the minimum halo mass to host a satellite galaxy. 

We use two sets of Sherwood hydrodynamic simulations: (i) one snapshot at redshift $z=2.8$ encompassing a comoving volume $V=160^3(\hinvMpc)^3$ where the number of CDM particles and gas particles in
the simulation is $2048^3$; and (ii) a series of snapshots at redshifts $z=2.0,\,2.4,\,2.8,\,3.2$ in a $V=80^3(\hinvMpc)^3$ volume with $1024^3$ particles keeping the resolution fixed across simulations. The initial conditions are generated using the \texttt{N-GenIC} code \cite{Springel05} with the corresponding number of either $2048^3$ or $1024^3$ particles. The fiducial cosmology is based on the best-fit \emph{Planck} 2013 cosmology with cosmological parameters set to $\Omega_m=0.308,\, \Omega_b=0.0482,\, h=0.678,\, \sigma_8=0.829,\, n_s=0.961$ \cite{Planck_first}. Additionally, the Sherwood simulations come with a halo catalog built using a friends-of-friends algorithm and with an available mass range of $10^9 \leq M_\odot \simlt 10^{14}$. For all simulations, redshift-space distortions have been applied along the $\zhat$-axis and particles have been assigned using a triangular-shaped cloud (TSC) algorithm. 

To explore cross-correlations of the \Lya forest with high-redshift galaxies we use hydrodynamic simulations of two types of star-forming galaxies in the state-of-the-art Astrid hydrodynamical simulations (see, e.g.,~\cite{2022MNRAS.512.3703B,2022MNRAS.513..670N,Obuljen:2022cjo, Ivanov:2024dgv, Sullivan:2025eei}). In particular, we use the samples consistent with the observed (angular) clustering and number density of \Lya emitters (LAEs) and Lyman-break galaxies (LBGs) at $z=3$. The samples are designed to approximately match the linear bias $b_1$ and number density of existing and future LSS surveys for LAEs (``ODIN'' \cite{White:2024lki} and S5 \cite{White:2024lki, Ebina:2024ojt,schlegel_megamapper_concept}) and similarly for LBGs (``CARS'' and S5; \cite{Hildrebrandt_CARS_2009,schlegel_megamapper_concept}).

\section{ Results}\label{sec:results}
In this section, we use the perturbative forward model introduced in Sec.~\ref{sec:model_setup} on the synthetic data presented in Sec.~\ref{sec:simulations}. We compare the \Lya forest auto-correlation as well as its cross-correlation with dark matter halos (used as proxies for high-redshift galaxies and quasars). We compare four metrics to assess the quality of the forward model: (i) the two-dimensional flux power spectrum, $P(k,\mu)$ which is the average of the squared norm of the Fourier transform of the flux decrement, $\td_F$, in bins of the Fourier wavenumber, $k$, with the cosine of the angle to the line-of-sight, $\mu=\kpar/k$. The corresponding error bars are computed from the square root of the diagonal of the Gaussian covariance based on the number of expected Fourier modes per bin $P(k,\mu)\sqrt{2/N(k,\mu)}$. (ii) The one-point probability density function which is a more stringent test of the model since we are effectively capturing information beyond the two-point function. (iii) The error power spectrum (or mean-squared model error), given in Eq.~\eqref{eq:Perr}, is a measure of the success of the model.
(iv) To investigate the agreement at the level of the phases and amplitudes, we compute the cross-correlation coefficient:
\be \label{eq:rcc}
r_{cc}(\delta_F^{\rm truth},\delta^{\rm model}_F)=\frac{\langle 
\delta^{\rm model}_F(\k)
[\delta_F^{\rm truth}(\k)]^*
\rangle}{
\left(\langle 
|\delta_F^{\rm truth}(\k)|^2\rangle
\langle
|\delta_F^{\rm model}(\k)|^2\rangle\right)^{1/2}
}
% = \sqrt{1-\frac{P_{\rm err}}{P_{\rm truth}}}
% = \sqrt{\frac{P_{\rm model}}{P_{\rm truth}}}
\,,
\ee 
for the simulated (`truth') and forward modeled (`model') fields. The cross-correlation coefficient allows us to investigate how spatially coherent the densities are in Fourier space and is a measure of the amount of cosmological information that can be extracted when using the model to describe \Lya forest measurements \cite{Schmittfull:2018yuk}. 

We show our results as a function of Fourier wavenumber $k$ and in three angular bins centered at $\mu = 0.17,\, 0.50,\, 0.83$. The maximum wavenumber shown is given by the Nyquist frequency which is $k_{\rm Ny}=1.61 \hMpcinv$ for Abacus with a resolution of $N_{\rm cell}=1024$ in a $2^3(\hinvGpc)^3$ volume and $k_{\rm Ny}=5.03,\, 10.05 \hMpcinv$ for Sherwood in $160^3 (\hinvMpc)^3$ and $80^3 (\hinvMpc)^3$ volumes with $N_{\rm cell}=256$, respectively. Throughout this work, we will discuss model I from Abacus and compare these results to the ones obtained from the Sherwood hydrodynamic simulations (if not already discussed in Ref.~\cite{deBelsunce:2025bqc}). In App.~\ref{app:model_dependence_abacus} we confirm the robustness of the perturbative forward model to different realizations (models II-IV) of the Abacus simulations stemming from different approaches to calibrating the FGPA procedure to plant a \Lya forest on top of an $N$-body simulation (see Ref.~\cite{Hadzhiyska:2023} for more details). 
%  kNy = Ncell * pi / Boxsize 

\subsection{\Lya forest forward model} \label{sec:lya_results}
\begin{figure*}
    \centering
    \includegraphics[width=0.49\linewidth]{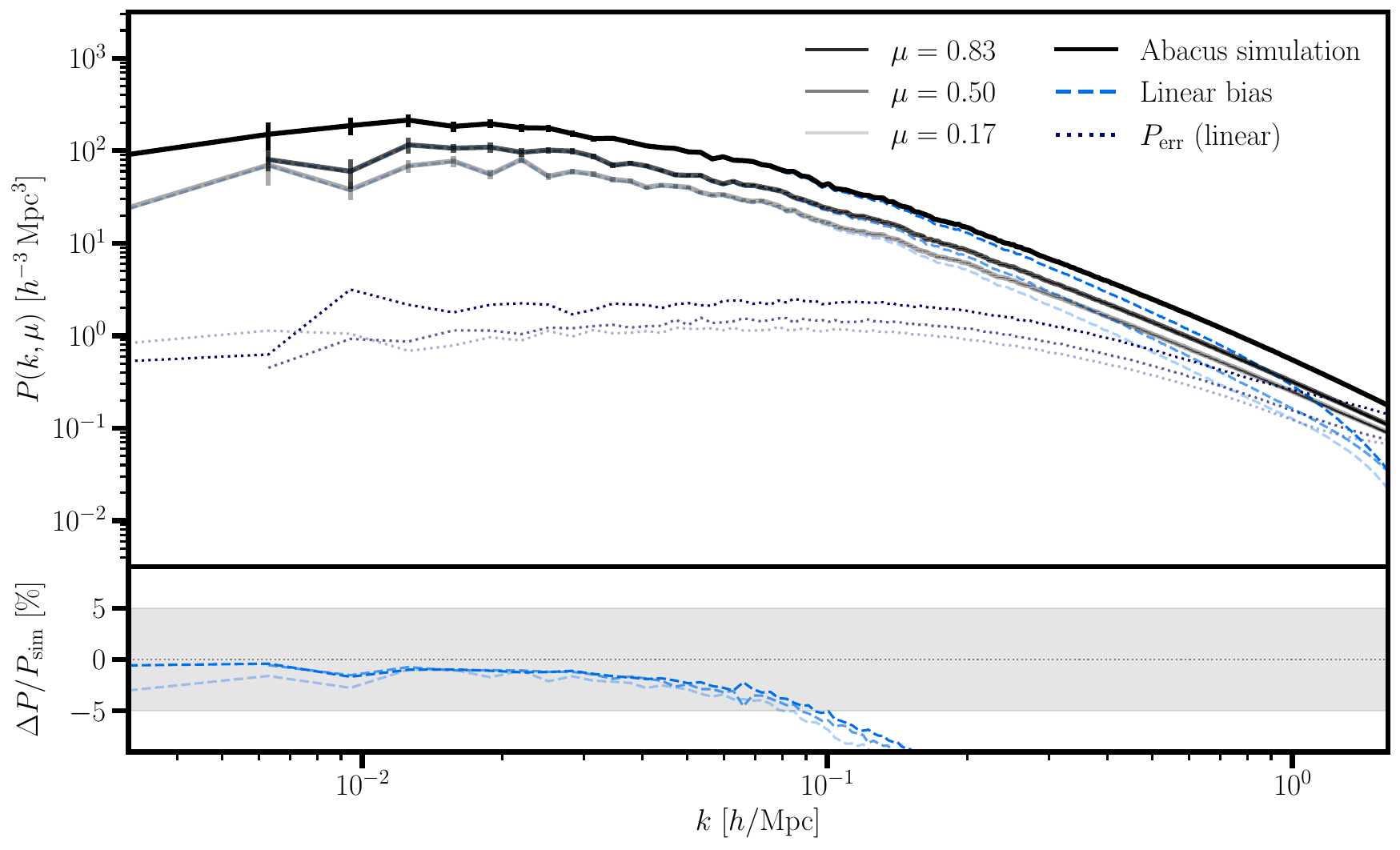}\hfill
    \includegraphics[width=0.49\linewidth]{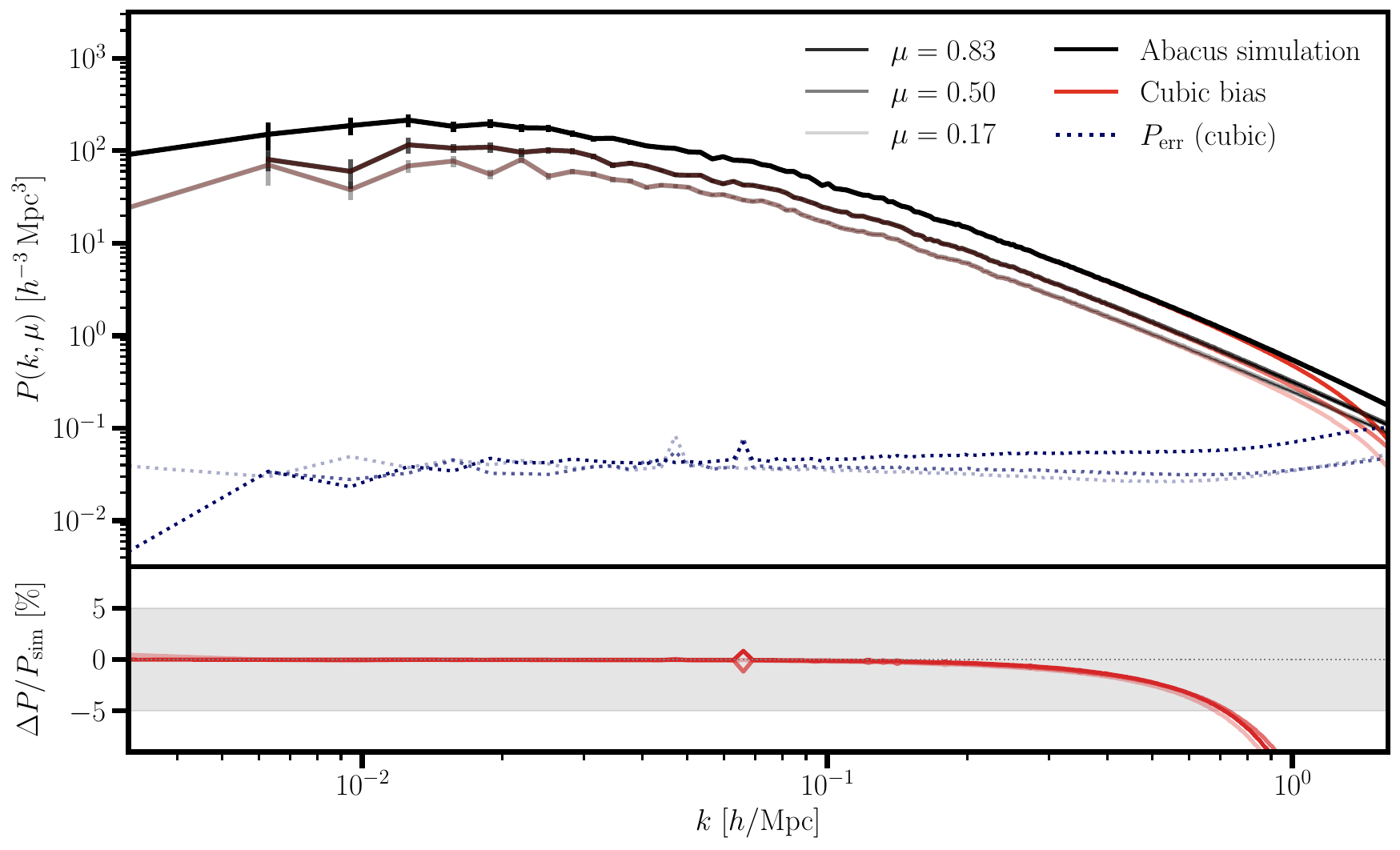}
    \vspace{-0.1in}
    \caption{\textbf{Best-Fit Abacus Power Spectra:}
    Comparison between the measured two-dimensional power spectrum from the Abacus simulation (black) and the best-fit forward model obtained from a linear and a cubic bias expansion in three angular wedges. \textit{Left}: Linear theory prediction (blue dashed line), which shows that the forward model breaks down beyond $\kmax\sim 0.1 \hMpcinv$ and a large error power spectrum $P_{\mathrm{err}}(\mathbf{k},\zhat) \equiv \langle |\tilde{\delta}_{\mathrm{sim}} - \tilde{\delta}_{\mathrm{EFT}}|^2 \rangle$ (blue dashed lines), indicating poor model performance. \textit{Right}: Cubic bias model prediction (red solid line), which matches the simulation much more closely, with substantially reduced residuals and a suppressed $P_{\mathrm{err}}$. In each panel, the power spectrum is shown in bins of Fourier wavenumber $k$ and angle to the line of sight, parametrized by $\mu = k_\parallel / k$, with darker (lighter) lines for $\mu = 0.83$ ($0.17$). The bottom panel displays the percent difference between the simulation and model power spectra. 
    A gray band highlights the $\pm5\%$ region in the bottom panel.
    }
    \label{fig:lya_pk}
\end{figure*}

In this section, we present the results of the \Lya forest forward model. In Fig.~\ref{fig:lya_pk} we compare the linear (left panel) and cubic (right panel) forward models at the power spectrum level for the Abacus simulation. 
Note that our linear theory model matches the phenomenological fitting functions employed in current \Lya forest analyses (see, e.g.,~\cite{McDonald:2001fe, Arinyo-i-Prats:2015, DESI_lya_2024}). We find that linear theory does not \textit{exactly} capture the largest scales $k\simlt 10^{-2} \hMpcinv$ and recovers the input power spectrum at the 5\% level only down to $\kmax \approx 0.1\hMpcinv$. Similarly, the linear error power spectrum, given in Eq.~\eqref{eq:Perr}, exhibits an orientation dependence and amplitude at the level of a few percent of the amplitude of the total power spectrum indicating poor model performance. For the cubic model, we find excellent agreement at the largest scales between the simulation and model power spectra. It captures  the power spectrum down to $\kmax \approx 0.8 \hMpcinv$ at the 5\% level.\footnote{Current DESI \Lya full-shape analyses use the correlation function down to a minimum scale of $r=25\hinvMpc$ \cite{Cuceu:2025nvl} which corresponds to $k \sim  \pi/25 = 0.13 \hMpcinv$ which is the range of validity where the present forward model is accurate at  the sub-percent level.}  We note that the error power spectrum exhibits a scale dependence above $k \simgt 0.6 \hMpcinv$ leading to increased stochasticity.\footnote{We refer the reader to Ref.~\cite{deBelsunce:2025gci} to further reduce the \Lya forest stochasticity by computing non-linear particle displacements using simulations in the context of  Hybrid effective field theory (HEFT). This extends the range of validity of the forward model to $k\approx 1\hMpcinv$.} 

\begin{figure}
    \centering
    \includegraphics[width=1\linewidth]{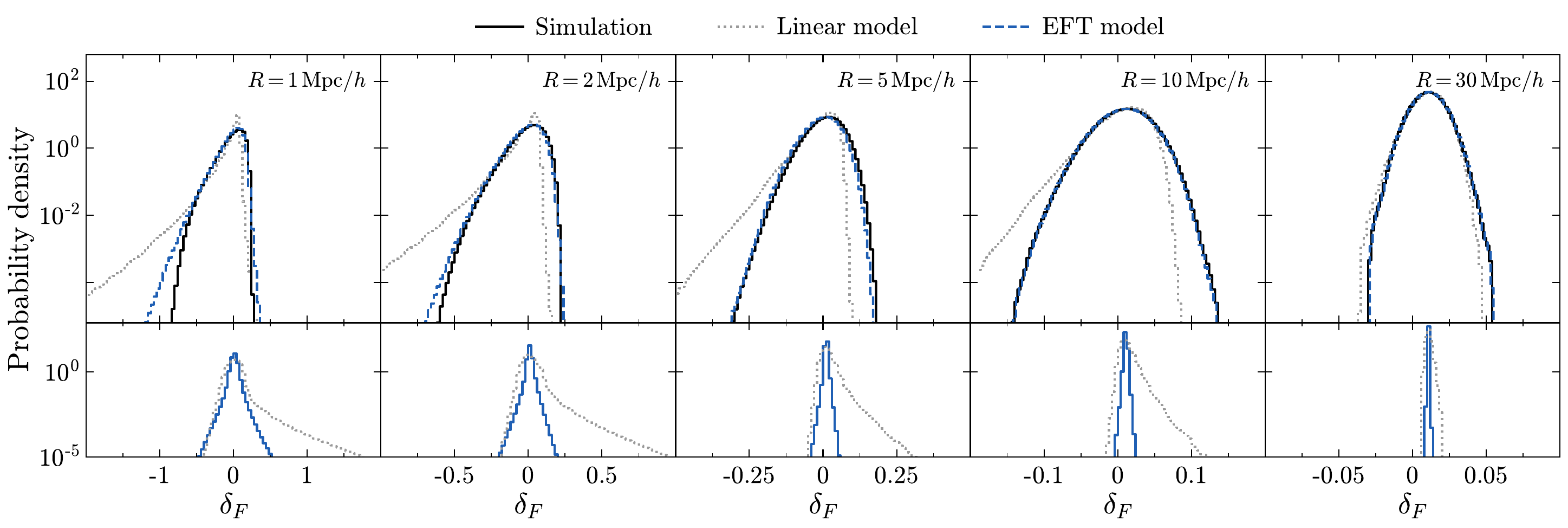}
    \caption{\textbf{One-point PDF Abacus:}
    Comparison of one-point probability density function from Abacus simulation (black) to perturbative forward model (dashed blue) and linear model (gray lines) with increasing degree of isotropic Gaussian smoothing applied to the fields $R=1 - 30 \hinvMpc$ from left to right. The bottom panel shows the residuals. The differences between the linear forward model and the simulation are visible for \textit{all} smoothing scales. This emphasizes the importance of the higher order terms in our bias expansion.
    }
    \label{fig:lya_pdf}
\end{figure}

In Fig.~\ref{fig:lya_pdf} we compare both \Lya forest fields at the field level using the one-point probability density function (PDF). The break down of the linear model is striking, even on large scales.\footnote{This is qualitatively similar, yet less pronounced than on the field-level methodology applied to the Sherwood hydrodynamic simulations, see~\cite{deBelsunce:2025bqc}, illustrating that the Abacus mocks can access larger scales and thus more quasi-linear modes.} We find excellent agreement between the EFT model and the simulations down to cell sizes of $\approx 2 \hinvMpc$ with the linear model breaking down for very large cell sizes of $10-30\hinvMpc$. Note that the number of pixels from the EFT forward model (blue dashed line) that are visibly deviating from the distribution obtained from the simulation (black line) is less than 0.01\%.  In Tab.~\ref{tab:moments_pdf} we quantify the agreement between the modeled and true field using the moments of the field with the same Gaussian kernel smoothing radii ($R=1\,, 2\,, 5\,, 10\,, 30\, \hinvMpc$) applied to them. Whilst this is not applicable to observed \Lya forest data \cite{Belsunce_Sullivan_skewspectrum}, it allows us to isolate larger scales. The residuals (denoted by $\Delta\td$ in Tab.~\ref{tab:moments_pdf}) for the best-fit EFT model are consistently smaller than the residuals from the linear model, highlighting the importance of higher order bias parameters. 

\begin{table}
\centering
\begin{tabular}{cccccccccccccccc}
\hline \hline
& \multicolumn{5}{c}{Variance} 
& \multicolumn{5}{c}{Skewness} 
& \multicolumn{5}{c}{Kurtosis} \\
\cmidrule(lr){2-6} \cmidrule(lr){7-11} \cmidrule(lr){12-16}
$R$ 
& $\td_{\rm truth}$ & $\td_{\rm lin.}$ &$\td_{\rm best-fit}$ & $\Delta\td_{\rm lin.}$ & $\Delta\td$ 
& $\td_{\rm truth}$ & $\td_{\rm lin.}$ &$\td_{\rm best-fit}$ & $\Delta\td_{\rm lin.}$ & $\Delta\td$
& $\td_{\rm truth}$ & $\td_{\rm lin.}$ &$\td_{\rm best-fit}$ & $\Delta\td_{\rm lin.}$ & $\Delta\td$ \\
\hline
1  & 0.0159 & 0.0109 & 0.0149 & 0.0051 & 0.0010 & -1.0635 & -3.7654 & -1.2825 & 0.8168 & -0.0412 & 1.4303 & 26.5017 & 2.5290 & 11.4647 & 4.0812 \\
2  & 0.0075 & 0.0060 & 0.0074 & 0.0015 & 0.0001 & -0.7810 & -2.7355 & -0.8410 & 1.1211 & 0.0031  & 0.8954 & 13.3336 & 1.0496 & 8.4781 & 7.0318 \\
5  & 0.0022 & 0.0020 & 0.0022 & 0.0002 & 0.0000 & -0.3764 & -1.4703 & -0.3882 & 1.2552 & 0.0624  & 0.2078 & 3.6621  & 0.1712 & 4.7769 & 4.4458 \\
10 & 0.0007 & 0.0007 & 0.0007 & 0.0000 & 0.0000 & -0.1635 & -0.8150 & -0.1646 & 1.1150 & 0.0772  & 0.0275 & 1.1027  & 0.0022 & 3.2269 & 1.9103 \\
30 & 0.0001 & 0.0001 & 0.0001 & 0.0000 & 0.0000 & -0.0161 & -0.2487 & -0.0124 & 0.4271 & 0.0299  & 0.0081 & 0.1016 & 0.0037 & 0.6529 & 0.2548 \\
\hline \hline
\end{tabular}
\caption{\textbf{Statistical Moments Abacus:} Statistical moments (variance, skewness, and kurtosis) of the Abacus simulation ($\td_{\rm sim}$) and the perturbative forward model ($\td_{\rm EFT}$) as well as their residuals for different isotropic smoothing scales $R$ (in $\hinvMpc$). Following baseline expectation, we find increasing agreement between the one-point probability density functions when removing small-scale modes. The mean is consistent with zero for all fields. Note that skewness and kurtosis vanish for Gaussian fields indicating that our forward model captures higher-order moments, i.e.,~non-Gaussian information, of the field.
}
\label{tab:moments_pdf}
\end{table}

\subsubsection{Impact of IR resummation}
\begin{figure}
    \centering
    \includegraphics[width=0.49\textwidth]{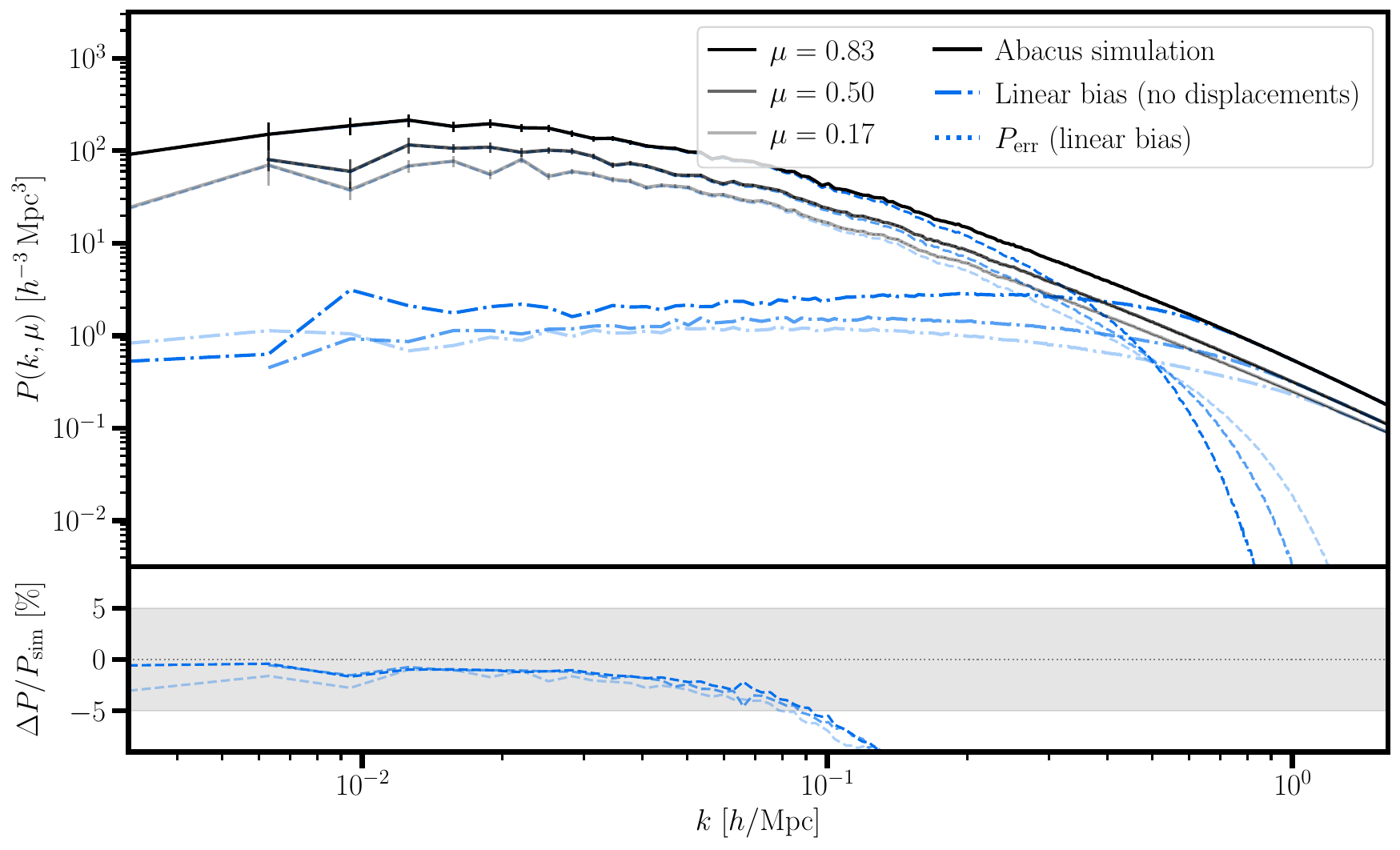}\hfill
    \includegraphics[width=0.49\textwidth]{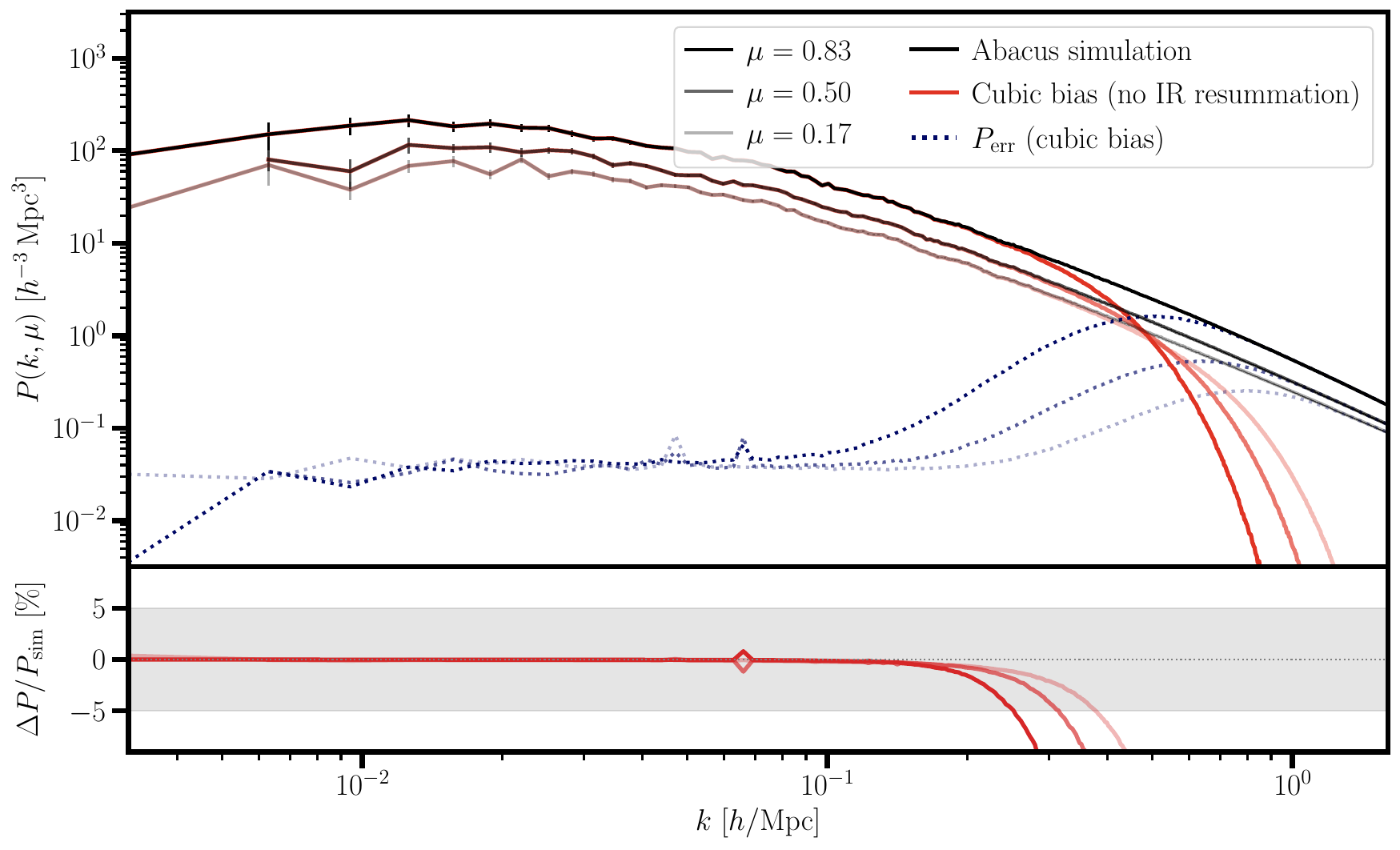}
    \caption{\textbf{Best-Fit Abacus Power Spectrum without IR Resummation:}
    Same as Fig.~\ref{fig:lya_pk} but with the bulk flows left unresummed for the
    linear (\textit{left panel}) and cubic (\textit{right panel}) forward model,
    to isolate the effect of IR resummation. Rather than discarding the
    displacements (which would also remove the displacement mode-coupling kernels
    and no longer constitute a consistent Eulerian PT), we Taylor-expand the shift
    operator to consistent order, retaining the kernels $\tilde{K}_2$,
    $\tilde{K}_3$ while omitting the exponential resummation (``EPT''; see text).
    The power spectrum is shown in three angular wedges $\mu=0.17,0.50,0.83$ for
    the \abacus simulation, with the dotted lines showing the error power spectrum
    $P_{\mathrm{err}}$. The broadband shape is preserved: the cubic residuals
    remain within $5\%$ down to $k\approx0.1\,h\,\mathrm{Mpc}^{-1}$ and the cubic
    $P_{\mathrm{err}}$ stays roughly an order of magnitude below the linear
    one. This confirms that the IR resummation does not affect the broadband shape (cf.\ the
    shifted/IR-resummed operator equivalence~\cite{Chen:2020fxs}). Near the BAO
    scale ($k\approx 0.1\,h\,\mathrm{Mpc}^{-1}$) the unresummed model fails
    to damp the acoustic feature, imprinting oscillatory, $\mu$-dependent
    structure on $P_{\mathrm{err}}$. The bottom panel shows the percent difference
    between model and simulation, with a gray band marking the $\pm5\%$ region.
    }
    \label{fig:perr_ratio_resummation}
\end{figure}
Standard Eulerian perturbation theory does not fully capture the non-linear damping of the baryon acoustic oscillation (BAO) feature. This stems from linear matter displacements, which are sensitive to  long-wavelength (infrared, IR) modes which are in turn responsible for the slow convergence of the perturbative expansion \citep{Crocce:2007dt}. While individual diagrams in standard perturbation theory exhibit IR-enhanced terms, these enhancements cancel out when all diagrams are summed, as required by the equivalence principle \citep{Blas:2013bpa,Blas:2015qsi}. However, the presence of the BAO feature prevents a complete cancellation, leaving residual IR sensitivity \citep{Baldauf:2015xfa,Blas:2016sfa}. IR resummation has been introduced to rigorously account for these large soft modes in the BAO signal \citep{Senatore:2014via,Baldauf:2015xfa,Vlah:2015zda,Blas:2015qsi,Blas:2016sfa,Ivanov:2018gjr,Vasudevan:2019ewf,2014JCAP...05..022P, 2015JCAP...09..014V,Chen20features,Chen:2020fxs, Chen:2020zjt,Chen24}. 

To isolate the effect of IR resummation we regenerate the field-level forward model
\emph{without} resumming the bulk flows. We stress that this is distinct from
discarding the displacement altogether: the shift operator
$e^{-i\mathbf{k}\cdot\boldsymbol{\Psi}}$ in Eq.~\eqref{eq:shifted_operators}
generates both the perturbative displacement mode-coupling and the all-orders
resummation of bulk flows. Setting $\boldsymbol{\Psi}\to0$ would remove the
former as well, yielding a model that is no longer a consistent Eulerian PT.
Instead, we Taylor-expand the shift to consistent order in the linear field,
retaining the displacement kernels ($\tilde{K}_2$, $\tilde{K}_3$) while omitting
the exponential resummation. Since shifted and IR-resummed Eulerian operators
are equivalent~\cite{Chen:2020fxs}, this expanded model agrees with the fully
resummed one on the smooth (no-wiggle) component of the power spectrum and
differs only in the treatment of the oscillatory (wiggle) part.

The result is shown in Fig.~\ref{fig:perr_ratio_resummation} for the linear
(\textit{left panel}) and cubic (\textit{right panel}) expansions. In contrast to
discarding the displacements altogether, the broadband shape is preserved: the
percent-level residuals remain within $5\%$ down to
$k\approx0.1\,h\,\mathrm{Mpc}^{-1}$ for the cubic model, which continues to
outperform linear theory, with the two error power spectra separated by roughly
an order of magnitude. The signature of the omitted resummation is instead
localized around the acoustic scale, 
where the unresummed model fails to damp the BAO feature: this produces a
residual mismatch between the modeled (undamped) and simulated (damped) wiggles
that imprints oscillatory, $\mu$-dependent structure on the error power spectrum
in this regime. This demonstrates that IR resummation is essential specifically
for recovering the BAO feature at the field level, while having a negligible
effect on the broadband shape.

\subsection{Cross-correlation of the \Lya forest with massive halos} \label{sec:lyaxhalo_results}
\begin{figure}
    \centering
    \includegraphics[width=0.49\linewidth]{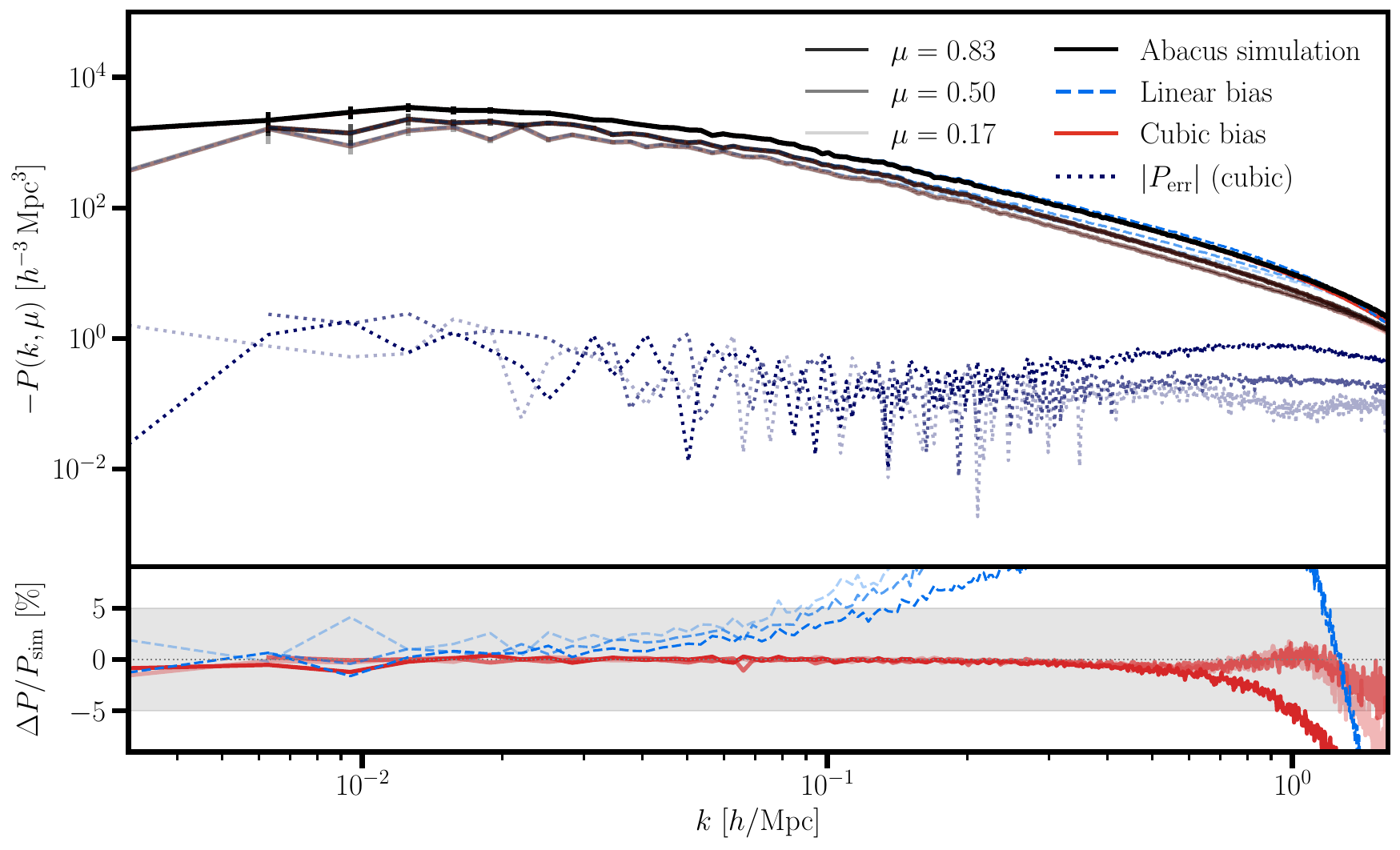}\hfill
    \includegraphics[width=0.49\linewidth]{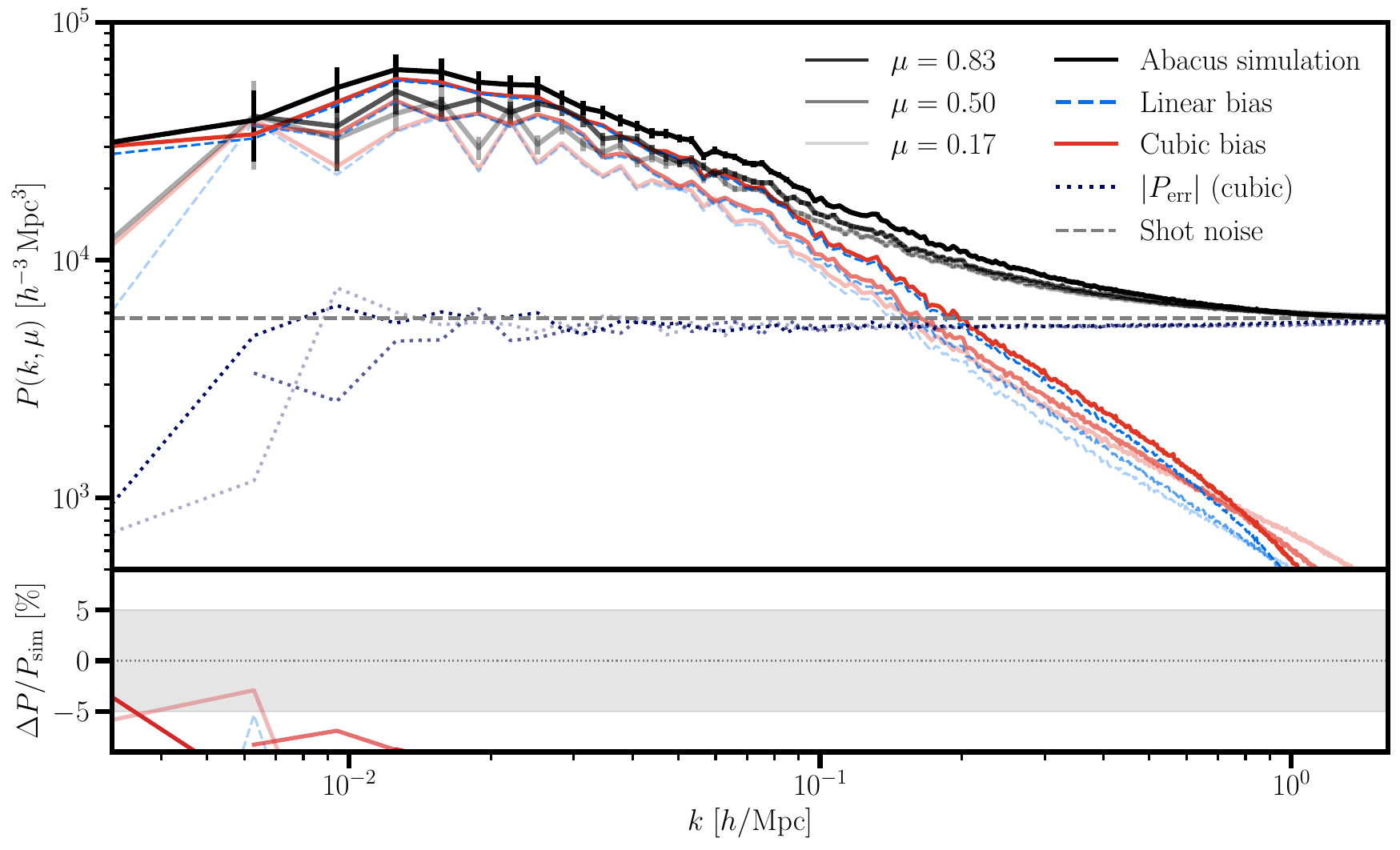}
    \vspace{-0.1in}
    \caption{\textbf{Best-Fit Abacus Power Spectra incl. Halos:}
    Same as Fig.~\ref{fig:lya_pk} for the cross-correlation of the \Lya forest with halos (\textit{left plot}) and the halo auto-correlation using all available halo masses (\textit{right plot}). The halo auto-correlation error power spectrum matches the shot noise ($\overline{n} = 1.75 \cdot 10^{-4}$). The model dependence of the cross spectrum is shown in Fig.~\ref{fig:abacus_pk_models_cross}.
    }
    \label{fig:lya_halo_pk}
\end{figure}

In this section, we extend the results from Sec.~\ref{sec:lya_results} to perform the field-level modeling in redshift space for galaxies \cite[see, e.g.,~][]{Schmittfull:2018yuk, Schmittfull:2020trd, Obuljen:2022cjo, Ivanov:2024xgb}. This allows to compute the cross correlation of the \Lya forest with massive halos (a proxy for high-redshift galaxies and quasars). Cross-correlations are a key source of cosmological information for \Lya analyses and help break the degeneracy of the growth rate $f$ with the (unknown) velocity gradient bias \cite{Font-Ribera:2013fha,Chudaykin:2025gsh, DESI_lya_2024}. In the left panel of Fig.~\ref{fig:lya_halo_pk} we show the cross spectrum of both tracers which matches remarkably well for the cubic model up to $\kmax \approx 1 \hMpcinv$ at the five per cent level. The linear theory model fails already  beyond $k\simgt 4\cdot10^{-2}\hMpcinv$. It is interesting to note that the mean-squared-error is reduced in the cross-correlation compared to the halo auto-correlation. This indicates that cross-correlation measurements, even with highly-biased tracers (as proposed by DESI-II), are a fruitful avenue to extract cosmological parameters.

The halo auto spectrum, shown in the right panel, shows a significantly larger error power spectrum which stems from the fact that the field-level model is only valid up to the shot noise. (Following baseline expectation, subtraction of the shot noise yields agreement at the per cent level down to $k\simlt 0.3\hMpcinv$.) The halo sample for the Abacus simulation consists of 1,403,076 objects which, in turn, yields a shot noise level of $1/\overline{n} = 5.7 \cdot 10^3$ $[\Mpch]^3$ with masses in the range $10.8 \leq \log_{10}(M/(h^{-1}M_\odot)) \leq 14.2$.\footnote{We verified that splitting the mass range to select massive or light halos does not affect our conclusion and we observer the same behavior. Further we verified that including additional cubic operators, such as $\mathcal{S}_3$, does not improve the error power spectrum.} 

\subsection{The error power spectrum}\label{sec:Perr}
To quantify the performance of our forward model we use the error (or noise) power spectrum, defined in Eq.~\eqref{eq:Perr},
% \be 
% P_{\rm err} (k,\mu) \equiv \langle |\delta^{\mathrm{truth}}_F(\k) - \delta^{\mathrm{model}}_F(\k)|^2 \rangle\,,
% \ee 
which reflects the agreement at the level of the phases. 
% \be \label{eq:perr_ns}
% P_{\rm err}=n_0(1+\alpha_1 k^2+
% \alpha_2 
% \mu^2 k^2)\quad \text{as}\quad k\to 0\,,
% \ee 
% for dimensional constants $n_0,\alpha_1,\alpha_2$ \cite{Ivanov:2024lya}.
In particular, a successful forward model should produce an error spectrum with a small amplitude and weak scale and orientation-dependence, i.e.,~EFT predicts the shape of the error power spectrum in the large scale limit given in Eq.~\eqref{eq:EFTPerr}. In this regime, we find the following values for the 
leading order stochasticity parameters
at $\kmax=0.6~\hMpc$:
\be 
\begin{split}
\text{Sherwood ($L=160\hinvMpc$):}\quad & n_0=0.16~[\Mpch]^3\,,\quad 
\alpha_1 = -0.22~[\Mpch]^2\,,\quad  
\alpha_2 = 0.75~[\Mpch]^2\,, \\
\text{Abacus ($L=2\hinvGpc$):}\quad & n_0=0.04~[\Mpch]^3\,,\quad 
\alpha_1 = -0.06~[\Mpch]^2\,,\quad  
\alpha_2 = 0.17~[\Mpch]^2\,.
\end{split}
\ee 
These measurements are consistent 
with the estimates 
given in Eq.~\eqref{eq:EFT_Perr_estimate}, 
pointing to a characteristic
stochasticity scale    
$R_{\rm stoch}\sim 0.5~\Mpch$ for Sherwood and
$R_{\rm stoch}\sim 0.3~\Mpch$ 
for Abacus. 
In addition, these values are consistent
with the \Lya stochasticity measurements
from the ACCEL$^2$ simulations~\cite{deBelsunce:2024rvv},
though the latter have significant errorbars. 
To the best of our knowledge, this 
is the first ever precise determination of the 
stochasticity of the Lyman-$\alpha$
forest field.

Interestingly, we see 
a high level of consistency between 
Sherwood and Abacus simulations  
in term of their stochasticity,
even though they represent very 
different small-scale models. 
Since the Abacus mocks 
capture dark matter physics only, 
the similarity between the 
Sherwood and Abacus results
suggest that the Lyman-$\alpha$
stochasticity may be driven by 
the stochasticity 
of the dark matter distribution.

In addition, we notice that 
the scale dependence of the stochastic power spectrum becomes more shallow
around $1~\hMpc$, where the error power
spectrum contributes significantly 
(by more than $5\%$) to the total 
power. This signals the breakdown
of the EFT gradient expansion 
for the stochastic effects. 
Theoretical modeling of the 
error power spectrum in this regime 
may represent a serious challenge.
The connection
to dark matter stochasticity
mentioned above opens up the possibility
to model the \Lya
stochasticity with the 
hybrid EFT technique where the 
Lagrangian bias expansion 
for the Lyman-$\alpha$
forest is supplemented
with the non-linear displacements
from an $N$-body simulation. 
This topic is further explored 
in Ref.~\cite{deBelsunce:2025gci}, where it has been shown 
that the incorporation of the non-linear 
displacements that include 
the stochastic components in them 
indeed allows one to reduce the 
scale and orientation
dependence of the error power 
spectrum extending the range of validity of EFT beyond $k\approx 1\hMpcinv$.

\begin{figure}
    \centering
    \includegraphics[width=0.49\linewidth]{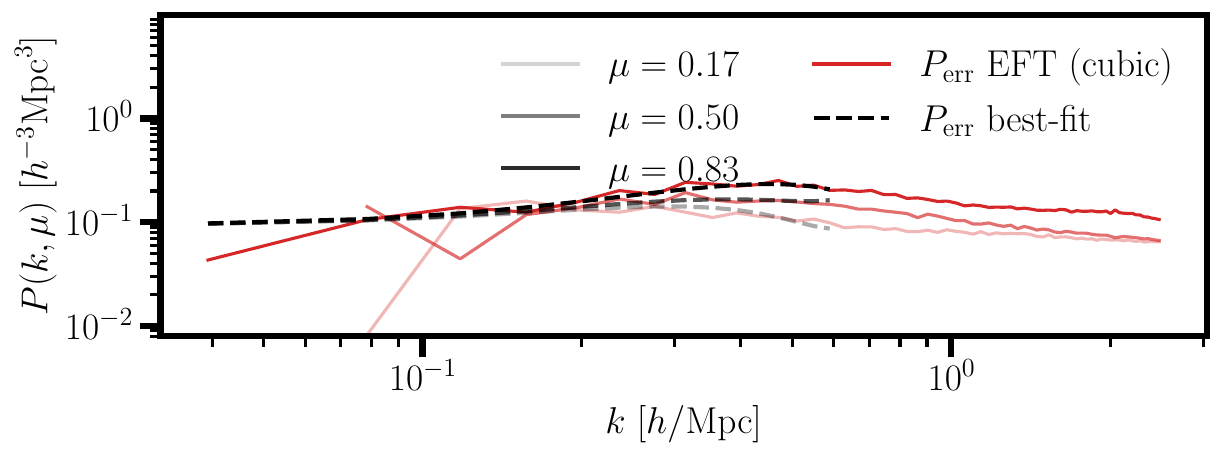}\hfill
    \includegraphics[width=0.49\linewidth]{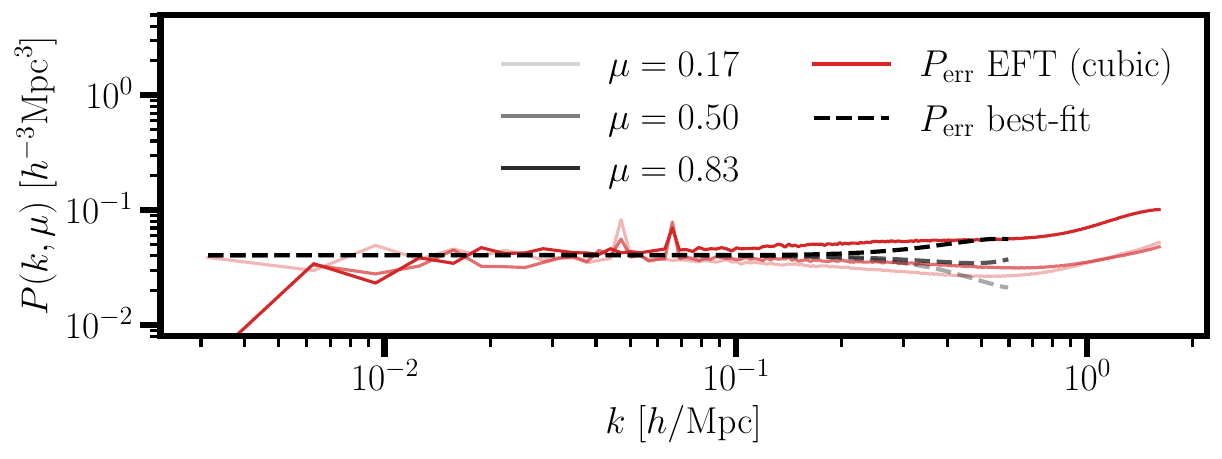}
    \caption{\textbf{Redshift-space error power spectrum} $P_{\rm err}$ corresponding to the best-fit transfer functions shown in Fig.~\ref{fig:transfer_func} for the Sherwood simulation (\textit{left panel}) and Abacus simulation (\textit{right panel}). $P_{\rm err}$ is computed in three angular bins with a maximum wavenumber of $k= 1 \hMpcinv$ for the fits. We show the polynomial fits to $P_{\rm err}$ up to $k=0.6 \hMpcinv$, the wavenumber up to which we can accurately reconstruct the power spectrum. The scale and angular dependence sets in for both simulations at around $k \approx 0.4 \hMpcinv$. The best-fit coefficients are tabulated in Tab.~\ref{tab:abacus_perr_results}. For Abacus the best-fit error power spectra on models two to four are shown in Fig.~\ref{fig:abacus_bestfit_perr_models}. }
    \label{fig:Perr}
\end{figure}

To investigate the error power spectrum beyond the $k\to 0$ limit, we fit the following functional form to it
\begin{equation} \label{eq:Perr_tf}
P_{\text{err}}(k, \mu) = a_0 + a_2 k^2 + a_3 k^3 + a_4 k^4 + \sum_{i=2,\,4} a_{ii} (k \mu)^i\,,
\end{equation}
where we have the $k^3$
term 
to account for 
the observed smooth 
scaling of the 
error power spectrum in the non-linear regime. 
Odd powers
of $k$ in the error power spectrum are forbidden in EFT; their inclusion here is purely phenomenological.
The best-fit parameters for both sets of simulations are tabulated in Tab.~\ref{tab:abacus_perr_results} and a subset are shown in Fig.~\ref{fig:Perr}. Following baseline expectation, the error power spectrum has a an amplitude that is more than two orders of magnitude smaller than the signal and constant on large scales. In particular, the onset of the scale-and orientation-dependence occurs around $k\approx 0.4-0.6\hMpcinv$ ultimately setting an upper limit on the applicability of EFT as a theory model (see Ref.~\cite{deBelsunce:2025gci}). We perform the same fits to the error power spectrum obtained from the cross-correlation of the \Lya forest with halos and find that the amplitude of $P_{\rm err}$ is not driven by the larger one obtained from the halos but rather by the one from the \Lya forest.

\begin{table}
\centering
\begin{tabular}{lccccccc}
\hline\hline
Data &  & $a_0$ & $a_2$ & $a_3$ & $a_4$ & $a_{22}$ & $a_{44}$ \\
\hline
Abacus & I & $\phantom{-}0.041$ & $-0.097$ & $\phantom{-}0.016$ & $\phantom{-}0.069$ & $\phantom{-}0.230$ & $-0.276$ \\
-- & II & $\phantom{-}0.048$ & $-0.195$ & $\phantom{-}0.433$ & $-0.345$ & $\phantom{-}0.046$ & $\phantom{-}0.235$ \\
-- & III & $\phantom{-}0.069$ & $-0.531$ & $\phantom{-}1.149$ & $-0.814$ & $\phantom{-}0.358$ & $-0.570$ \\
-- & IV & $\phantom{-}0.063$ & $-0.549$ & $\phantom{-}1.163$ & $-0.820$ & $\phantom{-}0.438$ & $-0.668$ \\
\hline
-- & I $\times$ halos & $\phantom{-}0.116$&$\phantom{-}0.061$&$\phantom{-}0.841$&$-1.486$&$\phantom{-}2.368$&$-0.945$ \\
-- & II $\times$ halos & $\phantom{-}0.121$&$-0.226$&$\phantom{-}1.461$&$-1.798$&$\phantom{-}2.251$&$-1.018$ \\
-- & III $\times$ halos & $\phantom{-}0.254$&$\phantom{-}1.897$&$-5.069$&$\phantom{-}2.916$&$\phantom{-}2.499$&$-2.814$ \\
-- & IV $\times$ halos & $\phantom{-}0.268$&$\phantom{-}1.694$&$-4.875$&$\phantom{-}2.833$&$\phantom{-}2.819$&$-2.958$ \\
\hline
Sherwood & $L=160 \hinvMpc$ & $\phantom{-}0.154$ & $-0.005$ & $-0.412$ & $\phantom{-}0.316$ & $\phantom{-}0.404$ & $-0.337$ \\
-- & $\times$ halos & $\phantom{-}0.034$&$\phantom{-}0.955$&$-3.074$&$\phantom{-}2.096$&$\phantom{-}1.441$&$-1.092$ \\
\hline\hline
\end{tabular}
\caption{\textbf{Best-fit $P_{\rm err}$ fits:} Coefficients for the polynomial fit to $P_{\rm err}$ given in Eq.~\eqref{eq:Perr_tf} for the Abacus simulations (\textit{top eight rows}) and the Sherwood simulation (\textit{bottom two row}), shown in Fig.~\ref{fig:Perr}. In addition we quote the best-fit parameters from fits to the error power spectrum of the cross-correlation with all available halos, denoted by ``$\times$ halos''. The corresponding transfer functions are tabulated in Tabs.~\ref{tab:beta_parameters} and \ref{tab:abacus_tf_results} and $a_n, a_{nn}$ are in units of ~$[\Mpch]^{3+n}$. The best-fit plots for Abacus models two to four are shown in Fig.~\ref{fig:abacus_bestfit_perr_models}.
}
 \label{tab:abacus_perr_results}
\end{table}

\subsection{Cross-correlation coefficient}\label{sec:rcc}
A second stringent test of the field-level model is provided by the cross-correlation coefficient, $r_{cc}$, defined in Eq.~\eqref{eq:rcc}. It relates the power spectrum of the simulated field to the mean-squared model error through $P_{\rm err}=P_{\rm truth}(1-r^2_{cc})$ and thus quantifies how faithfully the model reproduces the phases of the field. As such, $r_{cc}$ is our key metric for assessing the range of validity of the forward model. However, the scale at which the error power spectrum begins to rise with $r_{cc}$ correspondingly departing from unity does not by itself define the maximum scale of applicability. We therefore examine how well the error power spectrum can be captured by EFT stochastic counterterms~\cite{Ivanov:2024lya},
\begin{equation} \label{eq:P1D_stoch}
    P^{\rm stoch.}_{\rm 1D}(k_{\parallel}) = c_0 + c_1k_{\parallel}^2 + c_2 k_{\parallel}^4\, ,
\end{equation}
where the coefficient $c_n$ has units of $[\Mpch]^{(2n+1)}$ for $n=0,1,2$. As discussed in Ref.~\cite{deBelsunce:2025bqc}, the small-scale modeling is affected by non-Poisson stochasticity (the analog of the one-halo term in galaxy clustering), which renders the error power spectrum scale- and direction-dependent in a way that EFT cannot capture beyond the gradient expansion. This stochastic noise sets the limit of applicability of EFT~\cite{Baldauf:2015tla,Baldauf:2015zga,Schmittfull:2018yuk,Schmittfull:2020trd}, which we discuss in the following for the 3D power spectrum and in Sec.~\ref{sec:p1d} for the 1D power spectrum.

\begin{figure*}
    \centering
    \includegraphics[width=\linewidth]{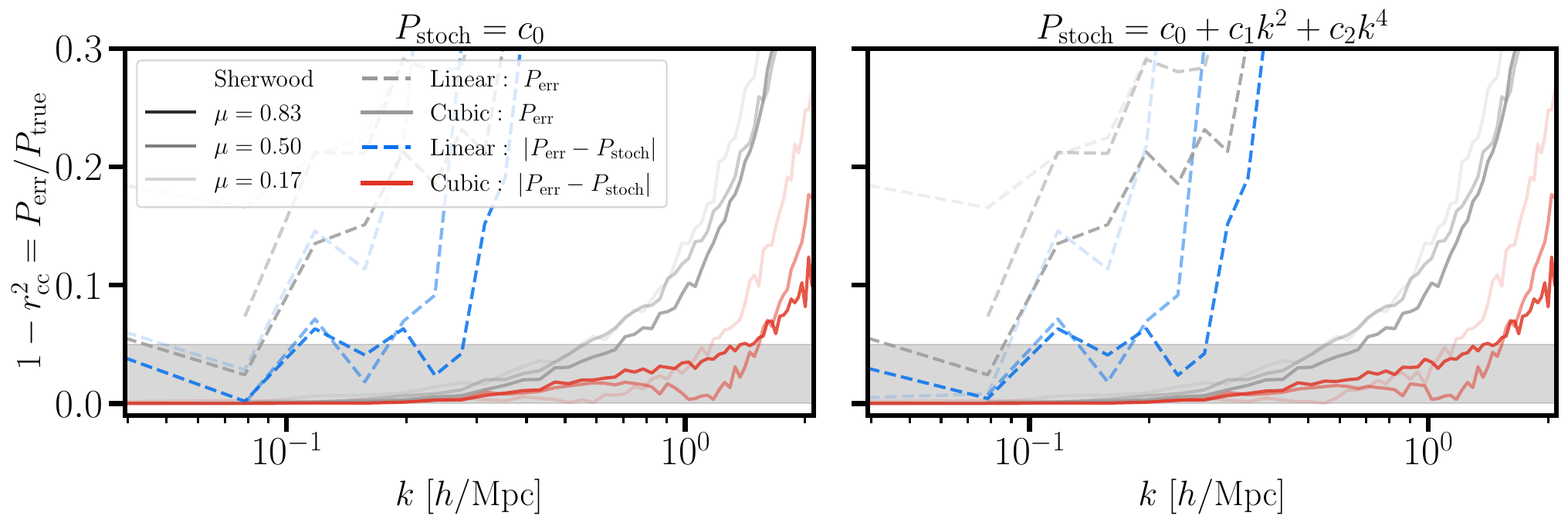}
    \caption{\textbf{Cross-Correlation Coefficients \Lya Sherwood:} Fractional mean-squared model error $1-r_{cc}^2 = |P_{\rm err}/P_{\rm true}|$ where $r_{cc}=r_{cc}(\td^{\rm truth},\td^{\rm model})$ is the cross-correlation coefficient between simulations and model. The results are shown for the Sherwood simulation using the linear (dashed lines) and cubic EFT model (solid lines) in three angular bins, $\mu$. The gray lines denote the ``raw'' cross-correlation coefficient and the colored lines illustrate the subtraction of the stochastic counterterms, given in Eq.~\eqref{eq:P1D_stoch}, from the error power spectrum. The gray shaded region indicates the 5\% error band.}
    \label{fig:rcc}
\end{figure*}

\begin{figure*}
    \centering
    \includegraphics[width=\linewidth]{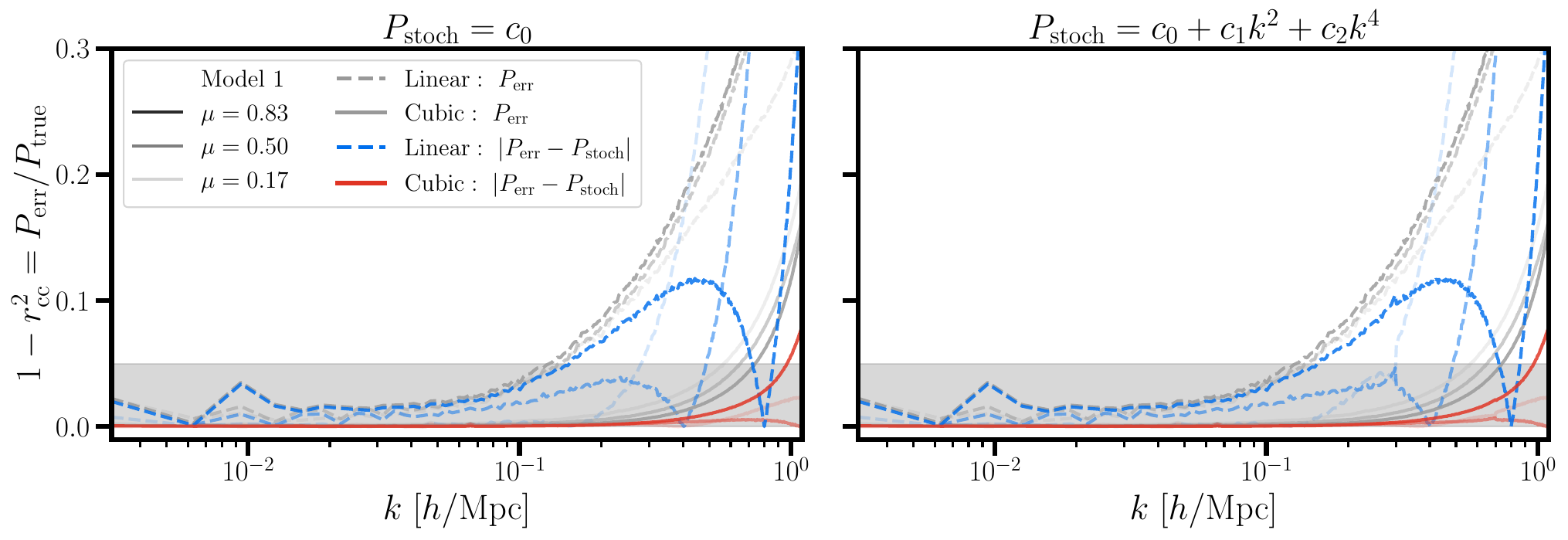}
    \caption{\textbf{Cross-Correlation Coefficients \Lya Abacus:} Same as Fig.~\ref{fig:rcc} for the Abacus simulation model I. The turnover in the linear model is an artifact from an over-subtraction at high $k$ and the result of plotting the norm.}
    \label{fig:rcc_abacus}
\end{figure*}

In Fig.~\ref{fig:rcc} we show the cross-correlation coefficients obtained for the linear (dashed lines) and cubic EFT model (solid lines) for the large \Lya Sherwood simulation at $z=2.8$. The gray lines use the ``raw'' error power spectrum, while the colored lines are obtained after subtracting the stochastic counterterms [see Eq.~\eqref{eq:P1D_stoch}] from it. Before subtraction, the cubic EFT model reproduces the phases in Fourier space to better than 1\% up to $k\simlt 0.3 \hMpcinv$ ($k\simlt 0.24 \hMpcinv$) for Abacus (Sherwood), with the 5\%-level reached at $k\simlt 0.7 \hMpcinv$ ($k\simlt 0.6 \hMpcinv$). Since Abacus covers a larger volume, it contains more quasi-linear modes and the linear model captures a larger range of scales than for Sherwood for which it captures the large-scale modes only at the $3\%$ on large scales.

After subtracting the stochastic contribution from the error power spectrum\footnote{This procedure is equivalent to the joint modeling 
of the deterministic and stochastic parts of the \Lya power spectrum.}, we extend the reach of both models: for Sherwood, the linear model reaches $k\simlt 0.3 \hMpcinv$ at the 5\% level and the cubic model reaches $k\simlt 0.5\, (1.3) \hMpcinv$ at the 1\% (5\%) level, with a similar performance for Abacus as shown in Fig.~\ref{fig:rcc_abacus}. From this analysis, the majority of the improvement stems from subtracting a constant offset with little weight stemming from the $c_1$ and $c_2$ terms. We note that going to higher redshifts, e.g.,~$z=3.2$ using the smaller Sherwood boxes, does not significantly improve the performance of the cubic model for the 3D power spectrum. We emphasize that, while both simulations use entirely different prescriptions to model the \Lya forest, they yield consistent results, demonstrating the robustness of the field-level methodology and setting the maximum scale for linear theory at $k\simlt 0.1-0.2\hMpcinv$.\footnote{It is interesting to compare the obtained cross-correlation coefficient to the one obtained from other approaches, e.g.~using the promising deep learning reconstruction presented in figure 6 in Ref.~\cite{Hafezianzadeh:2025ifw}. Whilst their cross-correlation coefficient significantly improves using their reconstruction approach, their model has a floor of approximately five per cent on large scales ($k\simlt 1\hMpcinv$). We leave the exploration of a hybrid approach combining EFT on large scales with a deep-learning-based reconstruction on small scales to future work.}

\begin{figure*}
    \centering
    \includegraphics[width=0.49\linewidth]{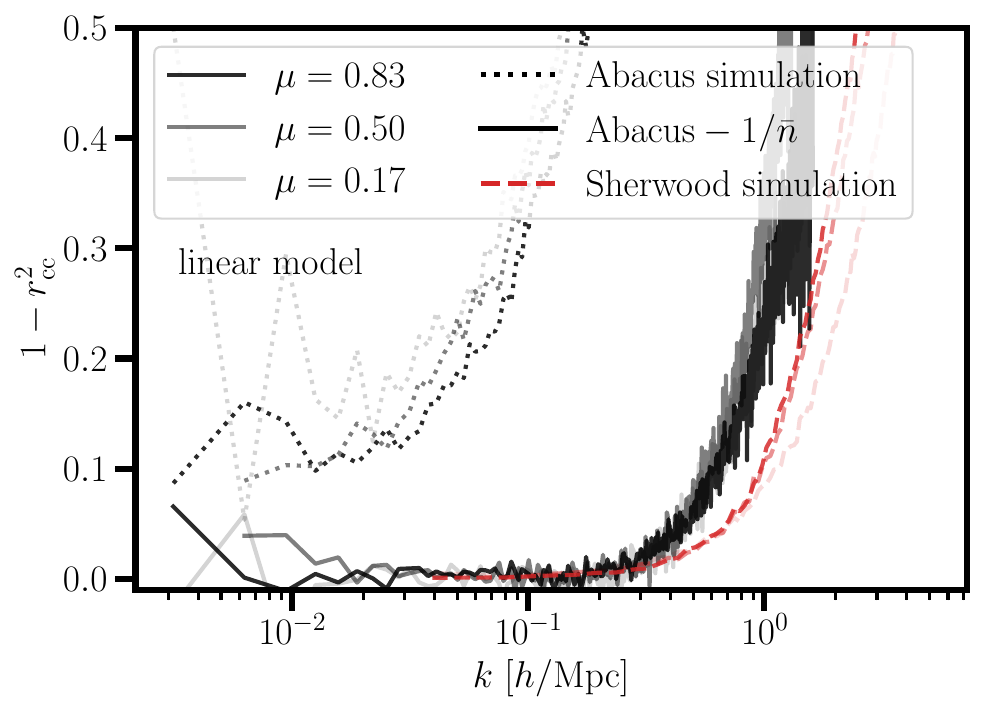}\hfill
    \includegraphics[width=0.49\linewidth]{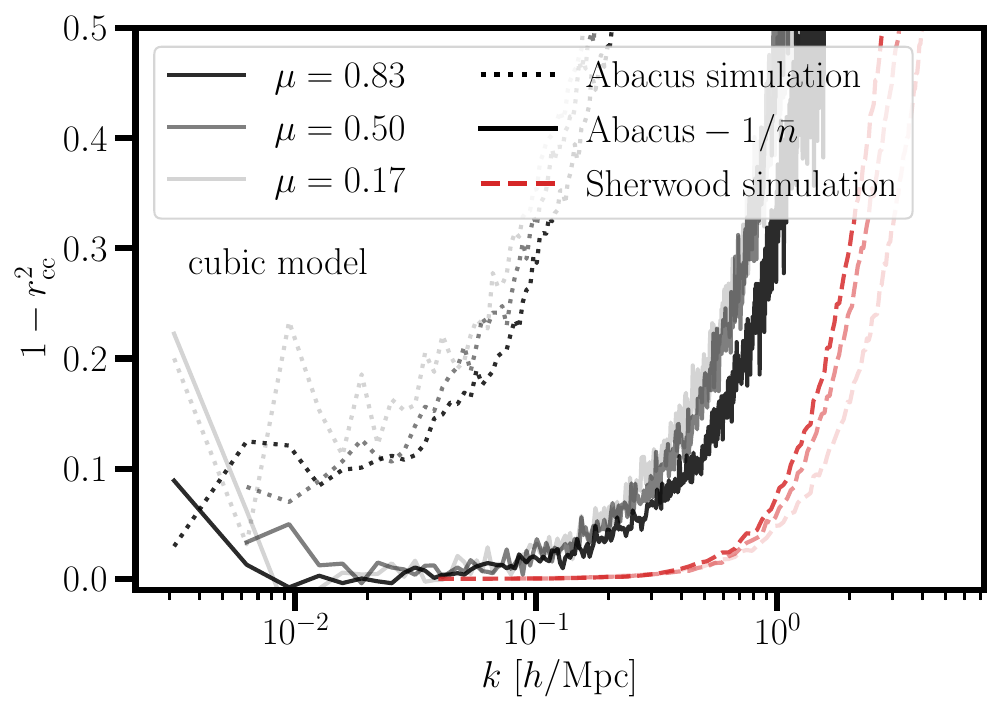}
    \caption{\textbf{Cross-Correlation Coefficients Halos:} Same as Fig.~\ref{fig:rcc} for the halo auto-correlation using all available halo masses for both simulations. As discussed in Fig.~\ref{fig:lya_halo_pk}, the error power spectrum for Abacus converges to the shot noise, setting a clear upper bound on the cross-correlation coefficient. Therefore, we show the shot noise subtracted version (solid black line) and $r_{cc}$ including the shot noise (dotted black line). This shows that for quasar (or galaxy) samples with lower shot noise, the field-level methodology extends its reach -- similar to the \Lya forest. For Sherwood the cross-correlation coefficient is very close to one since the provided halo catalogs are (unrealistically) light.
    }
    \label{fig:rcc_halo}
\end{figure*}

In addition to the \Lya forest, we show the fractional mean-squared model error for the halo auto-correlation in Fig.~\ref{fig:rcc_halo}, using the linear model (left panel) and the cubic model (right panel). Here the cubic model offers almost no improvement over the linear theory model, indicating that we are in the shot-noise limited regime for halos. Massive and light halos serve as proxies for quasars and high-redshift galaxies, respectively. In contrast to those from Abacus, the Sherwood halos are light and carry a very low shot noise; we therefore subtract the shot noise only from Abacus in Fig.~\ref{fig:rcc_halo}, moving from the dotted to the solid black lines for both models.

\begin{figure}
    \centering
    \includegraphics[width=\linewidth]{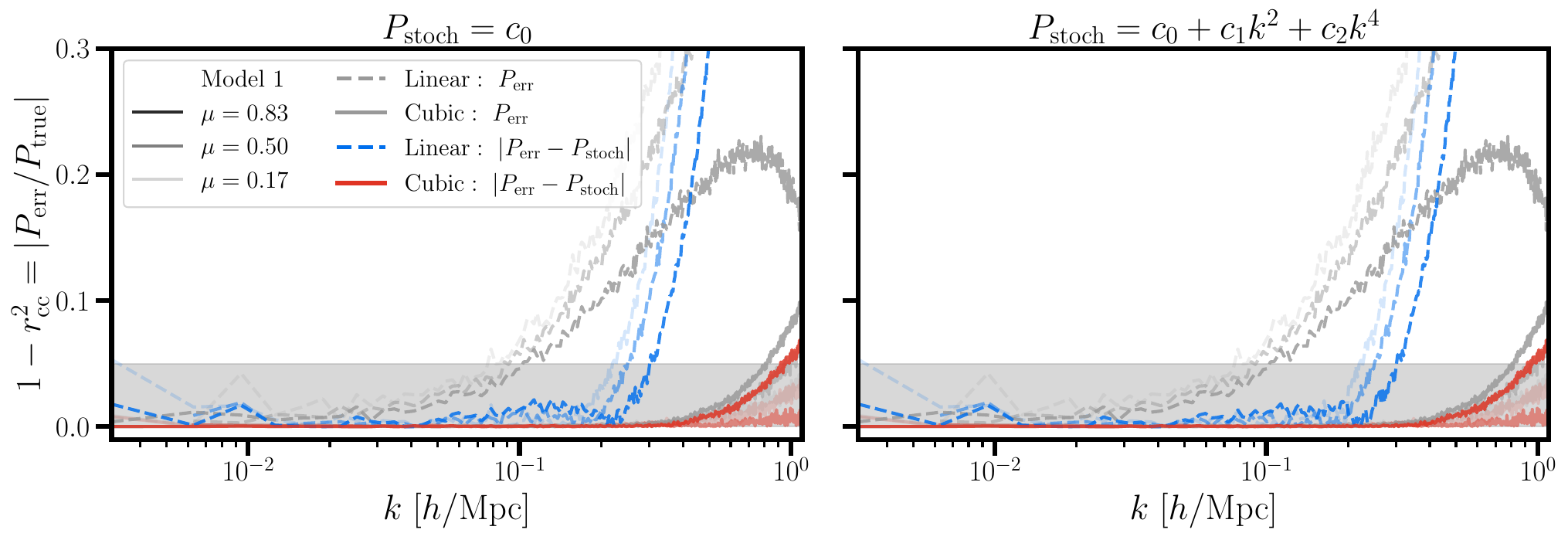}\\
    \includegraphics[width=\linewidth]{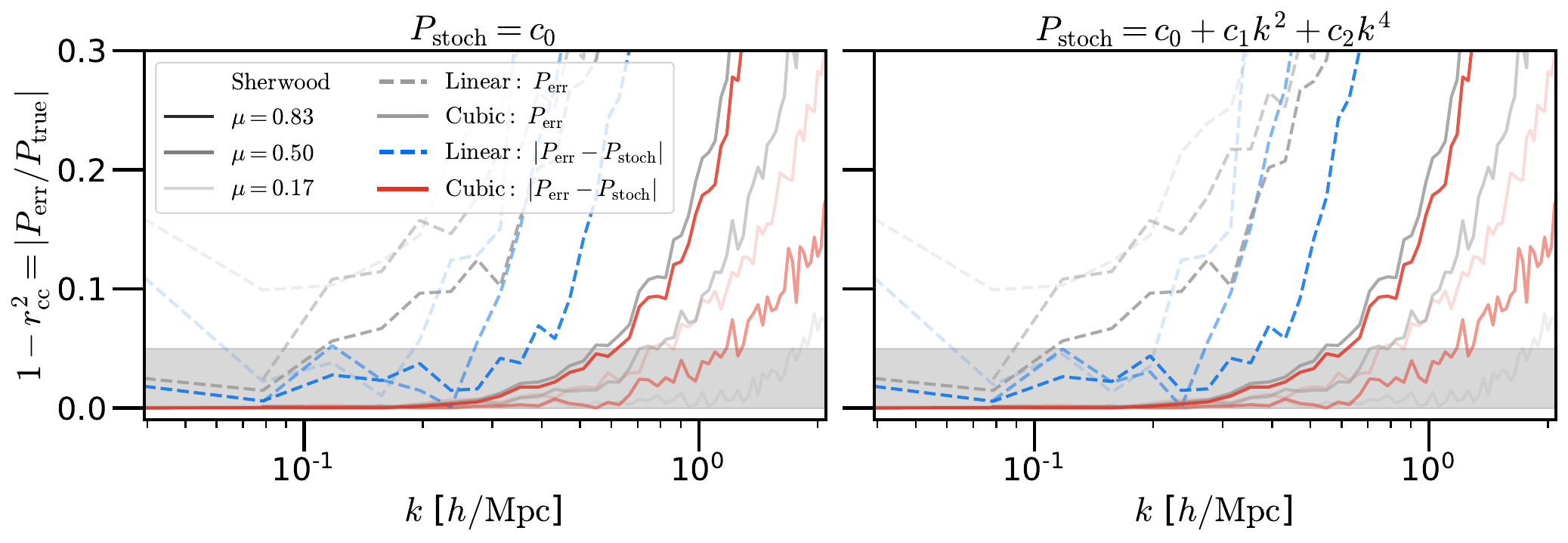}
    \vspace{-0.1in}
    \caption{\textbf{Cross-correlation coefficient \Lya $\times$ Halos:}
    Same as Figs.~\ref{fig:rcc} for the cross-correlation of the \Lya forest with halos for Abacus Model I (\textit{top row}) and Sherwood (\textit{bottom row}) using all available halo masses. The gray lines denote the ``raw'' cross-correlation coefficient and the colored lines illustrate the subtraction of the stochastic counterterms, given in Eq.~\eqref{eq:P1D_stoch}, from the error power spectrum. The gray shaded region indicates the 5\% error band.
    }
    \label{fig:rcc_cross}
\end{figure}

Finally, in Fig.~\ref{fig:rcc_cross} we show the 3D cross-correlation coefficient for the cross-correlation of the \Lya forest with halos, using all available halo masses for Abacus (top row) and Sherwood (bottom row). As before, we model the stochastic contributions to the error power spectrum: this yields a noticeable improvement for the linear model (gray to blue dashed lines), extending its reach to $k\simlt 0.1 \hMpcinv$, and a negligible improvement for the cubic model (gray to red solid lines), which reaches the 5\%-level at $k\simlt 1\hMpcinv$. Overall, the cubic forward model provides an excellent fit to the cross-correlation data.

\subsection{One-dimensional power spectrum} \label{sec:p1d}
\begin{figure*}
    \centering
    \includegraphics[width=0.49\linewidth]{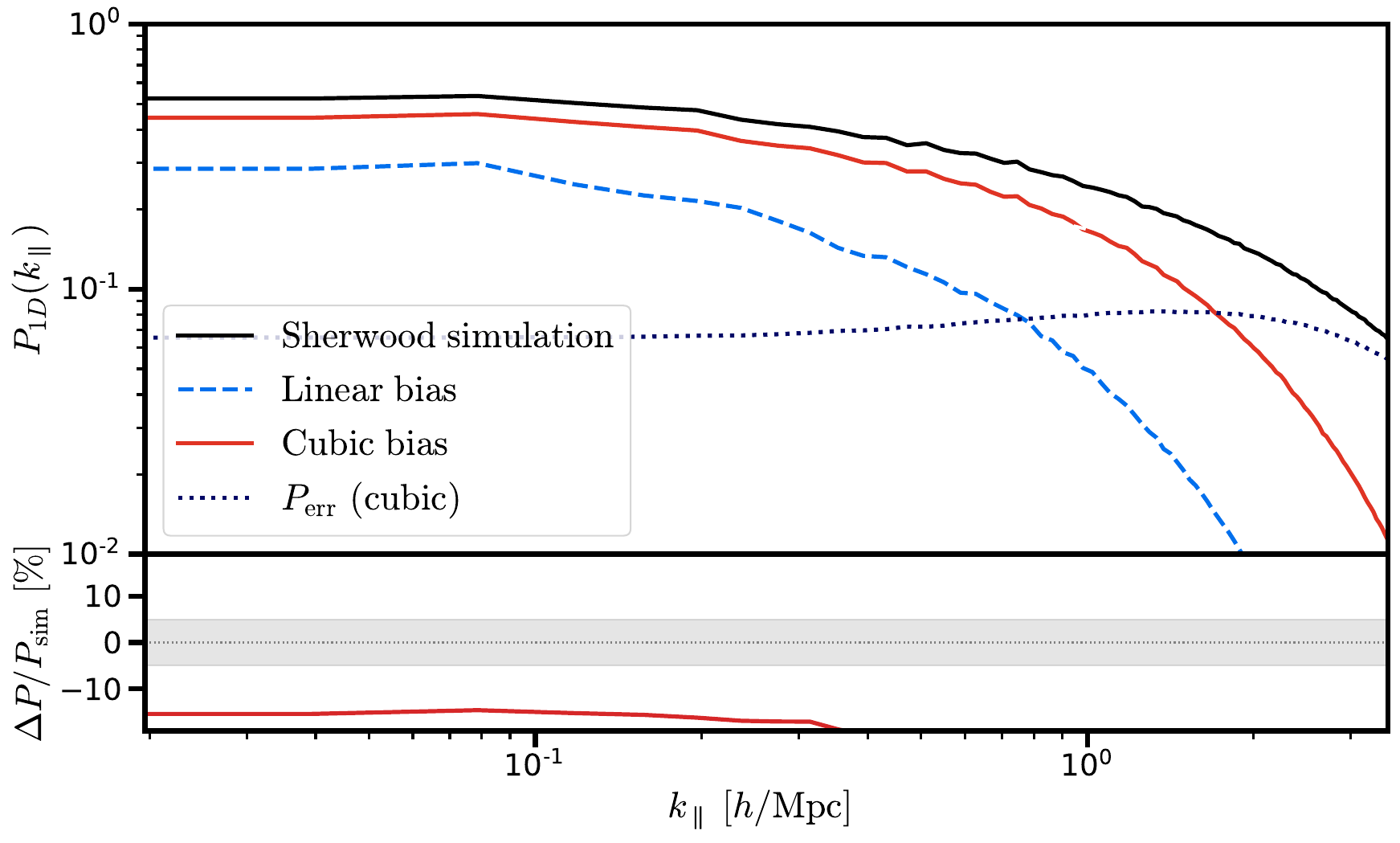}\hfill
    \includegraphics[width=0.49\linewidth]{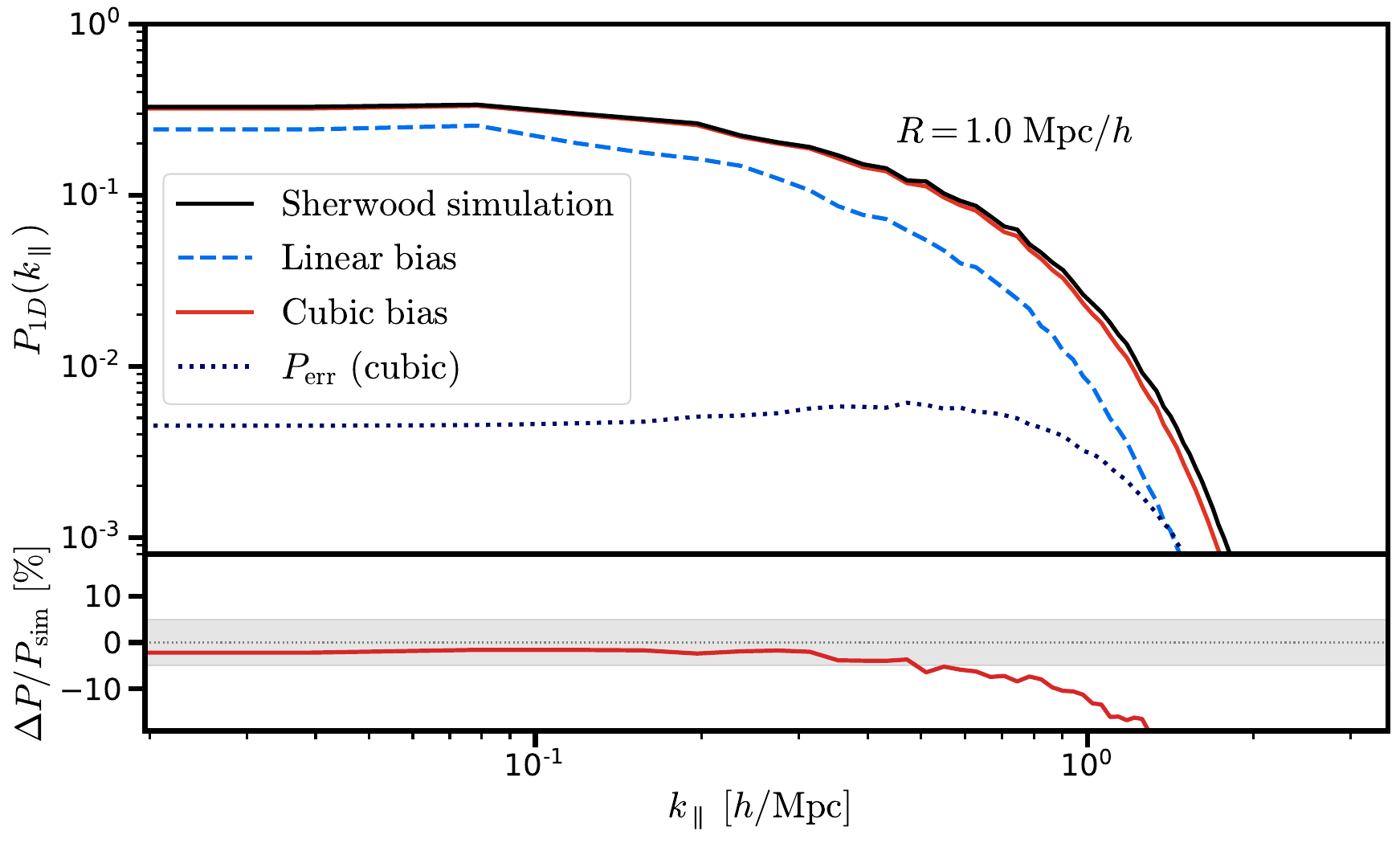}\hfill
    \vspace{-0.1in}
    \caption{\textbf{Best-Fit 1D Power Spectra Sherwood:}
    Comparison between the measured one-dimensional power spectrum from the Sherwood simulations with two different smoothing scales using a Gaussian kernel: left no smoothing, right $R=1 \hinvMpc$. The measured P1D is shown in black together with the best-fit forward model obtained from a linear (blue dashed) and a cubic bias (red solid) expansion, respectively. For the cubic bias expansion we show the error power spectrum (blue dotted line). For the baseline without smoothing, we find a 15\% offset between the cubic model and the true power spectrum and an even larger off set for the linear model. In both panels, the P1D is shown in bins of Fourier wavenumber $k_{\parallel}$ in units of $\hMpcinv$. The bottom panel displays the percent difference between the simulation and model power spectra. A gray band highlights the $\pm5\%$ region in the bottom panel.
    }
    \label{fig:lya_p1d_sherwood}
\end{figure*}

\begin{figure*}
    \centering
    \includegraphics[width=0.49\linewidth]{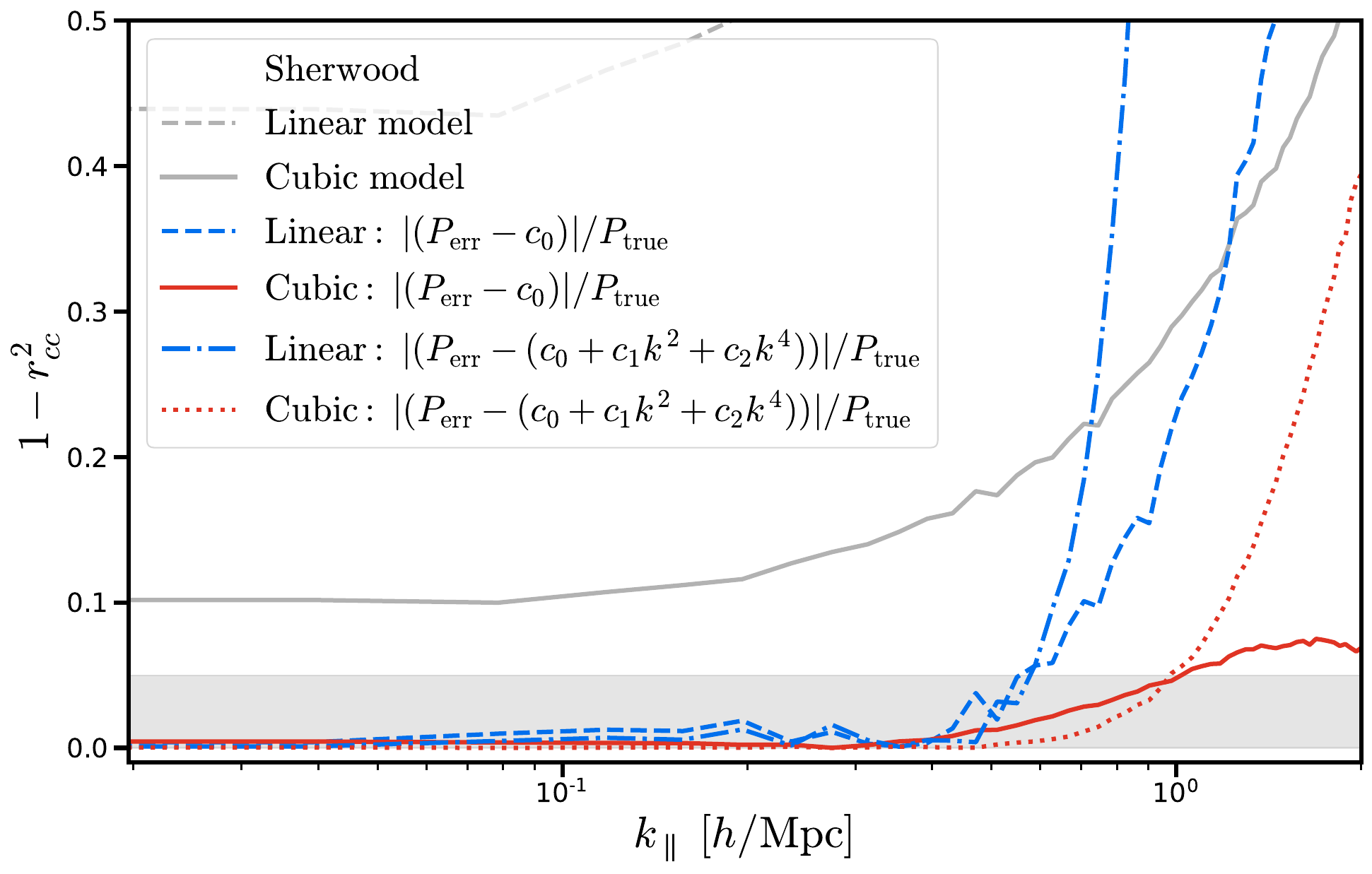}\hfill
    \includegraphics[width=0.49\linewidth]{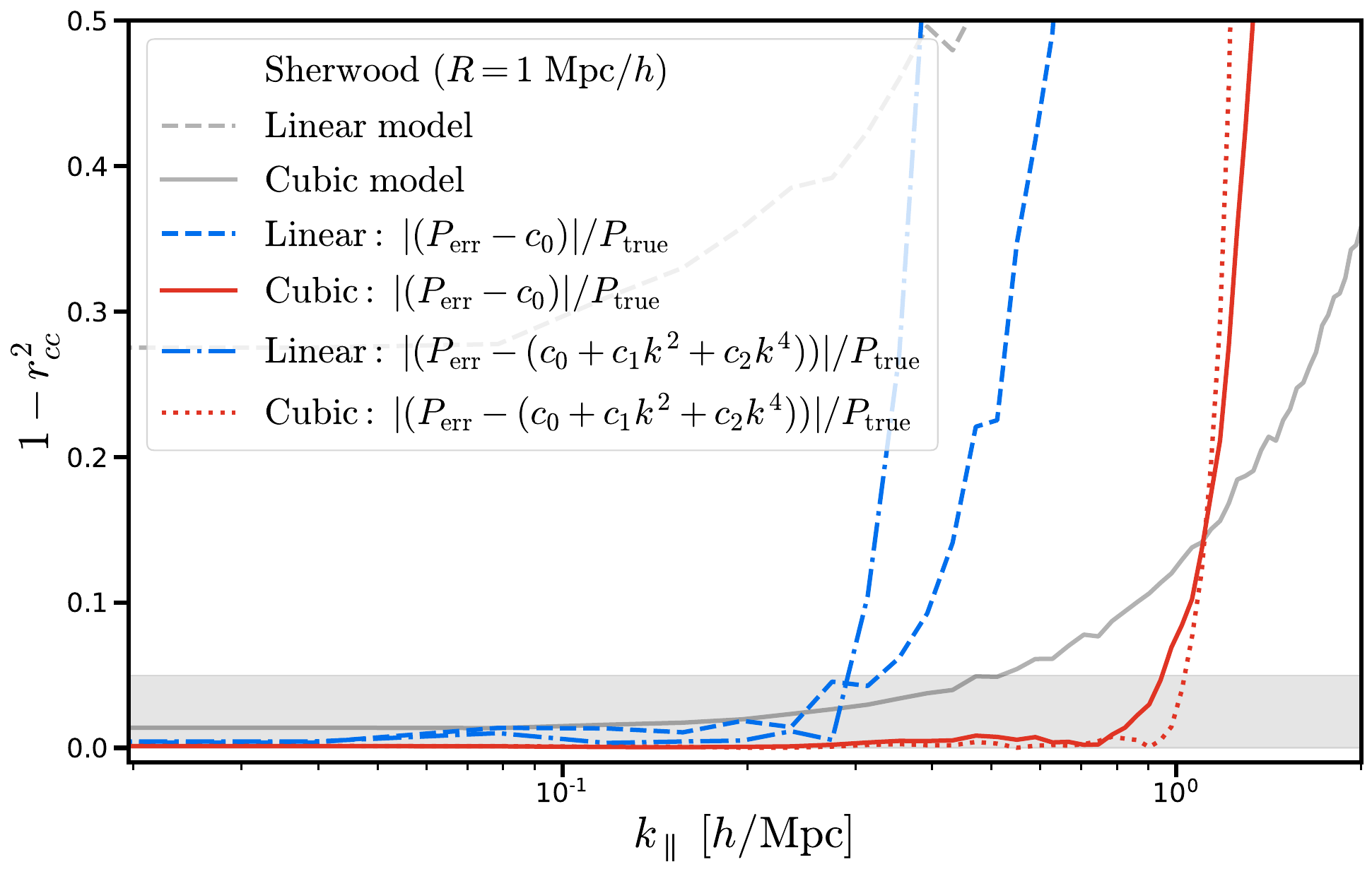}\hfill
    \vspace{-0.1in}
    \caption{\textbf{Cross-correlation coefficient P1D Sherwood:}
    Corresponding cross-correlation coefficient ($1-r_{\rm cc}^2$) to the one-dimensional power spectra shown in Fig.~\ref{fig:lya_p1d_sherwood}. The colored lines have, first, only the constant low-$k$ limit subtracted from the P1D denoted by $c_0$ and, second, a higher-order polynomial $c_0 + c_1k_{\parallel}^2 + c_2 k_{\parallel}^4$. This plot illustrates the importance of using a higher order biasing model compared to purely linear theory as the range of five-percent-level validity of the model is extended up to $k\approx 1 \hMpcinv$ for the baseline case without smoothing. 
    }
    \label{fig:lya_p1d_rcc_sherwood}
\end{figure*}

In Fig.~\ref{fig:lya_p1d_sherwood} we additionally compare the one-dimensional power spectrum -- a key summary statistic in the context of \Lya forest analyses (see, e.g.,~\cite{Seljak:2005, Viel:2005, McDonald06, PYB13, Chabanier:2019, Pedersen:2020, 2023MNRAS.526.5118R, 2024MNRAS.tmp..176K}). 
The left panel shows the model performance without any smoothing applied to the simulation and the cubic forward model recovers the shape of the true power spectrum well while showing a residual offset of $\sim 15\%$. In contrast, the linear theory model fails at recovering the P1D of the input simulation, even on large scales, to better than 40\% (here: outside of the residual plot range). The right panel includes a Gaussian smoothing kernel of $R=1\hinvMpc$ and removes small-scale information. This results in recovery of the input simulation power spectrum to better than 5\% (1\%) down to $k_\parallel\simlt 0.6\, (0.3)\hMpcinv$ for the cubic model and the linear theory results are off by more than $\approx 20 \%$. Our results suggest that there is more perturbative information in the P1D than discussed in Ref.~\cite{Irsic:2018hhg}. 
In fact, up to $k_\parallel\approx 1~\hMpc$
the P1D in simulations is dominated 
by perturbative modes. 
While we agree with~\cite{Irsic:2018hhg}
on the significant role of stochastic contributions
in P1D, we point out that 
this noise can be well modeled 
within EFT. 

The error power spectra of the cubic model (and for the linear model on large scales) are constant and do not show any scale dependence. The error power spectrum of the cubic model (dotted blue line in the left panel) of the Sherwood simulation is flat with an amplitude of $\approx 0.07\, [\hinvMpc]^3$. It can be modeled, as discussed in Sec.~\ref{sec:rcc}.
In Fig.~\ref{fig:lya_p1d_rcc_sherwood} we show the corresponding cross-correlation coefficients where we first only remove the constant low-$k_\parallel$ limit going from gray to solid colored lines (blue: linear theory, red: cubic model). The removal of the residual noise using Eq.~\eqref{eq:P1D_stoch} from the linear (cubic) model extends the range of validity (when investigating $1-r_{\rm cc}^2$) to $k_\parallel \approx 0.3\, (1.0) \hMpcinv$ resulting in the dash-dotted blue lines for the linear theory and dotted red lines for the cubic model, respectively. 
Note that the smoothing of the field is non-physical and only serves to illustrate the importance of using a higher-order bias expansion to better capture the small-scale clustering. We perform the same steps on the smoothed field and show the results in the right panel of the same figure. 
As visible from the cross-correlation coefficients, for the 3D power spectrum this sets in at $k\simgt 0.9-1.0\hMpcinv$ and for the 1D power spectrum at $k_\parallel\simgt 1.0\hMpcinv$, respectively. 

\begin{figure*}
    \centering
    \includegraphics[width=0.49\linewidth]{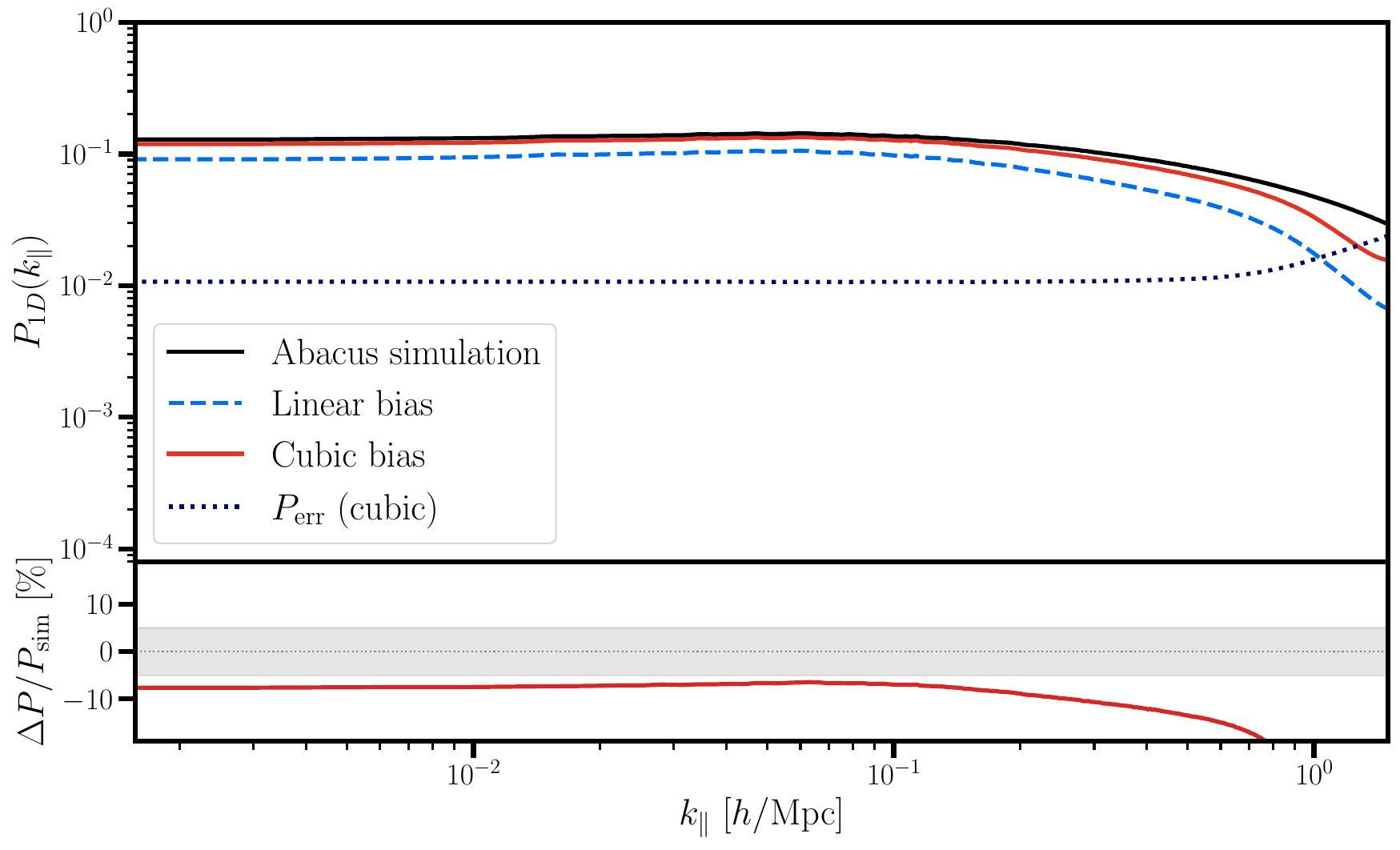}\hfill
    \includegraphics[width=0.49\linewidth]{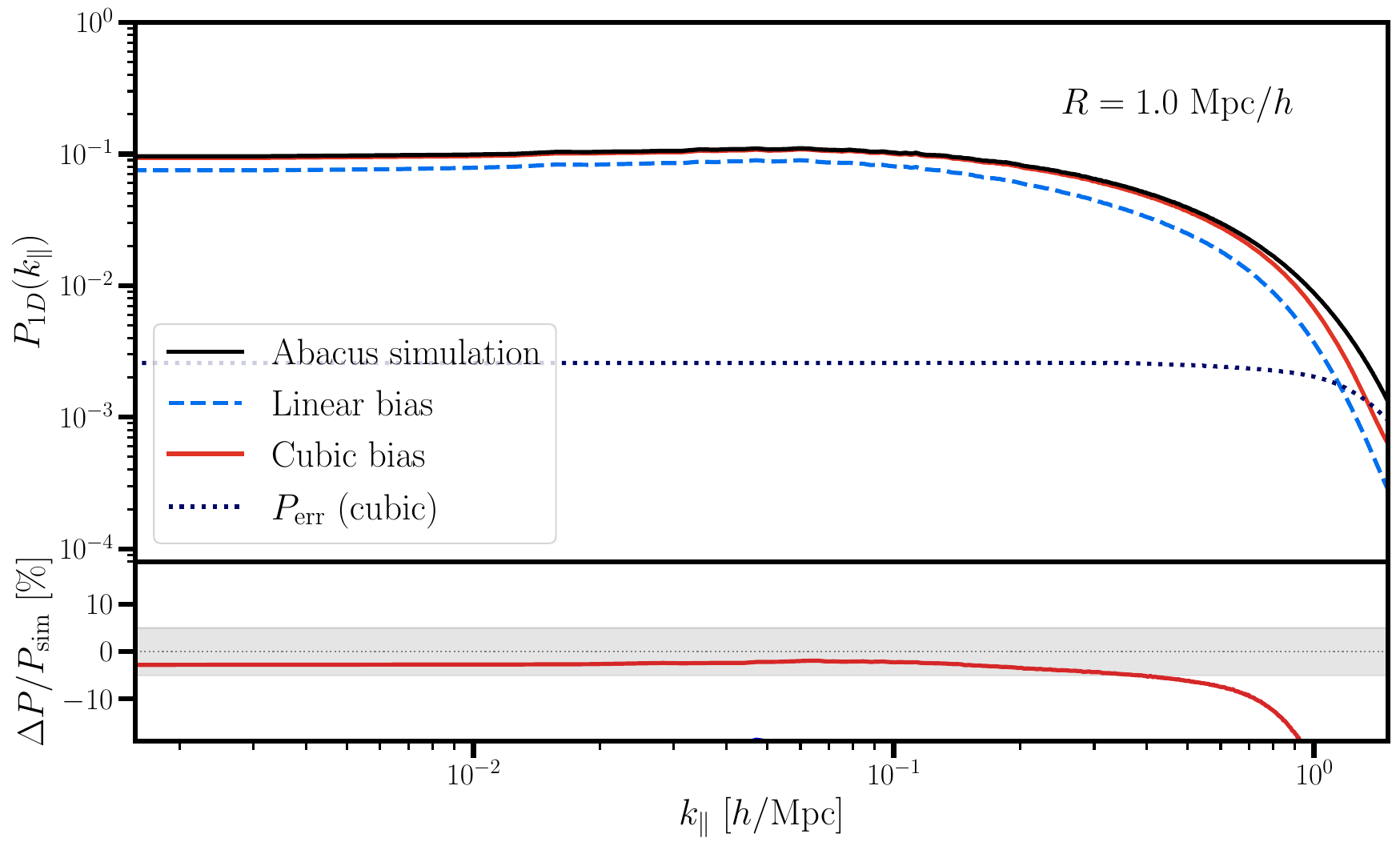}\hfill
    \vspace{-0.1in}
    \caption{\textbf{Best-Fit 1D Power Spectra Abacus:}
    Same as Fig.~\ref{fig:lya_p1d_sherwood} for the Abacus simulation and model I. 
    }
    \label{fig:lya_p1d_abacus}
\end{figure*}

\begin{figure*}
    \centering
    \includegraphics[width=0.49\linewidth]{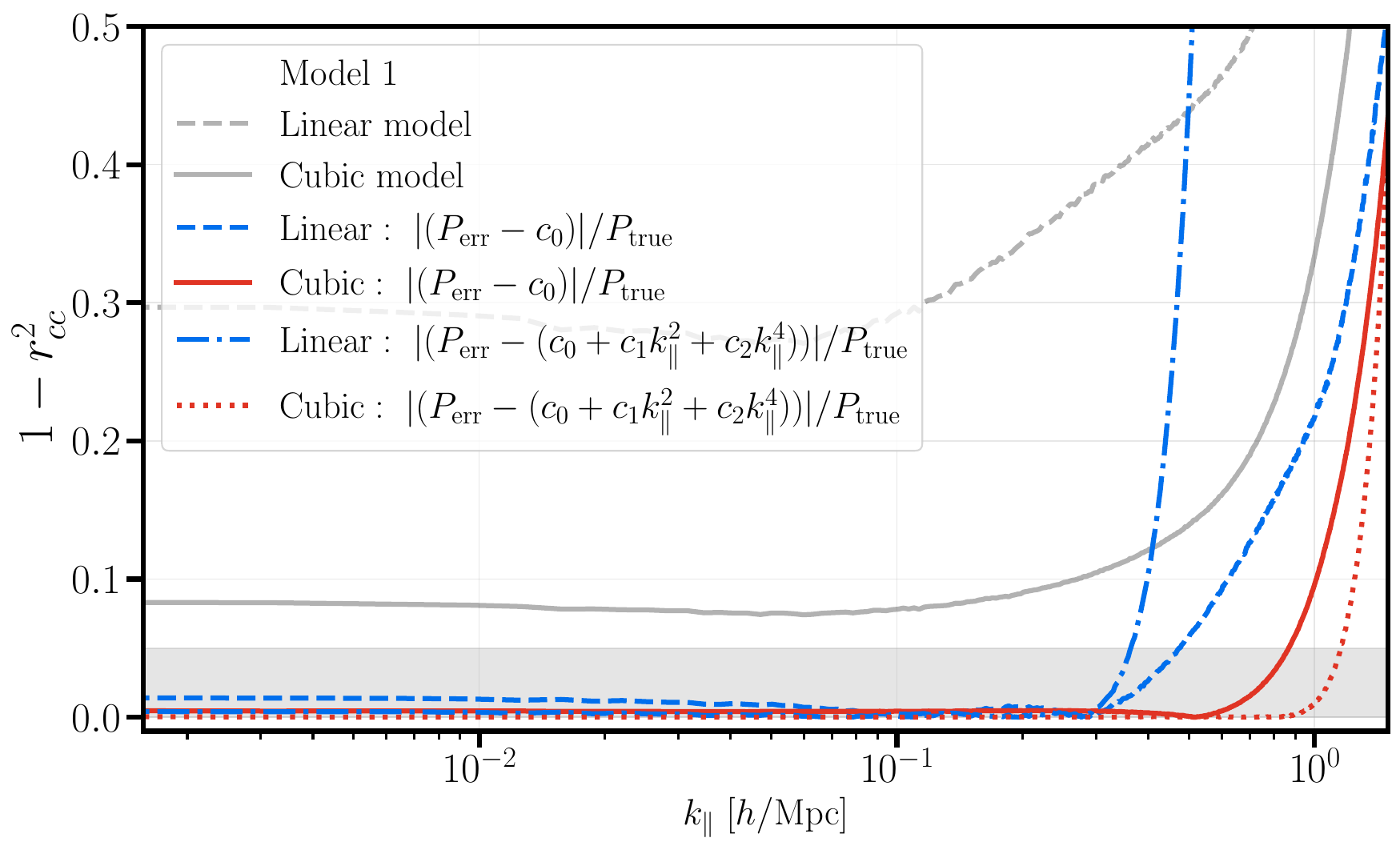}\hfill
    \includegraphics[width=0.49\linewidth]{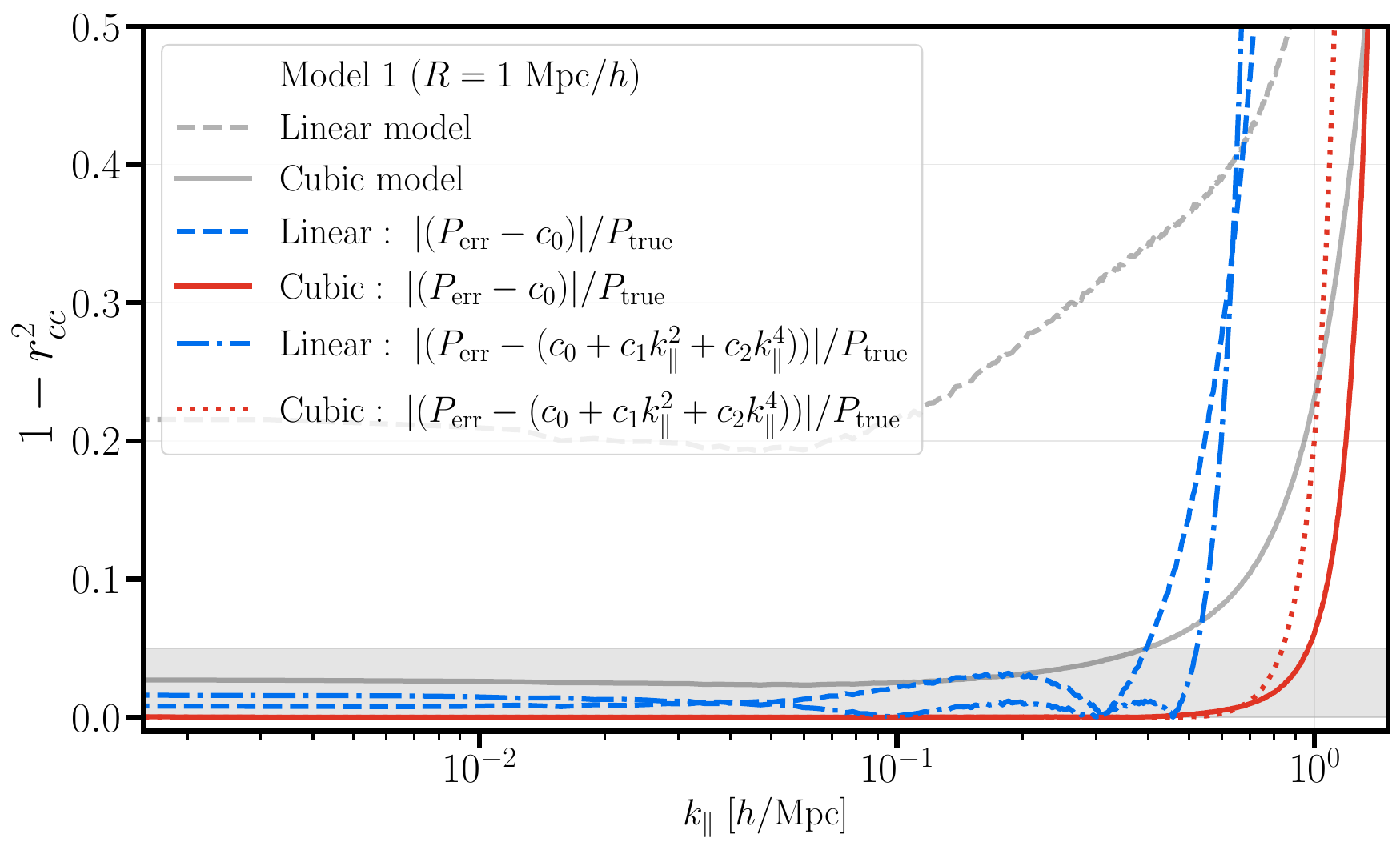}\hfill
    \vspace{-0.1in}
    \caption{\textbf{Cross-correlation coefficient P1D Abacus:}
    Cross-correlation coefficient ($1-r_{\rm cc}^2$) for the one-dimensional power spectrum of the Abacus simulation and model I where we model the residual noise floor using Eq.~\eqref{eq:P1D_stoch}; same as Fig.~\ref{fig:lya_p1d_rcc_sherwood}. 
    }
    \label{fig:lya_rcc_abacus}
\end{figure*}

In Fig.~\ref{fig:lya_p1d_abacus} we show the corresponding results on the P1D for the Abacus simulations and in Fig.~\ref{fig:lya_rcc_abacus} we show their cross-correlation coefficients. Qualitatively, the results agree with the ones obtained from Sherwood. Removing the stochastic terms we extend the range of validity of the cubic (linear) model to $k_\parallel\approx 1.0\, (0.4) \hMpcinv$ for the unsmoothed field. Analogously to the P3D results, the linear model has a bigger $k$-reach on Abacus compared to Sherwood given the larger box size and available number of quasi-linear modes. In Figs.~\ref{fig:abacus_p1d_models}-\ref{fig:abacus_rcc_models} we show the model dependence of our forward model in the context of the P1D.

\section{Transfer function fits} \label{sec:tranfer_func}
So far we have compared the best-fit perturbative forward model to a set of simulations and found an excellent agreement between the two. However, the best-fit transfer functions that minimize the error power spectrum (or mean-squared error)
have been completely free in each $k$-bin. To build intuition for the transfer functions from Eq.~\eqref{eqn:lya_model} we fit these as polynomial expansions in $k$ and $\mu$, following Ref.~\cite{deBelsunce:2025bqc}, given by
\begin{align} \label{eq:beta_tf}
\beta_i(k,\mu) = c_0 + c_{01}\mu^2 + 
                \left( c_1 + c_{12}\mu^2 + c_{14}\mu^4\right) \cdot k + 
                \left( c_4 k^2 + c_{22}\mu^2 + c_{44}k^2\mu^4\right)\cdot  k^2 \,,
\end{align}
which is different to the one used in Refs.~\cite{Schmittfull:2020trd, Obuljen:2022cjo} to capture the angular dependence of the transfer functions at low-$k$. In particular, we allow for higher order corrections, i.e.~the two-loop terms, by allowing for odd powers of $k$. Note that in the $k\to 0$ limit we recover the ``Kaiser''-type expression. The fits are performed jointly for all $\mu$-bins using non-linear least-squares minimization, weighting each data point by $k$ to account for
the different number of modes with a cut-off at $k=1\hMpcinv$.\footnote{We verified that reducing the maximum wavenumber chosen in the fit to $k=0.6 \hMpcinv$ does not change our conclusions.} 

\begin{figure*}
    \centering
    \includegraphics[width=1\linewidth]{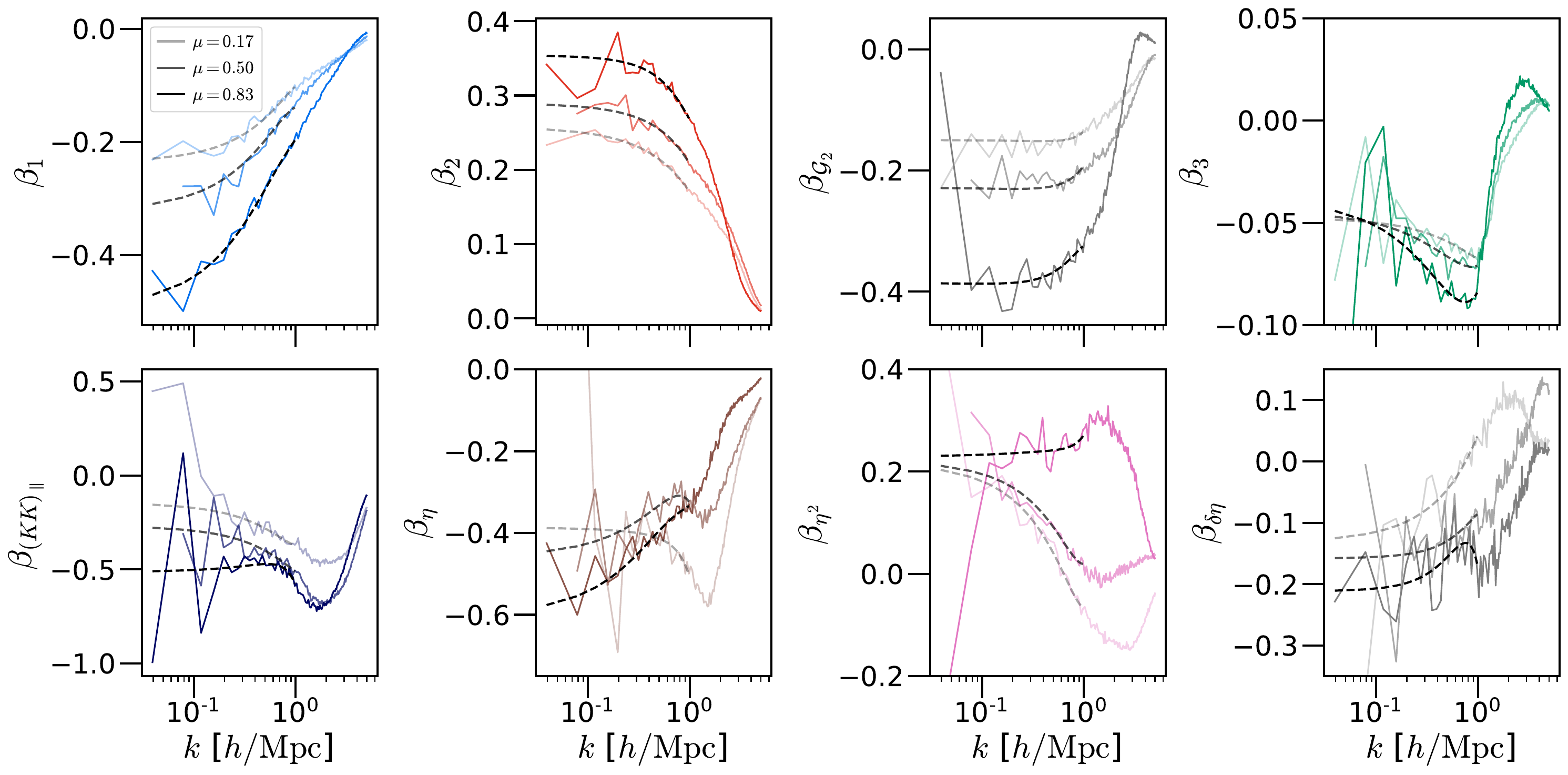}
    \caption{\textbf{Coefficients of Best-Fit Sherwood Transfer Functions:} Best-fit transfer functions $\beta_i(k,\mu)$ for the cubic EFT model obtained from fits to the Sherwood simulations. The corresponding polynomial model for the transfer functions $\beta(k,\mu)$ is given in Eq.~\eqref{eq:beta_tf} which reduces to the Kaiser model in the low-$k$ limit. The large fluctuations in the first two $k$-bins stem from the small number of available modes. Weighting each bin by its number of $k$ modes down weights very noisy bins. The coefficients are tabulated in Tab.~\ref{tab:beta_parameters}. }
    \label{fig:transfer_func}
\end{figure*}

\textbf{Sherwood simulation:} The resulting best-fit coefficients obtained for each transfer function are tabulated in Tab.~\ref{tab:beta_parameters} and compared to the measured transfer functions from Sherwood in Fig.~\ref{fig:transfer_func}. This illustrates that the transfer functions can be approximated by smooth functional forms.  We compare the coefficients of the polynomial fits in Tab.~\ref{tab:beta_parameters} to the bias parameters from fits of the one-loop power spectrum to the Sherwood simulation~\cite{Bolton17, Ivanov:2023yla} in the following. Note that the orthogonalization of the transfer functions mixes bias parameters and absorbs higher-order corrections, \textit{i.e.}~the present EFT model includes all contributions up to two-loop order (bar the $P_{33}$ term stemming from the cubic field $\td^3$). From $\beta_1$ the parameters $c_0$ and $c_{01}$ can directly be compared to $b_1$ and $fb_{\eta}$ as well as $c_0$ from $\beta_2$ to $b_2$. We find very good  agreement for $b_1$ and $b_2$, yet a larger difference for $b_\eta$ and $c_{01}(\beta_1)$. We emphasize, however, that fits of the one-loop power spectrum in the small hydrodynamic simulations can be strongly affected by cosmic variance and since the present fits do not have an associated covariance matrix, care should be taken when comparing the two. Further, the equivalence principle imposes that $c_0(\beta_\eta)$ matches $c_{01}(\beta_1)$ which is an independent confirmation of our theoretical model as we fit for the transfer functions completely independently of each other. 

\begin{table*}
\centering
\begin{tabular}{l|cccccccc}
\hline\hline
TF & $c_0$ & $c_{01}$ & $c_1$ & $c_{12}$ & $c_{14}$ & $c_4$ & $c_{22}$ & $c_{44}$ \\
\hline
$\beta_1$
& $-0.225$ & $-0.384$ & $\phantom{-}0.146$ & $\phantom{-}0.661$
& $-0.074$ & $-0.017$ & $-0.458$ & $\phantom{-}0.130$ \\

$\beta_2$
& $\phantom{-}0.254$ & $\phantom{-}0.145$ & $-0.090$ & $\phantom{-}0.158$
& $-0.091$ & $\phantom{-}0.001$ & $-0.121$ & $\phantom{-}0.041$ \\

$\beta_{G2}$
& $-0.140$ & $-0.354$ & $-0.006$ & $-0.073$
& $\phantom{-}0.062$ & $\phantom{-}0.017$ & $\phantom{-}0.195$ & $-0.133$ \\

$\beta_3$
& $-0.048$ & $\phantom{-}0.013$ & $-0.021$ & $-0.155$
& $-0.014$ & $\phantom{-}0.001$ & $\phantom{-}0.134$ & $-0.007$ \\

$\beta_{KK\parallel}$
& $-0.129$ & $-0.554$ & $-0.274$ & $-0.263$
& $\phantom{-}1.177$ & $\phantom{-}0.053$ & $-0.011$ & $-0.473$ \\

$\beta_{\eta}$
& $-0.378$ & $-0.319$ & $-0.093$ & $\phantom{-}2.083$
& $-1.546$ & $-0.068$ & $-0.599$ & $\phantom{-}0.297$ \\

$\beta_{\eta^2}$
& $\phantom{-}0.217$ & $\phantom{-}0.018$ & $-0.373$ & $\phantom{-}0.362$
& $\phantom{-}0.335$ & $\phantom{-}0.079$ & $-0.062$ & $-0.069$ \\

$\beta_{\delta\eta}$
& $-0.129$ & $-0.120$ & $\phantom{-}0.193$ & $-0.941$
& $\phantom{-}1.008$ & $-0.001$ & $\phantom{-}0.326$ & $-0.440$ \\
\hline\hline
\end{tabular}
\caption{\textbf{Coefficients of Best-Fit Sherwood Transfer Functions:} Best-fit parameters for the transfer function (TF) model, $\beta(k,\mu)$, given in Eq.~\eqref{eq:beta_tf} and illustrated in Fig.~\ref{fig:transfer_func} obtained from the Sherwood simulation. Each wavenumber bin is weighted by $k$ to down weight small scale modes with a cut off at $k_{\rm max}=1\hMpcinv$. The coefficients $c_n, c_{nm}$ are given in units of ~$[\Mpch]^{n}$.
}
\label{tab:beta_parameters}
\end{table*}

\textbf{Abacus simulation:}
Analogously to the analysis on Sherwood simulations, we fit the low-$k$ limit of the transfer functions obtained from the field-level technique applied to the Abacus simulations (see Sec.~\ref{sec:lya_results}). In Tab.~\ref{tab:abacus_tf_results} we tabulate the coefficients of the transfer functions. Here we only tabulate the values for simulation ``one'', Model I and line-of-sight $z$ which can be directly compared to the one-loop power spectrum fits presented in table 4 in Ref.~\cite{Hadzhiyska:2025cvk}. In particular, the coefficients $c_0$ and $a_{01}$ for $\beta_1$ are directly related to $b_1$ and $-fb_\eta$ and $c_0$ of $\beta_2$ to $b_2$. For all three we find reasonable agreement. In contrast to Sherwood for which we find a strong discrepancy between $c_{01}$ of $\beta_1$ indicating that the limited volume and the cosmic variance strongly affect constraints on the bias parameters of the one-loop power spectrum. 

\begin{figure*}
    \centering
    \includegraphics[width=1\linewidth]{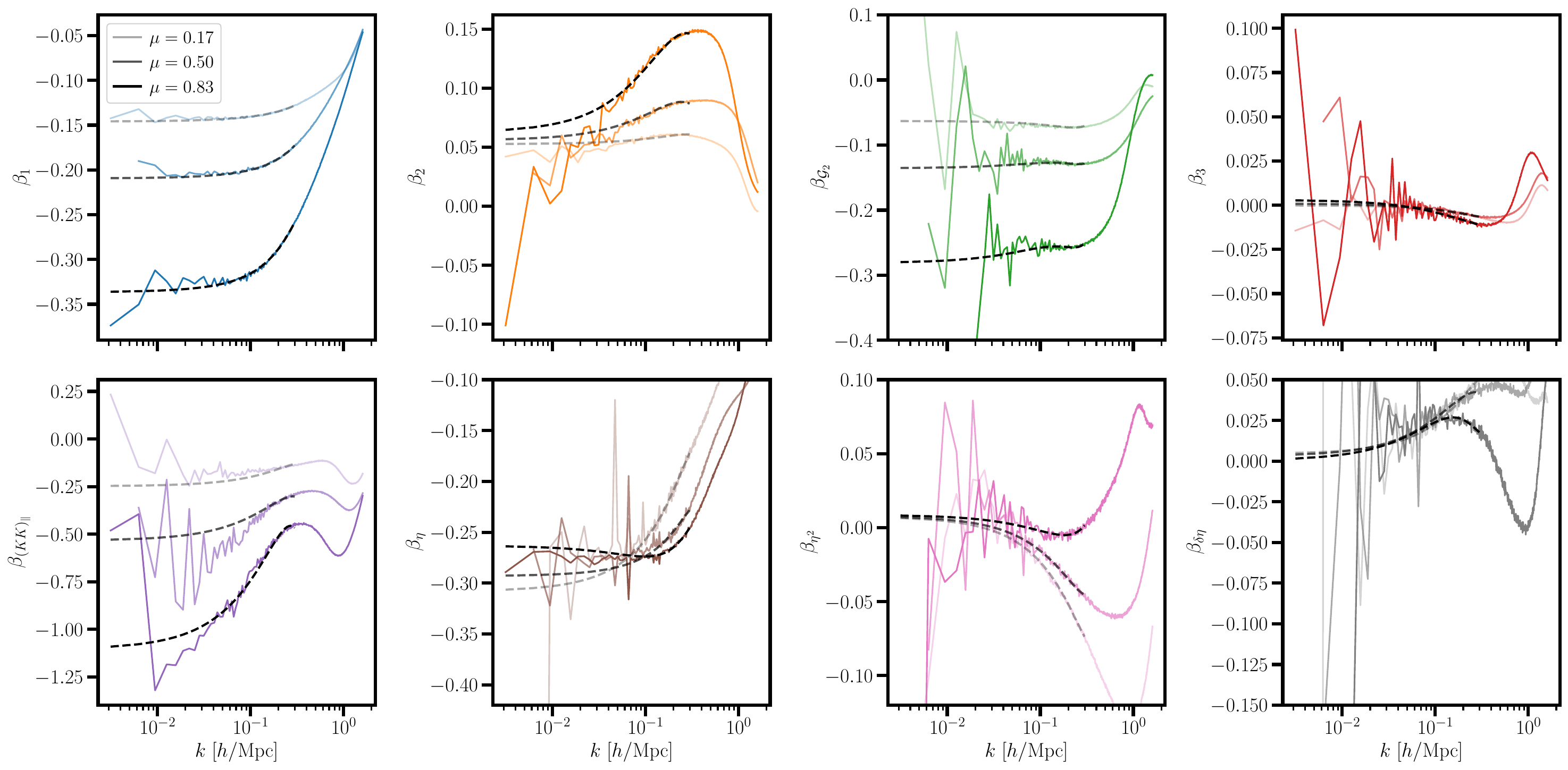}
    \caption{\textbf{Best-Fit Abacus Transfer Functions:} Same as Fig.~\ref{fig:transfer_func} but for Abacus model I. The coefficients are tabulated in Tab.~\ref{tab:abacus_tf_results}.
    }
    \label{fig:abacus_transfer_func}
\end{figure*}

In Fig.~\ref{fig:abacus_transfer_func} we show the fitted transfer functions and the resulting polynomial fits. Given the large box size, the transfer functions are computed out to large scales. Whilst the results agree qualitatively with Sherwood we note that the $\beta_2$ transfer functions exhibit a scale dependence already on large scales. The key takeaway from this section is that (i) we can measure each transfer function from the simulated data; and (ii) these can be approximated through simple polynomials. 

\begin{table*}
\centering
\begin{tabular}{l|cccccccc}
\hline\hline
TF & $c_0$ & $c_{01}$ & $c_1$ & $c_{12}$ & $c_{14}$ & $c_4$ & $c_{22}$ & $c_{44}$ \\
\hline
$\beta_1$
& $-0.152$ & $-0.156$ & $\phantom{-}0.082$ & $\phantom{-}0.283$
& $-0.059$ & $-0.009$ & $-0.176$ & $\phantom{-}0.062$ \\

$\beta_2$
& $\phantom{-}0.108$ & $\phantom{-}0.067$ & $-0.017$ & $\phantom{-}0.086$
& $-0.089$ & $-0.021$ & $-0.066$ & $-0.052$ \\

$\beta_{\mathcal{G}_2}$
& $-0.087$ & $-0.167$ & $\phantom{-}0.021$ & $\phantom{-}0.044$
& $\phantom{-}0.086$ & $\phantom{-}0.020$ & $\phantom{-}0.005$ & $\phantom{-}0.110$ \\

$\beta_3$
& $-0.003$ & $-0.012$ & $-0.020$ & $-0.117$
& $\phantom{-}0.095$ & $\phantom{-}0.017$ & $\phantom{-}0.097$ & $\phantom{-}0.012$ \\

$\beta_{KK_\parallel}$
& $-0.026$ & $-0.154$ & $-0.207$ & $\phantom{-}0.640$
& $-0.755$ & $-0.060$ & $-0.997$ & $\phantom{-}0.922$ \\

$\beta_{\eta}$
& $-0.213$ & $\phantom{-}0.064$ & $\phantom{-}0.124$ & $-0.275$
& $\phantom{-}0.121$ & $-0.011$ & $-0.036$ & $\phantom{-}0.187$ \\

$\beta_{\eta^2}$
& $\phantom{-}0.073$ & $-0.043$ & $-0.111$ & $\phantom{-}0.104$
& $\phantom{-}0.092$ & $\phantom{-}0.018$ & $-0.009$ & $\phantom{-}0.058$ \\

$\beta_{\delta\eta}$
& $-0.070$ & $\phantom{-}0.007$ & $-0.016$ & $-0.318$
& $\phantom{-}0.017$ & $-0.011$ & $\phantom{-}0.581$ & $-0.347$ \\
\hline\hline
\end{tabular}
\caption{\textbf{Coefficients of Best-Fit Abacus Transfer Functions:} Coefficients of the fitted transfer functions (TF) for Abacus simulation ``one'', line-of-sight $z$ and Model I for the polynomial given in Eq.~\eqref{eq:beta_tf} and plotted in Fig.~\ref{fig:abacus_transfer_func}. The coefficients $c_n, c_{nm}$ are given in units of ~$[\Mpch]^{n}$.} \label{tab:abacus_tf_results}
\end{table*}

\subsection{High-redshift galaxy transfer functions}
In this work, we use a set of transfer functions matching the clustering properties of observations at $z=3$ obtained from calibrations on state-of-the-art Astrid hydrodynamical simulations (see, e.g.,~\cite{Ivanov:2024dgv, Sullivan:2025eei}), representing \Lya emitters (LAEs) and Lyman-break galaxies (LBGs). Here, we follow Ref.~\cite{Sullivan:2025eei} to which the reader is referred for a fuller presentation of the employed  high-redshift galaxy samples. These are designed to approximately match the linear bias, $b_1$, and number density for LBGs (``CARS''; \cite{Hildrebrandt_CARS_2009}) and LAEs (``ODIN'' \cite{White:2024lki}) and a  futuristic sample based on projections for Stage-V spectroscopy (denoted by S5; \cite{schlegel_megamapper_concept}), summarized as LBG and LAE S5 in table II of Ref.~\cite{Sullivan:2025eei}. 

\begin{figure*}
    \centering
    \includegraphics[width=1\linewidth]{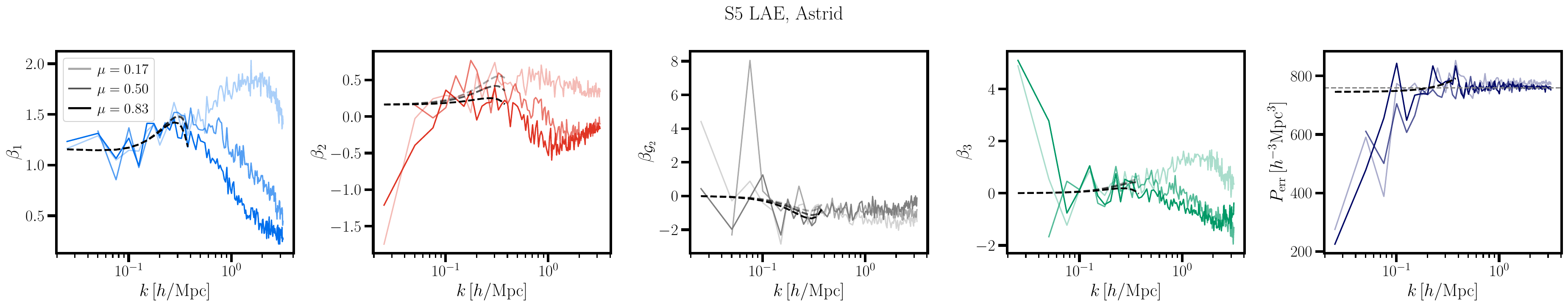}\\
    \includegraphics[width=1\linewidth]{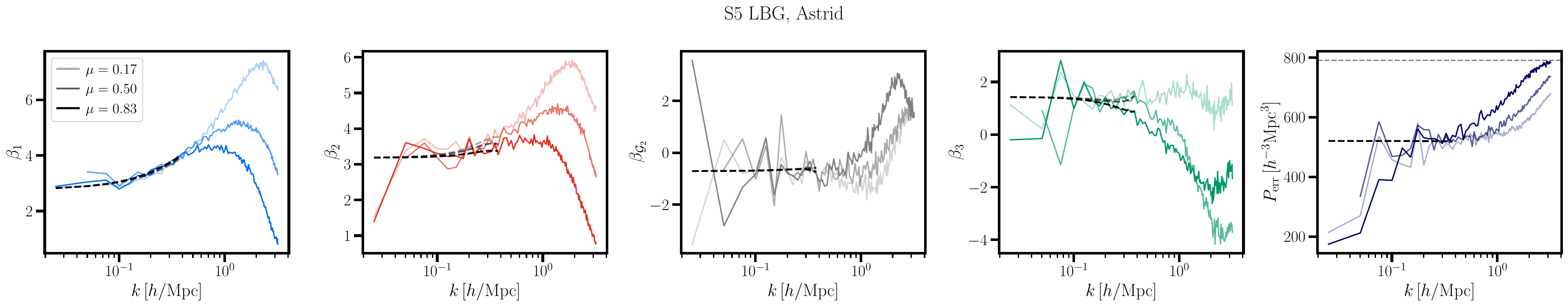}\\
    \includegraphics[width=1\linewidth]{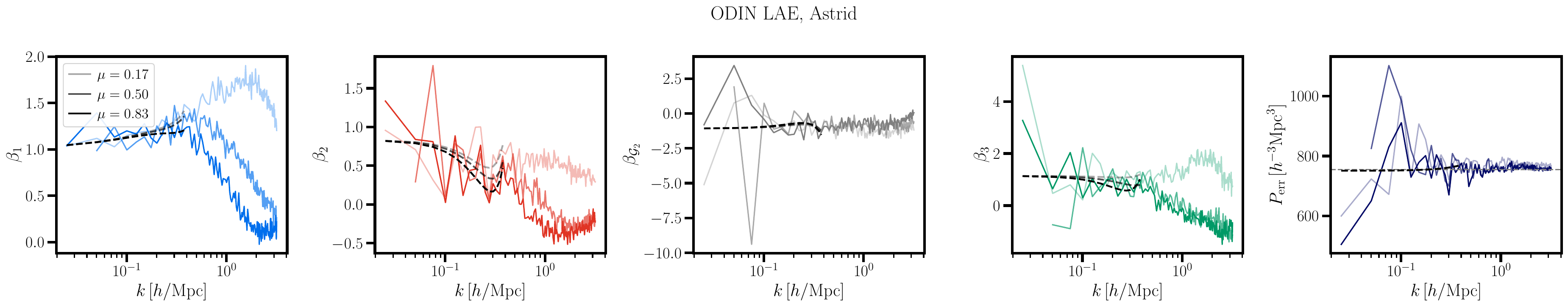}\\
    \includegraphics[width=1\linewidth]{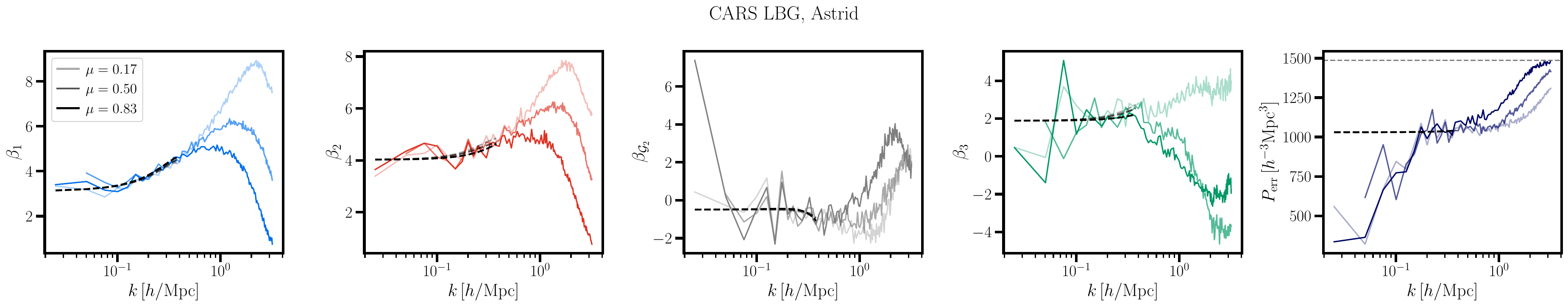}\\
    \caption{Best-fit transfer functions $\beta_i(k,\mu)$ for the cubic EFT model obtained from fits to the Astrid simulation tuned to an ``S5'' and ``ODIN'' sample in the two top rows and both bottom rows show the same for an LBG-type sample calibrated on Astrid simulations. The first four columns show the redshift-space transfer functions for galaxies $\beta_1$, $\beta_2$, $\beta_{\mathcal{G}_2}$, $\beta_3$, and the last column shows the error power spectrum $P_{\rm err}\equiv \langle |\delta^{\mathrm{truth}}_F(\k) - \delta^{\mathrm{model}}_F(\k)|^2 \rangle$ a quantitative measure of the model performance. Note that for the error spectrum we use the constant large-scale theoretical limit as input for the field-level mocks. The intensity of the lines denotes the angular wedges $\mu$ given in three bins. The dashed black curves are fits to the measured transfer functions from Ref.~\cite{Ivanov:2024dgv} using the polynomial model given in Eqs.~\eqref{eq:beta1_tf_LAE}-\eqref{eq:perr_tf_LAE}. 
    The horizontal dashed gray lines in the last column indicates the effective shot noise of the sample ($1/\overline{n}$).
    }
    \label{fig:Astrid_TF}
\end{figure*}

We fit the redshift space transfer functions for both galaxy samples as polynomial expansions in $k$ and $\mu$, using a functional form inspired by the  one used in Ref.~\cite{Obuljen:2022cjo}
\begin{align} 
\beta_1(k,\mu) &= c_0 + c_{1}k + c_2 k^2 + c_4k^4 \label{eq:beta1_tf_LAE} + c_{22}(k\mu)^2 + c_{44}(k\mu)^4\,, \\
\beta_{i\neq 1}(k,\mu) &= c_0 + c_2 k^2 + c_4k^4 \label{eq:beta2_tf_LAE}  + c_{22}(k\mu)^2 + c_{44}(k\mu)^4\,, \\
P_{\rm err}(k,\mu) &= a_0 + a_2 k^2 
% +a_3k^3 + a_4k^4 
\label{eq:perr_tf_LAE} 
% + a_{22}(k\mu)^2 
% +a_{33}(k\mu)^3
% + a_{44}(k\mu)^4
\,.
\end{align}
The fits are performed jointly for all $\mu$-bins using nonlinear least-squares minimization, weighting each data point by $k$ with a cut off at $k=0.4\hMpcinv$.\footnote{We verified that varying the maximum used wavenumber by $\Delta k=0.1\hMpcinv$ does not affect our conclusions.} 
The resulting best-fit transfer functions are shown in Fig.~\ref{fig:Astrid_TF}.

\section{Transfer functions in perturbation theory} \label{sec:theory_transfer_func}

In this section we will 
develop the perturbation theory 
modeling for the field-level transfer functions and use
them to measure EFT 
parameters at the field-level, connecting the fitted transfer functions to a theory prediction. In particular, addressing the question whether the large freedom in $k$ and $\mu$ when minimizing the mean-squared-error is justified. 
The key idea of this method
is that in the absence of orthogonalization
the transfer functions 
should be constant on large-scales --
by construction. These constants then can be matched to the 
EFT parameters. However, 
this constancy
of transfer functions 
in the $k\to 0$
limit 
is violated by the Gram-Schmidt orthogonalization, 
which on large scales 
removes the operators which 
can be represented as 
other operators times 
a $\mu$-dependent transfer function.
Specifically, it 
removes the 
contributions of the $\eta_{\rm new}$, given in Eq.~\eqref{eq:eta_new},
operator and the $\Pi_\parallel^{[2]}$
operators 
at the linear 
and quadratic orders, respectively. As a result 
of the Gram-Schmidt
procedure, we end up with 
$\mu$-dependent, but $k$-independent transfer functions
on large scales. 
We will deal with this 
by forward modeling the 
orthogonalization procedure. 

The first step of our 
fitting procedure is to specify 
the EFT perturbation theory basis appropriate for the transfer function computations. 
It is natural
to use the EFT formulation
used to compute the 
one-loop predictions
with the \texttt{CLASS-PT} code~\cite{Chudaykin:2020aoj,Ivanov:2023yla}.
However, it turns out that it is convenient 
to use an equivalent, 
but slightly different 
EFT model for the purpose 
of the matching, which we discuss
here.

To start off, we introduce 
the field level EFT at the 
linear level, i.e.~we ignore all quadratic operators  
and the orthogonalization procedure. 
Therefore, we take the present EFT model 
and assume all transfer functions
to be constants, yielding
% Our model then reads
\be 
\delta_F = 
\beta_1\tilde{\delta}_1 + 
\beta_\eta \delta_Z = (\beta_1+\beta_\eta+ \beta_\eta f\mu^2 )\delta_1(\k)+\mathcal{O}(\delta_1^2)\,.
\ee 
In
the $k\to 0$ limit, where 
the linear theory prediction
is dominant, our model simply reproduces the Kaiser formula
provided that the transfer
functions $\beta_1$
and $\beta_\eta$ take 
constant values. These values
then can be converted
into the usual \Lya EFT linear bias parameters
from \texttt{CLASS-PT}
via 
\be 
\label{eq:conv_b1}
b_1 = \beta_1 +\beta_\eta\,,\quad b_\eta =-\beta_\eta\,.
\ee 
What we actually measure though
are orthogonalized transfer functions, i.e. the model
we use to fit the 
simulation snapshots 
takes the following 
form in the linear approximation
\be 
\delta^{\rm model}_F= \beta^F_1(k,\mu)\tilde{\delta}_1 + 
\beta_\eta(k,\mu) \delta^\perp_Z +\mathcal{O}(\delta_1^2)\,.
\ee 
Since the Zel'dovich density field
is 100\% correlated with the 
shifted field $\tilde{\delta}_1$
in linear theory, we have
\be 
\delta^\perp_Z = \delta_Z -\frac{\langle \delta_Z  \tilde{\delta}_1\rangle'}{\langle |\tilde{\delta}_1|^2\rangle'}\tilde{\delta}_1
=(1+f\mu^2)\delta_1-(1+f\mu^2)\delta_1 +\mathcal{O}(\delta_1^2)
=0+\mathcal{O}(\delta_1^2)\,,
\ee 
i.e. at the linear order we simply have 
\be 
\delta_F= \beta^F_1(k,\mu)\tilde{\delta}_1 +\mathcal{O}(\delta_1^2)\,,
\ee 
so that both coefficients of our
constant transfer function model 
can be matched from the $\mu$-dependence of $\beta_1$. 
Using the definition
of $\beta_1^F$
and our new EFT model we get 
\be 
\beta_1^F (k,\mu)=\frac{\langle\delta_F
\tilde{\delta}_1 \rangle'}{\langle|\tilde{\delta}_1|^2\rangle'}
% \ee 
% \be 
% \beta_1^F (k,\mu)
\Bigg|_{k\to 0}=\beta_1+\beta_\eta + \beta_\eta f\mu^2\,.
\ee 
Once we have $\beta_1,\beta_\eta$
measurements from fits to the $\beta_1^F (k,\mu)$ data, we can convert
them into the usual EFT 
parameters using 
Eq.~\eqref{eq:conv_b1}.
This matching exercise shows
that it is easier  
to match the transfer functions
to our new EFT model 
which is directly equivalent to the 
forward model up to
the Gram-Schmidt process. 
These coefficents then can be converted into 
the usual bias parameters. 
% that the EFT transfer function
% $\beta_1$ is expected to be
Let us now take a look at this
constant EFT transfer function model
at the quadratic order.
The corresponding \Lya
kernel then reads
\be
\begin{split}
\label{eqn:lya_kernel}
  K_2(\k_1,\k_2) = &
   \beta_1 \tilde K_2 
   +\beta_{\eta} \left(
   F_2^{\rm ZA}-\frac{3}{7} f \mu^2F_{\mathcal{G}_2}\right)
   +\beta_2 
2F_{\delta^2}
+\beta_3 6F_{\delta^3}
\\
&+\beta_{\mathcal{G}_2}
F_{\mathcal{G}_2} +\beta_{\delta \eta} 
F_{\delta \eta}
% ^\perp 
 +\beta_{\eta^2} 
F_{{\eta}^{2}}  
+\beta_{KK_\parallel}
F_{{(KK)}_\parallel}
+\b_{\Pi_\parallel^{(2)}}
F_{\Pi_\parallel^{(2)}}
% ^\perp 
% \,, \nonumber 
% \end{align}
% \be
% \label{eq:K2full_3}
% \begin{split}
\\
\equiv & (\b_1+\b_\eta) F_{\delta}
-\b_1 F_{\delta \eta}
+\b_\eta \left(
-F_\eta +F_{\eta^2}-F_{\delta \eta}
\right)
+ 2\b_2 F_{\delta^2}
+\left(-\frac{2}{7}\b_1 
+\frac{3}{14}\b_\eta 
+\b_{\G}\right)F_{\G} \\
&+\b_{\delta \eta}F_{\delta \eta}+
\b_{\eta^2}  F_{\eta^2}
+\b_{\Pi^{[2]}_\parallel}
\Pi^{[2]}_\parallel
+\b_{(KK)_\parallel}
F_{(KK)_\parallel}\\
= & (\b_1+\b_\eta) F_{\delta}
-\b_\eta F_\eta 
+ 2\b_2 F_{\delta^2}
+\left(-\frac{2}{7}\b_1 
+\frac{3}{14}\b_\eta 
+\b_{\G}\right)F_{\G} \\
&+(-\b_1-\b_\eta+\b_{\delta \eta})F_{\delta \eta}+
(\b_{\eta^2} +\b_\eta) F_{\eta^2}
+\b_{\Pi^{[2]}_\parallel}
F_{\Pi^{[2]}_\parallel}
+\b_{(KK)_\parallel}
F_{(KK)_\parallel}
\end{split}
\ee
Note that this model 
explicitly has the $\Pi_\parallel^{[2]}$
operator. 
The above can be compared with the usual EFT kernel in Eq.~\eqref{eq:K2full}. 
Matching the coefficients we get:
\be 
\label{eq:eft_rel}
\begin{split}
& b_1 = \b_1+\b_\eta \,,
\quad b_\eta = - \b_\eta\,,
\quad b_2 = 2\beta_2\,, 
\quad b_3=6\beta_3 \,,
\quad 
b_{\G}=-\frac{2}{7}\b_1 
+\frac{3}{14}\b_\eta 
+\b_{\G}\,,\\
& b_{\delta \eta}=-\b_1-\b_\eta+\b_{\delta \eta}\,,\quad b_{\eta^2}=\b_{\eta^2}+\b_{\eta}\,,\quad b_{\Pi^{[2]}_\parallel}=\b_{\Pi^{[2]}_\parallel}\,, \quad 
b_{(KK)_\parallel}=
\b_{(KK)_\parallel}~\,.
\end{split}
\ee 
The above matching allows one to determine
linear and quadratic EFT parameters
from the large-scale
limits of the EFT transfer functions.
Importantly, the above matching is consistent
with the linear theory matching. 
Let us also note that while 
the $\delta^3$ operator 
is present in the forward model, 
its correlator with $\delta_1$
is redundant 
at the one-loop order, while its correlations with the quadratic operators start only at the two-loop order which is beyond the scope of our work. Therefore, we will ignore this operator in our
EFT transfer 
function model. 

The correlators between 
relevant operators
$\langle \mathcal{O}_A \mathcal{O}_B\rangle$
can be readily extracted from \texttt{CLASS-PT}~\cite{Chudaykin:2020aoj}, which computes
all possible cross-correlations above
as part of the 
$P_{22}$ computation routine. 
In practice, however, this 
problem is complicated by
four effects. First, 
as discussed before, 
we measure the transfer 
functions of the orthogonal
operators, 
and thus we need to 
account for the 
Gram-Schmidt procedure
in our forward model. 
Second, there are non-linear 
corrections that introduce
scale-dependence of the transfer functions, making it hard to extract their 
constant parts from small
simulation boxes. Third,
large-scales are affected 
by residual cosmic variance,
which is non-negligible 
for quadratic operators. 
And fourth, related to the previous points, we need to know the covariance
between the transfer function data.
In the following, we will address each point.

\begin{figure}
    \centering
\includegraphics[width=1\linewidth]{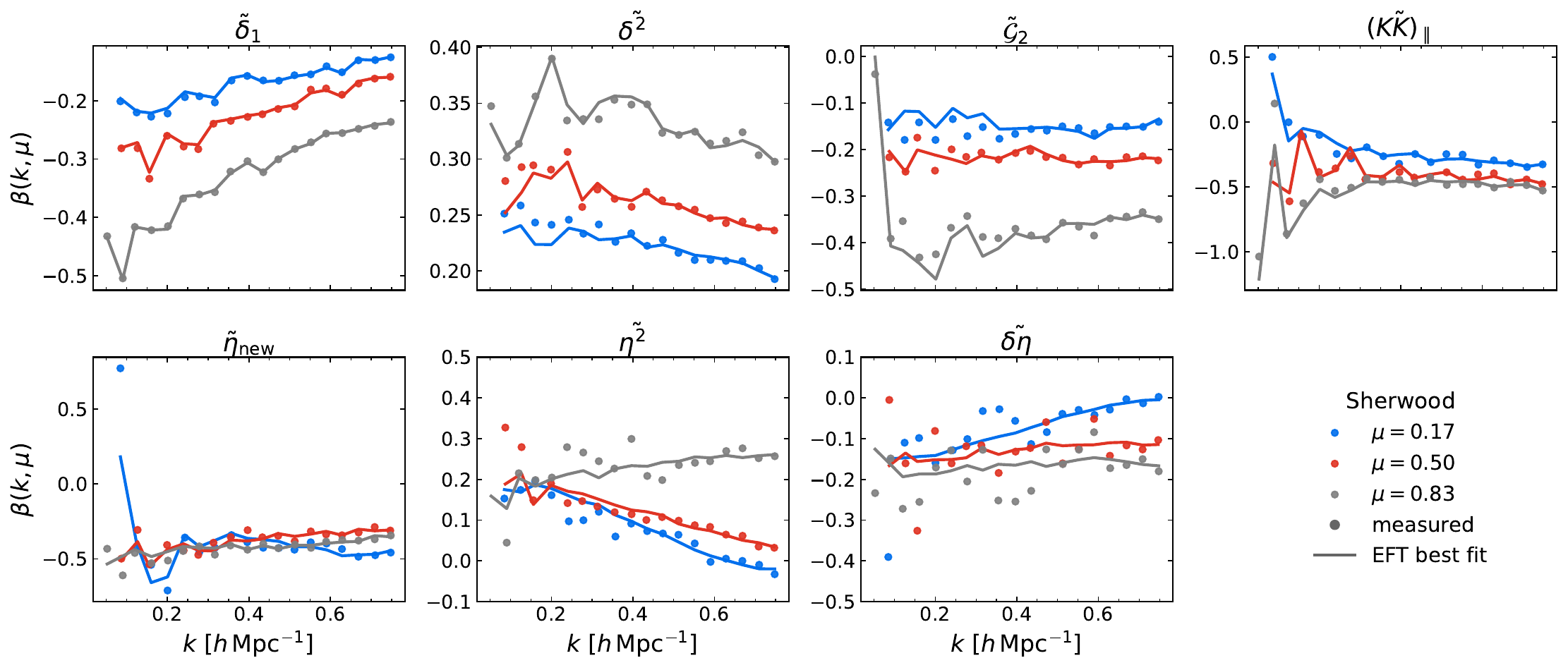}
\includegraphics[width=1\linewidth]{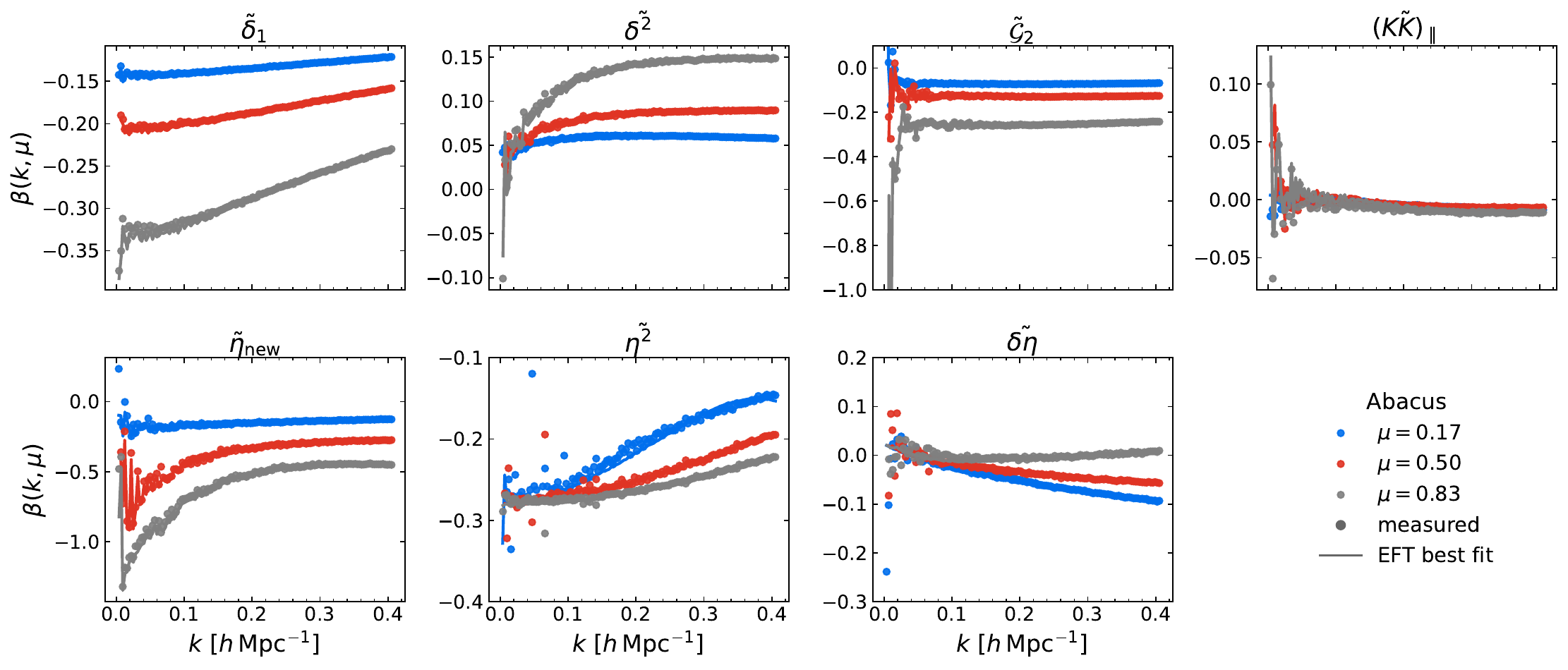}
\vspace{-0.1in}
    \caption{\textbf{Transfer Functions:} Best-fitting EFT models for the seven transfer functions obtained from the Sherwood simulations (\textit{top panel}) and the Abacus FGPA mocks (\textit{bottom panel}). We compare the measured transfer functions (filled circles) to the EFT predictions (solid lines) in three angular bins $\mu$ (red where the angular bin is centered at $\mu=0.83$ is close to the line-of-sight) and Fourier wavenumber $k$. Given the smaller statistical error bars of the Sherwood simulation, two-loop effects set in at a lower $k$ compared to Abacus. Since the small-scale physics in the Sherwood simulations is much more accurate (by construction) than in Abacus, we measure and predict the transfer functions down to $k=0.8\hMpcinv$ for Sherwood and to $k=0.4 \hMpcinv$ for Abacus. Note that we remove $\delta^3$ from the fits to the measured transfer functions, as discussed in the main text.
    }
    \label{fig:bf_tr}
    \vspace{-0.1in}
\end{figure}
% \newpage

First, we forward model the Gram-Schmidt process. Using
the non-orthogonalized bias
expansion $\delta_F=\sum_{A} \beta_A \mathcal{O}_A$,
the measured transfer function
of an operator 
$\mathcal{O}_a$
follows the model
\be 
\beta^F_a(k,\mu) = \frac{\langle \delta_F \mathcal{O}^\perp_a\rangle'}{\langle |\mathcal{O}^\perp_a|^2\rangle'}=
\frac{\langle \delta_F M_{aB}\mathcal{O}_B\rangle'}{\langle |\mathcal{O}^\perp_a|^2\rangle'}=
\frac{1}{\langle |\mathcal{O}^\perp_a|^2\rangle'}M_{aB}\langle \delta_F \mathcal{O}_B\rangle'=\frac{1}{\langle |\mathcal{O}^\perp_a|^2\rangle'}\sum_{A}M_{aB}\beta_A\langle \mathcal{O}_A \mathcal{O}_B\rangle'\,,
\ee 
where indices $A,B$ run over all the eight
operators including $\Pi^{[2]}_\parallel$, in contrast 
to the $a$ index which runs only over seven operators. 
$M_{aB}$ is the $\k$-dependent rotation matrix built from the cross spectra
of the original, non-orthogonal operators. 
If we had access to only one bin, the inversion of the above equation would not be possible. However, using data from multiple bins, 
we can obtain the solution to the above problem as 
a least squares problem. 

Second, to account for 
scale-dependent non-linearities, we 
introduce additional polynomials in our 
transfer function model with free fitting parameters for every $\mu$-bin, 
\be 
\beta^F_a(k,\mu)\Big|_{\rm model} = \frac{1}{\langle |\mathcal{O}^\perp_a|^2\rangle'}\sum_{A}M_{aB}\beta_A\langle \mathcal{O}_A(\k) \mathcal{O}_B(-\k)\rangle'+a_{a,\mu}k^2+b_{a,\mu}k^4\,.
\ee

\begin{figure}
    \centering
\includegraphics[width=0.329\linewidth]{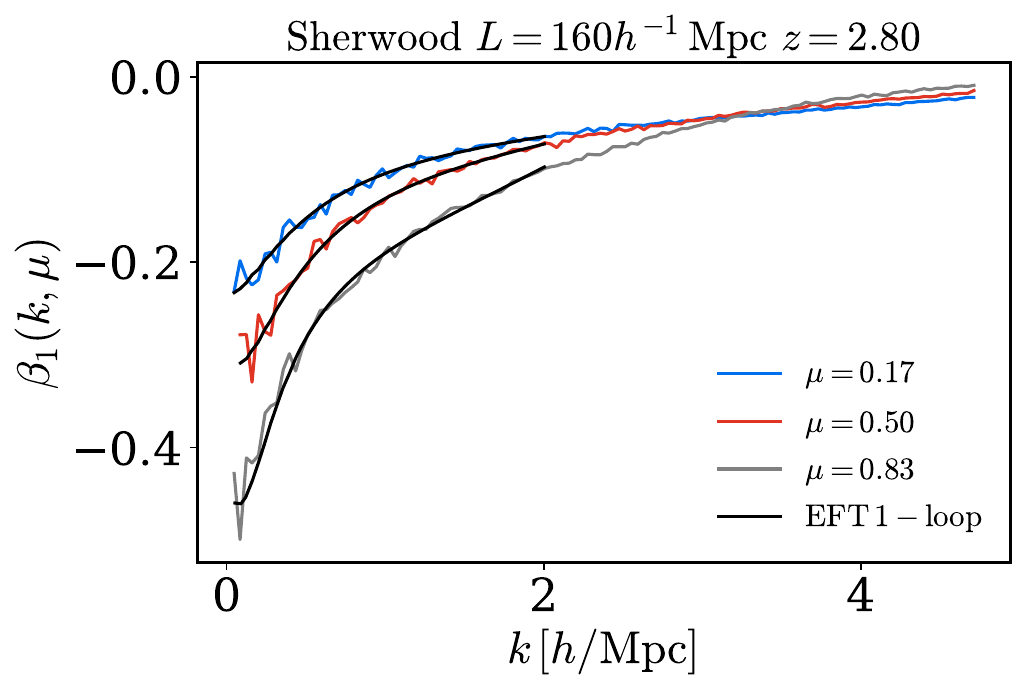}\hfill
\includegraphics[width=0.329\linewidth]{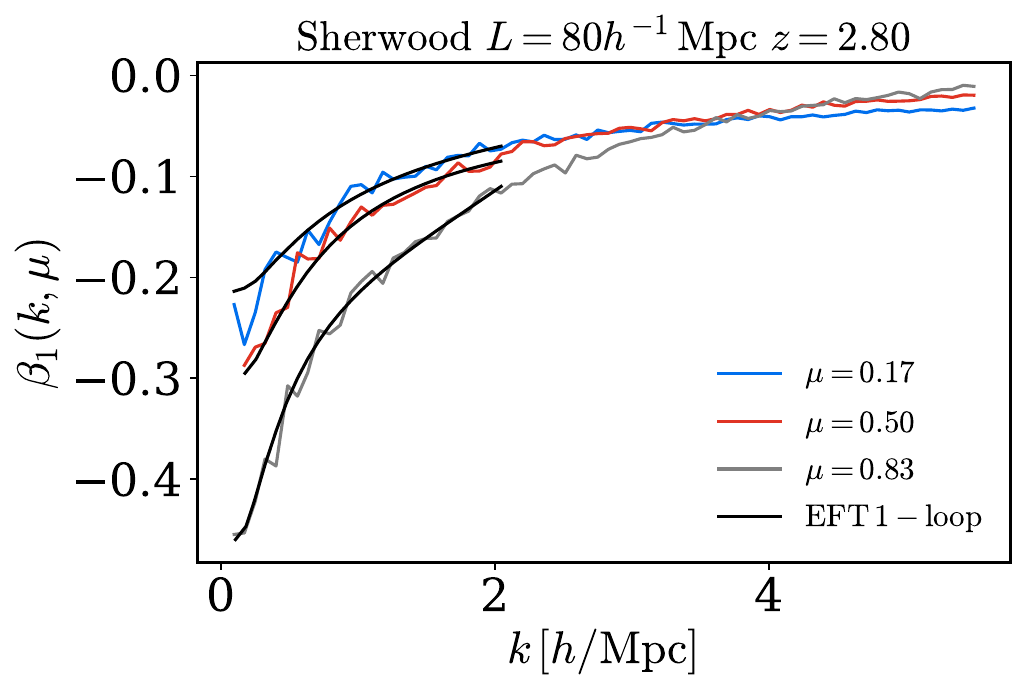}\hfill
\includegraphics[width=0.329\linewidth]{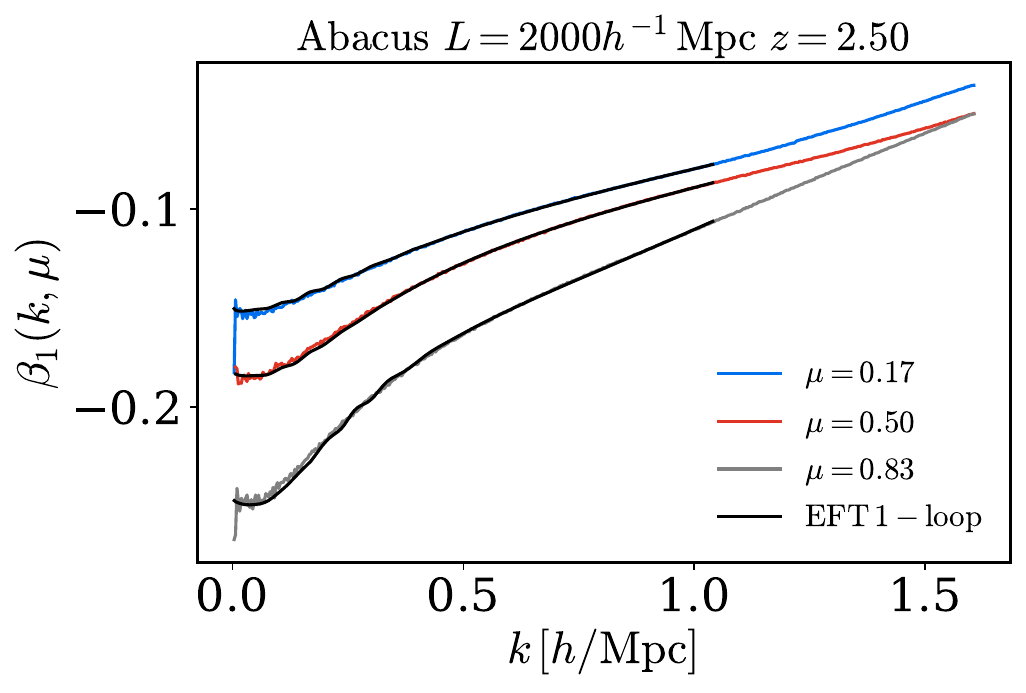}\hfill
\vspace{-0.1in}
    \caption{\textbf{Best-fit EFT TF:} We show the best-fitting EFT models for
    the $\beta_1$ transfer function for the Sherwood (\textit{left and center panel}) and the Abacus FGPA (\textit{right panel}) mock. The transfer functions are shown in three angular bins $\mu$ (blue: $\mu=0.17$, red: $\mu=0.50$, gray: $\mu=0.83$) and as a function of Fourier wavenumber $k$. The EFT 1-loop best-fit transfer functions are shown as solid black lines up to $\kmax = 2.0 \hMpcinv$ ($\kmax=1.0\hMpcinv$) for Sherwood (Abacus) which provide a good fit to the data. See Fig.~\ref{fig:bf_tr_b1_abacus} for the other Abacus models.}\label{fig:bf_tr_b1}
    \vspace{-0.1in}
\end{figure}

Third, to reduce the residual cosmic variance we use 
the correlators
$\langle \mathcal{O}_A \mathcal{O}_B\rangle$
computed directly at the field level 
using the modes
available in the 
simulation box. Fourth,
we assume 
the transfer function
errors $\propto 1/k$, as follows from
mode counting. Since we do not have access to many simulations from which to extract the errors and the covariance, we apply the following procedure to account for the fact that $\beta_1$ must have the smallest errors 
of all transfer functions
because it is 
a linear operator. 
We first fit $\beta_1$ only using the linear theory model corrected by the polynomials, 
\be 
\beta^F_1(k,\mu) = \beta_1+\beta_\eta(1+f\mu^2) + a_{1\mu}k^2 + b_{1\mu}k^4\,,
\ee 
and then use the extracted values  
of $\b_1$
and $\b_\eta$
as priors in the full fit including quadratic operators.
We use $\kmax=0.5 ~(0.2)~\hMpc$
at the first stage (fitting $\beta_1$) 
and $\kmax=0.8 ~(0.4)~\hMpc$
at the second stage for Sherwood (Abacus).
The results are shown
in Fig.~\ref{fig:bf_tr} and we highlight a key aspects: Given the larger number of quasi-linear modes in Abacus, our transfer function fits match better the measured ones. This is particularly noticeable for $\tilde{\eta}_{\rm new}$, $\tilde{\eta}^2$, and $\tilde{\td\eta}$ where we get shape mismatches for Sherwood. The obtained best-fit values 
$\beta_A$
then can be easily converted 
into the usual EFT parameters via Eq.~\eqref{eq:eft_rel}, yielding the tabulated best-fit parameters in Tab.~\ref{tab:sherwood_eft_biases_nod3}. 
% \be 
% \begin{split}
% \text{Sherwood:}\quad & b_1=-0.23\,,\quad 
% b_2 = 0.45\,,\quad  
% b_{\G} = 0.08\,,\quad  
% b_3=-0.06\,,\quad  b_{(KK)_\parallel}=-0.51\,,\quad  \\
% & b_\eta = 0.32\,,\quad 
% b_{\eta^2}=-0.27\,,\quad
% b_{\delta\eta}=0.29\,,\quad
% b_{\Pi^{[2]}_\parallel}=0.28\,.\\
% \text{Abacus:}\quad & b_1=-0.15\,,\quad 
% b_2 = 0.13\,,\quad  
% b_{\G} = -0.016\,,\quad  
% b_3=0.003\,,\quad  b_{(KK)_\parallel}=-0.046\,,\quad  \\
% & b_\eta = 0.15\,,\quad 
% b_{\eta^2}=-0.20\,,\quad
% b_{\delta\eta}=0.07\,,\quad
% b_{\Pi^{[2]}_\parallel}=0.024\,.
% \end{split}
% \ee 
The obtained linear bias parameters are fully consistent with full-shape fits obtained from the one-loop power spectrum to Sherwood \cite{Ivanov:2024lya} and the Abacus simulations \cite{Hadzhiyska:2025cvk}. 
The cubic operators 
and counterterms 
can be extracted from the scale
and orientation 
dependence of the 
$\beta_1$
transfer function.
For that we use the 
full one-loop EFT model
\be 
\beta^F_1(k,\mu) =\frac{P_{F\tilde{1}}}{P_{\tilde{1}\tilde{1}}}=
\frac{K^{\rm lin+ctr}_{1}(\k)P_{11}+P^{F\tilde{1}}_{22}+P^{F\tilde{1}}_{13}}{P_{11}+\tilde{P}_{22}+2\tilde{P}_{13}}~\,,
\ee 
built from the 
non-linear kernels
of the field-level shifted operators computed in~\cite{Ivanov:2024hgq,Ivanov:2024xgb},
\be 
\begin{split}
& K^{\rm lin+ctr}_1
=
b_1 -b_\eta f\mu^2 + k^2(c_0+c_1\mu^2+c_2\mu^4)~\,,\\
& \tilde{P}_{22}=2\int_{\q} [\tilde{K}_2(\k-\q,\q)]^2P_{\rm lin}(q)P_{\rm lin}(|\k-\q|)\,,\quad \tilde{P}_{13} = 3 P_{\rm lin}(k)\int_{\q}
\tilde{K}_3(\k,-\q,\q)P_{\rm lin}(q)\,,\\
& {P}^{F\tilde{1}}_{22}=2\int_{\q} \tilde{K}_2(\k-\q,\q)
K_2(-\k+\q,-\q)
P_{\rm lin}(q)P_{\rm lin}(|\k-\q|)\,,\\
&{P}^{F\tilde{1}}_{13} = 3 K_1(\k) P_{\rm lin}(k)\int_{\q}
\tilde{K}_3(\k,-\q,\q)P_{\rm lin}(q)+3  P_{\rm lin}(k)\int_{\q}
K_3(\k,-\q,\q)P_{\rm lin}(q)~\,,
\end{split}
\ee 
and impose the linear and quadratic bias parameter measurements as priors. 
This procedure is then applied at $\kmax=2~\hMpc$
$(1~\hMpc)$
for the 
Sherwood (Abacus) simulations
yielding the bias parameters tabulated in Tab.~\ref{tab:sherwood_eft_biases_nod3}
% \be 
% \begin{split}
% \text{Sherwood:}\quad & b_{\Gamma_3}  = 0.99 \,,\quad 
% b_{\Pi^{[3]}_\parallel}  = 4.69\,,\quad 
% b_{\delta\Pi^{[2]}_\parallel}  = 7.1\,,\quad
% b_{K\Pi^{[2]}_\parallel}    = -2.58\,,
% \quad
% b_{\eta\Pi^{[2]}_\parallel}    = 17.4\,,\\
% & c_0    = 0.001~[\Mpch]^2\,,\quad 
% c_2    = -0.002~[\Mpch]^2\,,\quad 
% c_4    = 0.030~[\Mpch]^2\,,\\
% \text{Abacus:}\quad & b_{\Gamma_3}  = 0.81 \,,\quad 
% b_{\Pi^{[3]}_\parallel}  = 1.60\,,\quad 
% b_{\delta\Pi^{[2]}_\parallel}  = 3.14\,,\quad
% b_{K\Pi^{[2]}_\parallel}    = -1.23\,,
% \quad
% b_{\eta\Pi^{[2]}_\parallel}    = 8.09\,,\\
% & c_0    = 0.015~[\Mpch]^2\,,\quad 
% c_2    = -0.020~[\Mpch]^2\,,\quad 
% c_4    = 0.083~[\Mpch]^2\,.
% \end{split}
% \ee 
with $b_1$ and $b_\eta$ being similar to the fitting value
from the previous stage. The best-fit 
$\beta_1$
transfer functions
from these fits
are displayed in Fig.~\ref{fig:bf_tr_b1}. 

\begin{table*}
\centering
\begin{tabular}{l|c|cccc|cccc}
\hline
\hline
 & \multicolumn{5}{c|}{\textbf{Sherwood}} 
 & \multicolumn{4}{c}{\textbf{Abacus}} \\
$L$
& $160\hinvMpc$
& \multicolumn{4}{c|}{$80\hinvMpc$}
& \multicolumn{4}{c}{$2\hinvGpc$}
\\
$z$
& $2.8$
& $2.0$
& $2.4$
& $2.8$
& $3.2$
& \multicolumn{4}{c}{$2.5$}
\\
% Particles
% & $2048^3$
% & \multicolumn{4}{c|}{$1024^3$}
% & \multicolumn{4}{c}{$6912^3$}
Models
& --
& --
& --
& --
& --
& I
& II
& III
& IV
\\[0.5ex]
\hline
\hline
$b_1$                          
& $-0.210$
& $-0.104$ & $-0.148$ & $-0.202$ & $-0.270$
& $-0.150$
& $-0.133$ & $-0.136$ & $-0.132$ \\
$b_{\eta}$                   
& $\phantom{-}0.323$
& $\phantom{-}0.199$ & $\phantom{-}0.262$ & $\phantom{-}0.336$ & $\phantom{-}0.410$
& $\phantom{-}0.149$
& $\phantom{-}0.136$ & $\phantom{-}0.287$ & $\phantom{-}0.321$ \\
$b_2$                        
& $\phantom{-}0.392$
& $\phantom{-}0.170$ & $\phantom{-}0.265$ & $\phantom{-}0.415$ & $\phantom{-}0.619$
& $\phantom{-}0.128$
& $\phantom{-}0.127$ & $\phantom{-}0.100$ & $\phantom{-}0.107$ \\
$b_{G_2}$                    
& $-0.045$
& $-0.075$ & $-0.089$ & $-0.087$ & $-0.066$
& $-0.016$
& $-0.012$ & $-0.007$ & $-0.007$ \\
$b_{(KK)_\parallel}$         
& $-0.423$
& $-0.269$ & $-0.415$ & $-0.636$ & $-0.852$
& $-0.046$
& $-0.061$ & $-0.016$ & $-0.030$ \\
$b_{\eta^2}$                 
& $-0.250$
& $-0.055$ & $-0.036$ & $-0.001$ & $-0.061$
& $-0.203$
& $-0.164$ & $-0.268$ & $-0.284$ \\
$b_{\delta\eta}$             
& $\phantom{-}0.079$
& $-0.114$ & $-0.130$ & $-0.104$ & $-0.022$
& $\phantom{-}0.073$
& $\phantom{-}0.064$ & $\phantom{-}0.106$ & $\phantom{-}0.106$ \\
$b_{\Pi^{[2]}_\parallel}$    
& $\phantom{-}0.347$
& $\phantom{-}0.207$ & $\phantom{-}0.288$ & $\phantom{-}0.386$ & $\phantom{-}0.590$
& $\phantom{-}0.024$
& $\phantom{-}0.008$ & $-0.074$ & $-0.066$ \\
\hline
$b_{\Gamma_3}$               
& $-0.154$
& $-0.754$ & $-0.855$ & $-0.627$ & $\phantom{-}0.736$
& $\phantom{-}0.814$
& $\phantom{-}0.599$ & $\phantom{-}1.253$ & $\phantom{-}1.223$ \\
$b_{\Pi^{[3]}_\parallel}$    
& $\phantom{-}1.544$
& $\phantom{-}0.043$ & $\phantom{-}0.109$ & $\phantom{-}0.214$ & $\phantom{-}0.512$
& $\phantom{-}1.603$
& $\phantom{-}1.349$ & $\phantom{-}3.795$ & $\phantom{-}3.908$ \\
$b_{\delta\Pi^{[2]}_\parallel}$  
& $\phantom{-}0.239$
& $-4.141$ & $-6.325$ & $-9.348$ & $-12.373$
& $\phantom{-}3.135$
& $\phantom{-}2.615$ & $\phantom{-}6.035$ & $\phantom{-}6.091$ \\
$b_{K\Pi^{[2]}_\parallel}$   
& $\phantom{-}0.841$
& $\phantom{-}1.279$ & $\phantom{-}1.009$ & $-0.353$ & $-3.930$
& $-1.225$
& $-0.814$ & $-3.901$ & $-3.939$ \\
$b_{\eta\Pi^{[2]}_\parallel}$
& $-2.157$
& $-12.177$ & $-17.132$ & $-22.932$ & $-26.187$
& $\phantom{-}8.093$
& $\phantom{-}6.490$ & $\phantom{-}15.917$ & $\phantom{-}15.954$ \\
\hline
$c_0~[\Mpch]^2$              
& $\phantom{-}0.002$
& $-0.007$ & $-0.013$ & $-0.024$ & $-0.039$
& $\phantom{-}0.015$
& $\phantom{-}0.014$ & $\phantom{-}0.019$ & $\phantom{-}0.018$ \\
$c_2~[\Mpch]^2$              
& $-0.130$
& $-0.011$ & $-0.008$ & $-0.006$ & $-0.005$
& $-0.020$
& $-0.018$ & $\phantom{-}0.022$ & $\phantom{-}0.015$ \\
$c_4~[\Mpch]^2$              
& $\phantom{-}0.219$
& $\phantom{-}0.048$ & $\phantom{-}0.057$ & $\phantom{-}0.078$ & $\phantom{-}0.115$
& $\phantom{-}0.083$
& $\phantom{-}0.075$ & $\phantom{-}0.061$ & $\phantom{-}0.075$ \\
\hline
\hline
\end{tabular}
\caption{\textbf{Best-fit EFT bias parameters from TFs:} Best-fit EFT parameters obtained from the transfer functions, introduced in Sec.~\ref{sec:theory_transfer_func}, via Eq.~\eqref{eq:eft_rel} for both sets of simulations analyzed in the present work: (i) from the $L=160\hinvMpc$  and $L=80\hinvMpc$ Sherwood simulations at four different redshifts using the same effective resolution; and (ii) the additional Abacus models I-IV obtained from a box of length $L=2\hinvGpc$ using a different resolution. The top block quotes the linear and quadratic bias operators, the middle block the cubic ones and the bottom block are the counterterms. 
We emphasize that these values should be interpreted with caution. The quoted bias parameters are best-fit values for which uncertainties cannot yet be reliably estimated, as the covariance of the transfer functions has not been measured (and is expected to be very small because of the Gram-Schmidt orthogonalization). Furthermore, the cubic bias parameters are significantly degenerate because they are inferred from a global fit to $\beta_1$ rather than measured directly at the field level. We leave a more robust determination of these parameters to future work.
}
\label{tab:sherwood_eft_biases_nod3}
\end{table*}

\begin{figure}
    \centering
    \includegraphics[width=0.9\linewidth]{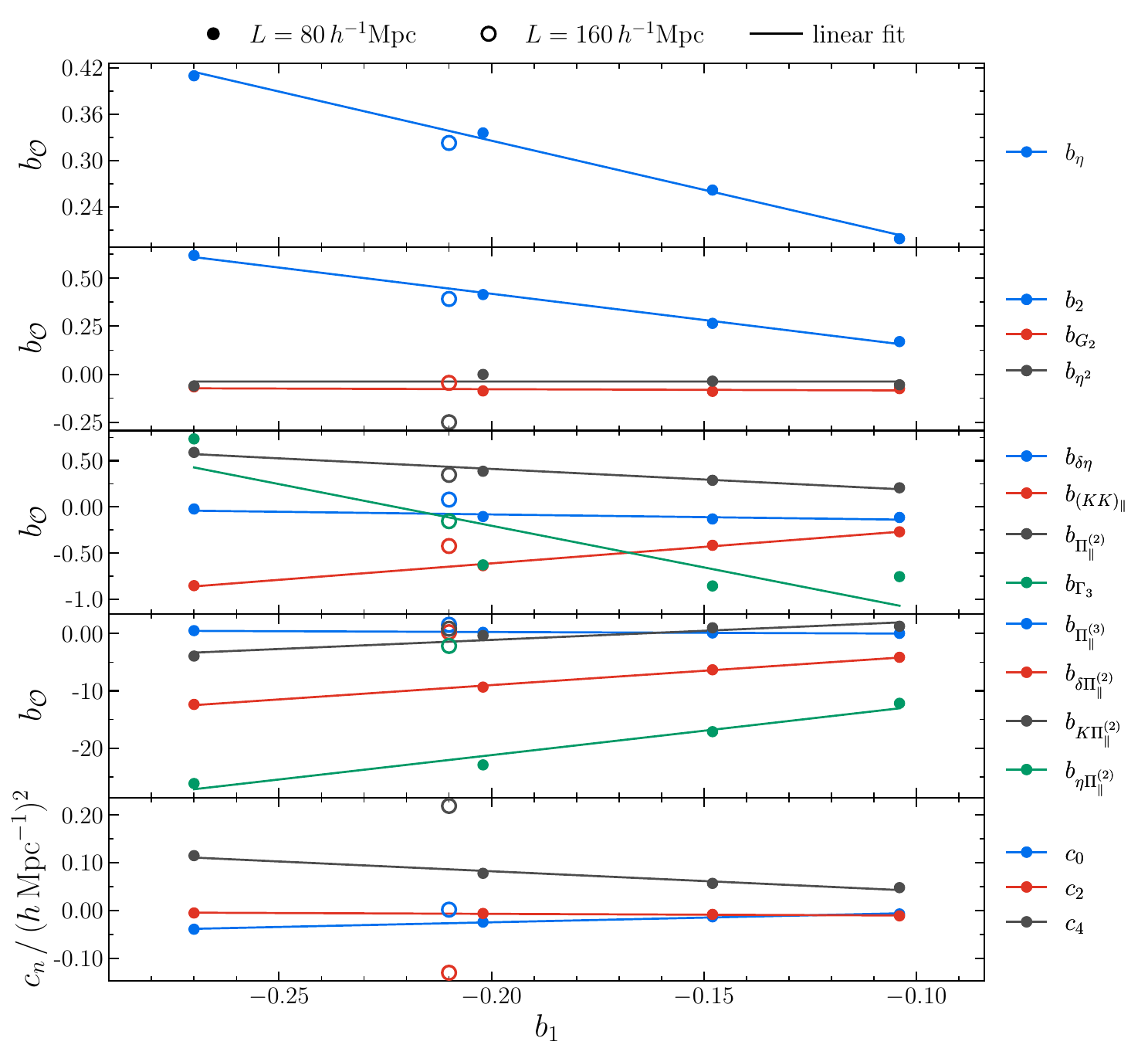}
    \caption{\textbf{Bias parameter relation:} Comparison of best-fit bias parameters and counterterms obtained from field-level fits to the Sherwood $L=160\hinvMpc$ (empty circles) at $z=2.8$ and $L=80\hinvMpc$ boxes at four redshifts (from left to right for $z=3.2,\, 2.8,\, 2.4,\, 2.0$ as filled circles). Note that we do not show error bars as the field-level fits are best-fit values performed for a single realization. The solid lines are linear fits to the evolution of the bias parameters and counter terms of the smaller box. We remind the reader that our convention of $b_\eta$ is related to the literature by a negative sign. The coefficients for the linear fits are tabulated in Tab.~\ref{tab:bias_b1_fits}.}
    \label{fig:eftparam}
\end{figure}

To illustrate the results further and compare both hydrodynamic simulations, we show in Fig.~\ref{fig:eftparam} the relation of 
the non-linear bias parameters $b_{\mathcal{O}}$ as a function
of the linear bias $b_1$ and linear fits to the $b_1-b_{\mathcal{O}}$ and $b_1-c_n$ relations, tabulated in Tab.~\ref{tab:bias_b1_fits}.  It is interesting to note that within the $80\hinvMpc$ Sherwood boxes the results follow the bias relation but for the larger volume we do see a difference in the linear bias parameters even though the resolution is kept constant for both Sherwood boxes. Note that we do not have error bars since we are performing best-fits and are not exploring the parameter space using e.g.~MCMC chains. 
Note that the cubic parameters and counterterms
have a significant
degree of correlation between them. We believe
that the unbroken degeneracies
between these parameters 
are responsible 
for certain 
large values 
of cubic parameters, 
like $b_{\eta \Pi^{[2]}_\parallel}$. A more robust way to measure these parameters will be 
to explicitly add
all necessary 
operators to the forward
model, and fit their Wilson coefficients
from the appropriate 
transfer 
functions. 

We conclude that the EFT successfully
predicts the shape of the transfer functions, from which the bias parameters can be measured. In future our fitting procedure can be improved with
accurate covariances, 
two-loop computations, and 
the inclusion of 
cubic operators at the field level. We leave these for future investigation. 

\begin{table*}
\centering
\begin{tabular}{lcc @{\hskip 2em} | lcc @{\hskip 2em} | lcc}
\hline
\hline
$b_{\mathcal{O}}$ & $p_1$ & $p_0$ & $b_{\mathcal{O}}$ & $p_1$ & $p_0$ & $b_{\mathcal{O}}$ & $p_1$ & $p_0$ \\
\hline
$b_\eta$        &  -1.273 &  \phantom{-}0.071 & $b_{(KK)_\parallel}$           &   \phantom{-}3.562 &  \phantom{-}0.102 & $b_{K\Pi^{[2]}_\parallel}$    &  31.792 &  \phantom{-}5.256 \\
$b_2$           &  -2.729 & -0.127 & $b_{\Pi^{[2]}_\parallel}$      &  -2.287 & -0.046 & $b_{\eta\Pi^{[2]}_\parallel}$ &  85.125 & -4.199 \\
$b_{G_2}$       &  -0.066 & -0.091 & $b_{\Gamma_3}$                 &  -9.012 & -2.006 & $c_0$                         &   \phantom{-}0.196 &  \phantom{-}0.015 \\
$b_{\eta^2}$    &   \phantom{-}0.002 & -0.038 & $b_{\Pi^{[3]}_\parallel}$      &  -2.806 & -0.288 & $c_2$                         &  -0.035 & -0.014 \\
$b_{\delta\eta}$ & -0.580 & -0.198 & $b_{\delta\Pi^{[2]}_\parallel}$ & 50.060 &  1.014 & $c_4$                         &  -0.409 &  \phantom{-}0.000 \\ [2ex]
\hline
\end{tabular}
\caption{\textbf{Linear relation between $b_1$ and $b_{\mathcal{O}}$: }Coefficients of the linear fits of the EFT bias parameters and counterterms to the linear bias $b_1$, $b_{\mathcal{O}}(b_1)=p_1\,b_1+p_0$, obtained from the four Sherwood snapshots at $z=2.0,2.4,2.8,3.2$ ($L=80\,h^{-1}\mathrm{Mpc}$) and shown in Fig.~\ref{fig:eftparam}. Counterterms $c_n$ are in units of $(h^{-1}\mathrm{Mpc})^{2}$.}
\label{tab:bias_b1_fits}
\end{table*}

\section{Large-scale clustering mocks} \label{sec:EFT_mocks}

One of the main bottlenecks in cosmological analyses of the \Lya forest is the generation of large-volume high-resolution mocks. Therefore, approximate prescriptions such as log-normal mocks \cite{Farr20}, or augmented LPT mocks \cite{Sinigaglia:2023lem} as well as the present Abacus FPGA mocks \cite{Hadzhiyska:2023} have been used in the past. We present and validate a new  approach to generate (in principle, arbitrarily) large \Lya mocks calibrated on small-scale hydrodynamic simulations which can be used for pipeline testing and covariance matrix estimation. As a first proof-of-principle we generate large-scale clustering mocks with a box length of $V=2^3(\hinvGpc)^3$ and a cell size of $\approx 1.95 \hinvMpc$ calibrated on Sherwood simulations.\footnote{We find consistent results when changing the resolution by a factor of two.} 

In summary, we perform the following steps to create the three-dimensional \Lya forest flux decrement or dark matter halo density fields. 
\begin{enumerate}
    \item We fit the transfer functions given in Eq.~\eqref{eqn:lya_model} for the \Lya forest and Eq.~\eqref{eqn:g_model} for dark matter halos at the field level on hydrodynamic simulations. These fits benefit from cosmic variance cancellation as we use the same set of initial conditions for the forward model as for the simulations. 
    \item Next, we compute (and fit) the perturbative $\beta_1$ transfer function predicted by the EFT one-loop model obtained from \texttt{ClASS-PT} and fit a polynomial to the remaining $\beta_{\mathcal{O}}$ transfer functions. Similarly, we fit a polynomial to the measured mean-square model error ($P_{\rm err}$). 
    \item To generate large-scale clustering mocks we first generate a new set of initial conditions with a given box size (here: $L=2\hinvGpc$) and grid resolution (here: $1.95\hinvMpc$). These are used to construct the shifted fields $\tilde{\td}$ which are subsequently orthogonalized using the Gram-Schmidt procedure. Note that since the dependence on the specific ICs is captured by the shifted fields, we can apply the transfer functions and their smooth polynomial fits to realizations with different ICs. 
    \item We multiply each orthogonal shifted field $\tilde{\delta}^{\perp}$ by the corresponding best-fit transfer function obtained in step (1) which yields the part of the \Lya forest (or halo) field that is correlated with the ICs.
    \item In addition to the signal, we compute the stochastic part by generating a field with the same box size and resolution with a power spectrum matching the mean-squared model error, given by the polynomial fits from step (2). 
\end{enumerate}
The final mock is the sum of the signal and noise component. We generate these mocks for both the \Lya forest, calibrated on Sherwood simulations, and high-redshift LAE and LBG samples, calibrated on Astrid simulations, respectively. We validate our methodology to construct large-scale clustering mocks by performing full-shape and BAO analyses at fixed cosmology for (i) the \Lya forest alone, and (ii) a joint analysis of the \Lya forest with LAEs/LBGs. 

We stress that the mocks we generate 
contain the deterministic non-Gaussian 
information. 
We use a Gaussian approximation for the stochastic noise only, because it allows for a  straightforward 
implementation of the 
scale- and angle-dependence
of the error
power spectrum.
In principle, one may consider a more complicated model for the distribution
of the noise field, 
which may include a combination
of Poissonian 
and Gaussian components.
However, using a scale-dependent Gaussian noise 
is sufficient 
for the modeling of the two-point statistics,
which the main goal
of our current study.

\subsection{Validation procedure: Full-shape \& BAO fits}\label{sec:validation_BAO_fits_theory}

The key quantity of interest of DESI \Lya and galaxy clustering analyses is to measure the position of the BAO peak \cite{DESI_BAO_2024, DESI_lya_2024, DESI_lya_dr2}. The BAO scaling parameters are obtained by setting constraints on distortions created from a difference between the fiducial and the true cosmology when converting observed to 3D coordinates, the so-called Alcock-Paczy\'nski effect~\citep{Alcock:1979mp}, parameterized by $\boldsymbol{\alpha}\equiv \{\apar, \aperp\}$. The radial and transverse parameters encode the Hubble parameter $H(z)$, the sound horizon at the redshift of decoupling $r_s(z_d)$ and the angular diameter distance $D_A(z)$ at the effective redshift $z$ of the sample by re-mapping true to observed positions $k \rightarrow k'\,,  \mu \rightarrow \mu'$ with
\begin{align}\label{eq:coord-rescaling}
    k' = \frac{k}{\alpha_\perp} \left[ 1 + \mu^2 \left( \frac{1}{F^2} - 1 \right) \right]^{1/2}\,, \qquad
    \mu' = \frac{\mu}{F} \left[ 1 + \mu^2 \left( \frac{1}{F^2} - 1 \right) \right]^{-1/2}\,,
\end{align}
where we define $F\equiv\apar/\aperp$. Now, the BAO scaling factors are defined as \cite{Padmanabhan:2009}
\begin{align} 
\alpha_{\parallel} \equiv \frac{H^{\text{fid}}(z) r_s^{\text{fid}}(z_d)}{H(z) r_s(z_d)}\,,\quad \alpha_{\perp} \equiv \frac{D_A(z) r_s^{\text{fid}}(z_d)}{D_A^{\text{fid}}(z) r_s(z_d)}\,,
\end{align}
with $r_s^{\text{fid}}(z_d)$ being the fiducial value of the sound horizon scale at the drag epoch. We ``observe'' unprimed quantities which are evaluated at the redshift of the mock. The superscripts \textit{fid} and \textit{tem} refer to quantities in the fiducial and template (here: Sherwood) cosmology. The final expression for the fits is a function of the rescaled coordinates $k'$ and $\mu'$ and an additional normalization (volume) factor $P(k,\mu)= P_m(k',\mu')/(\aperp^2\apar)$.

For the model of the power spectrum $P(k,\mu)$ we employ the one-loop EFT model of the \Lya forest which consists of four  key components
\begin{equation} \label{eq:Pmodel}
    P^{\rm th.}(k,\mu) = P^{\rm tree}(k,\mu) + P^{\rm 1-loop}(k,\mu) + P^{\rm ct}(k,\mu) + P^{\rm st}(k,\mu) \,,
\end{equation}
where $k$ is the Fourier wavenumber and $\mu$ the angle of $k=\{\kpar,\kvperp\}$ to the line-of-sight, $\mu \equiv \kpar/k$. The infrared resummed linear theory power spectrum\footnote{For the fits we use time-sliced perturbation theory to decompose the power spectrum into a smooth (nw) and oscillatory (w) component~\citep{Blas:2016sfa,Chudaykin:2020aoj}. Both components are treated independently throughout the modeling procedure and are combined only at the final stage.} is given by 
\beq
P^{\rm tree}(k,\mu) = K^2_1(\kvec) P_{\rm lin}(k)\,, \qquad K_1(\kvec)\equiv (b_1-b_{\eta}f\mu^2)
\eeq
where $f$ is the (linear) growth rate. The first higher-order correction to the tree-level power spectrum is 
\begin{align}
P^{\rm 1-loop}& (k,\mu) 
=2\int_{\qvec} K_2^2(\qvec,\kvec-\qvec)
P_{\text{lin}}(|\kvec-\qvec|)P_{\text{lin}}(q)  + 6 K_1(\kvec)P_{\rm lin}(k)\int_{\qvec} K_3(\kvec,-\qvec,\qvec)P_{\text{lin}}(q)\,.
\end{align} with higher order redshift-space kernels, $K_{2,3}$ \citep[see Eq.~(3.19) in][]{Ivanov:2024lya} and we use the notation $\int_{\qvec}\equiv\int \frac{d^3q}{(2\pi)^3}$ to denote the three-dimensional integral over $\qvec$. The counter terms 
\begin{align}
\label{eq:shoch}
P^{\rm ct}(k,\mu) = -2(c_0+c_2\mu^2+c_4\mu^4)K_1(\kvec)k^2 P_{\rm lin}(k)\,,
\end{align} scale as $k^2P_{\rm lin}(k)$ and the stochastic contributions 
\begin{align}
P^{\rm st}(k,\mu) = P_{\text{shot}}+a_0\frac{k^2}{\knl^2}+a_2\frac{k^2\mu^2}{\knl^2}\,,
\end{align}
scale as a constant shot noise piece and a scale-and angle-dependent term  capturing small-scale clustering. We refer the reader to Refs.~\cite{deBelsunce:2024rvv,Chudaykin:2025gsh, Hadzhiyska:2025cvk} for a fuller presentation of the methodology. 

To jointly fit the \Lya forest and high-redshift galaxy samples (here: LBGs and LAEs) we use a one-loop EFT model, following the procedure described in Refs.~\citep{Chudaykin:2025gsh, Hadzhiyska:2025cvk}. In linear theory, this corresponds to computing the geometric mean of both tree-level tracers, $P^{\rm tree}_{\times}(k, \mu) = (b_1 - b_{\eta}f \mu^2) (b_1^q + f \mu^2) P_{\rm lin}(k)$, where the subscript $\times$ denotes the cross-correlation, and superscript $q$ represents the quasar (or halo) tracer, i.e.~$b_1^q$ is the linear bias  parameter for quasars. In practice, we remove the line-of-sight dependent terms by remapping the bias parameters (see, e.g.,~\cite{Desjacques:2016bnm, Ivanov:2024lya, Belsunce_Sullivan_skewspectrum}) which yields for the cross-correlation the following form
\begin{align}\label{P1loopX}
P_{\times}^{\rm th.}(k,\mu) = &K_1(\textbf{k})K_1^{\rm q}(\textbf{k})P_{\rm lin}(k)  +2\int_\qvec K_2(\qvec,\textbf{k}-\qvec)K_2^{\rm q}(\qvec,\textbf{k}-\qvec)
P_{\text{lin}}(|\textbf{k}-\qvec|)P_{\text{lin}}(q)  \\
&+ 3 P_{\rm lin}(k)\int_\qvec [K_1(\textbf{k}) K_3^{\rm q}(\textbf{k},-\qvec,\qvec)
+ K_1^{\rm q}(\textbf{k}) K_3(\textbf{k},-\qvec,\qvec)]P_{\text{lin}}(q)
-(c_0+c_2\mu^2+c_4\mu^4)K_1^{\rm q}(\textbf{k})k^2P_{\rm lin}(k)\nonumber\\
&-(c^{\rm q}_0+c^{\rm q}_2\mu^2+c_4^q\mu^4)K_1(\textbf{k})k^2P_{\rm lin}(k)
 -c_x^q(f\mu k)^4(K^q_1(\kvec))^2 P_{\rm lin}(k)\,,\nonumber
\end{align}
where $K_1^q(\kvec)\equiv (b_1^q+f\mu^2)$ and $K_{2,3}^q$ are the standard redshift space kernels for galaxies (see, e.g., \cite{Bernardeau:2001CPT}), and $c_x$ is the next-to-leading order $k^4$ redshift-space counter term \cite{Chudaykin:2025gsh}. 

We fit for  the bias and BAO scaling parameters by using a Gaussian likelihood to fit the one-loop \Lya forest power spectrum, denoted by $P^{\rm model}$, to the measured 2D power spectra, $P_i^{\rm data}$. Therefore, we sample a $\chi^2$ function
\begin{equation} \label{eq:chi2}
    \chi^2 = \sum_i \frac{\left[P_i^{\rm data}-P^{\rm model}(k_i,\mu_i)\right]^2}{2 \left(P_i^{\rm data}\right)^2/N_i}\,,
\end{equation}
where $N_i$ are the Fourier modes per bin, $k_i$ the Fourier wavenumbers with the cosine of the angle to the line-of-sight, $\mu=\kpar/k$. The fits are done using five $\mu$ bins and a wavenumber spacing of $\Delta k = 0.003\hMpcinv$. 

\subsection{\Lya forest mocks at fixed redshift}
\begin{figure}
    \centering
    \includegraphics[width=\linewidth]{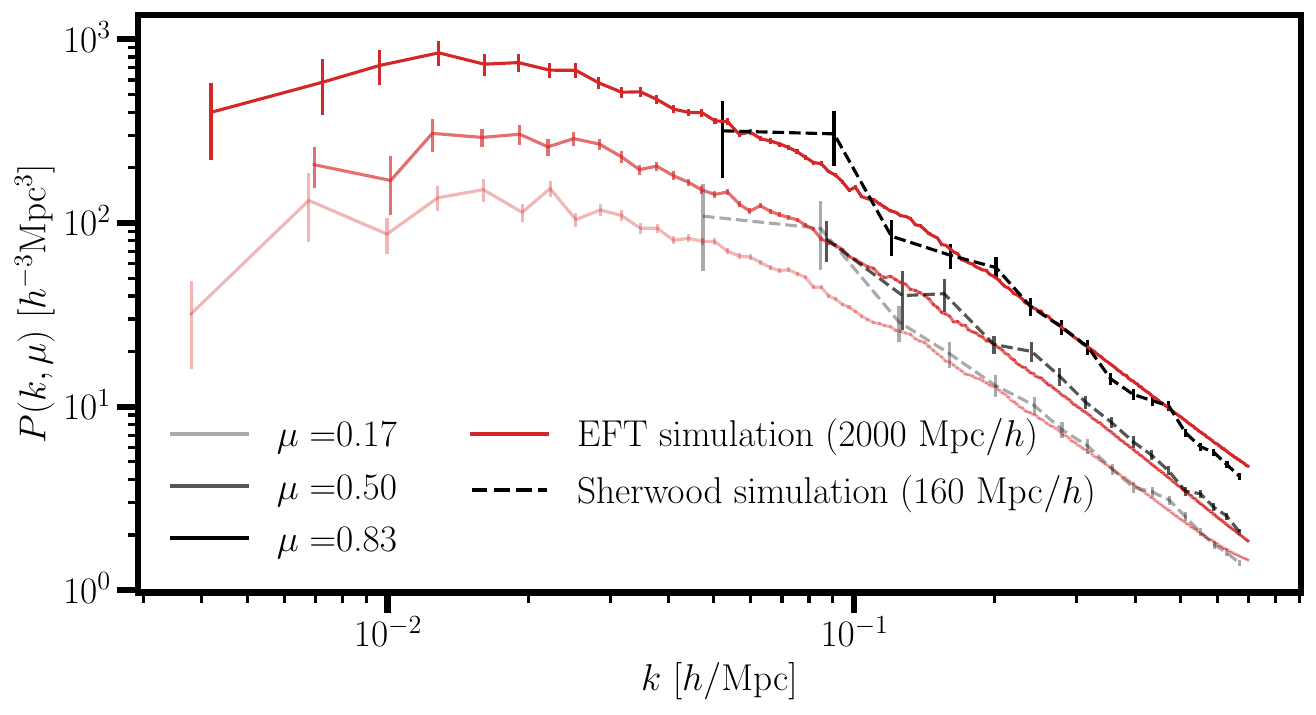}
    \caption{\textbf{Large-scale Clustering \Lya Mock:} Two-dimensional Power spectrum $P(k,\mu)$ in three angular bins of the measured power spectrum from the Sherwood hydrodynamic simulation (dashed black lines) compared to the measured power spectrum of the new \Lya mock using the Sherwood transfer functions applied to the Abacus initial conditions. This demonstrates that the field-level methodology is able to generate large-scale clustering mocks with arbitrary box sizes (here: $L=2 \hinvGpc$) calibrated on small hydrodynamic simulations (Sherwood: $L=160 \hinvMpc$). The EFT simulation is generated at a resolution with $N_{\rm cell}=1024$. }
    \label{fig:DESI_lya_mock}
\end{figure}
First, we take the best-fit transfer functions from the Sherwood simulations obtained from $L=160\hinvMpc$ and $L=80\hinvMpc$ boxes. To generate a large \Lya forest mock, we use the readily available ICs from the Abacus simulations in a box of size $L=2\hinvGpc$ with $N_{\rm grid}=1024$, which is comparable to, yet slightly higher resolution than, current state-of-the-art simulations for cosmological analyses of the \Lya forest in DESI ($\sim 2.5\hinvMpc$) \cite{Farr20}.\footnote{Note that the DESI \Lya forest mocks require a volume of $V=10^3(\hinvGpc)^3$. We leave this engineering challenge to future work.} The resulting mock is at redshift $z=2.8$ and shown in Fig.~\ref{fig:DESI_lya_mock}. This shows a qualitative agreement between the power spectrum of the field-level simulation and the input high-resolution Sherwood simulation. For illustration purposes, we show for this case three angular wedges for which we find good qualitative agreement down to scales of $k\approx 0.7 \hMpcinv$ between the power spectrum measured on the small-scale high-resolution simulation compared to the one measured from the field-level simulation.

\subsubsection{Validation of \Lya forest mocks at fixed redshift of $z=2.8$}\label{sec:EFT_mock_fix_z}
\begin{figure}
    \centering
    \includegraphics[width=0.49\linewidth]{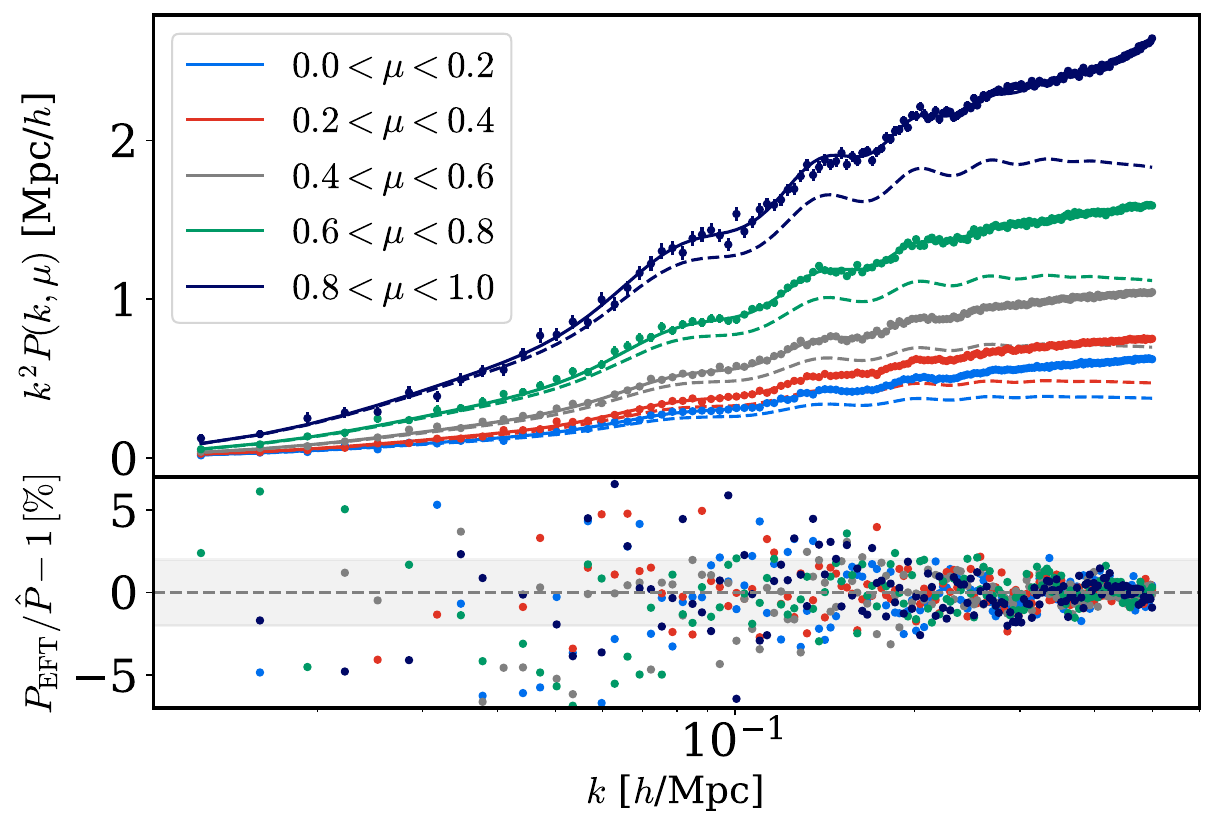}\hfill
    \includegraphics[width=0.49\linewidth]{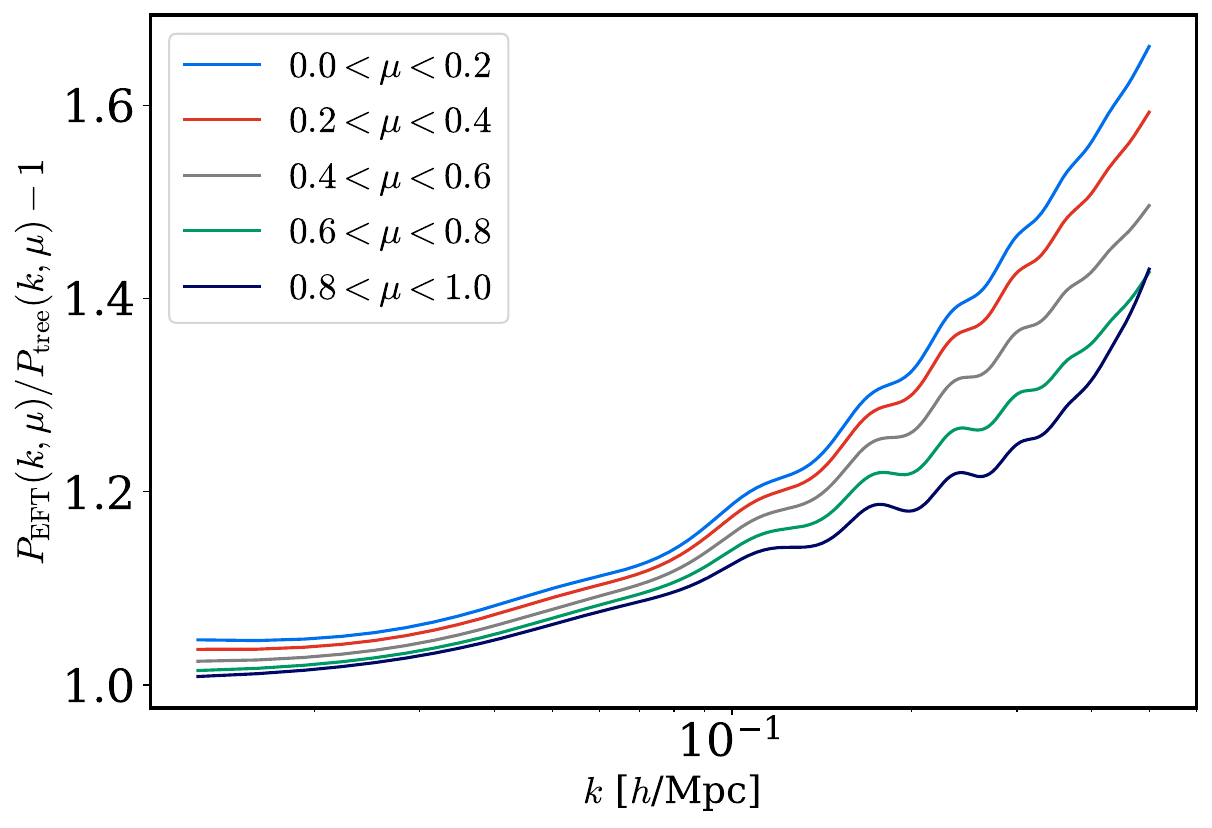}\hfill \\
    \includegraphics[width=0.49\linewidth]{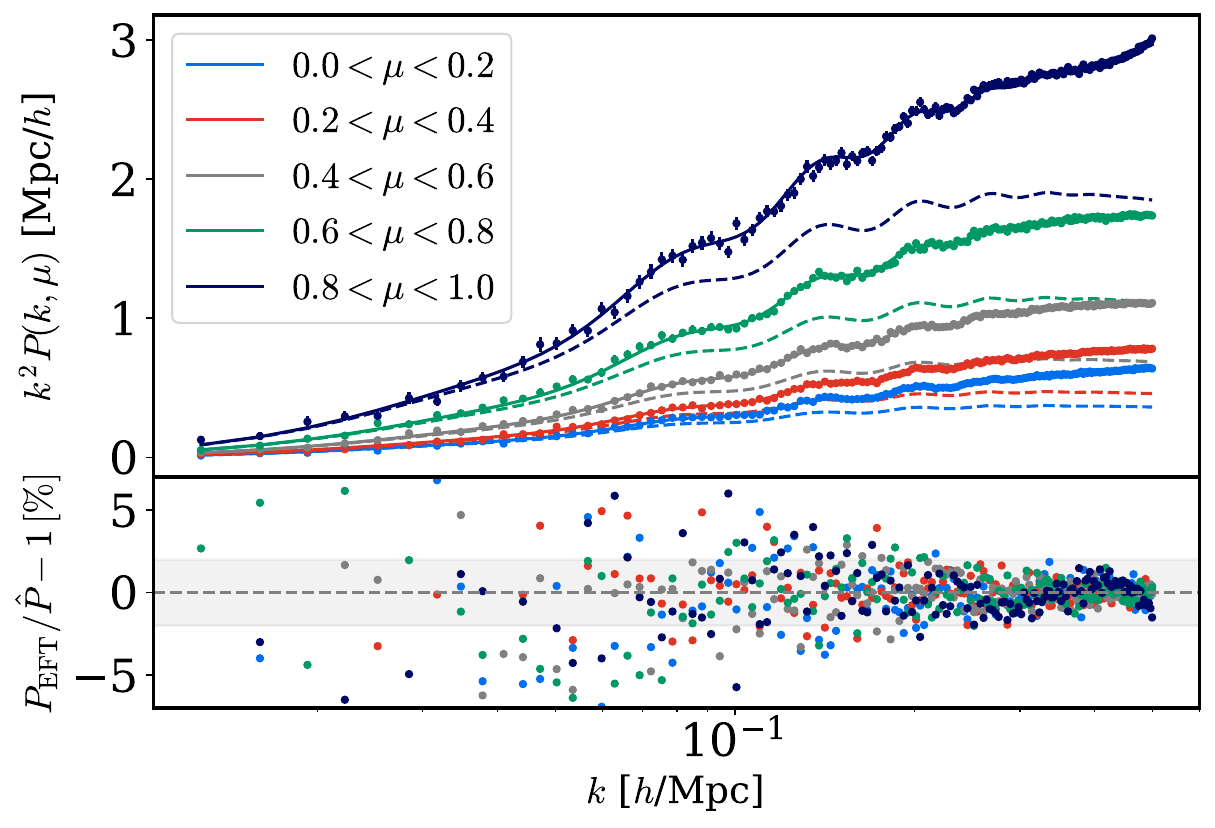}\hfill
    \includegraphics[width=0.49\linewidth]{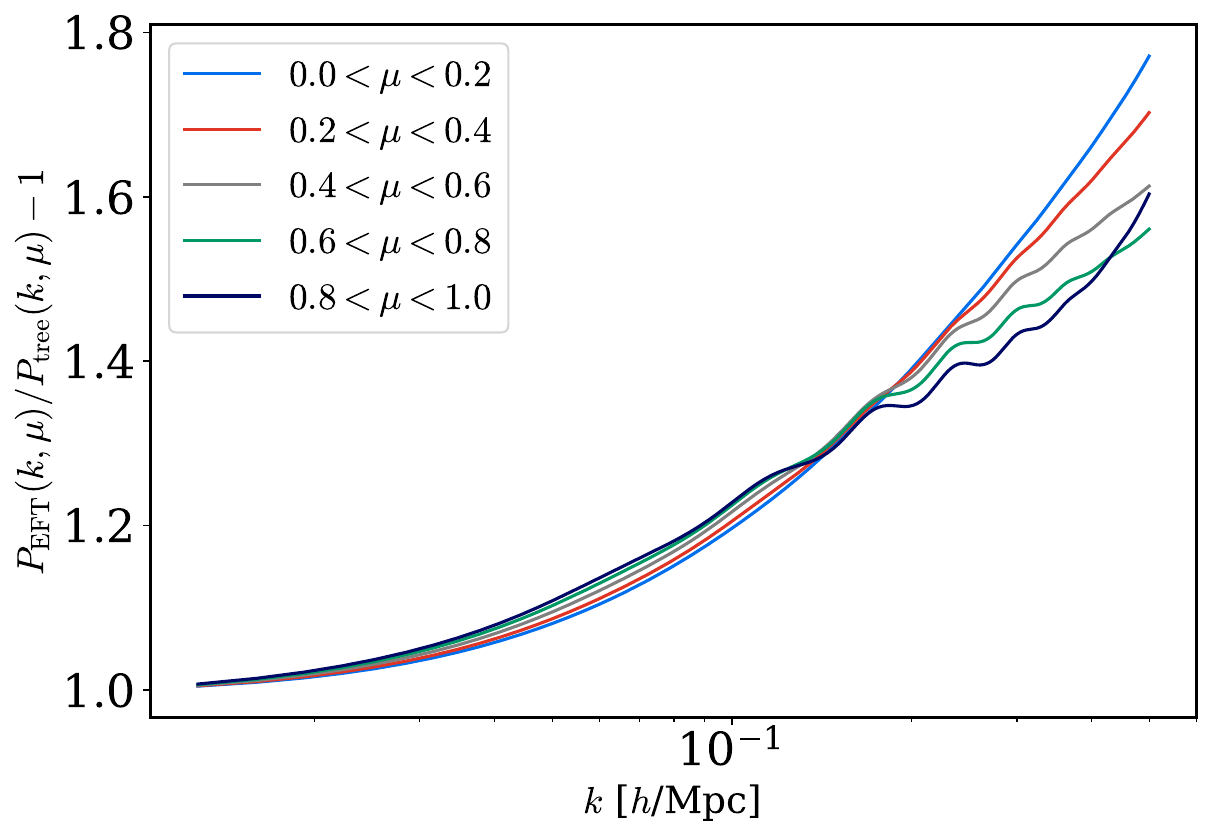}\hfill \\ 
    \caption{\textbf{Validation of \Lya Mocks:} EFT fits to the \Lya auto-correlation (\textit{left panel}) and corresponding loop-correction (\textit{right panel}) shown in five angular bins, $\mu$, with $\kmax =0.5\hMpcinv$ for a single realization. The top (bottom) panel uses the $L=160\,(80)\, \hinvMpc$ box as input. The best-fit EFT model (solid line) is compared to $P^{\rm tree}$ (dashed line) and the data points show error bars which are obtained assuming a diagonal Gaussian covariance based on the number of expected Fourier modes per bin $P(k,\mu)\sqrt{2/N(k,\mu)}$. The bottom panel shows the ratio between model and data of which one is subtracted in addition to a gray band indicating the 2\% error band to guide the eye.}
    \label{fig:EFT_mock_bestfit}
\end{figure}

We validate the field-level mock, shown in Fig.~\ref{fig:DESI_lya_mock}, by performing BAO fits to the measured power spectra, as detailed in Sec.~\ref{sec:validation_BAO_fits_theory}. In Fig.~\ref{fig:EFT_mock_bestfit} we show the resulting best-fit power spectra using $\kmax=0.5\hMpcinv$ (left panel) and the corresponding loop correction plot (right panel) for the large-scale clustering mock obtained from the transfer functions fitted to the $L=160\hinvMpc$ in the top row and the $L=80\hinvMpc$ in the bottom row. The model yields an excellent fit to the simulation at all scales (solid line) and linear bias (dashed lines) shows strong deviations on quasi-linear scales. The tree-level model deviation in the loop correction plot approximately matches our previous findings, i.e.~beyond $k\approx 0.08\hMpcinv$ the corrections surpass the 5\% threshold.

\begin{figure}
    \centering
    \includegraphics[width=1\linewidth]{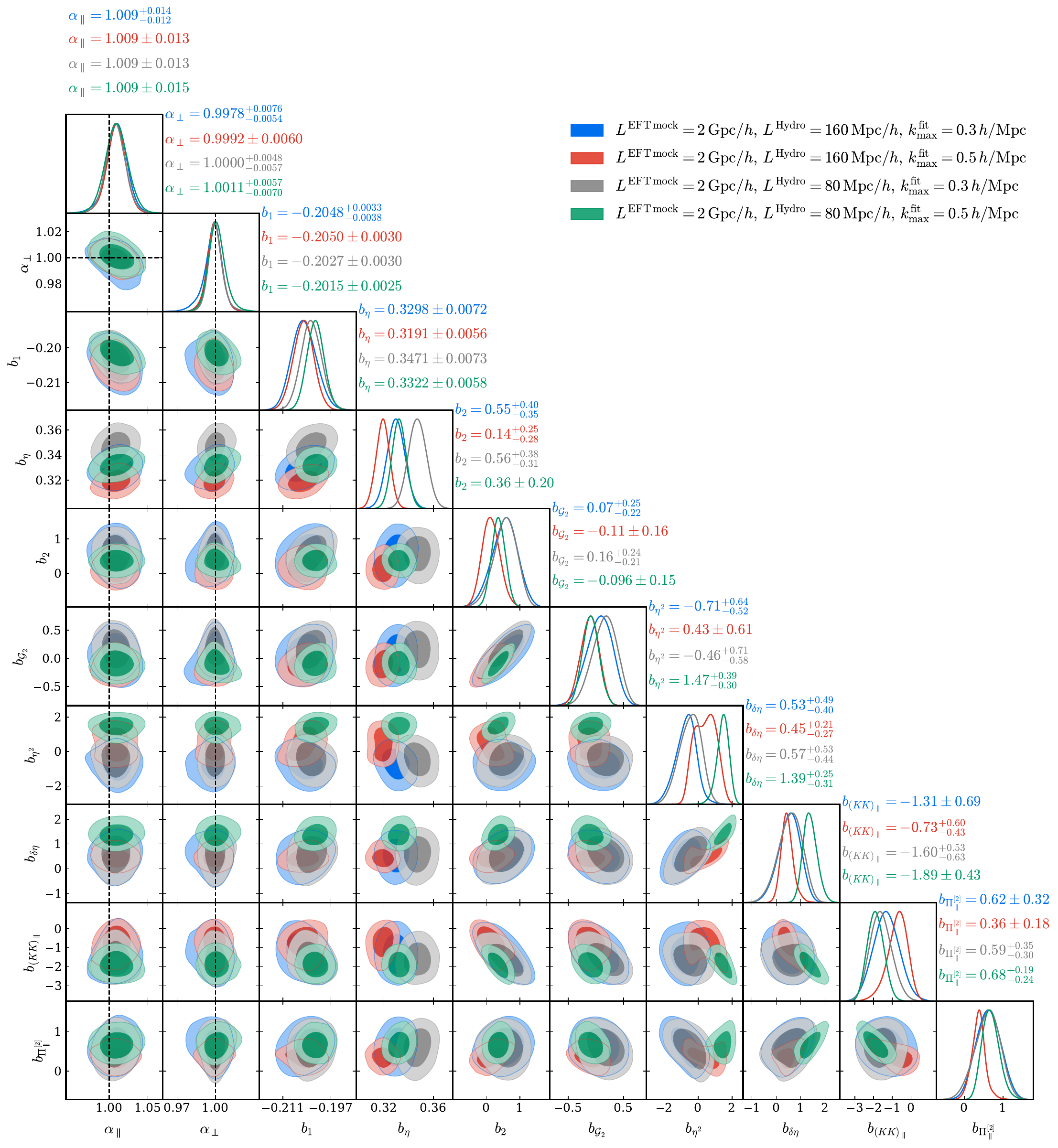}
    \caption{\textbf{Drift plot of \Lya Mock at $z=2.8$:} Best-fit 1D and 2D marginalized posteriors for the two Sherwood simulations with different box sizes and for two different maximum wavenumbers $\kmax=0.3,\,0.5 \hMpcinv$ used in the fits centered at $z=2.8$. We find consistent results within each set of simulations. Additionally, we recover unbiased BAO scaling parameters that are consistent with the fiducial values at the $1\sigma$-level -- the key parameter of interest for DESI analyses. Note that the box sizes denote the Sherwood simulations that have been used to perform the field-level fits of the transfer functions and that the BAO fits have been performed on power spectra measured from boxes of length $L=2 \hinvGpc$.
    }
    \label{fig:EFT_mock_drift_plot}
\end{figure}

In Fig.~\ref{fig:EFT_mock_drift_plot}, we present the corresponding drift plots to Fig.~\ref{fig:EFT_mock_bestfit} obtained by fitting each box using two values of $\kmax = 0.3,\,0.5\hMpcinv$ and obtain excellent agreement between the fits for the same realization. As expected, the one- and two-dimensional marginalized posteriors systematically shrink with increasing $\kmax$. The linear bias parameters $b_1$ and $b_\eta$ values are fully consistent with our field-level fits discussed further in Sec.~\ref{sec:tranfer_func} and tabulated in Tab.~\ref{tab:sherwood_eft_biases_nod3}. We highlight two key results from these fits: First, we recover unbiased BAO scaling parameters at the $1\sigma$ level, demonstrating how these mocks can be used to validate cosmological fitting pipelines for DESI \Lya forest analyses (see, e.g.,~\cite{DESI_lya_2024, Cuceu:2025nvl}). Second, the recovered bias parameters from the field-level mock are consistent with both, the field-level measurements (see Sec.~\ref{sec:theory_transfer_func}) and the P3D measurements presented in Ref.~\cite{Ivanov:2024lya}. The remaining differences with the latter  are primarily driven by analysis choices: (i) we perform an anisotropic one-loop IR resummation instead of an isotropic tree-level treatment, and (ii) we fit for the BAO scaling parameters, which the $L=160\hinvMpc$ box can only marginally resolve; and (iii) the fitted transfer functions $\beta(k,\mu)$ obtained from field-level analyses benefit from cosmic variance cancellation, and can therefore differ from P3D fits derived from small-volume simulations that are strongly affected by cosmic variance.

In the following we quantify the agreement between the recovered bias parameters from fits to the large-volume mock (Tab.~\ref{tab:EFT_bias_lya_auto}) compared to the transfer function fits mapped onto the CLASS-PT basis (Tab.~\ref{tab:sherwood_eft_biases_nod3}). The well-constrained linear and
velocity-bias parameters, $b_1$ and $b_\eta$, agree at the $\lesssim 1.7\sigma$
level, the only mildly elevated values arising for $b_1$ at $L=160\,\hinvMpc$
($1.4$--$1.7\,\sigma$). The quadratic bias $b_2$ is recovered within $1\sigma$ in every case. We thus find no significant tension between the two analyses. Note that we recover unbiased BAO parameters and that the model captures the shift \textit{and} smearing of the BAO peak through the one-loop corrections and the IR resummation, respectively.

\subsubsection{\Lya clustering mocks at $z_{\rm eff}=2.33$}

\begin{figure}
    \centering
    \includegraphics[width=0.49\linewidth]{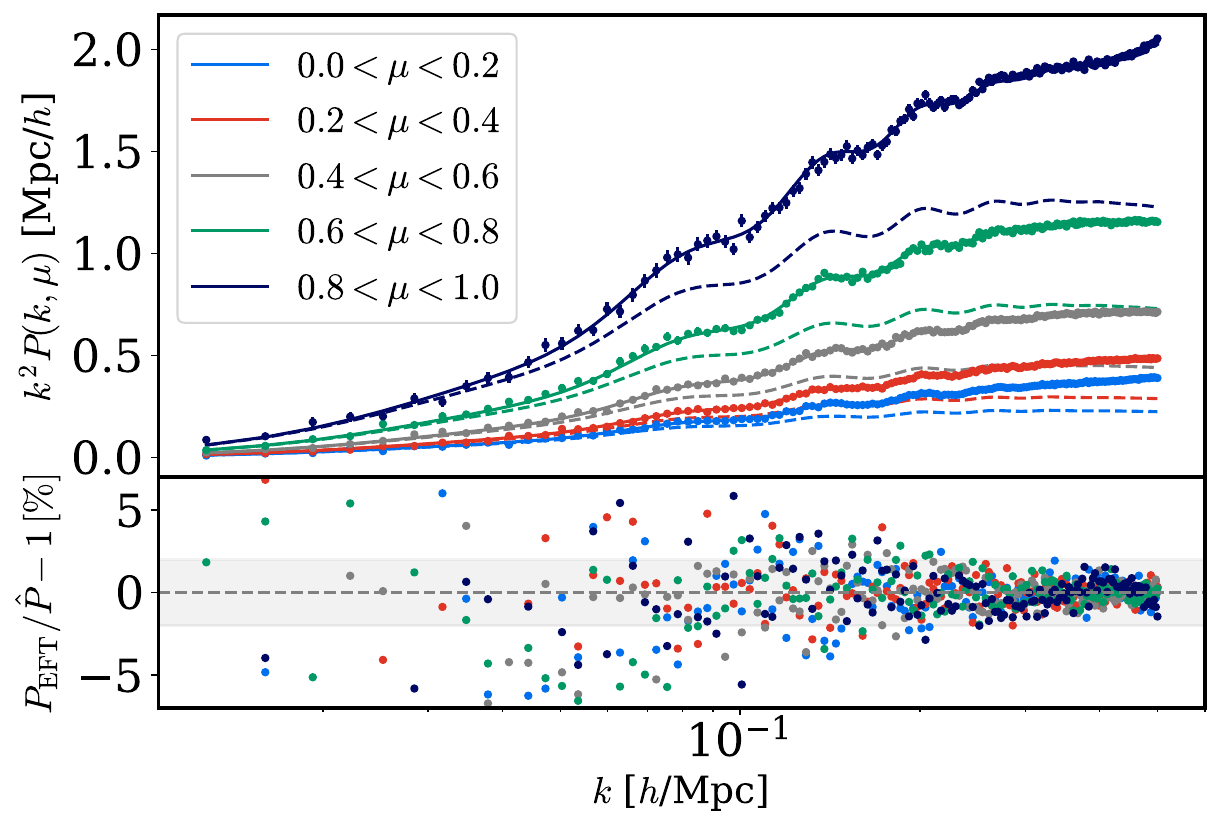}\hfill
    \includegraphics[width=0.49\linewidth]{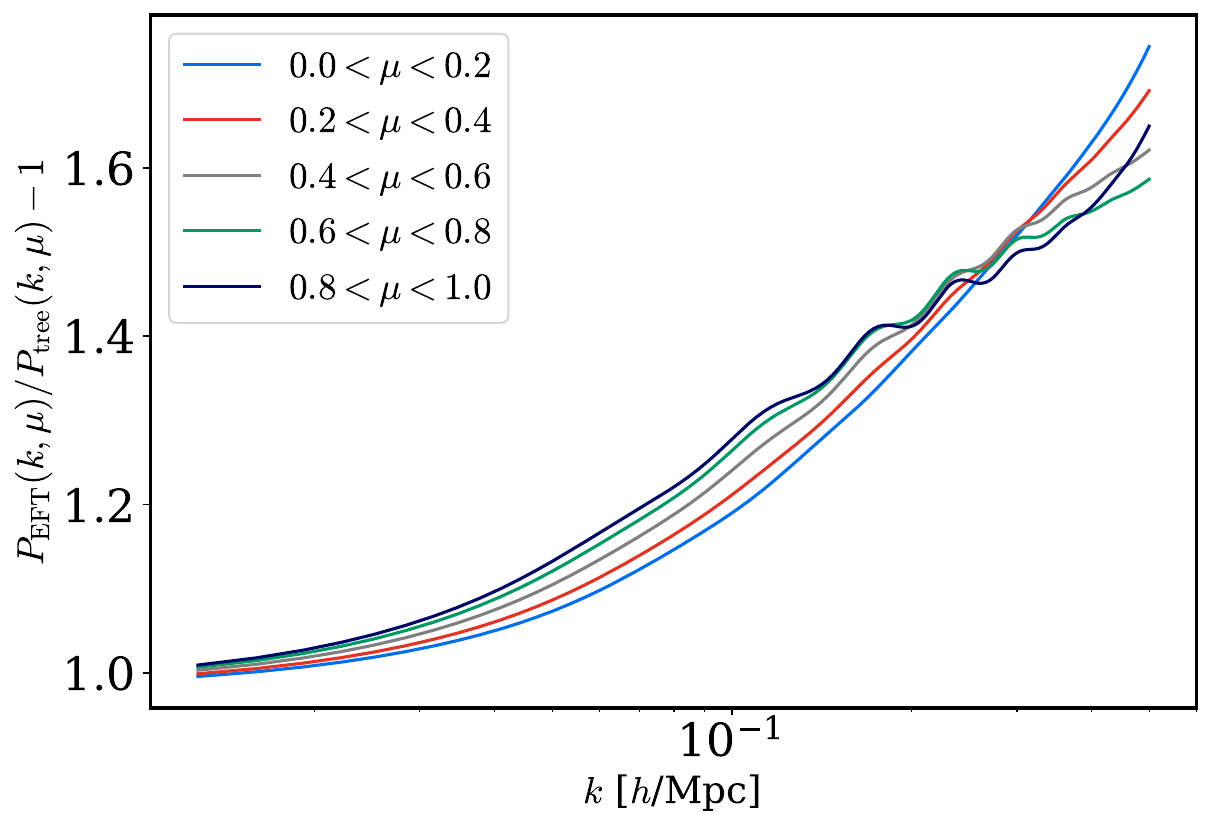}\hfill \\ 
    \caption{\textbf{Validation of \Lya Mocks at DESI redshift:} Same as Fig.~\ref{fig:EFT_mock_bestfit} but at the effective redshift of DESI for the \Lya forest using a $k_{\rm max}=0.5 \hMpcinv$ for the fits.}
    \label{fig:EFT_mock_bestfit_z233}
\end{figure}

Similarly, we perform a BAO fit of the \Lya forest EFT mocks at $z=2.33$, the effective redshift of the \Lya forest sample of DESI DR1 \cite{DESI_lya_2024}. The best-fit spectra are shown in Fig.~\ref{fig:EFT_mock_bestfit_z233} and the marginalized EFT nuisance and BAO parameters are tabulated in the last two columns of Tab.~\ref{tab:EFT_bias_lya_auto}. We highlight two key findings for the $z=2.33$ mock. First, we find excellent agreement between our linear bias parameters and the ones measured by DESI DR1 \cite{DESI_lya_2024}. Second, We find unbiased constraints on the BAO scaling parameters:
\begin{align}
& \alpha_{\parallel} = 1.011 \pm 0.018,\qquad
\alpha_{\perp} = 1.0021^{+0.0060}_{-0.0080}\,.
\end{align}
The resulting best-fit spectra are shown in Fig.~\ref{fig:EFT_mock_bestfit_z233}. Importantly, the loop corrections are exceeding 20\% at quasi linear scales of $k\approx0.1 \hMpcinv$ emphasizing the importance of higher order bias parameters.

\begin{figure}
    \centering
    \includegraphics[width=1\linewidth]{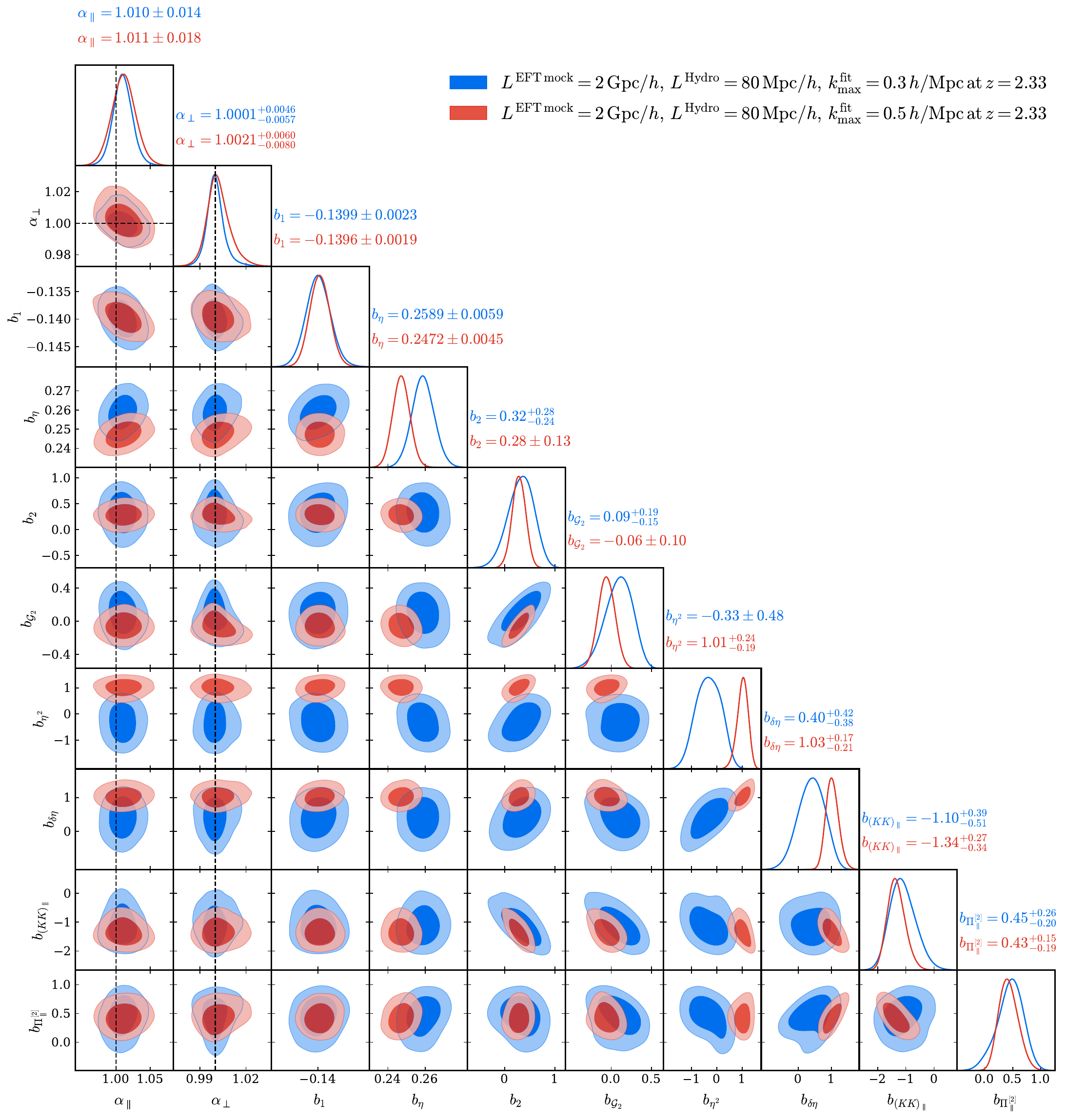}
    \caption{\textbf{Drift plot of \Lya Mock at $z=2.33$:} Best-fit 1D and 2D marginalized posteriors for the $l=80\hinvMpc$ Sherwood simulation using two different maximum wavenumbers for the full-shape fits $\kmax=0.3,\,0.5 \hMpcinv$. Crucially, we recover \textit{unbiased BAO scaling parameters}. 
    }
    \label{fig:EFT_mock_drift_plot_z233}
\end{figure}

The resulting 1D and 2D marginalized posteriors for the sampled parameters are shown in Fig.~\ref{fig:EFT_mock_drift_plot_z233}. 

\begin{table*}
    \centering
    \renewcommand{\arraystretch}{1.3}
    \setlength{\tabcolsep}{2pt}
    \begin{tabular}{l | cc cc | cc}
    \hline
    \hline
    Redshift
    & \multicolumn{4}{c}{\boldmath$z=2.8$}
    & \multicolumn{2}{c}{\boldmath$z=2.33$} \\
    \cmidrule(lr){2-5}\cmidrule(lr){6-7}
    Sherwood
    & \multicolumn{2}{c}{$L=160\,\hinvMpc$}
    & \multicolumn{2}{c}{$L=80\,\hinvMpc$}
    & \multicolumn{2}{c}{$L=80\,\hinvMpc$} \\
    \cmidrule(lr){2-3}\cmidrule(lr){4-5}\cmidrule(lr){6-7}
    $k_{\rm max}^{\rm fit}$
    & $0.3$ & $0.5$
    & $0.3$ & $0.5$
    & $0.3$ & $0.5$ \\[0.5ex]
    \hline
    \multicolumn{7}{@{}l}{\textit{Sampled parameters}}\\
    $b_1$
      & $-0.2048^{+0.0033}_{-0.0038}$ & $-0.2050 \pm 0.0030$ & $-0.2027 \pm 0.0030$ & $-0.2015 \pm 0.0025$ & $-0.1399 \pm 0.0023$ & $-0.1396 \pm 0.0019$ \\
    $b_{\eta}$
      & $\phantom{-}0.3298 \pm 0.0072$ & $\phantom{-}0.3191 \pm 0.0056$ & $\phantom{-}0.3471 \pm 0.0073$ & $\phantom{-}0.3322 \pm 0.0058$ & $\phantom{-}0.2589 \pm 0.0059$ & $\phantom{-}0.2472 \pm 0.0045$ \\
    $b_2$
      & $\phantom{-}0.55^{+0.40}_{-0.35}$ & $\phantom{-}0.14^{+0.25}_{-0.28}$ & $\phantom{-}0.56^{+0.38}_{-0.31}$ & $\phantom{-}0.36 \pm 0.20$ & $\phantom{-}0.32^{+0.28}_{-0.24}$ & $\phantom{-}0.28 \pm 0.13$ \\
    $b_{\mathcal{G}_2}$
      & $\phantom{-}0.07^{+0.25}_{-0.22}$ & $-0.11 \pm 0.16$ & $\phantom{-}0.16^{+0.24}_{-0.21}$ & $-0.096 \pm 0.15$ & $\phantom{-}0.09^{+0.19}_{-0.15}$ & $-0.06 \pm 0.10$ \\
    $b_{\eta^2}$
      & $-0.71^{+0.64}_{-0.52}$ & $\phantom{-}0.43 \pm 0.61$ & $-0.46^{+0.71}_{-0.58}$ & $\phantom{-}1.47^{+0.39}_{-0.30}$ & $-0.33 \pm 0.48$ & $\phantom{-}1.01^{+0.24}_{-0.19}$ \\
    $b_{\delta \eta}$
      & $\phantom{-}0.53^{+0.49}_{-0.40}$ & $\phantom{-}0.45^{+0.21}_{-0.27}$ & $\phantom{-}0.57^{+0.53}_{-0.44}$ & $\phantom{-}1.39^{+0.25}_{-0.31}$ & $\phantom{-}0.40^{+0.42}_{-0.38}$ & $\phantom{-}1.03^{+0.17}_{-0.21}$ \\
    $b_{(KK)_\parallel}$
      & $-1.31 \pm 0.69$ & $-0.73^{+0.60}_{-0.43}$ & $-1.60^{+0.53}_{-0.63}$ & $-1.89 \pm 0.43$ & $-1.10^{+0.39}_{-0.51}$ & $-1.34^{+0.27}_{-0.34}$ \\
    $b_{\Pi^{[2]}_\parallel}$
      & $\phantom{-}0.62 \pm 0.32$ & $\phantom{-}0.36 \pm 0.18$ & $\phantom{-}0.59^{+0.35}_{-0.30}$ & $\phantom{-}0.68^{+0.19}_{-0.24}$ & $\phantom{-}0.45^{+0.26}_{-0.20}$ & $\phantom{-}0.43^{+0.15}_{-0.19}$ \\
    $\apar$
      & $\phantom{-}1.009^{+0.014}_{-0.012}$ & $\phantom{-}1.009 \pm 0.013$ & $\phantom{-}1.009 \pm 0.013$ & $\phantom{-}1.009 \pm 0.015$ & $\phantom{-}1.010 \pm 0.014$ & $\phantom{-}1.011 \pm 0.018$ \\
    $\aperp$
      & $\phantom{-}0.9978^{+0.0076}_{-0.0054}$ & $\phantom{-}0.9992 \pm 0.0060$ & $\phantom{-}1.0000^{+0.0048}_{-0.0057}$ & $\phantom{-}1.0011^{+0.0057}_{-0.0070}$ & $\phantom{-}1.0001^{+0.0046}_{-0.0057}$ & $\phantom{-}1.0021^{+0.0060}_{-0.0080}$ \\[0.5ex]
    \midrule
    \multicolumn{7}{@{}l}{\textit{Analytically marginalized}}\\
    $b_{\Pi^{[3]}_\parallel}$
      & $\phantom{-}1.3854 \pm 0.3592$
      & $\phantom{-}2.2579 \pm 0.1161$
      & $\phantom{-}2.3390 \pm 0.3818$
      & $\phantom{-}3.9789 \pm 0.1261$
      & $\phantom{-}1.6803 \pm 0.2466$
      & $\phantom{-}2.7891 \pm 0.0766$ \\
    
    $b_{\delta\Pi^{[2]}_\parallel}$
      & $-0.2006 \pm 0.8437$
      & $-0.1935 \pm 0.4518$
      & $-0.7482 \pm 0.8638$
      & $-1.5778 \pm 0.4703$
      & $-1.1724 \pm 0.7020$
      & $-0.7901 \pm 0.3159$ \\
    
    $b_{(K\Pi^{[2]})_\parallel}$
      & $-1.2460 \pm 0.8426$
      & $-3.3381 \pm 0.3855$
      & $-2.9227 \pm 0.8766$
      & $-5.5891 \pm 0.3947$
      & $-2.4507 \pm 0.5353$
      & $-3.4554 \pm 0.2444$ \\
    
    $b_{\eta\Pi^{[2]}_\parallel}$
      & $-0.4614 \pm 1.7829$
      & $-0.4338 \pm 1.2428$
      & $-0.4876 \pm 1.8040$
      & $-0.5198 \pm 1.2934$
      & $-0.4975 \pm 1.7575$
      & $\phantom{-}0.7903 \pm 0.9370$ \\
    
    $b_{\Gamma_3}$
      & $\phantom{-}0.5259 \pm 0.3569$
      & $\phantom{-}0.2269 \pm 0.2086$
      & $\phantom{-}0.5012 \pm 0.3592$
      & $\phantom{-}0.6851 \pm 0.2144$
      & $\phantom{-}0.6367 \pm 0.2737$
      & $\phantom{-}0.4511 \pm 0.1472$ \\
    
    $P_{\rm shot}$
      & $-0.7920 \pm 0.5576$
      & $-0.9231 \pm 0.2003$
      & $-2.8145 \pm 0.5607$
      & $-2.4783 \pm 0.1957$
      & $-2.8691 \pm 0.3647$
      & $-2.2519 \pm 0.1173$ \\
    
    $a_0$
      & $\phantom{-}0.7054 \pm 5.0299$
      & $\phantom{-}12.2297 \pm 3.4545$
      & $\phantom{-}0.9571 \pm 4.9795$
      & $\phantom{-}16.1193 \pm 3.4851$
      & $\phantom{-}1.2038 \pm 4.9625$
      & $\phantom{-}17.7688 \pm 2.6652$ \\
    
    $a_2$
      & $\phantom{-}0.2653 \pm 5.0186$
      & $\phantom{-}2.7929 \pm 4.8289$
      & $\phantom{-}0.0609 \pm 4.9918$
      & $\phantom{-}3.6760 \pm 4.8137$
      & $-0.0266 \pm 5.0211$
      & $\phantom{-}1.9947 \pm 4.7306$ \\
    
    $c_0$
      & $\phantom{-}0.3195 \pm 0.2075$
      & $\phantom{-}0.1045 \pm 0.0756$
      & $\phantom{-}0.7356 \pm 0.2084$
      & $\phantom{-}0.6500 \pm 0.0754$
      & $\phantom{-}0.9325 \pm 0.1623$
      & $\phantom{-}0.6460 \pm 0.0535$ \\
    
    $c_2$
      & $\phantom{-}0.0144 \pm 0.4208$
      & $-0.0773 \pm 0.1190$
      & $-0.8157 \pm 0.4414$
      & $-1.3573 \pm 0.1221$
      & $-1.1771 \pm 0.3688$
      & $-1.3204 \pm 0.1009$ \\
    
    $c_4$
      & $-0.1286 \pm 0.1724$
      & $\phantom{-}1.0108 \pm 0.0797$
      & $\phantom{-}0.3731 \pm 0.1792$
      & $\phantom{-}2.3128 \pm 0.0826$
      & $\phantom{-}0.3669 \pm 0.1878$
      & $\phantom{-}1.9862 \pm 0.0731$ \\
    
    $\chi^2_{\rm marg}$
      & \phantom{-}341.2
      & \phantom{-}590.7
      & \phantom{-}341.8
      & \phantom{-}581.0
      & \phantom{-}338.1
      & \phantom{-}592.8 \\[0.5ex]
    \hline
    \end{tabular}
    \caption{\textbf{\Lya forest auto-correlation fits:} Marginalized best-fit EFT parameters obtained from fits to the \Lya forest auto-power spectrum of large-volume ($L=2{,}000\,\hinvMpc$) mocks calibrated on Sherwood hydrodynamic simulations, for hydrodynamic box sizes $L=160$ and $80\,\hinvMpc$ at redshifts $z=2.8$ (\textit{first four columns}) and interpolated to $z=2.33$ (\textit{last two columns}). The top section lists the sampled parameters; the bottom section lists the parameters that we analytically marginalize over and recover \textit{a posteriori} from the chains. The counterterms $c_0, c_2, c_4$ are quoted in units of $[\Mpch]^2$ and $\kmax$ is in $\hMpcinv$. The $\kmax=0.3\,(0.5)\,\hMpcinv$ fits use 460 (780) data points. The linear bias parameters are consistent at the $1$--$2\sigma$ level with fits from the three-dimensional power spectrum \cite{Ivanov:2024lya}.
    }
    \label{tab:EFT_bias_lya_auto}
\end{table*}

\subsection{Cross-correlations between the \Lya forest and high-redshift galaxies} \label{sec:lya_x_LBG_LAE}
\begin{figure}
    \centering
    \includegraphics[width=0.32\linewidth]{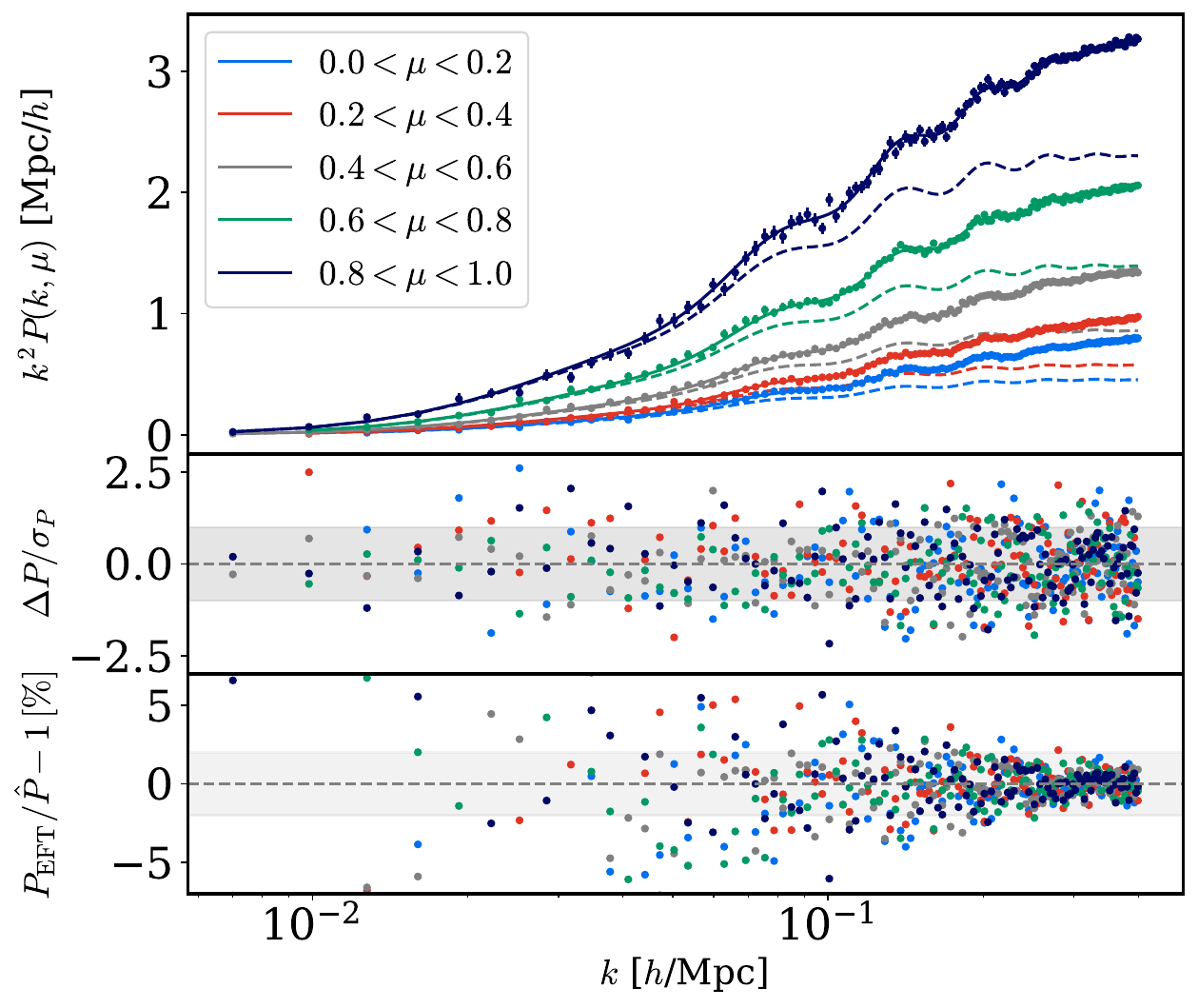}\hfill
    \includegraphics[width=0.32\linewidth]{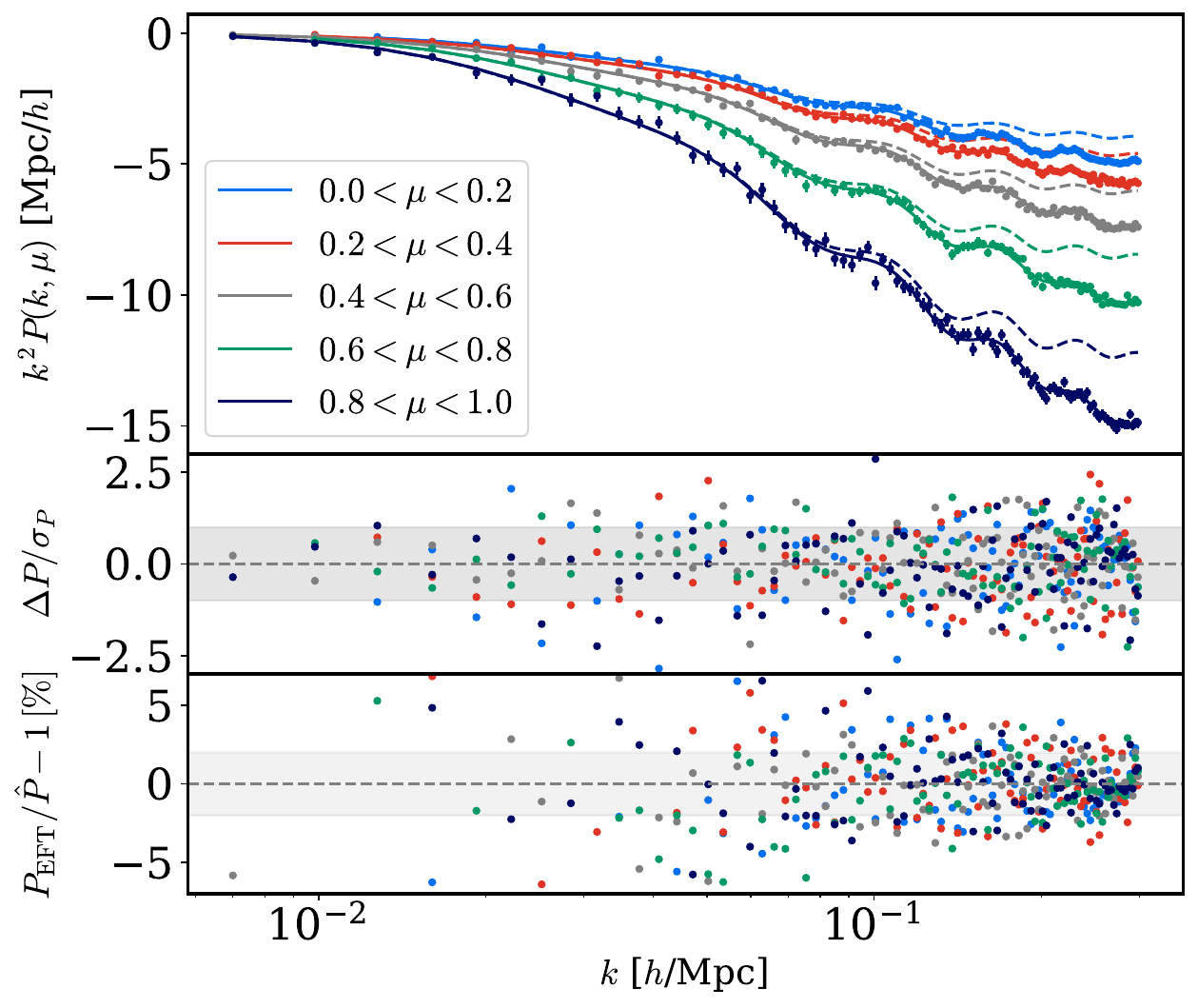}\hfill
    \includegraphics[width=0.32\linewidth]{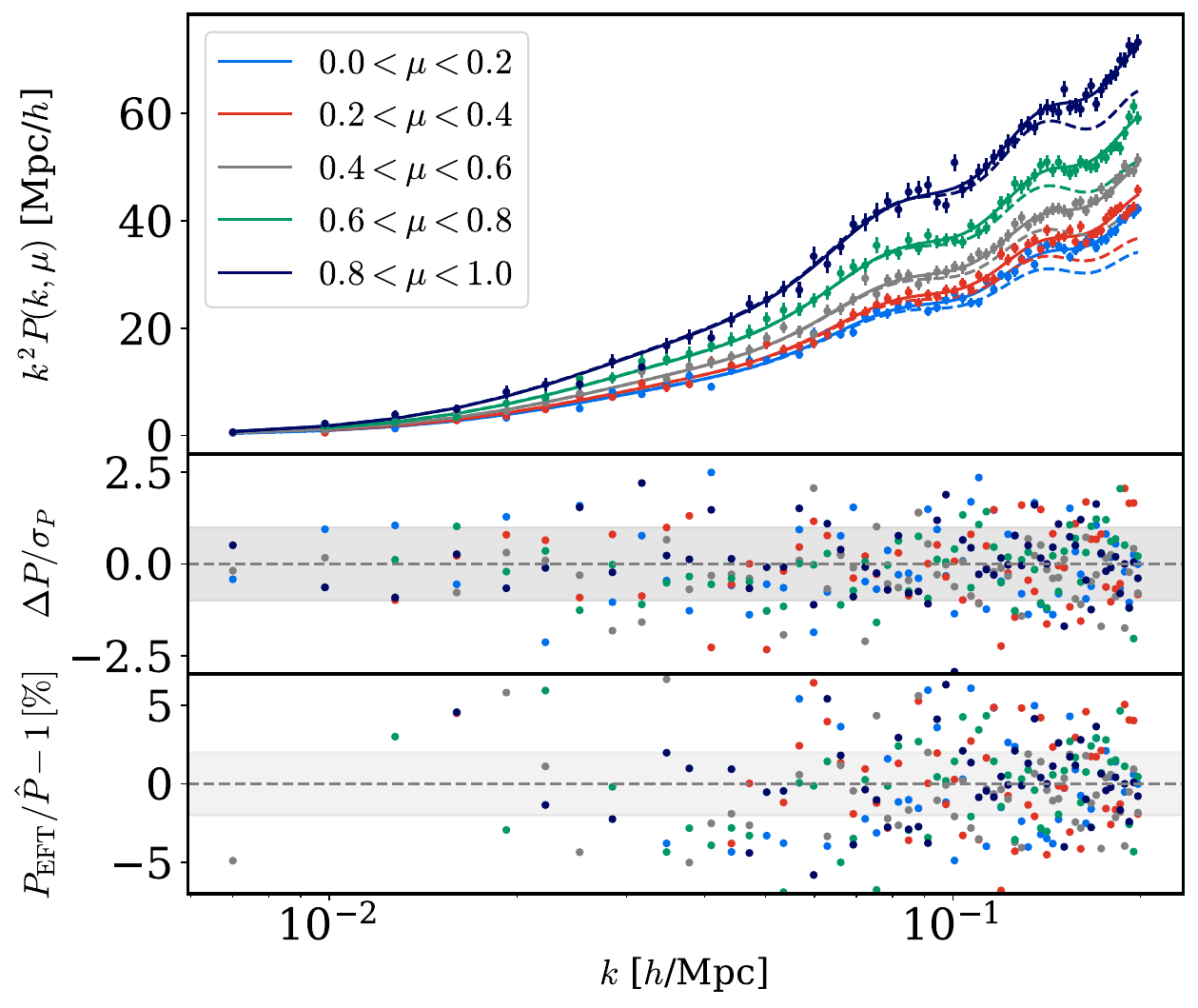}\hfill \\
    \includegraphics[width=0.32\linewidth]{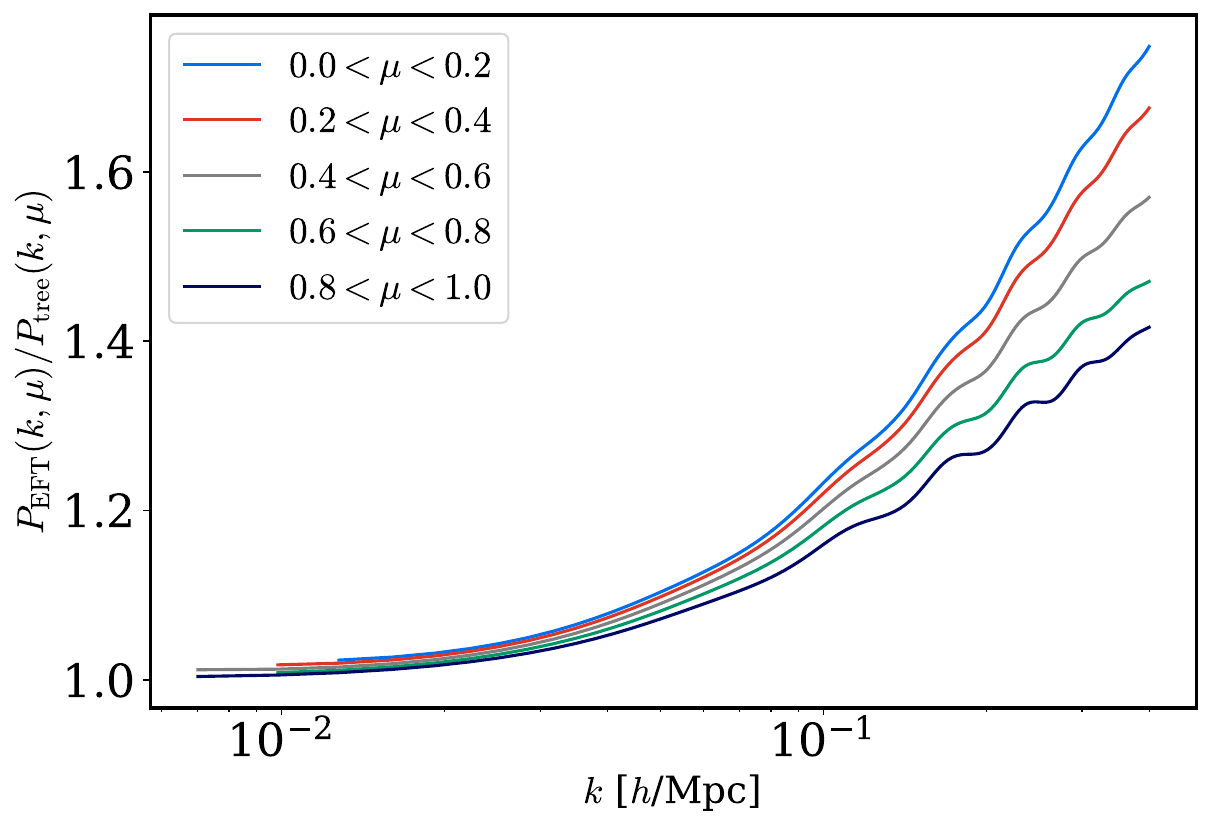}\hfill
    \includegraphics[width=0.32\linewidth]{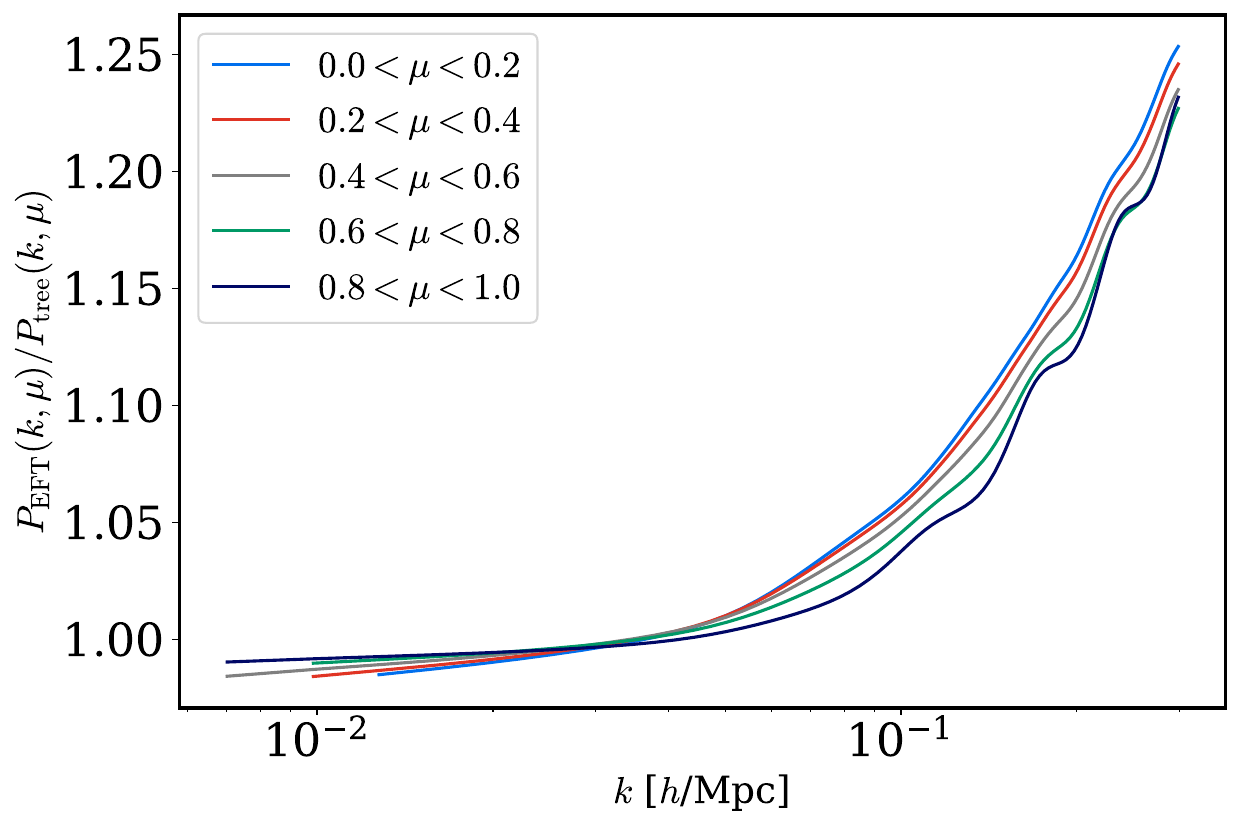}\hfill
    \includegraphics[width=0.32\linewidth]{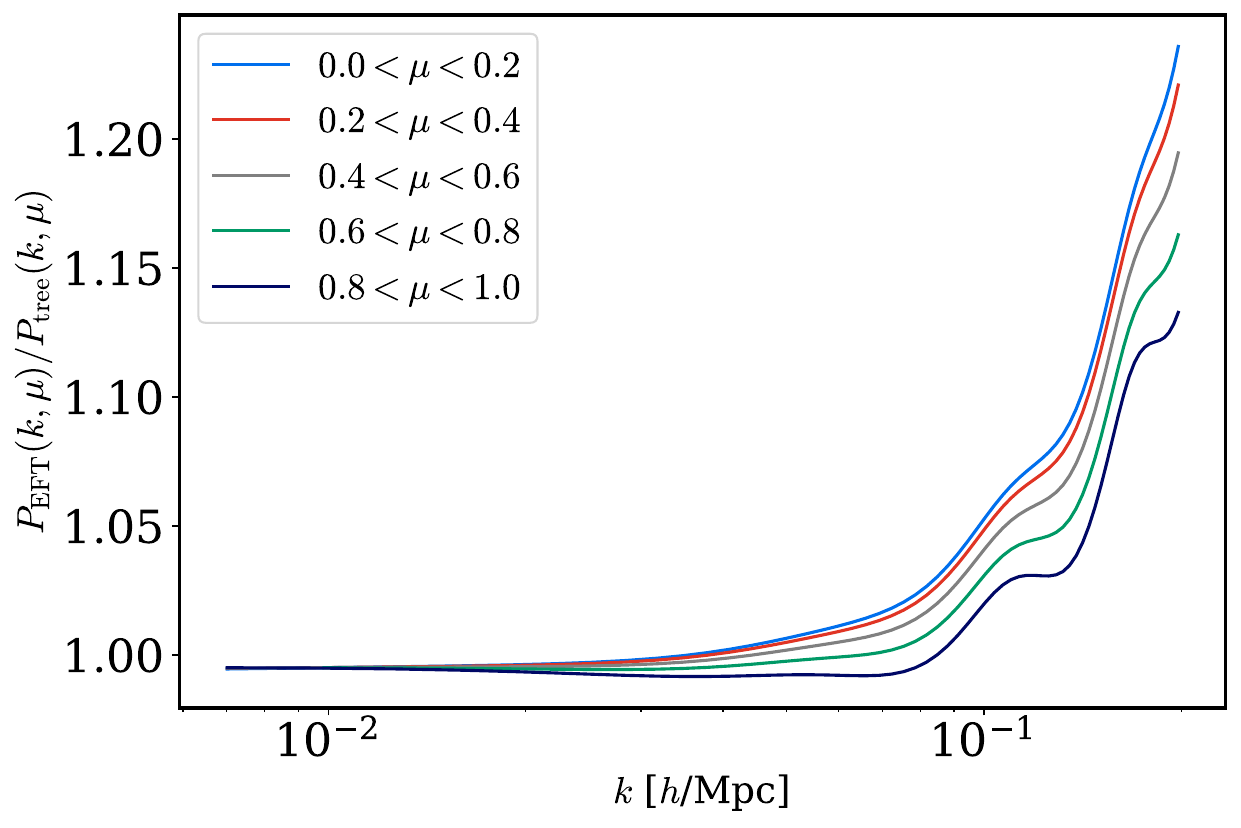}\hfill \\
    \caption{\textbf{Validation of \Lya $\times$ LAE S5 Mock:} 
    Same as Fig.~\ref{fig:EFT_mock_bestfit} but for the joint fit of the \Lya auto-correlation (\textit{left column}) with the cross-correlation of \Lya and galaxies (\textit{center column}) and the galaxy auto-correlation (\textit{right column}) using LAEs of the mimicked ``S5'' sample. The top row shows the best-fit EFT model (\textit{solid line}) which is compared to $P^{\rm tree}$ (\textit{dashed line}) and the data points with corresponding error bars. The two bottom panels show, first, the normalized and fractional residuals. For both, a gray band indicating the 2\% (or 1$\sigma$) band is shown to guide the eye. The bottom row shows the loop corrections which are the ratio between the best-fit EFT one-loop power spectrum and the tree-level power spectrum. The joint fits use the following maximum wavenumbers for the fits: $k_{\rm max}^{\rm FF} =0.40\hMpcinv$, $k_{\rm max}^{\rm Fg} =0.3\hMpcinv$, and $k_{\rm max}^{\rm gg} =0.2\hMpcinv$ for each realization. Note that the \Lya forest transfer functions are computed using a Sherwood simulation with box size $L=80\hinvMpc$ and the mock is centered at $z=3$.}
    \label{fig:EFT_mock_bestfit_cross_LAE}
    \vspace{-0.1in}
\end{figure}

\begin{figure}
    \centering
    \includegraphics[width=0.32\linewidth]{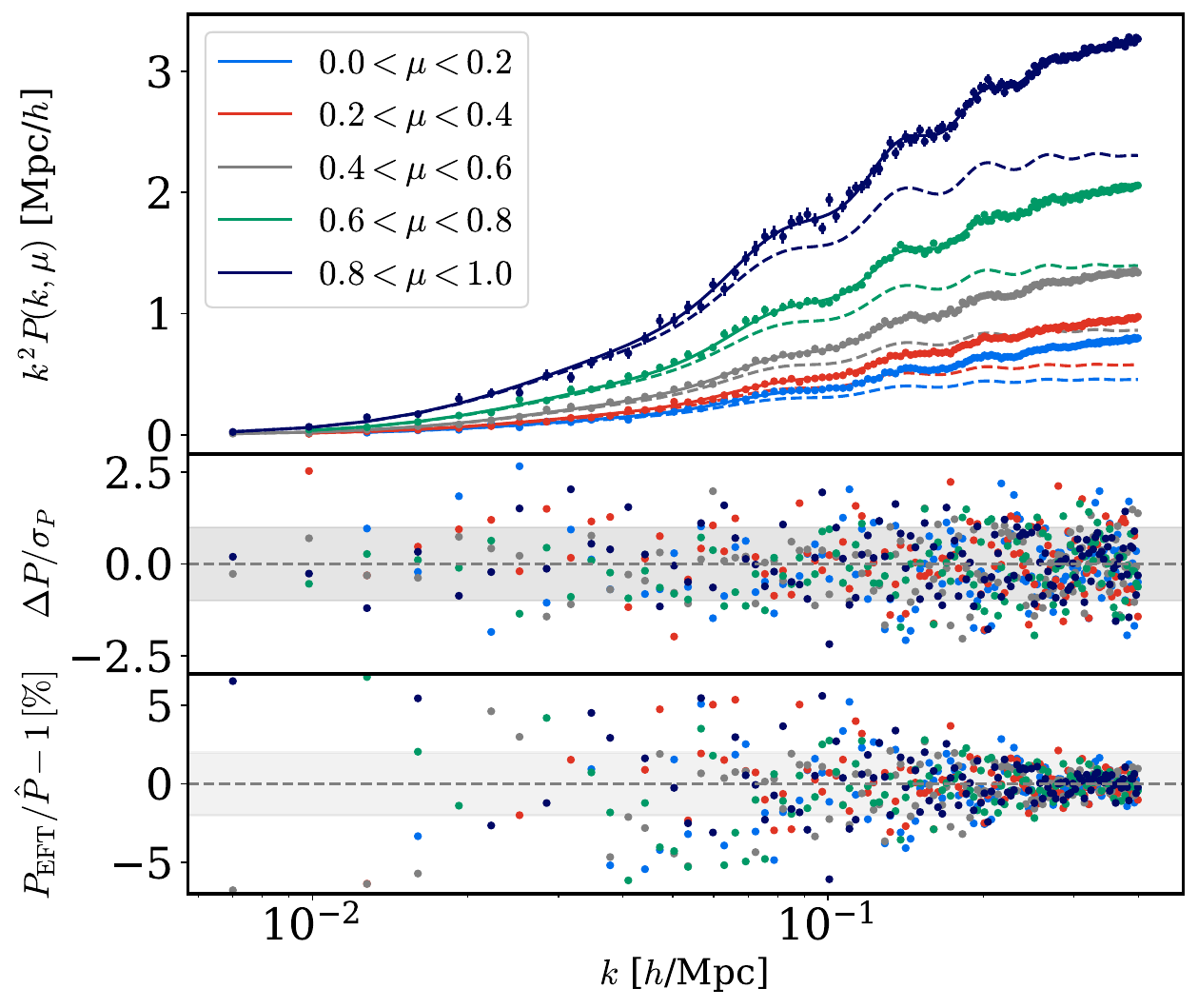}\hfill
    \includegraphics[width=0.32\linewidth]{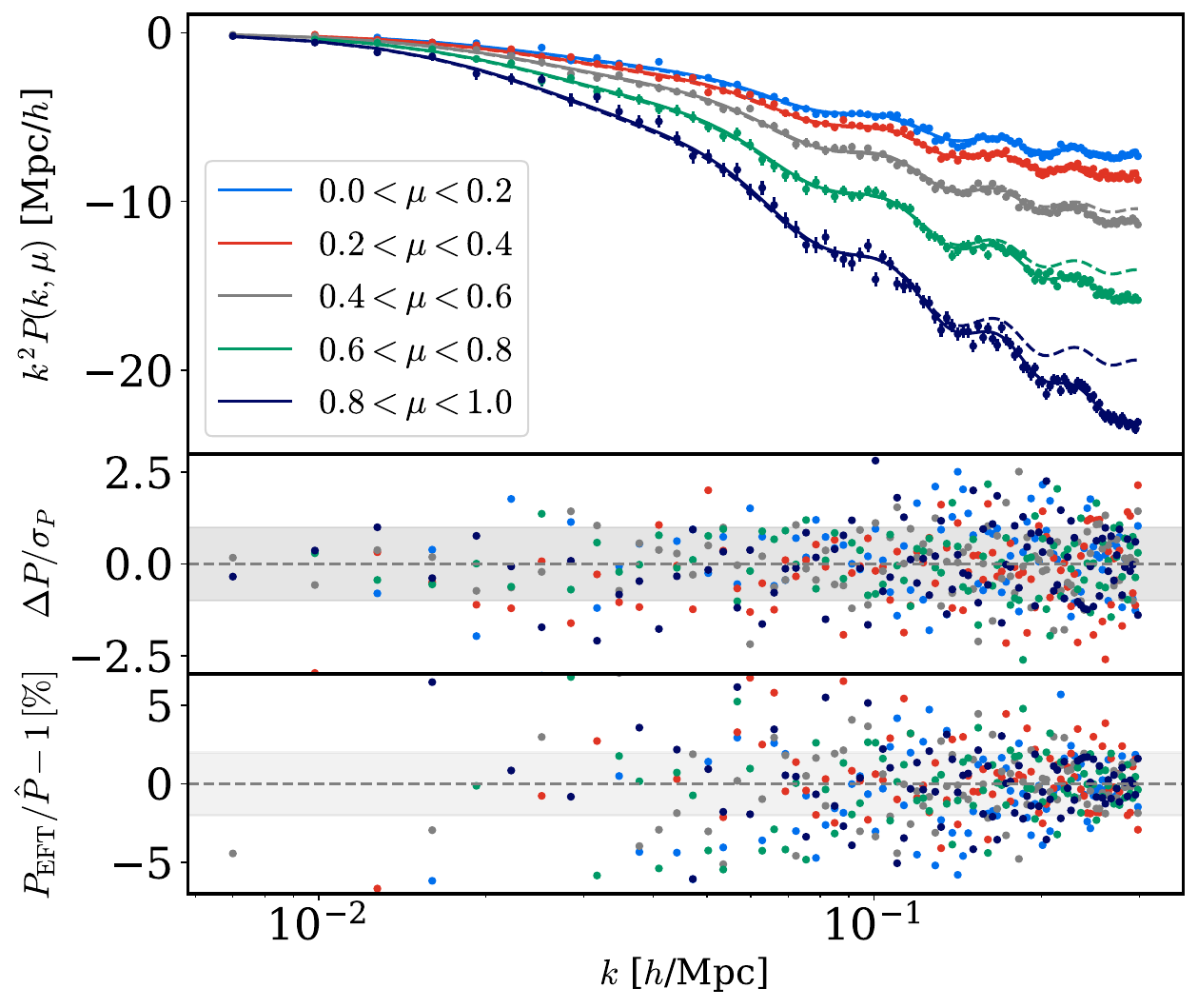}\hfill
    \includegraphics[width=0.32\linewidth]{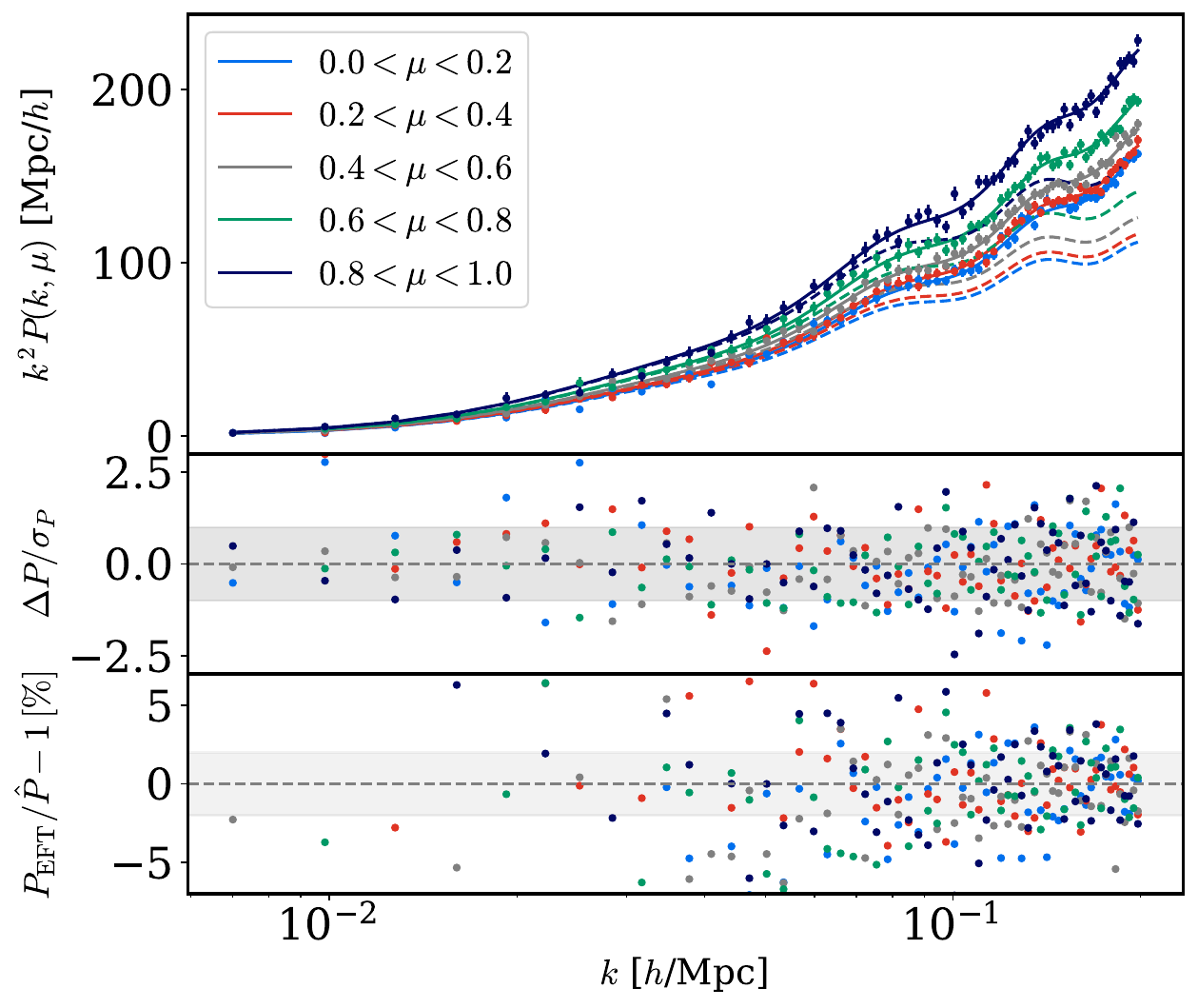}\hfill \\
    \includegraphics[width=0.32\linewidth]{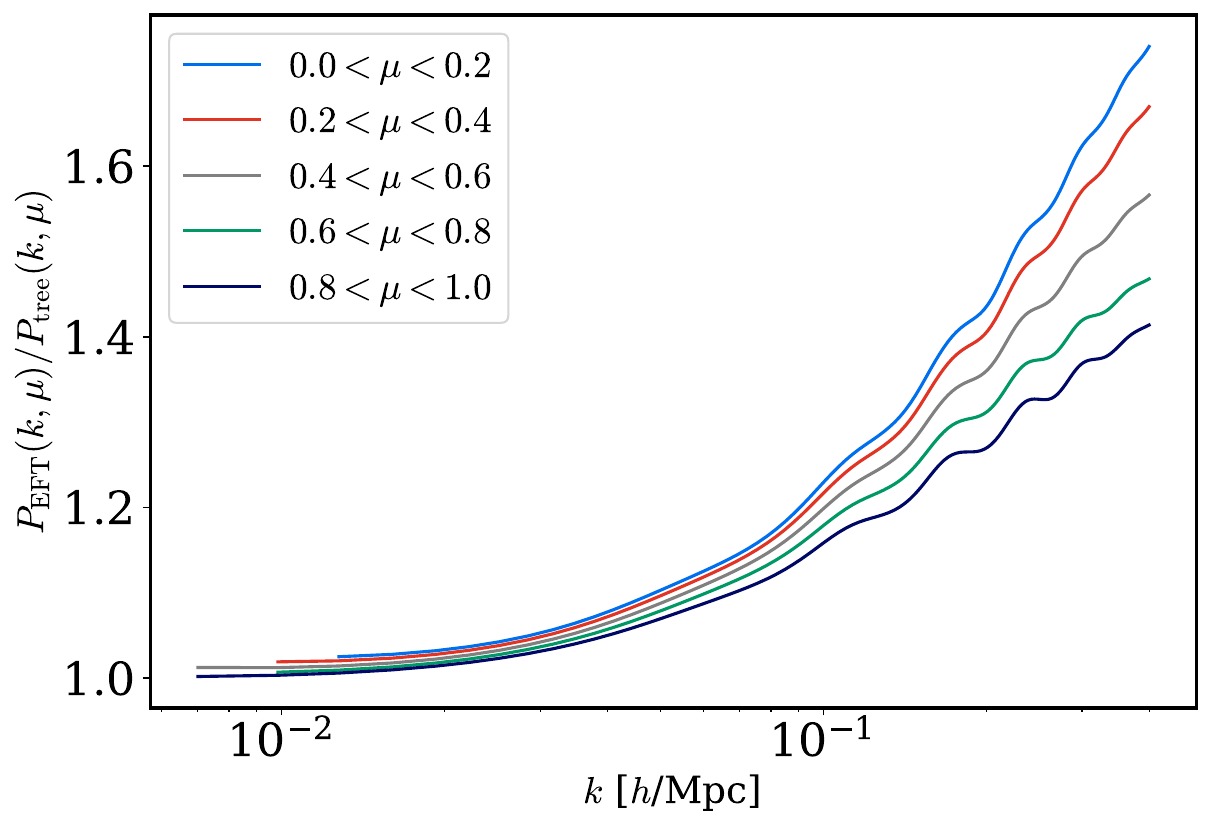}\hfill
    \includegraphics[width=0.32\linewidth]{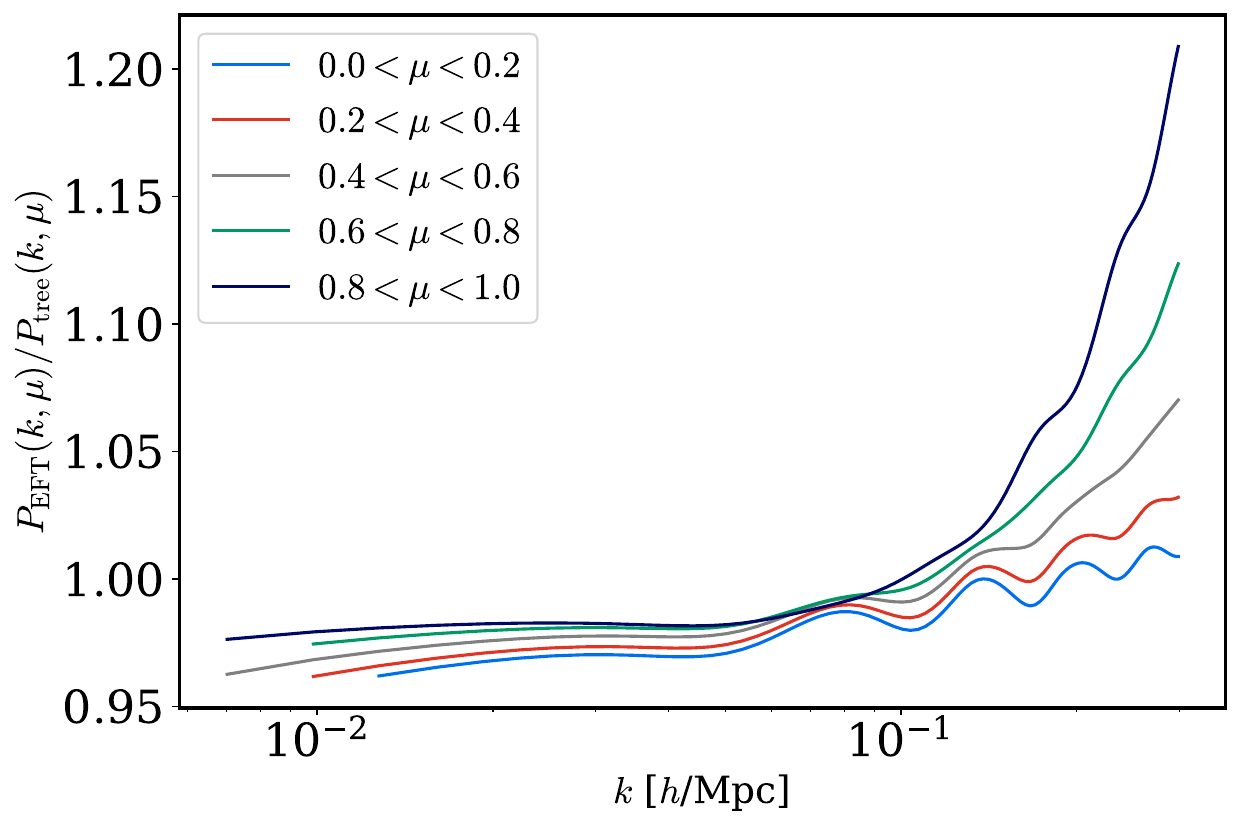}\hfill
    \includegraphics[width=0.32\linewidth]{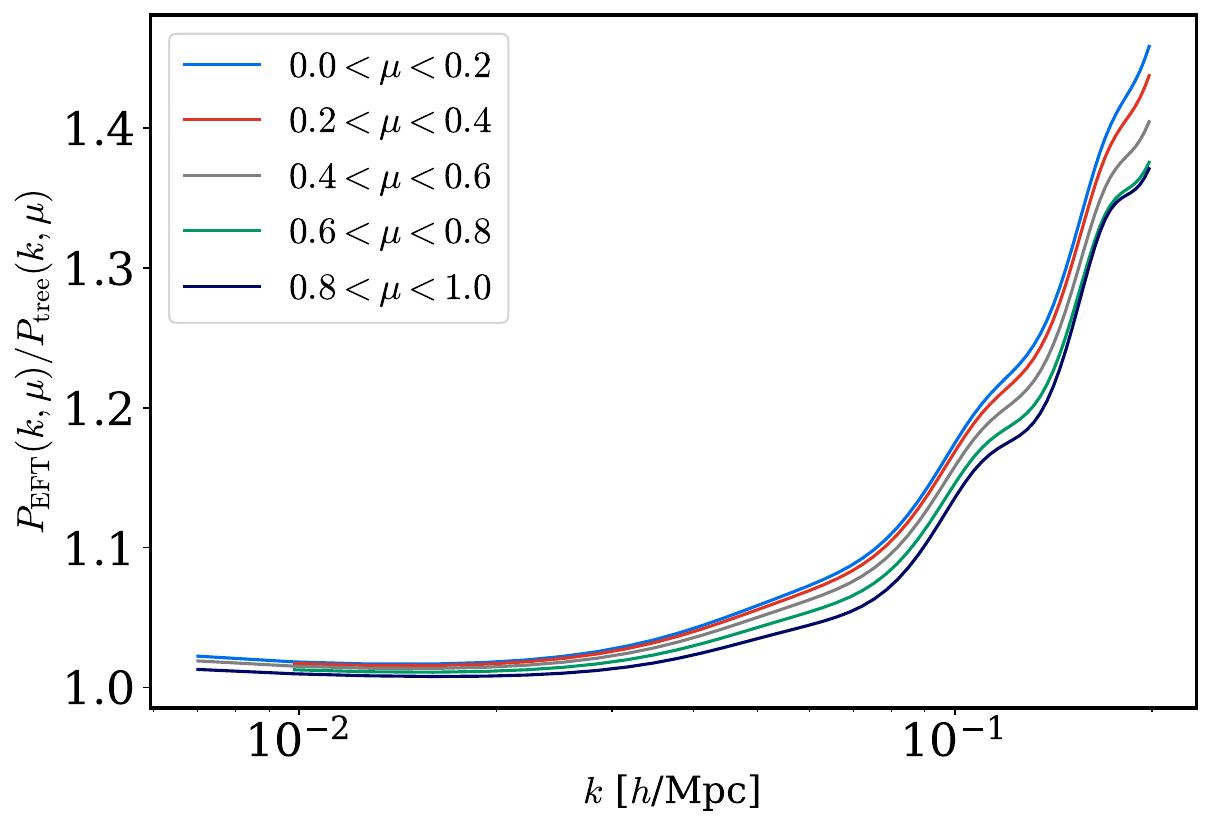}\hfill \\
    \caption{\textbf{Validation of \Lya $\times$ LBG S5 Mock:} Same as Fig.~\ref{fig:EFT_mock_bestfit_cross_LAE} but for S5 LBG sample.}
    \label{fig:EFT_mock_bestfit_cross_LBG}
    \vspace{-0.1in}
\end{figure}

A key source of cosmological information in \Lya forest analyses stems from cross-correlating the \Lya forest with quasar positions (see, e.g.,~\cite{Font-Ribera:2013fha, DESI_lya_2024, Cuceu:2025nvl, DESI_lya_dr2}). Near-future surveys such as DESI-II will measure large samples of high-redshift galaxies such as Lyman-break galaxies (LBGs) and 
\Lya emitters (LAEs). Yet, their cosmological analyses in combination with the \Lya forest is barely explored~\cite{Herrera-Alcantar:2025yry}. Here, we create large-scale clustering mocks of high-redshift galaxy samples, introduced in Sec.~\ref{sec:simulations}. To this end, we use the transfer functions shown in Fig.~\ref{fig:Astrid_TF} with a  scale cut of $\kmax=0.3 \hMpcinv$ which is driven by the scale and angular dependence of the error power spectrum \cite{Sullivan:2025eei}. We extend the previous section to create large-scale clustering mocks for LAEs and LBGs and perform full-shape and BAO fits for the cross-correlation of the \Lya forest with mocks of these galaxy samples. 

% \subsubsection{Validation of \Lya forest and high-redshift galaxy cross-correlation mocks}\label{sec:validation_lya_x_lae}

In Figs.~\ref{fig:EFT_mock_bestfit_cross_LAE}-\ref{fig:EFT_mock_bestfit_cross_LBG} we show the resulting best-fit spectra of the joint power spectrum fits to the \Lya forest and the high-redshift galaxy samples (LAEs and LBGs).\footnote{We leave the inclusion of the cross-covariance between the two tracers to future work.} We show the best-fit spectra in the top row with the corresponding residuals and the loop corrections in the bottom row of each figure for the LBG S5 and the LAE S5 samples. The joint fits use $k_{\rm max}^{\rm FF} =0.40\hMpcinv$, $k_{\rm max}^{\rm Fg} =0.30\hMpcinv$, and $k_{\rm max}^{\rm gg} =0.20\hMpcinv$.  Following baseline expectation, the loop corrections vanish on large scales and exceed the 5\% threshold beyond $k=0.1\hMpcinv$ when including the point tracers. We find good agreement between the best-fit parameters obtained from the joint fits given in Tab.~\ref{tab:EFT_bias_lae} to the ones summarized in table III in \cite{Sullivan:2025eei}. In addition to this, the BAO fits confirm, first, the usefulness of the presented mocks as we find unbiased constraints on the BAO scaling parameters from the joint fits of the \Lya forest and high-redshift galaxies at the $1\sigma$ level which can be used for inference pipeline validation. Second, the EFT model is applicable for cross-correlations with high-$z$ galaxies, even for highly-biased samples such as LBGs.

In this section we have demonstrated a new procedure to generate large-scale clustering mocks that can be used to validate cosmological inference pipelines. Note that the generated perturbative mocks are at fixed redshifts. To put these simulations on a light cone one would evolve all the fields and the gravitational potential with redshift and place an observer at, e.g.,~the center of the cartesian grid. Since the $160\hinvMpc$ Sherwood simulation only has a single redshift, we would again use the $L=80 \hinvMpc$ suite of simulations and fit to each snapshot the transfer functions at the field-level, see App.~\ref{app:Sherwood_tf_80}. We would interpolate the resulting polynomial fits to the desired redshift allowing us to compute a realization of the \Lya field at arbitrary redshift (within the redshift coverage of the snapshots) similar to what we have done in Sec.~\ref{sec:lya_x_LBG_LAE}. We leave the construction of light cone mocks to future work.

\begin{table*}
    \begin{center}
    \begin{tabular}{c c c c c}
     \hline\hline 
    Sample
    & \multicolumn{2}{c}{\textbf{LBG}}
    & \multicolumn{2}{c}{\textbf{LAE}}\\
    \cmidrule(lr){2-3}\cmidrule(lr){4-5}
    $b_{\mathcal{O}}$ 
    & CARS 
    & S5 
    & S5 
    & ODIN \\
     \hline
    $b_1^{\rm q}$ 
    & $\phantom{-}4.073 \pm 0.031$ 
    & $\phantom{-}3.798^{+0.027}_{-0.028}$ 
    & $\phantom{-}2.094^{+0.022}_{-0.017}$ 
    & $\phantom{-}1.969^{+0.023}_{-0.038}$ \\
    $b_2^{\rm q}$ 
    & $\phantom{-}11.71^{+2.89}_{-2.83}$ 
    & $\phantom{-}10.400^{+2.930}_{-2.938}$ 
    & $\phantom{-}2.745^{+2.947}_{-1.615}$ 
    & $-5.436^{+0.266}_{-3.694}$ \\
    $b_{\mathcal{G}_2}^{\rm q}$ 
    & $-0.7115^{+0.8136}_{-0.8095}$ 
    & $-1.4360^{+0.7184}_{-0.7140}$ 
    & $-1.1070^{+0.5132}_{-0.5050}$ 
    & $-1.0540^{+0.6946}_{-0.7010}$ \\
    $b_1$ 
    & $-0.236 \pm 0.002$ 
    & $-0.2359 \pm 0.0020$ 
    & $-0.2354 \pm 0.0020$ 
    & $-0.2369 \pm 0.0020$ \\
    $b_{\eta}$ 
    & $\phantom{-}0.3810 \pm 0.0052$ 
    & $\phantom{-}0.3818 \pm 0.0052$ 
    & $\phantom{-}0.3820 \pm 0.0053$ 
    & $\phantom{-}0.3882^{+0.0056}_{-0.0057}$ \\
    $b_2$ 
    & $\phantom{-}0.5299^{+0.1680}_{-0.1355}$ 
    & $\phantom{-}0.5349^{+0.1690}_{-0.1304}$ 
    & $\phantom{-}0.3768^{+0.2397}_{-0.1390}$ 
    & $\phantom{-}0.24820^{+0.21780}_{-0.18153}$ \\
    $b_{\mathcal{G}_2}$ 
    & $-0.17110^{+0.09106}_{-0.09010}$ 
    & $-0.2119^{+0.0905}_{-0.0903}$ 
    & $-0.122400^{+0.119932}_{-0.088200}$ 
    & $-0.08786^{+0.17645}_{-0.13754}$ \\
    $b_{\eta^2}$ 
    & $-0.8555^{+0.2447}_{-0.2035}$ 
    & $-0.8688^{+0.2247}_{-0.1972}$ 
    & $-0.8933^{+0.3710}_{-0.2787}$ 
    & $-0.3790^{+0.3311}_{-0.2789}$ \\
    $b_{\delta \eta}$ 
    & $\phantom{-}0.8217^{+0.2643}_{-0.2159}$ 
    & $\phantom{-}0.8855^{+0.2635}_{-0.2066}$ 
    & $\phantom{-}0.7630^{+0.2920}_{-0.2158}$ 
    & $\phantom{-}0.8819^{+0.2681}_{-0.2205}$ \\
    $b_{(KK)_\parallel}$ 
    & $-0.05146^{+0.15616}_{-0.15654}$ 
    & $-0.09012^{+0.15319}_{-0.15138}$ 
    & $\phantom{-}0.02497^{+0.22013}_{-0.16717}$ 
    & $-0.3620^{+0.2957}_{-0.2878}$ \\
    $b_{\Pi^{[2]}_\parallel}$ 
    & $\phantom{-}0.6957^{+0.1911}_{-0.2289}$ 
    & $\phantom{-}0.7214^{+0.1875}_{-0.2204}$ 
    & $\phantom{-}0.7328^{+0.2263}_{-0.2257}$ 
    & $\phantom{-}0.5261^{+0.1542}_{-0.2165}$ \\
    $\apar$ 
    & $\phantom{-}1.005^{+0.006}_{-0.005}$ 
    & $\phantom{-}1.0040^{+0.0050}_{-0.0052}$ 
    & $\phantom{-}1.007^{+0.005}_{-0.006}$ 
    & $\phantom{-}1.0030^{+0.0050}_{-0.0057}$ \\
    $\aperp$ 
    & $\phantom{-}1.0010^{+0.0040}_{-0.0027}$ 
    & $\phantom{-}1.0010^{+0.0040}_{-0.0026}$ 
    & $\phantom{-}0.9985^{+0.0035}_{-0.0032}$ 
    & $\phantom{-}1.000 \pm 0.003$ \\
    \hline
    $b_{\Pi^{[3]}_\parallel}$ 
    & $\phantom{-}2.526 \pm 0.22$ 
    & $\phantom{-}2.438 \pm 0.22$ 
    & $\phantom{-}2.416 \pm 0.218$ 
    & $\phantom{-}1.917 \pm 0.227$ \\
    $b_{\delta\Pi^{[2]}_\parallel}$ 
    & $-0.1866 \pm 0.673$ 
    & $-0.3883 \pm 0.625$ 
    & $-0.578 \pm 0.673$ 
    & $-1.455 \pm 0.662$ \\
    $b_{(K\Pi^{[2]})_\parallel}$ 
    & $-5.57 \pm 0.515$ 
    & $-5.59 \pm 0.498$ 
    & $-6.077 \pm 0.52$ 
    & $-5.323 \pm 0.511$ \\
    $b_{\eta\Pi^{[2]}_\parallel}$ 
    & $\phantom{-}0.2743 \pm 1.72$ 
    & $-0.06847 \pm 1.65$ 
    & $-0.3481 \pm 1.74$ 
    & $-0.2983 \pm 1.68$ \\
    $b_{\Gamma_3}$ 
    & $\phantom{-}1.67 \pm 0.254$ 
    & $\phantom{-}1.808 \pm 0.256$ 
    & $\phantom{-}1.786 \pm 0.265$ 
    & $\phantom{-}2.104 \pm 0.267$ \\
    $P_{\rm shot}$ 
    & $-0.2746 \pm 0.3$ 
    & $-0.2858 \pm 0.292$ 
    & $-0.182 \pm 0.334$ 
    & $-0.5699 \pm 0.338$ \\
    $a_0$ 
    & $\phantom{-}0.165 \pm 0.73$ 
    & $\phantom{-}0.1384 \pm 0.729$ 
    & $\phantom{-}0.1246 \pm 0.754$ 
    & $\phantom{-}0.136 \pm 0.715$ \\
    $a_2$ 
    & $-0.007086 \pm 0.74$ 
    & $-0.04853 \pm 0.736$ 
    & $\phantom{-}0.008977 \pm 0.749$ 
    & $\phantom{-}0.0487 \pm 0.741$ \\
    $c_0$ 
    & $-0.04172 \pm 0.0924$ 
    & $-0.03431 \pm 0.0896$ 
    & $-0.04338 \pm 0.105$ 
    & $\phantom{-}0.08935 \pm 0.107$ \\
    $c_{2}$ 
    & $\phantom{-}0.1545 \pm 0.156$ 
    & $\phantom{-}0.153 \pm 0.159$ 
    & $\phantom{-}0.07084 \pm 0.178$ 
    & $-0.5302 \pm 0.184$ \\
    $c_{4}$ 
    & $-0.1949 \pm 0.11$ 
    & $-0.2521 \pm 0.106$ 
    & $-0.1554 \pm 0.117$ 
    & $\phantom{-}0.1098 \pm 0.117$ \\
    $b_4^{\times}$ 
    & $-11.26 \pm 3.81$ 
    & $-3.662 \pm 3.68$ 
    & $\phantom{-}4.088 \pm 3.62$ 
    & $\phantom{-}11.2 \pm 3.65$ \\
    \hline
    $P_{\rm shot}^{\rm q}$ 
    & $-1.011 \pm 0.0809$ 
    & $-2.773 \pm 0.126$ 
    & $-0.2253 \pm 0.0332$ 
    & $-0.001836 \pm 0.0273$ \\
    $a_0^{\rm q}$ 
    & $\phantom{-}71.65 \pm 19.2$ 
    & $\phantom{-}87.29 \pm 31.3$ 
    & $\phantom{-}38.71 \pm 7.31$ 
    & $\phantom{-}6.731 \pm 6.55$ \\
    $a_2^{\rm q}$ 
    & $-166.6 \pm 67.1$ 
    & $-208.1 \pm 105$ 
    & $-0.06663 \pm 24.7$ 
    & $-7.069 \pm 21$ \\
    $c_{0}^{\rm q}$ 
    & $-2.699 \pm 1.7$ 
    & $-2.007 \pm 1.53$ 
    & $\phantom{-}3.008 \pm 1.02$ 
    & $\phantom{-}1.615 \pm 0.954$ \\
    $c_{2}^{\rm q}$ 
    & $\phantom{-}10.28 \pm 3.37$ 
    & $\phantom{-}12.15 \pm 3.13$ 
    & $\phantom{-}4.394 \pm 2.04$ 
    & $-2.152 \pm 1.89$ \\
    $c_{4}^{\rm q}$ 
    & $\phantom{-}2.279 \pm 2.73$ 
    & $-0.0832 \pm 2.53$ 
    & $\phantom{-}1.991 \pm 1.64$ 
    & $-1.039 \pm 1.54$ \\
    $b_{\Gamma_3}^{\rm q}$ 
    & $\phantom{-}2.252 \pm 0.0336$ 
    & $\phantom{-}2.38 \pm 0.033$ 
    & $\phantom{-}2.038 \pm 0.0354$ 
    & $\phantom{-}1.28 \pm 0.0354$ \\
    $b_4^{\rm q}$ 
    & $-78.02 \pm 43.2$ 
    & $-38.43 \pm 38.5$ 
    & $\phantom{-}6.535 \pm 35.6$ 
    & $\phantom{-}63.54 \pm 37.1$ \\
    $\chi^2_{\rm marg}$ 
    & $1204.7$
    & $1242.2$
    & $1179.3$
    & $1226.4$ \\
    \hline \hline
    \end{tabular}
    \end{center}
    \caption{\textbf{Joint \Lya and LAE/LBG fits:} Marginalized best-fit EFT parameters obtained from joint fits of the auto- and cross-correlation of the \Lya forest and the LAE/LBG samples. The top section shows the sampled parameters and the middle (bottom) section the ones that we analytically marginalize over for the auto- (cross-) correlation. We use 1400 data points, sample over 13 parameters explicitly (\textit{top block}) and analytically marginalize over 20 (\textit{bottom two blocks}) yielding a marginalized $\chi^2$ that is $\approx 0.85$ for each fit.}
    \label{tab:EFT_bias_lae}
\end{table*}

The galaxy bias parameters recovered from our joint fits agree well
with the field-level measurements of \cite{Sullivan:2025eei} (their
Table~III), who fit the same LBG and LAE selections in the Astrid
hydrodynamical simulations at $z=3$. 
For the linear bias we find
$b_1^{\rm q}\simeq3.8$--$4.1$ (LBGs) and $b_1^{\rm q}\simeq2.0$--$2.1$ (LAEs),
reproducing the expected LBG/LAE hierarchy and lying within or just below their
Astrid range ($b_1\simeq3.9$--$4.2$ and $1.9$--$2.7$, respectively); our
ODIN value $b_1^{\rm q}=1.97$ is moreover consistent with the observed
$b_1^{\rm q}\approx 2.0$. The tidal bias is similarly consistent: we obtain order-unity,
negative values, $b_{\mathcal{G}_2}^{\rm q}\in[-1.44,-0.71]$, in agreement with
\cite{Sullivan:2025eei} given the large recovered uncertainties on these parameters. The quadratic bias shows the largest offset: for the LBG samples we measure $b_2^{\rm q}\simeq10$--$12$,
about twice the reported $\simeq6$. 
Note that the shotnoise value for the galaxies is very small as we already subtract it prior to performing the fits. 

\section{\label{sec:conclusions} Summary and Conclusions}

The \Lya forest is a powerful tracer of the large-scale structure of our Universe. It probes a cosmological volume that remains inaccessible to galaxy surveys until Stage-V spectroscopy \cite{schlegel_megamapper_concept}, thereby extending the reach of perturbative methods. Owing to its higher redshift range ($2 \leq z \leq 5$), the \Lya forest also exhibits rotated degeneracy directions among cosmological parameters. A key advantage of the \Lya forest is the well-understood physics relating the neutral hydrogen distribution to the underlying dark matter, which has enabled accurate hydrodynamical simulations down to sub-kpc scales \cite{McDonald:2001fe, 2013ApJ...765...39A, Lukic:2015, Bolton17, Bird:2023evb}. These simulations are commonly used to calibrate linear-theory-based models augmented by phenomenological fitting functions for the power spectrum (or two-point correlation function) down to scales of a few $\hMpcinv$ \cite{McDonald:2001fe, Arinyo-i-Prats:2015vqa, Givans:2022qgb, mcquinn2011}, resulting in remarkably robust cosmological constraints (see, e.g.,~\cite{Slosar2013, dMdB:2020, DESI_lya_2024}). However, fits to the BAO scaling parameters -- the primary observables for constraining the cosmic expansion history with DESI -- using quasi-linear theory models are biased at the $\approx 0.3\%$ level \cite{deBelsunce:2024rvv, Hadzhiyska:2025cvk}. This is comparable to the forecasted cumulative precision of $\simlt 0.2\%$ when combining all tracers and redshift bins \cite{DESI:2016} indicating that existing modeling frameworks are approaching their limits.

Recent advances in theoretical modeling using the effective field theory (EFT) of large-scale structure, extended to the \Lya forest \cite{Ivanov:2024lya, Chudaykin:2025gsh, Belsunce_Sullivan_skewspectrum}, enable a consistent description of the \Lya forest from large to intermediate scales. In a companion \textit{Letter} \cite{deBelsunce:2025bqc}, we demonstrated that the same bias expansion can be used to successfully model the \Lya forest flux decrement and halo densities at the field level. In this work, we present the theoretical framework underlying this approach and summarize our main conclusions as follows.

\begin{itemize}

\item \textbf{Field-level modeling of the \Lya forest.}  
The EFT framework can accurately model the \Lya forest flux decrement at the field level. Using an
 EFT perturbative model, we find agreement at the $\leq 5\%$ level between the modeled (and fitted) and simulated power spectra down to $\kmax \leq 1.0\hMpcinv$ for the 3D and 1D power spectra, for a range of redshifts $z=2.0-3.2$. The corresponding one-point probability distribution function agrees down to cell sizes of $\approx 2\hinvMpc$, capturing information beyond the two-point function. Importantly, this forward model reproduces \emph{all} amplitudes and phases across \emph{all} modes, representing a substantially more stringent test of accuracy than comparisons based solely on mode-averaged statistics. More quantitatively, the forward-modeled and simulated 1D and 3D fields are correlated at the $>99\%$ level for $k \simlt 0.3\hMpcinv$ and at the $>95\%$ level down to $k \simlt 1.0\hMpcinv$.
 A high precision description
 of the \Lya data for $k\gtrsim 1\hMpcinv$
 must include a consistent modeling of the \Lya stochasticity (the analog of the one-halo term),
 which we have 
 quantified 
 in detail for the first time. 
 Note that the redshifts studied in our current analysis are lower than those currently utilized 
 in the actual EFT-based 
 analyses of the \Lya P1D data~\cite{Ivanov:2024jtl,He:2025jwp,Ivanov:2025pbu,Parashari:2026dxo}. A steep  
 increase of the non-linear 
 wavenumber (proxy to the 
 EFT cutoff) at 
 $z > 3$ suggests that 
 the field-level EFT 
 modeling should be 
 applicable to 
 an even wider range of scales 
 at these redshifts~\cite{Ivanov:2023yla,Ivanov:2025pbu}.

\item \textbf{Cross-correlations with halos.}  
The same EFT framework -- without the line-of-sight-dependent operators specific to the \Lya forest -- can accurately model the cross-correlation with massive dark matter halos, used here as proxies for high-redshift galaxies, up to the shot-noise limit. For both simulations, we find agreement at the $\leq 5\%$ level between the modeled and simulated power spectra down to $\kmax \leq 1.0\hMpcinv$ for the 3D cross power, using all available halo masses. 
Abacus uses a mass range of $10.8 \leq \log_{10}(M/(h^{-1}M_\odot)) \leq 14.2$ approximately yielding a linear bias matching DESI observations $b_q\approx 3.3$ and a number density of $\sim 1.75\times 10^{-4} \ (\hinvMpc)^{-3}$. Sherwood uses halo in a mass range of $10^9 \leq M_\odot \simlt 10^{14}$.

\item \textbf{Linear theory modeling at the field level.}  
We find that a linear theory forward model fails to generate the \Lya forest flux decrement at the field level. While the model and simulation power spectra agree at the $5$--$10\%$ level down to $k \simlt 0.1-0.3\hMpcinv$, the corresponding error power spectrum only captures these large scales at the $3-5\%$ level in this regime. This implies that fits to the power spectrum (or two-point correlation function) using a linear theory model effectively fit noise. The discrepancy is even more pronounced for the one-point probability distribution functions, which differ drastically between the linear theory fields and the simulations, even for large cell sizes of $R = 10$ and $30\hinvMpc$. Indeed, one would obtain a formally perfect match to the power spectrum using the forward model $\delta_F^{\rm model} = (P_{\rm F}(k,\mu)/P_{\rm lin}(k))^{1/2} \delta_1(\k)$ where $P_{\rm F}(k,\mu)$ is the non-linear \Lya forest power spectrum. At the field level, the form of this construction is equivalent to 
that of the
linear model employed here, yet it fails beyond scales of $k \simgt 0.1\hMpcinv$. 
These findings highlight the necessity of a higher-order bias expansion to capture non-linearities in the data already on quasi-linear scales.

\item \textbf{Large-volume mock generation and DESI-II relevance.}  
We construct large-volume ($V = 2^3(\hinvGpc)^3$) clustering mocks for the \Lya forest and its cross-correlation with high-redshift galaxies, such as \Lya emitters (LAEs), a key tracer for the near-future DESI-II survey. These mocks are calibrated on hydrodynamical simulations and exploit the computational efficiency of the perturbative forward model presented here, enabling the generation of large ensembles over cosmological volumes. This capability is particularly relevant given that current \Lya forest analyses \cite{DESI_lya_2024, Cuceu:2025nvl} rely on approximate lognormal mocks \cite{Ramirez-Perez:2021cpq} that fail to capture the physics shaping the BAO feature \cite{Chen:2024tfp, deBelsunce:2024rvv}. To validate these semi-analytic mocks, we perform a BAO analysis at fixed cosmology and recover unbiased BAO scaling parameters, making them applicable for validating inference pipelines. 
\end{itemize}

Our approach bridges the gap between theory and simulation by enabling efficient modeling of the flux decrement and the dark matter halo  field directly at the field level, rather than restricting the analysis to summary statistics. This framework naturally opens several promising avenues for future work: To extend the perturbative reach of our approach, transfer functions can be computed fully within the EFT framework; beyond $\beta_1$, this will require two-loop calculations along the lines developed for halos in Ref.~\cite{Abidi:2018eyd}. The present semi-analytic mocks are well suited for validating inference pipelines and constructing simulation-based covariance matrices for DESI analyses. Extending these mocks to the light cone is therefore a natural next step paired with a detailed investigation of their properties, as well as data-driven forecasts of effects such as radiative transfer measured via the cross-correlation between the \Lya forest and high-redshift galaxies.  Further directions include the development of simulation-based priors for EFT-based full-shape analyses along the lines of~\cite{Ivanov:2024hgq,Ivanov:2024xgb,Ivanov:2024dgv,Cabass:2024wob,Akitsu:2023eqa,Akitsu:2024lyt,Sullivan:2025eei,Chudaykin:2026nls}, and the 
application of 
our technique to the modeling of the \Lya
field at high redshifts, $3.2\lesssim z\lesssim 5.4$. 
The latter is especially important 
since the \Lya data at these redshifts
is especially  
powerful at constraining new physics with EFT,
making the development of 
robust simulation-based
priors for such analysis 
a priority
~\cite{Ivanov:2024jtl,He:2025jwp,Ivanov:2025pbu,Parashari:2026dxo}.
Finally, it would be interesting to study 
cosmological parameter
inference directly from Fourier modes~\cite{2019A&A...625A..64J,Lavaux:2019fjr,Nguyen:2020hxe,Millea:2020cpw,Porqueres:2020qgy,Tsaprazi:2021mft,Andrews:2022nvv,Babic:2022dws,Bayer:2022vid,Boruah:2022lsu,Zhou:2023ezg,Stadler:2023hea,Porqueres:2023drp,Beyond-2pt:2024mqz,Stadler:2024aff,Stadler:2024fui,Babic:2025fgv,Peron:2025lgh,Nguyen:2024yth,Akitsu:2025boy}. 
While the latter remains debated for galaxy surveys, particularly regarding the information content beyond the power spectrum and bispectrum~\cite{Spezzati:2025zsb,Akitsu:2025boy},
it is well motivated and worth investigating for the \Lya forest. We leave these exciting research directions for future investigation.

\acknowledgments
We thank Martin White and Pat McDonald for fruitful discussions. We are also grateful to Jamie Bolton, Jon{\'a}s Chaves-Montero, Jahmour Givans, Julien Guy, Naim G\"oksel Kara{\c{c}}ayl{\i}, and Andreu Font-Ribera for valuable discussions and assistance with the Sherwood files.

This work is supported by the National Science Foundation under Cooperative Agreement PHY-2019786 (The NSF AI Institute for Artificial Intelligence and Fundamental Interactions, \url{http://iaifi.org/}). 
This research used resources of the National Energy Research Scientific Computing Center (NERSC), a U.S. Department of Energy Office of Science User Facility operated under Contract No. DE-AC02-05CH11231.
JMS acknowledges that, in part, support for this work was provided by The Brinson Foundation through a Brinson Prize.
KA acknowledges supports from Fostering Joint International Research (B) under Contract No. 21KK0050 and the Japan Society for the Promotion of Science (JSPS) KAKENHI Grant No. JP24K17056. Support for
this work was provided by NASA through the NASA
Hubble Fellowship grant HST-HF2-51572.001 awarded
by the Space Telescope Science Institute, which is operated by the Association of Universities for Research in
Astronomy, Inc., for NASA, under contract NAS5-26555.
\appendix

\section{Transfer functions from Sherwood simulations} \label{app:Sherwood_tf_80}
For the mocks correlating the \Lya forest with high redshift galaxies in Sec.~\ref{sec:lya_x_LBG_LAE}, we use the $L=80\hinvMpc$ snapshots since these are available at a series of redshifts $z=2.0,\,2.4,\,2.8,\,3.2$ which we interpolate between redshifts.

\begin{table*}
\centering
\begin{tabular}{l|cccccccc}
\hline\hline
TF & $c_0$ & $c_{01}$ & $c_1$ & $c_{12}$ & $c_{14}$ & $c_4$ & $c_{22}$ & $c_{44}$ \\
\hline $\mathbf{z=2.0}$ &&&&&&\\
$\beta_1$ & $-0.123$ & $-0.265$ & $\phantom{-}0.077$ & $\phantom{-}0.733$ & $-0.290$ & $-0.005$ & $-0.546$ & $\phantom{-}0.298$ \\
$\beta_2$ & $\phantom{-}0.104$ & $\phantom{-}0.124$ & $-0.028$ & $-0.076$ & $\phantom{-}0.014$ & $-0.001$ & $\phantom{-}0.026$ & $-0.039$ \\
$\beta_{\mathcal{G}_2}$ & $-0.088$ & $-0.360$ & $\phantom{-}0.026$ & $\phantom{-}0.531$ & $-0.153$ & $-0.023$ & $-0.336$ & $\phantom{-}0.139$ \\
$\beta_3$ & $-0.018$ & $-0.007$ & $\phantom{-}0.032$ & $\phantom{-}0.090$ & $-0.108$ & $-0.010$ & $-0.046$ & $\phantom{-}0.106$ \\
$\beta_{KK_\parallel}$ & $-0.115$ & $-0.127$ & $-0.126$ & $-0.206$ & $\phantom{-}0.362$ & $\phantom{-}0.064$ & $\phantom{-}0.123$ & $-0.190$ \\
$\beta_{\eta}$ & $-0.172$ & $-0.215$ & $-0.046$ & $\phantom{-}1.351$ & $-0.599$ & $\phantom{-}0.046$ & $-0.816$ & $\phantom{-}0.233$ \\
$\beta_{\eta^2}$ & $\phantom{-}0.043$ & $\phantom{-}0.038$ & $-0.122$ & $\phantom{-}0.300$ & $-0.001$ & $\phantom{-}0.046$ & $-0.272$ & $\phantom{-}0.095$ \\
$\beta_{\delta\eta}$ & $\phantom{-}0.051$ & $-0.346$ & $-0.077$ & $\phantom{-}1.878$ & $-0.604$ & $\phantom{-}0.094$ & $-1.957$ & $\phantom{-}1.188$\\
\hline $\mathbf{z=2.4}$ &&&&&&\\
$\beta_1$ & $-0.171$ & $-0.385$ & $\phantom{-}0.090$ & $\phantom{-}1.042$ & $-0.350$ & $\phantom{-}0.009$ & $-0.777$ & $\phantom{-}0.368$ \\
$\beta_2$ & $\phantom{-}0.158$ & $\phantom{-}0.180$ & $-0.040$ & $-0.078$ & $\phantom{-}0.006$ & $-0.001$ & $-0.016$ & $\phantom{-}0.012$ \\
$\beta_{\mathcal{G}_2}$ & $-0.118$ & $-0.500$ & $\phantom{-}0.017$ & $\phantom{-}0.899$ & $-0.314$ & $-0.020$ & $-0.618$ & $\phantom{-}0.324$ \\
$\beta_3$ & $-0.036$ & $\phantom{-}0.006$ & $\phantom{-}0.053$ & $\phantom{-}0.043$ & $-0.110$ & $-0.012$ & $-0.009$ & $\phantom{-}0.133$ \\
$\beta_{KK_\parallel}$ & $-0.157$ & $-0.215$ & $-0.259$ & $-0.340$ & $\phantom{-}0.642$ & $\phantom{-}0.121$ & $\phantom{-}0.110$ & $-0.216$ \\
$\beta_{\eta}$ & $-0.331$ & $-0.254$ & $-0.083$ & $\phantom{-}2.338$ & $-1.127$ & $\phantom{-}0.069$ & $-1.384$ & $\phantom{-}0.420$ \\
$\beta_{\eta^2}$ & $\phantom{-}0.103$ & $\phantom{-}0.024$ & $-0.235$ & $\phantom{-}0.696$ & $\phantom{-}0.059$ & $\phantom{-}0.086$ & $-0.710$ & $\phantom{-}0.329$ \\
$\beta_{\delta\eta}$ & $\phantom{-}0.013$ & $-0.492$ & $-0.033$ & $\phantom{-}2.802$ & $-0.873$ & $\phantom{-}0.091$ & $-2.998$ & $\phantom{-}1.818$ \\
\hline $\mathbf{z=2.8}$ &&&&&&\\
$\beta_1$
& $-0.231$ & $-0.536$ & $\phantom{-}0.100$ & $\phantom{-}1.483$
& $-0.422$ & $\phantom{-}0.033$ & $-1.113$ & $\phantom{-}0.451$ \\

$\beta_2$
& $\phantom{-}0.247$ & $\phantom{-}0.250$ & $-0.061$ & $-0.080$
& $-0.013$ & $\phantom{-}0.001$ & $-0.069$ & $\phantom{-}0.074$ \\

$\beta_{G2}$
& $-0.168$ & $-0.591$ & $\phantom{-}0.024$ & $\phantom{-}1.149$
& $-0.494$ & $-0.030$ & $-0.822$ & $\phantom{-}0.556$ \\

$\beta_3$
& $-0.075$ & $\phantom{-}0.007$ & $\phantom{-}0.090$ & $-0.005$
& $-0.050$ & $-0.017$ & $\phantom{-}0.051$ & $\phantom{-}0.098$ \\

$\beta_{KK\parallel}$
& $-0.156$ & $-0.380$ & $-0.553$ & $-0.046$
& $\phantom{-}0.803$ & $\phantom{-}0.234$ & $-0.373$ & $\phantom{-}0.090$ \\

$\beta_{\eta}$
& $-0.607$ & $-0.239$ & $-0.194$ & $\phantom{-}4.289$
& $-2.365$ & $\phantom{-}0.097$ & $-2.494$ & $\phantom{-}0.937$ \\

$\beta_{\eta^2}$
& $\phantom{-}0.230$ & $-0.077$ & $-0.406$ & $\phantom{-}1.504$
& $\phantom{-}0.220$ & $\phantom{-}0.144$ & $-1.533$ & $\phantom{-}0.704$ \\

$\beta_{\delta\eta}$
& $-0.105$ & $-0.589$ & $\phantom{-}0.083$ & $\phantom{-}3.623$
& $-1.154$ & $\phantom{-}0.062$ & $-4.100$ & $\phantom{-}2.611$ \\
\hline $\mathbf{z=3.2}$ &&&&&&\\
$\beta_1$ & $-0.310$ & $-0.714$ & $\phantom{-}0.113$ & $\phantom{-}2.108$ & $-0.521$ & $\phantom{-}0.069$ & $-1.602$ & $\phantom{-}0.566$ \\
$\beta_2$ & $\phantom{-}0.393$ & $\phantom{-}0.322$ & $-0.103$ & $-0.038$ & $-0.063$ & $\phantom{-}0.012$ & $-0.168$ & $\phantom{-}0.145$ \\
$\beta_{\mathcal{G}_2}$ & $-0.240$ & $-0.587$ & $\phantom{-}0.046$ & $\phantom{-}1.117$ & $-0.622$ & $-0.058$ & $-0.785$ & $\phantom{-}0.726$ \\
$\beta_3$ & $-0.169$ & $\phantom{-}0.035$ & $\phantom{-}0.192$ & $-0.252$ & $\phantom{-}0.147$ & $-0.051$ & $\phantom{-}0.312$ & $-0.086$ \\
$\beta_{KK_\parallel}$ & $-0.050$ & $-0.703$ & $-1.146$ & $\phantom{-}1.645$ & $\phantom{-}0.393$ & $\phantom{-}0.476$ & $-2.141$ & $\phantom{-}1.082$ \\
$\beta_{\eta}$ & $-0.968$ & $-0.132$ & $-0.627$ & $\phantom{-}7.547$ & $-4.960$ & $\phantom{-}0.190$ & $-3.648$ & $\phantom{-}1.495$ \\
$\beta_{\eta^2}$ & $\phantom{-}0.461$ & $-0.335$ & $-0.639$ & $\phantom{-}2.958$ & $\phantom{-}0.543$ & $\phantom{-}0.210$ & $-2.974$ & $\phantom{-}1.356$ \\
$\beta_{\delta\eta}$ & $-0.332$ & $-0.650$ & $\phantom{-}0.260$ & $\phantom{-}4.658$ & $-1.677$ & $\phantom{-}0.053$ & $-5.621$ & $\phantom{-}3.806$ \\
\hline\hline
\end{tabular}\label{tab:sherwood_80_z20}
\caption{\textbf{Best-Fit Sherwood Transfer Function:} Sherwood snapshots at $z=2.0,\,2.4,\,2.8$ and $z=3.2$ for the $80\hinvMpc$ box.}
\end{table*}

\begin{table}
\centering
\begin{tabular}{lcccccc}
\hline\hline
$z$& $a_0$ & $a_2$ & $a_3$ & $a_4$ & $a_{22}$ &  $a_{44}$ \\
\hline
$2.0$ & $\phantom{-}0.018$ & $\phantom{-}0.590$ & $-1.908$ & $\phantom{-}1.482$ & $\phantom{-}0.229$ & $\phantom{-}0.013$ \\
$2.4$ & $\phantom{-}0.028$ & $\phantom{-}1.181$ & $-3.816$ & $\phantom{-}3.037$ & $\phantom{-}0.295$ & $\phantom{-}0.058$ \\
$2.8$ & $\phantom{-}0.089$ & $\phantom{-}0.422$ & $-1.158$ & $\phantom{-}0.694$ & $\phantom{-}0.272$ & $-0.265$ \\
$3.2$ & $\phantom{-}0.030$ & $\phantom{-}5.860$ & $-17.706$ & $\phantom{-}14.405$ & $\phantom{-}0.835$ & $-1.620$ \\
\hline \hline
\end{tabular}
\caption{\textbf{Error Power Spectrum Sherwood:} Fitted parameters for $P_{\rm err}$ for the different snapshots in the $L=80\hinvMpc$ Sherwood simulation.}
\end{table}

\section{Field-level fits on different Abacus FGPA mocks}\label{app:model_dependence_abacus}
\begin{figure*}
    \centering
    \includegraphics[width=0.329\linewidth]{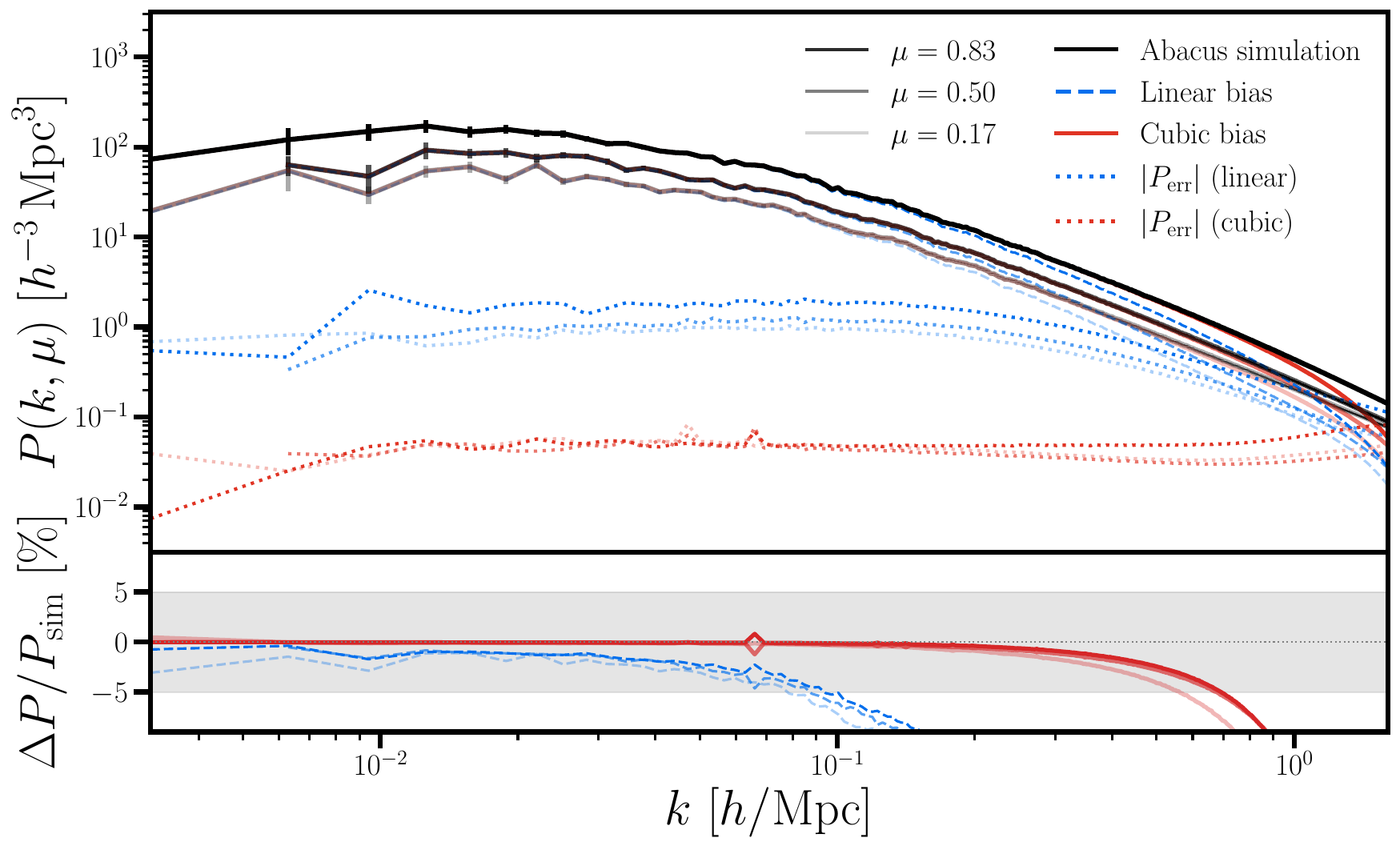}\hfill
    \includegraphics[width=0.329\linewidth]{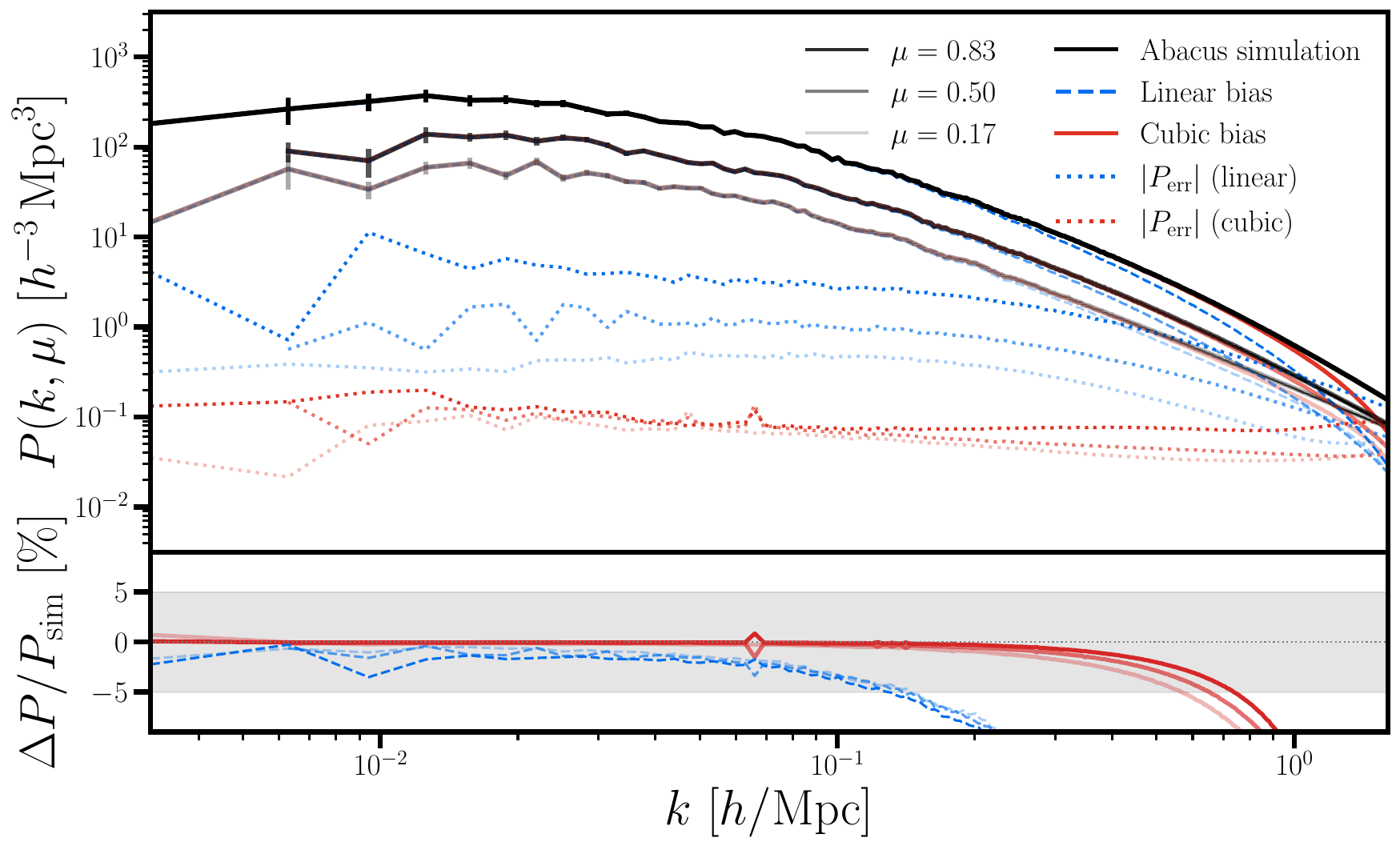}\hfill
    \includegraphics[width=0.329\linewidth]{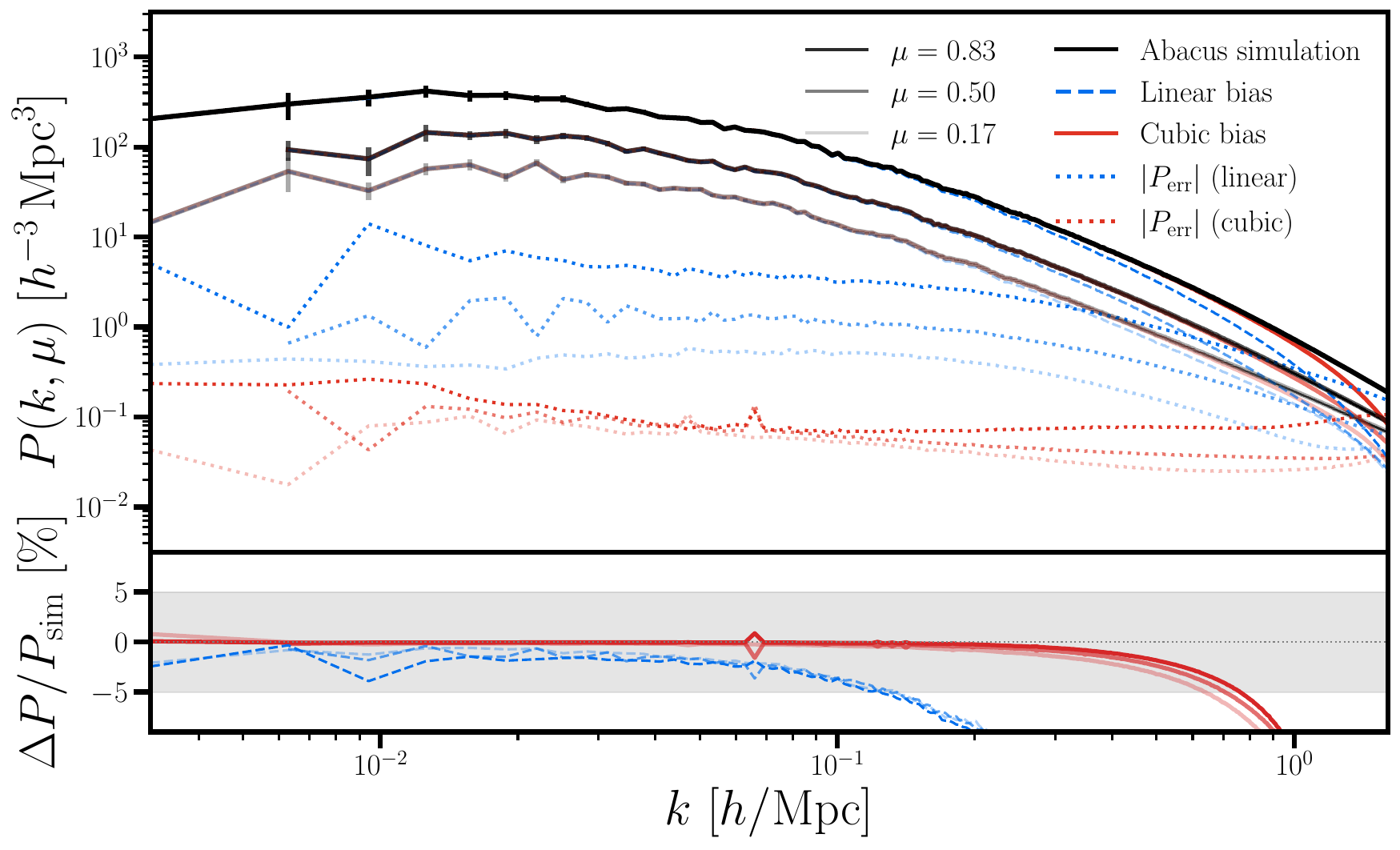}\hfill
    \vspace{-0.1in}
    \caption{\textbf{Model-dependence of Power Spectra:}
    Same as Fig.~\ref{fig:lya_pk} for three different FGPA implementations. Models one, two and three are shown from left to right comparing the linear to the cubic model as well as their error power spectra. Our perturbative forward model is robust to the details of the Abacus simulations, finding a consistent agreement at the 5\% level between the forward model and the input simulation. The scale and orientation dependence of the error power spectrum does, however, change with increasing $b_\eta$ (which we use as proxy for a larger amount of RSD in the simulations). The bottom panel displays the percent difference between the simulation and model power spectra. 
    A gray band highlights the $\pm5\%$ region in the bottom panel.
    }
    \label{fig:abacus_pk_models}
\end{figure*}

To assess the robustness of the outlined analytical forward model, we apply it to three additional implementations of the FGPA procedure (see \cite{Hadzhiyska:2023} for more details) representing different combinations of bias parameters. These have been fitted using linear theory in Ref.~\cite{Hadzhiyska:2023} and using the one-loop power spectrum in Ref.~\cite{Hadzhiyska:2025cvk}. Whilst we have so far used model `one' which is similar to model `two', i.e. having a low value of $b_\eta$ we also compare to realizations (model `three' and `four') with large values of $b_\eta$. Qualitatively the results look very similar and are shown for the \Lya power spectrum in Fig.~\ref{fig:abacus_pk_models} and their cross-correlation coefficient in Fig.~\ref{fig:abacus_rcc_models}. The values obtained from the polynomial fits are tabulated in Tab.~\ref{tab:abacus_tf_results_2}. In Fig.~\ref{fig:abacus_p1d_models} we show the model dependence for the P1D performance. Note that we find the largest differences between the four models for the P1D performance. In particular, for models III and IV where the velocity bias value is larger (and closer to the data) the linear theory model performs better than for both models with low $b_\eta$. 

\begin{figure*}
    \centering
    \includegraphics[width=0.329\linewidth]{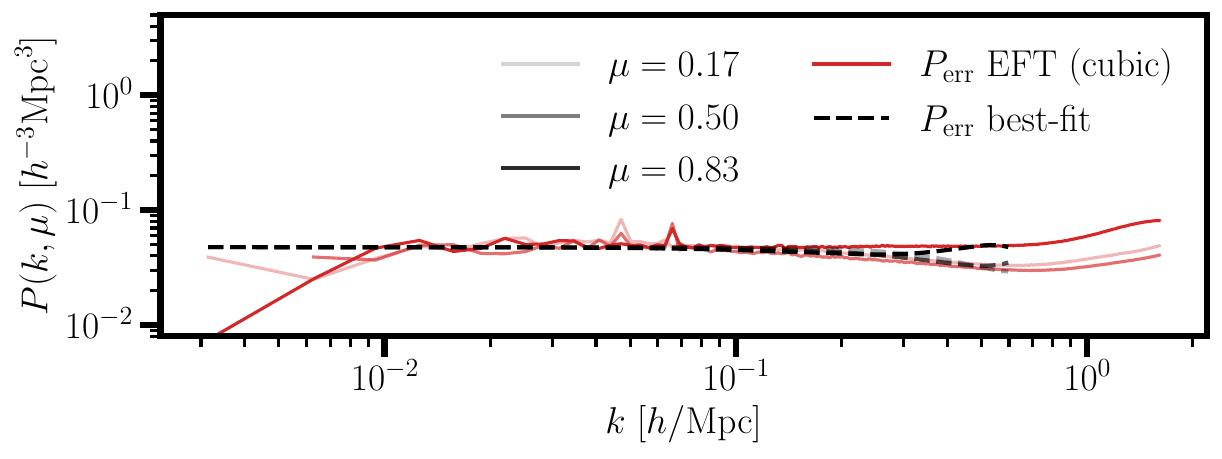}\hfill
    \includegraphics[width=0.329\linewidth]{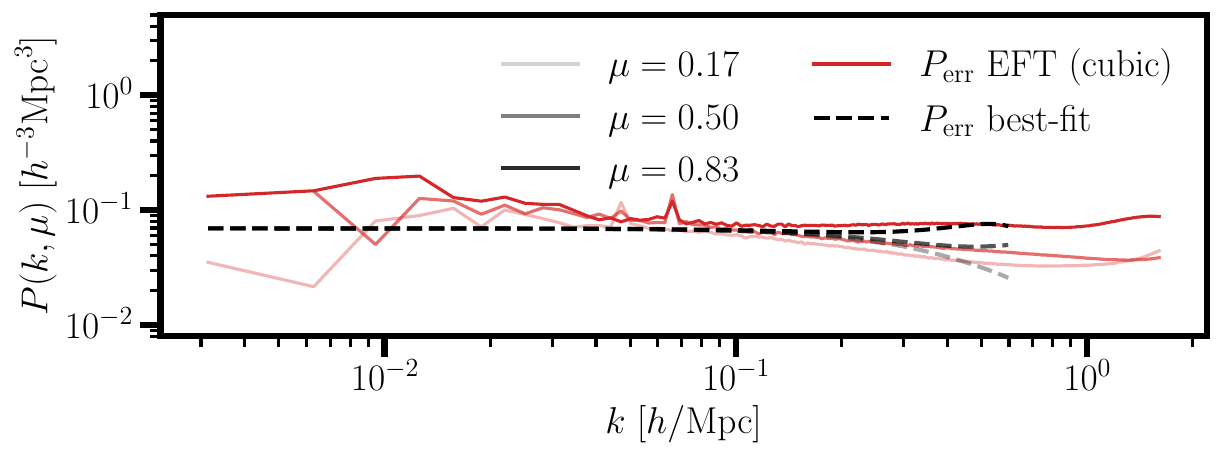}\hfill
    \includegraphics[width=0.329\linewidth]{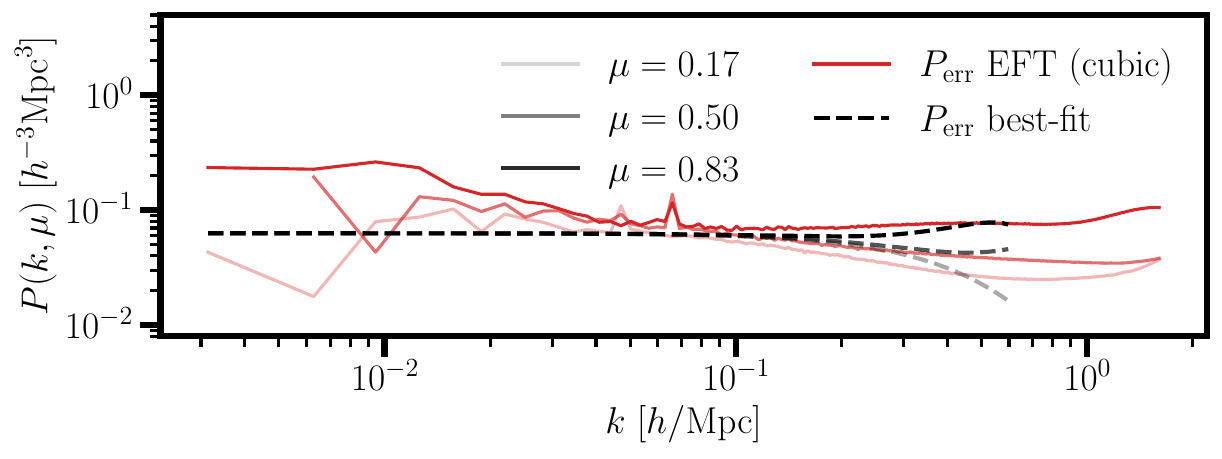}\hfill
    \vspace{-0.1in}
    \caption{\textbf{Model-dependence of best-fit $P_{\rm err}$:}
    Same as Fig.~\ref{fig:abacus_pk_models} showing the best-fit model for the error power spectrum for models two, three and four. The fits use $\kmax=0.6\hMpcinv$ and show a mild onset of scale and orientation dependence of the error power spectrum at around $0.2-0.3\hMpcinv$. 
    }
    \label{fig:abacus_bestfit_perr_models}
\end{figure*}

\begin{figure*}
    \centering
    \includegraphics[width=0.329\linewidth]{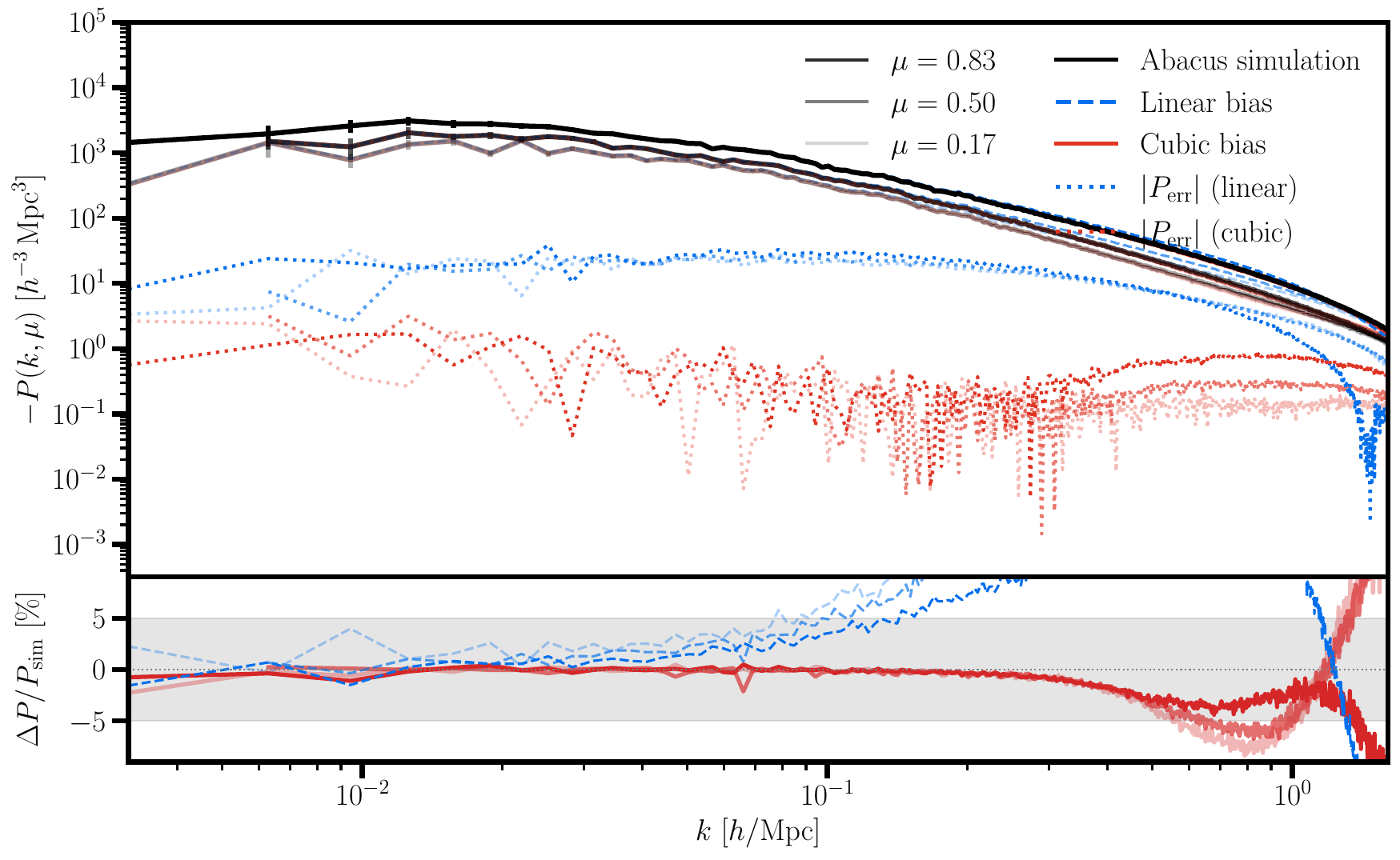}\hfill
    \includegraphics[width=0.329\linewidth]{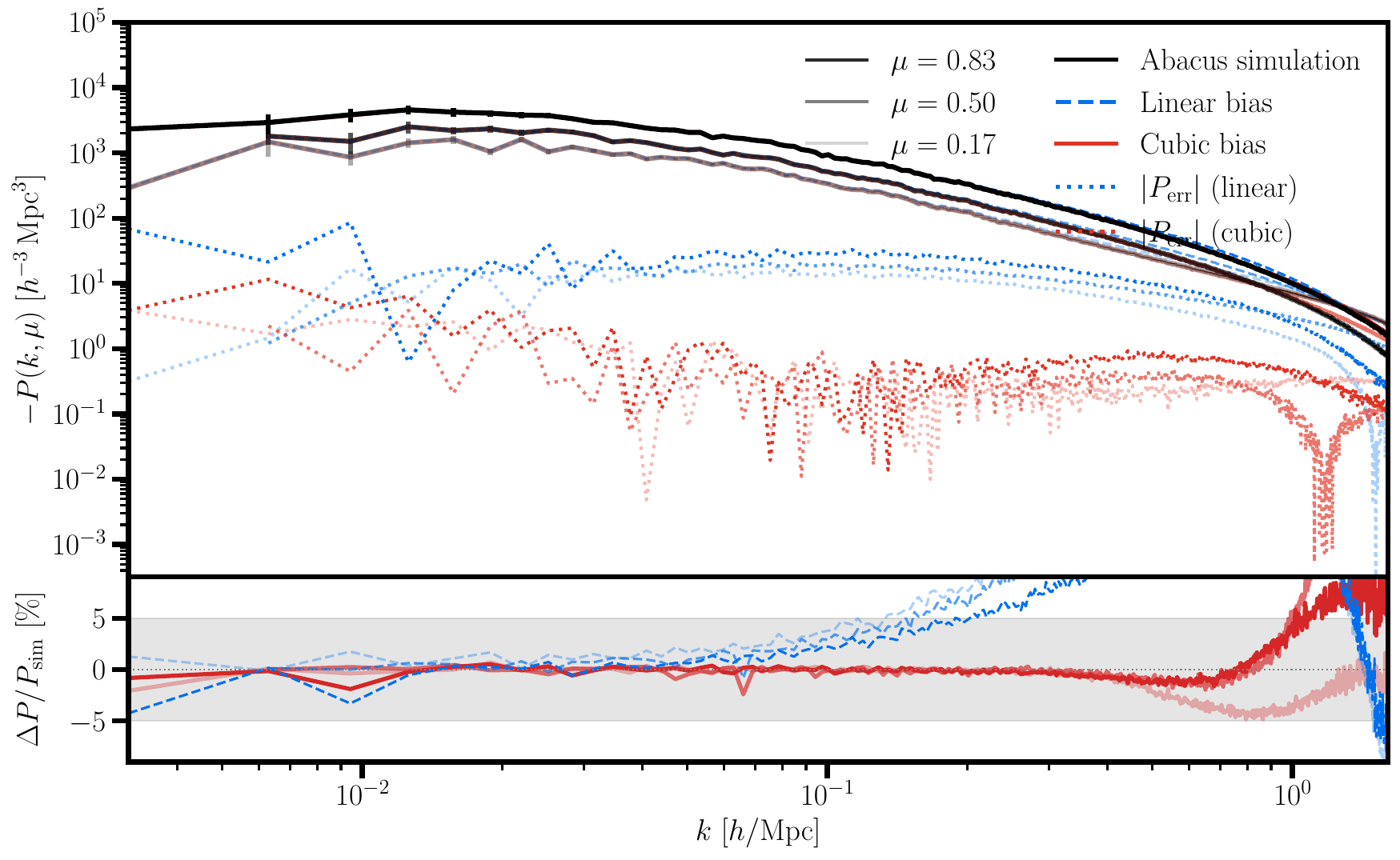}\hfill
    \includegraphics[width=0.329\linewidth]{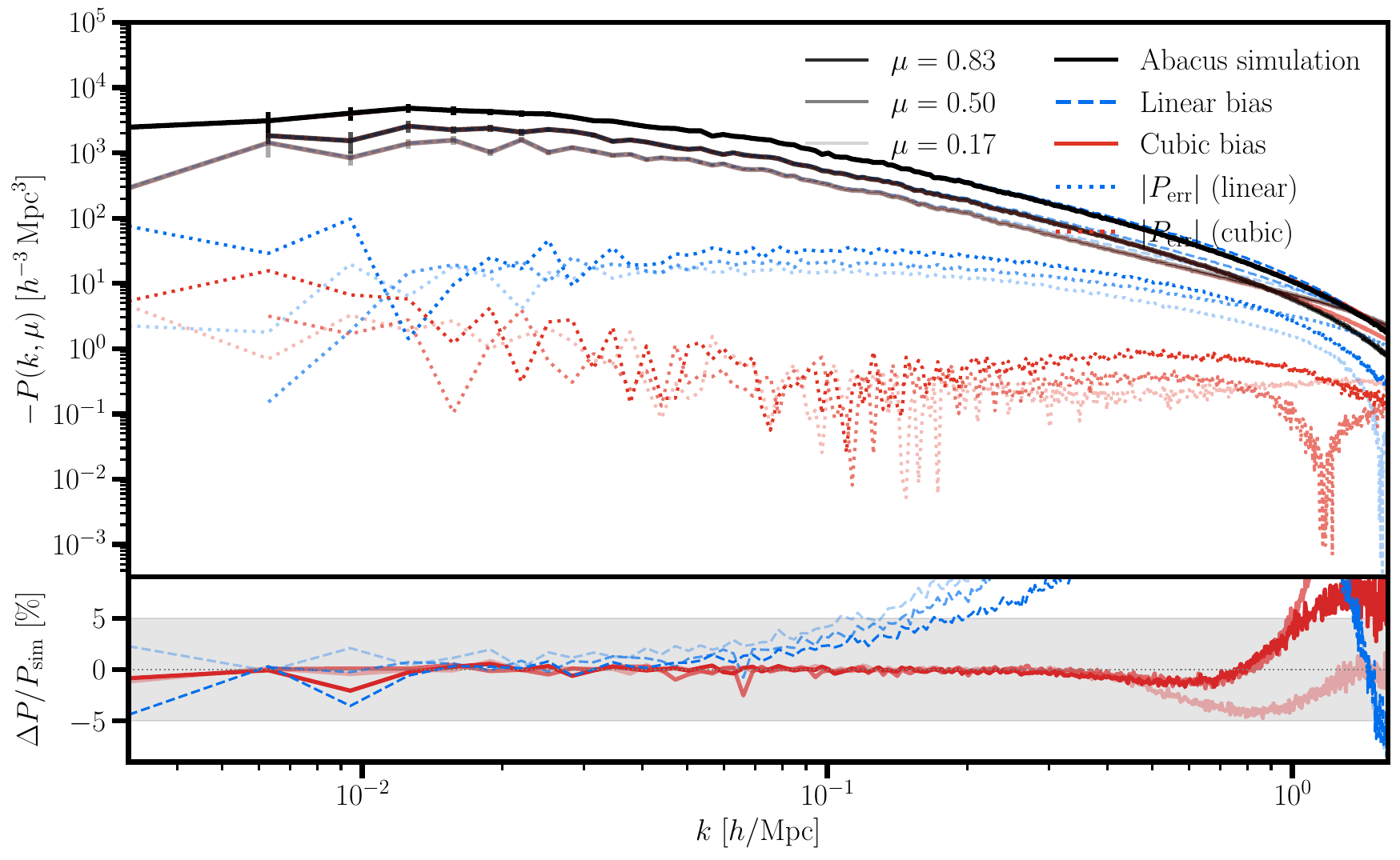}\hfill
    \vspace{-0.1in}
    \caption{\textbf{Model-dependence of Power Spectra:}
    Same as Fig.~\ref{fig:lya_halo_pk} for three different FGPA implementations, denoted by models II-IV.
    }
    \label{fig:abacus_pk_models_cross}
\end{figure*}

\begin{figure*}
    \centering
    \includegraphics[width=\linewidth]{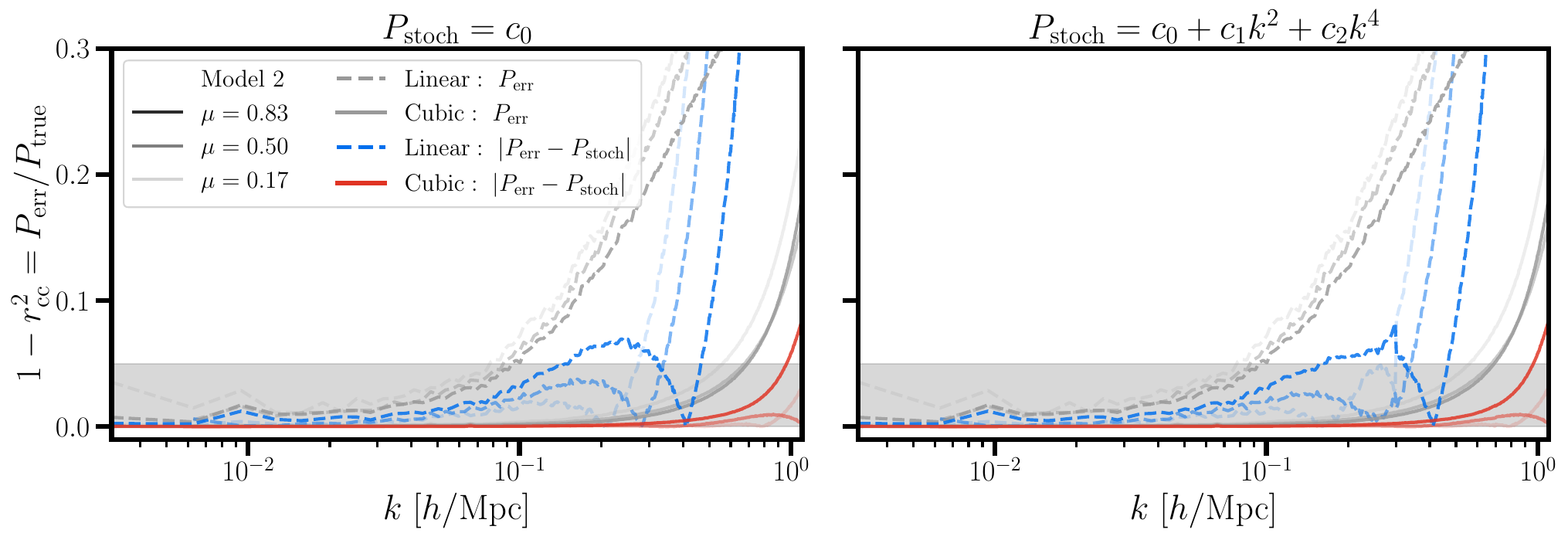}\\
    \includegraphics[width=\linewidth]{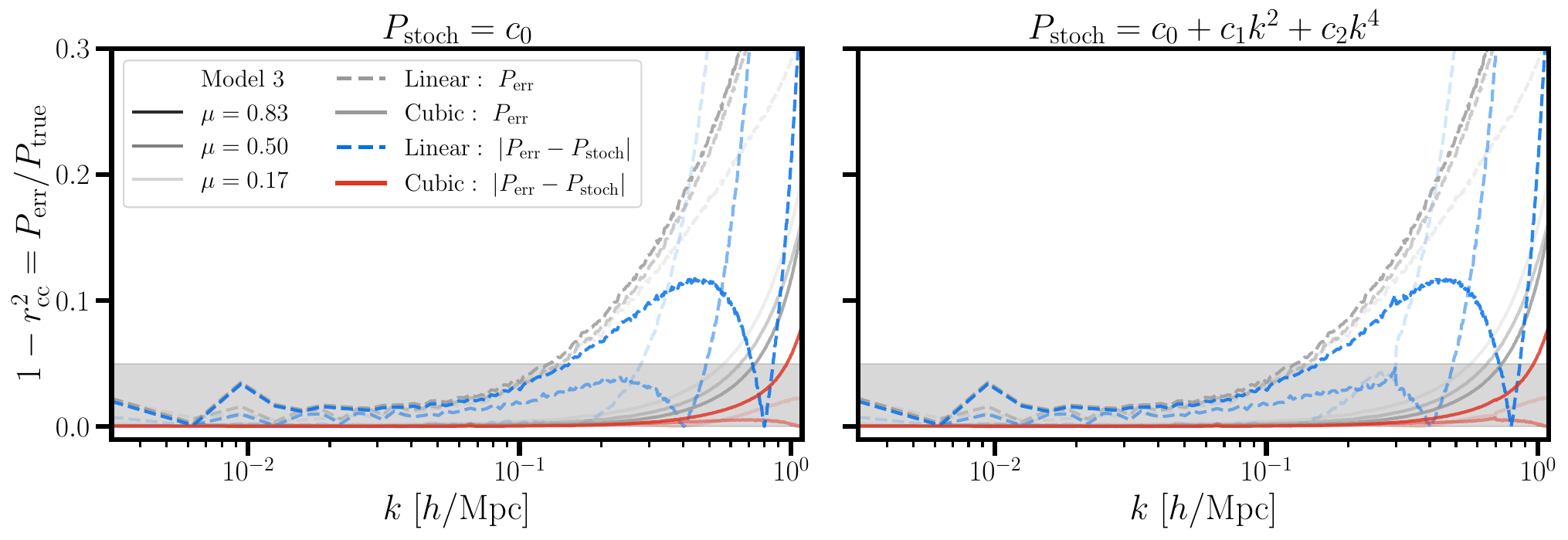}\\
    \includegraphics[width=\linewidth]{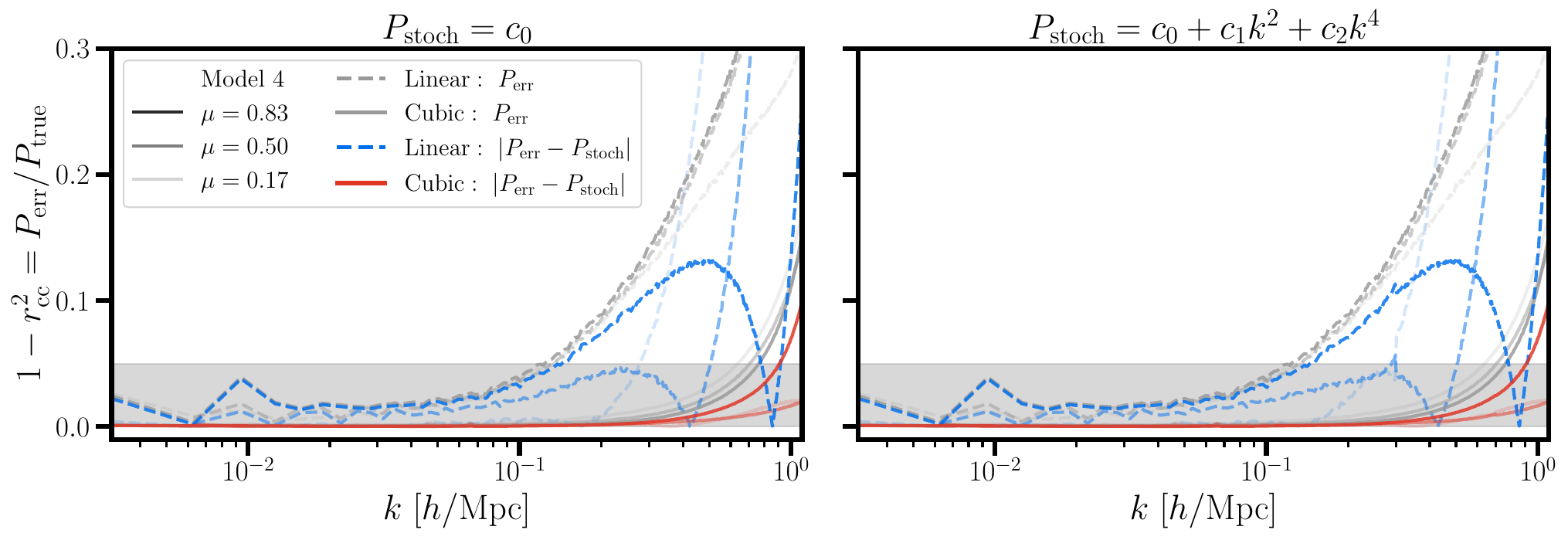}\\
    \vspace{-0.1in}
    \caption{\textbf{Model-dependence of $1-r_{\rm cc}^2$:}
    Same as Fig.~\ref{fig:rcc_cross} showing the cross-correlation coefficient for models II-IV. 
    }
    \label{fig:abacus_bestfit_perr_models_cross}
\end{figure*}

\begin{figure*}
    \centering
    \includegraphics[width=0.329\linewidth]{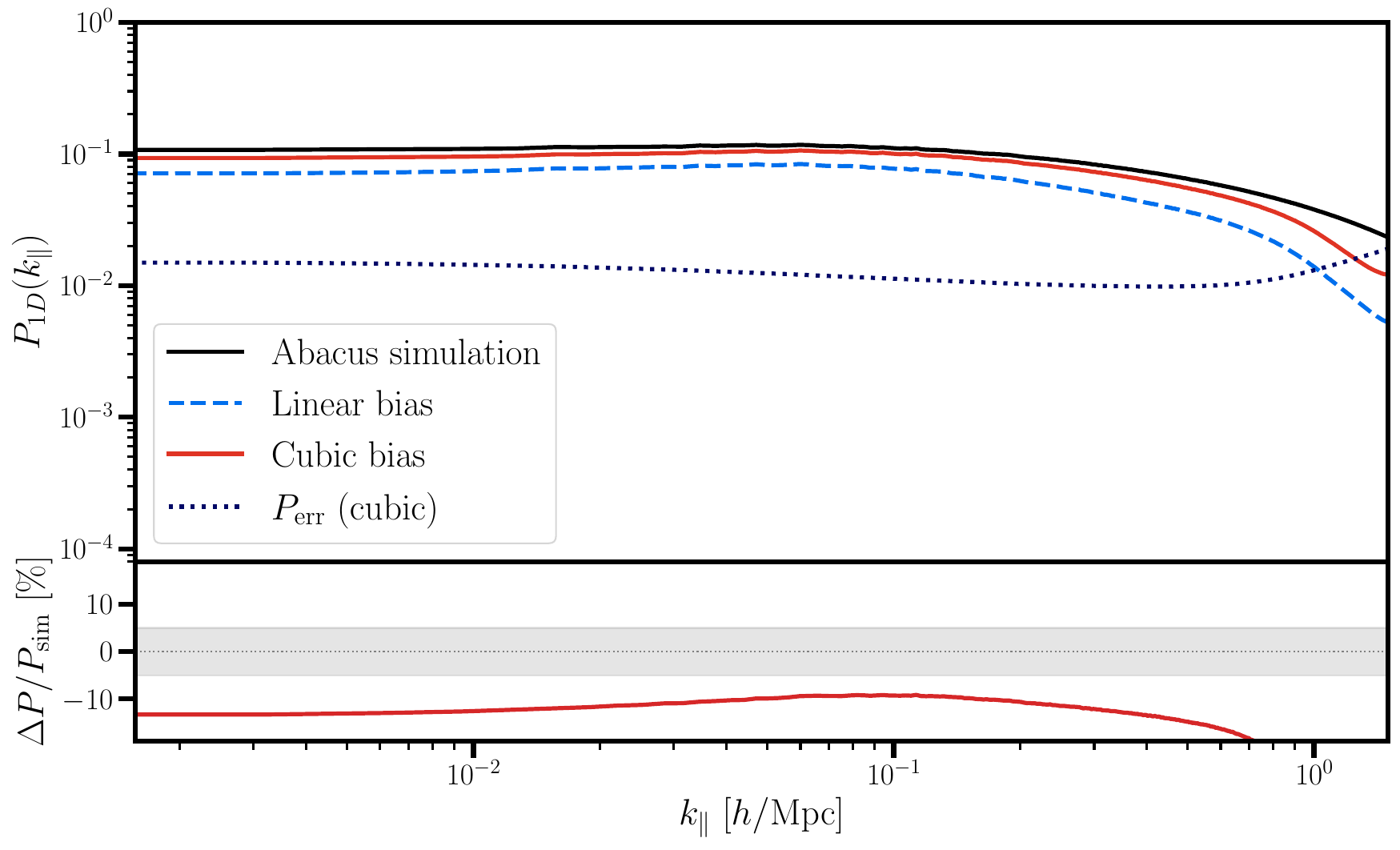}\hfill
    \includegraphics[width=0.329\linewidth]{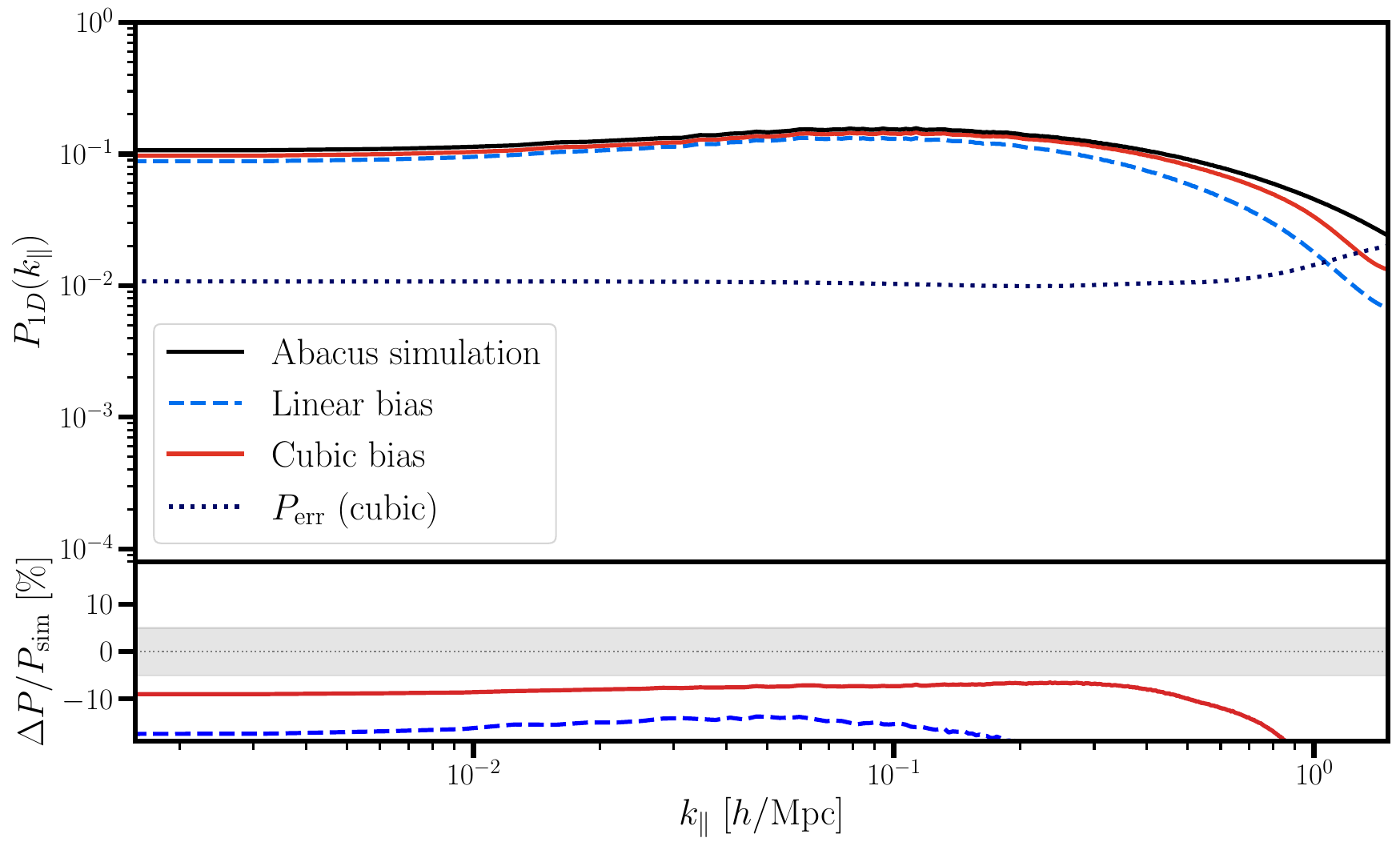}\hfill
    \includegraphics[width=0.329\linewidth]{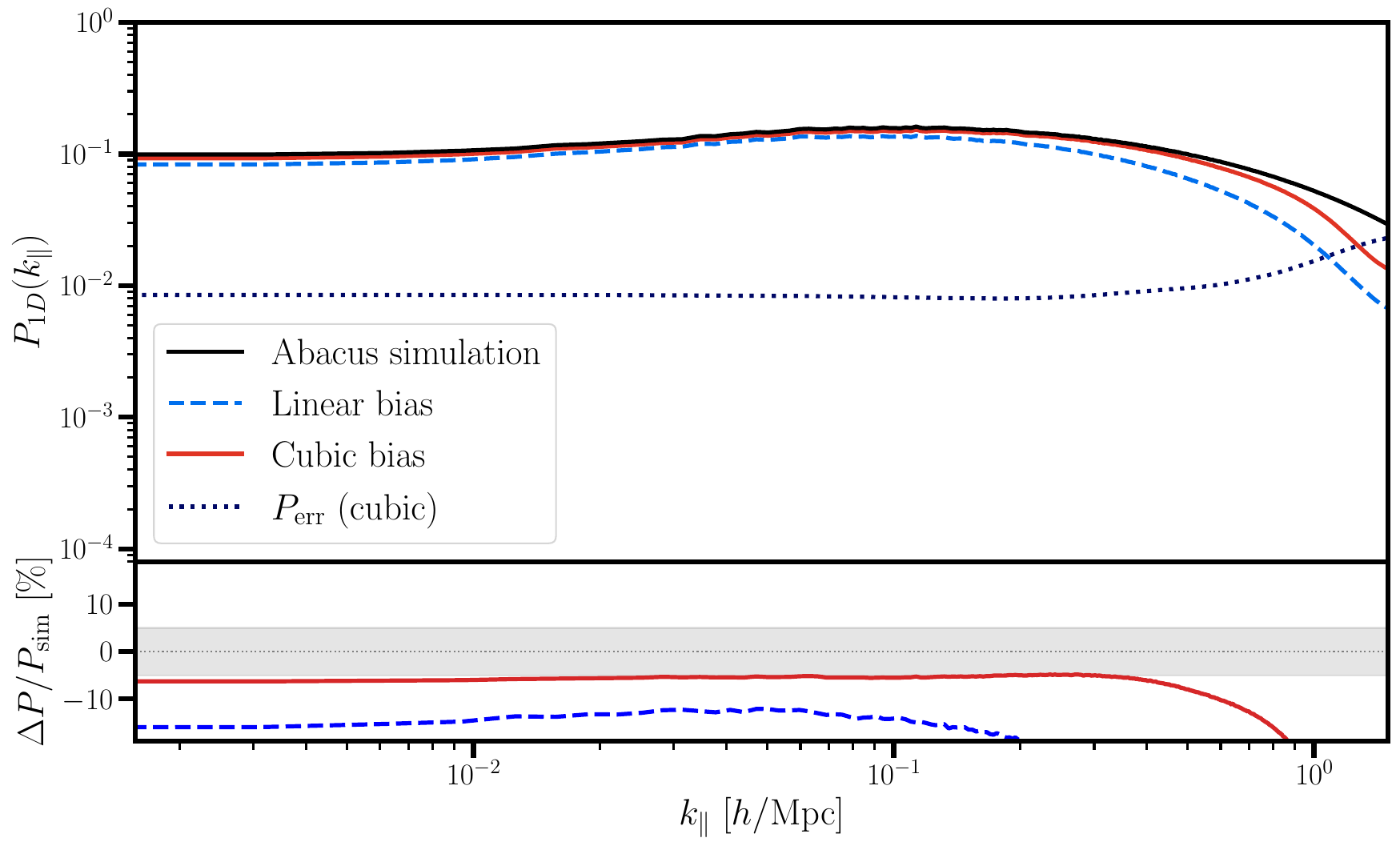}\hfill \\
    \includegraphics[width=0.329\linewidth]{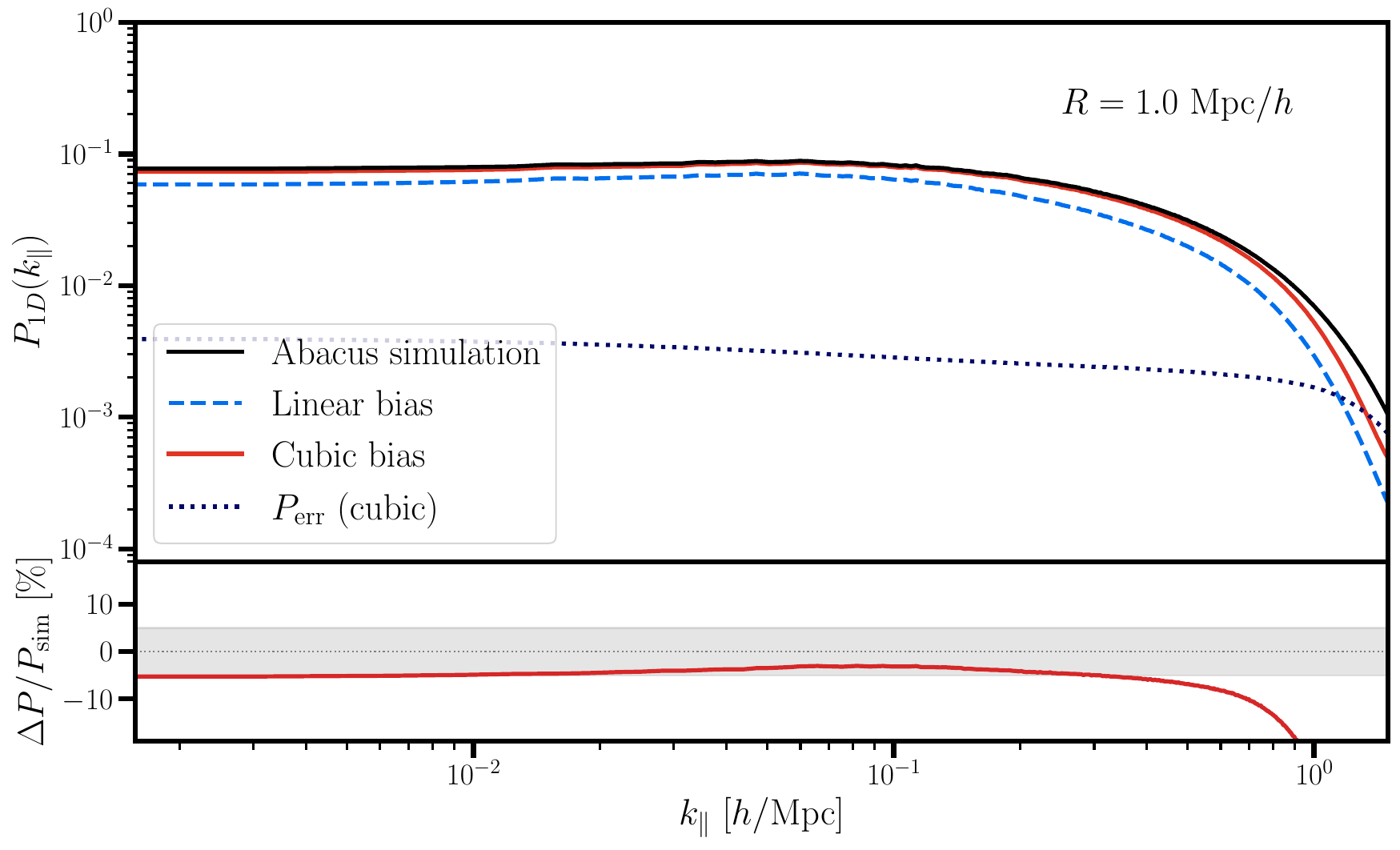}\hfill
    \includegraphics[width=0.329\linewidth]{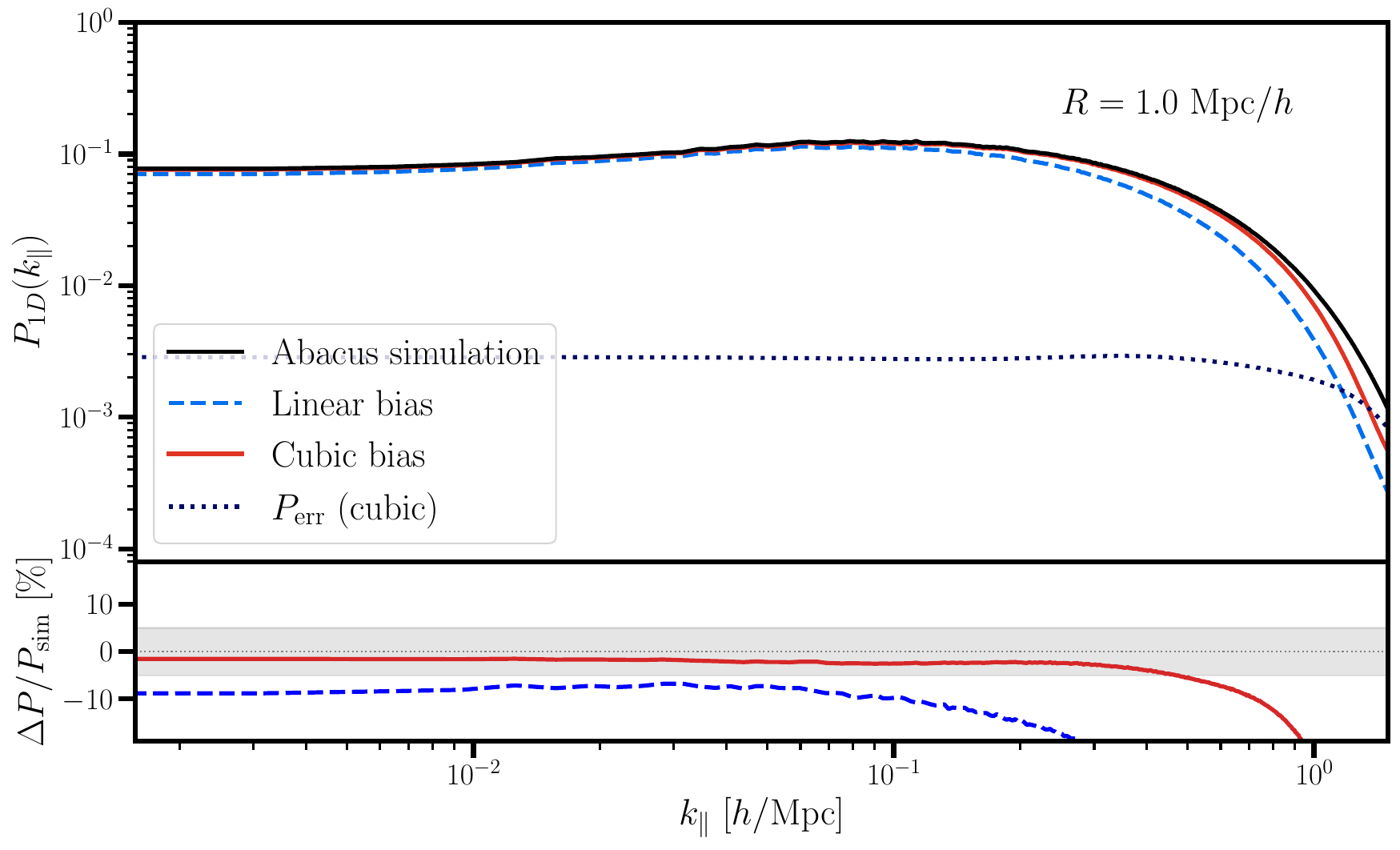}\hfill
    \includegraphics[width=0.329\linewidth]{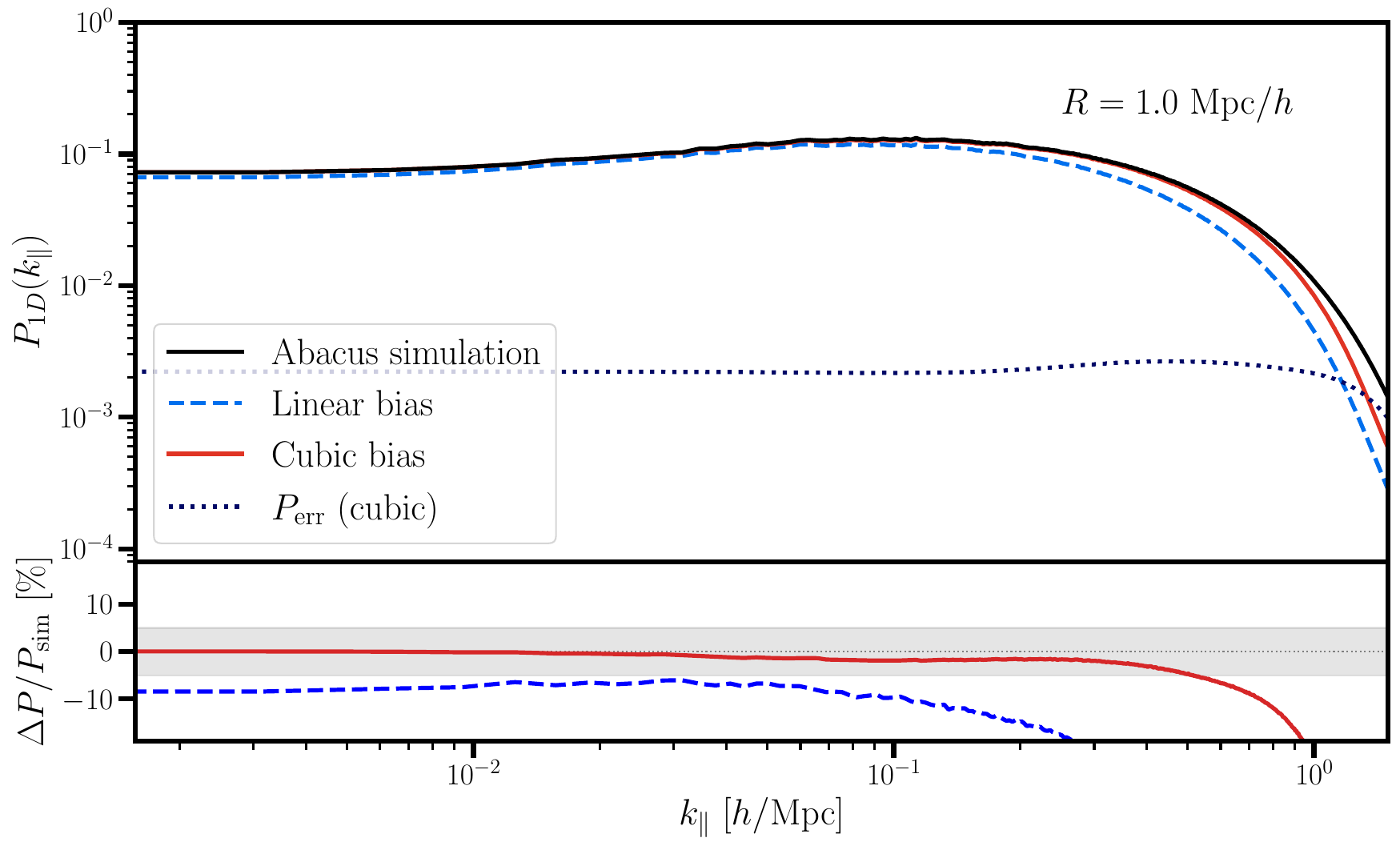}\hfill
    \vspace{-0.1in}
    \caption{\textbf{Model-dependence of P1D:}
    Same as Fig.~\ref{fig:abacus_pk_models} showing the one-dimensional power spectrum (P1D) for models II, III and IV (from left to right panel) comparing the smoothing scale $R=1\hinvMpc$ with the case of no smoothing. 
    }
    \label{fig:abacus_p1d_models}
\end{figure*}

\begin{figure*}
    \centering
    \includegraphics[width=0.329\linewidth]{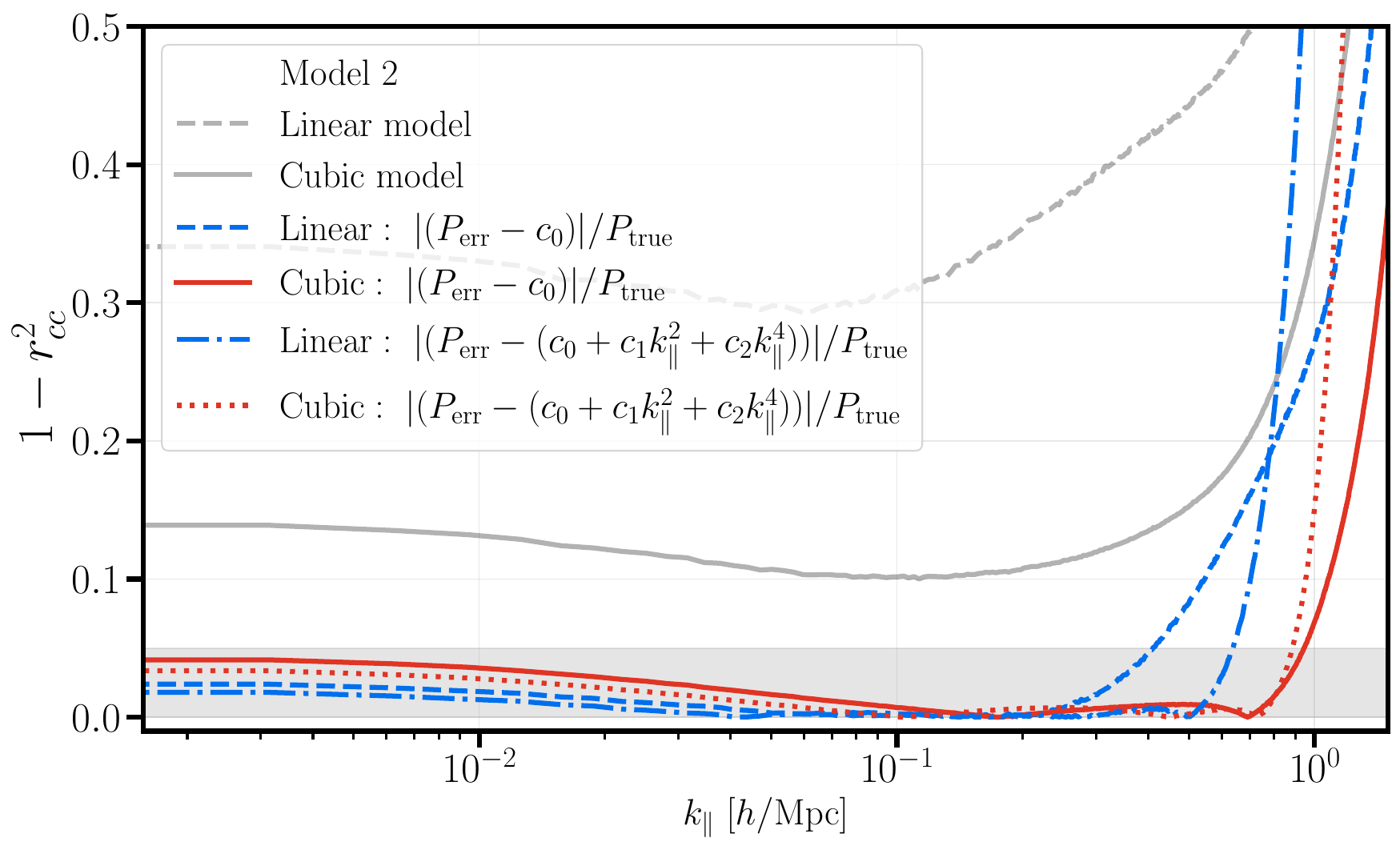}\hfill
    \includegraphics[width=0.329\linewidth]{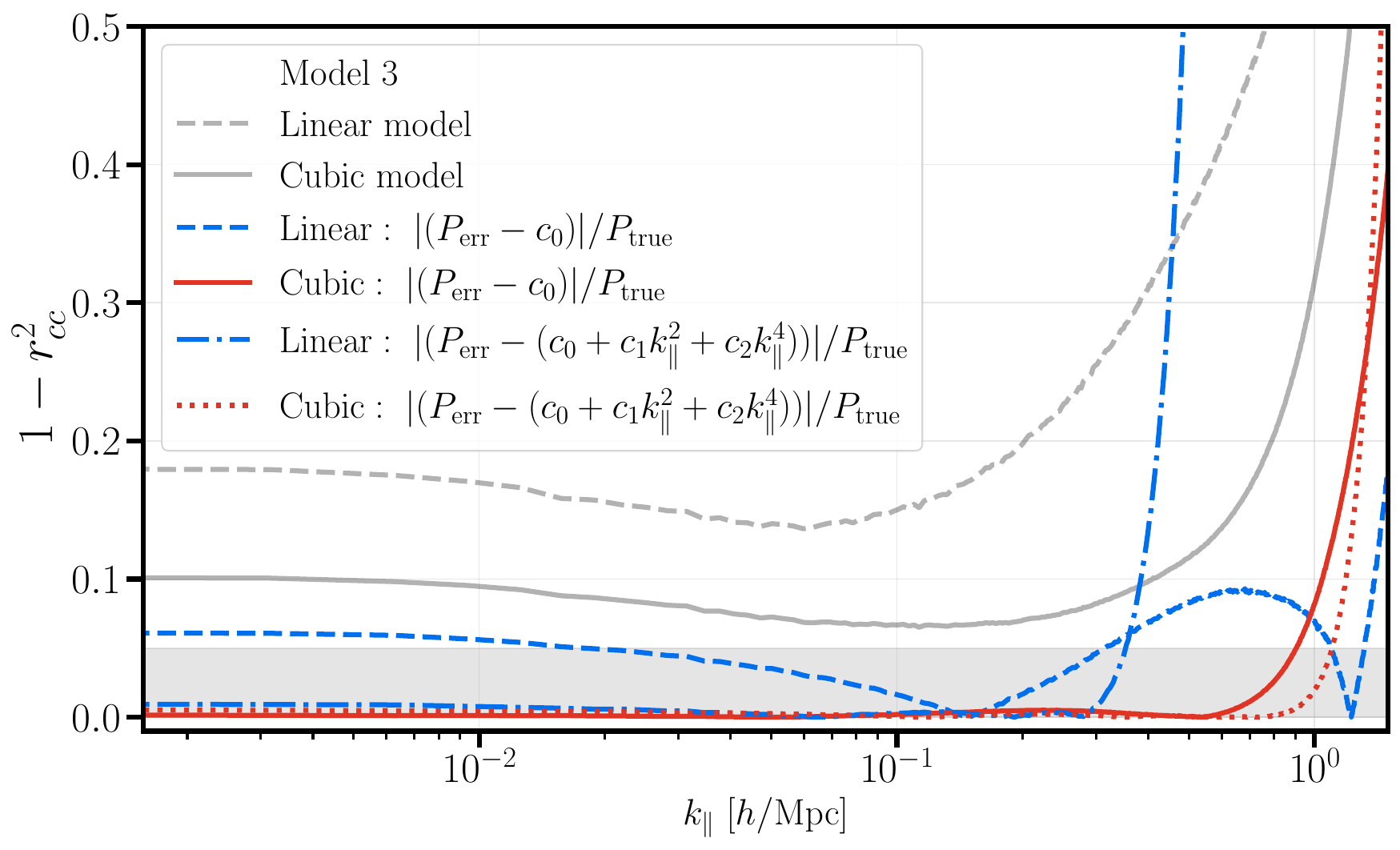}\hfill
    \includegraphics[width=0.329\linewidth]{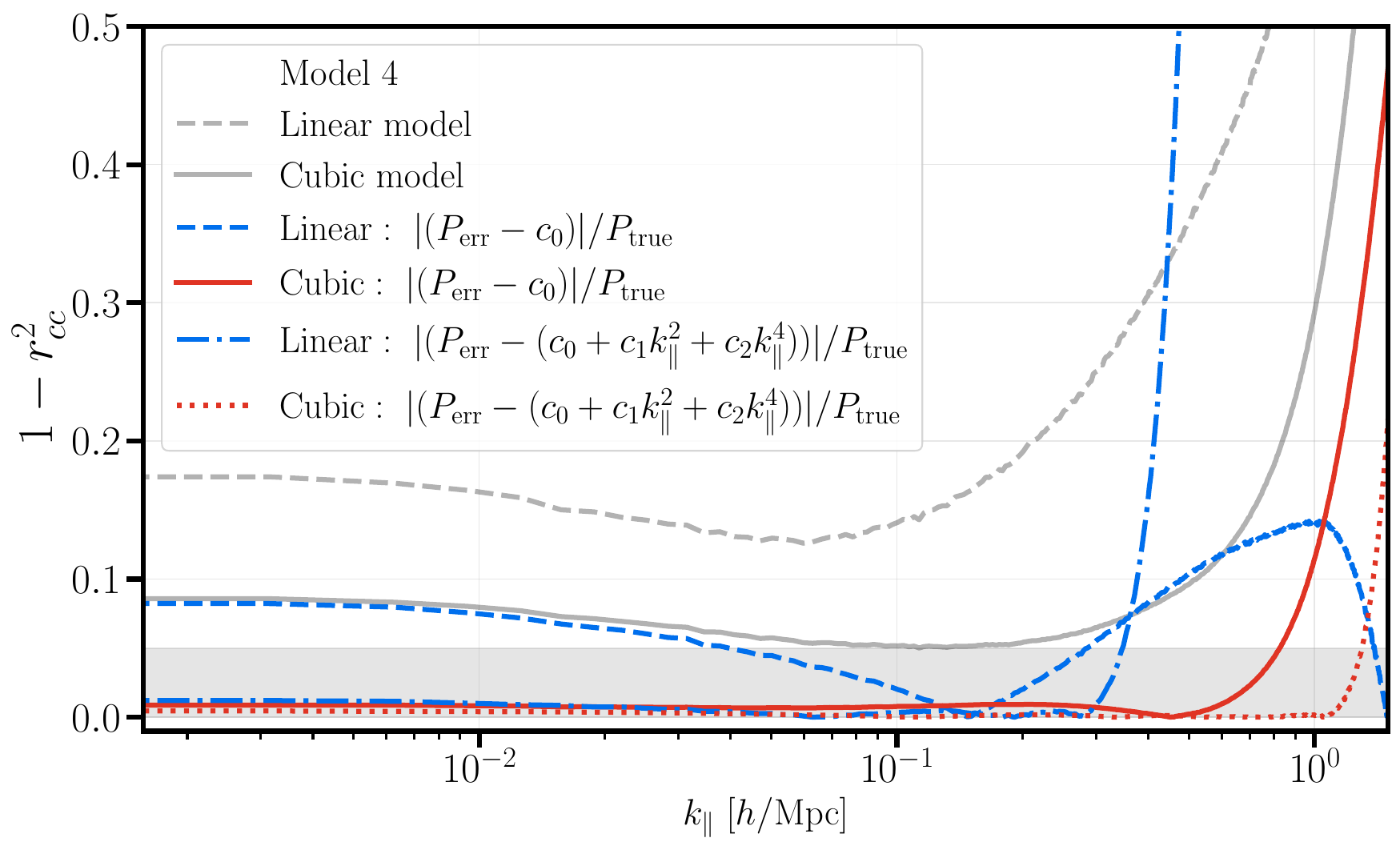}\hfill\\
    \includegraphics[width=0.329\linewidth]{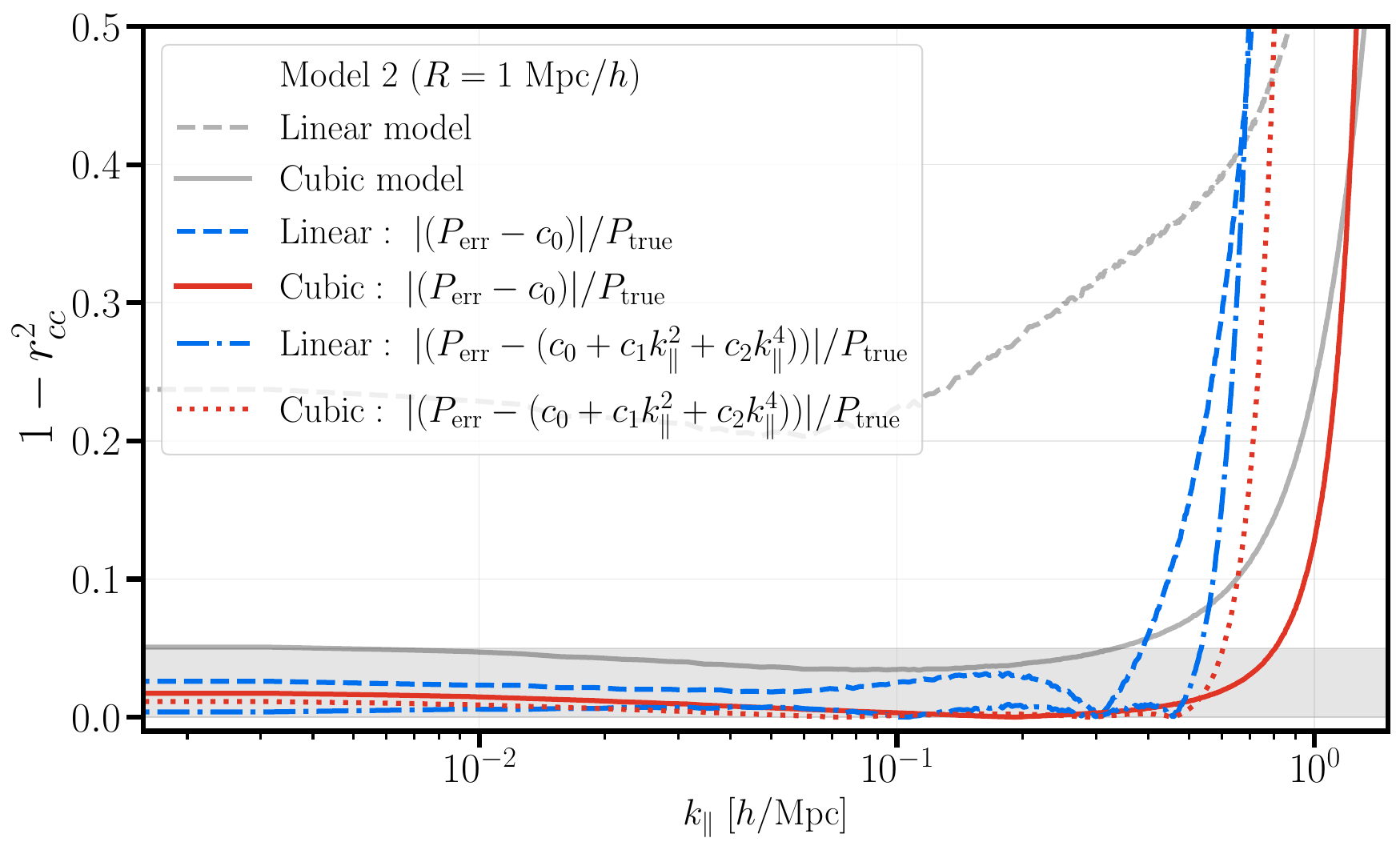}\hfill
    \includegraphics[width=0.329\linewidth]{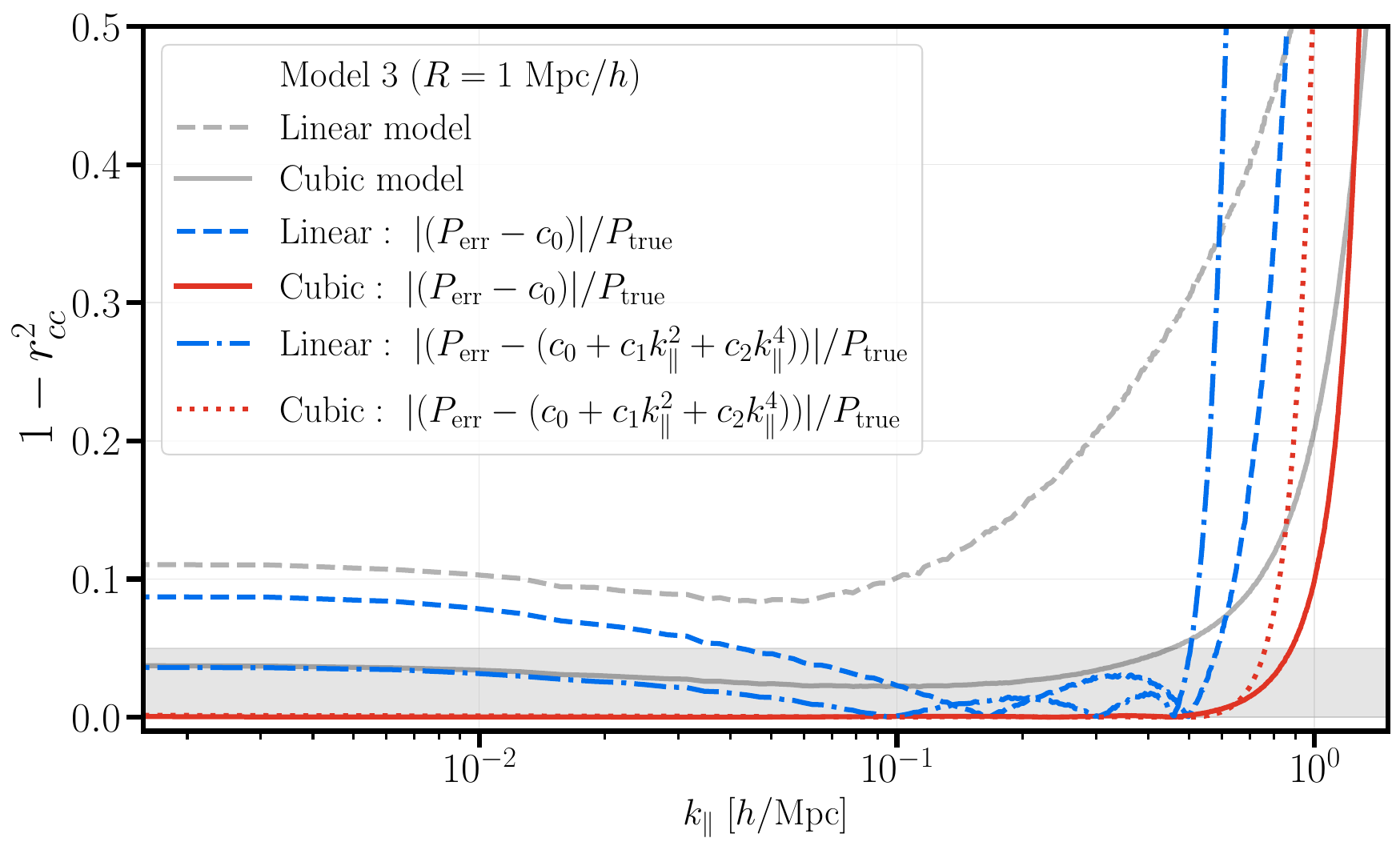}\hfill
    \includegraphics[width=0.329\linewidth]{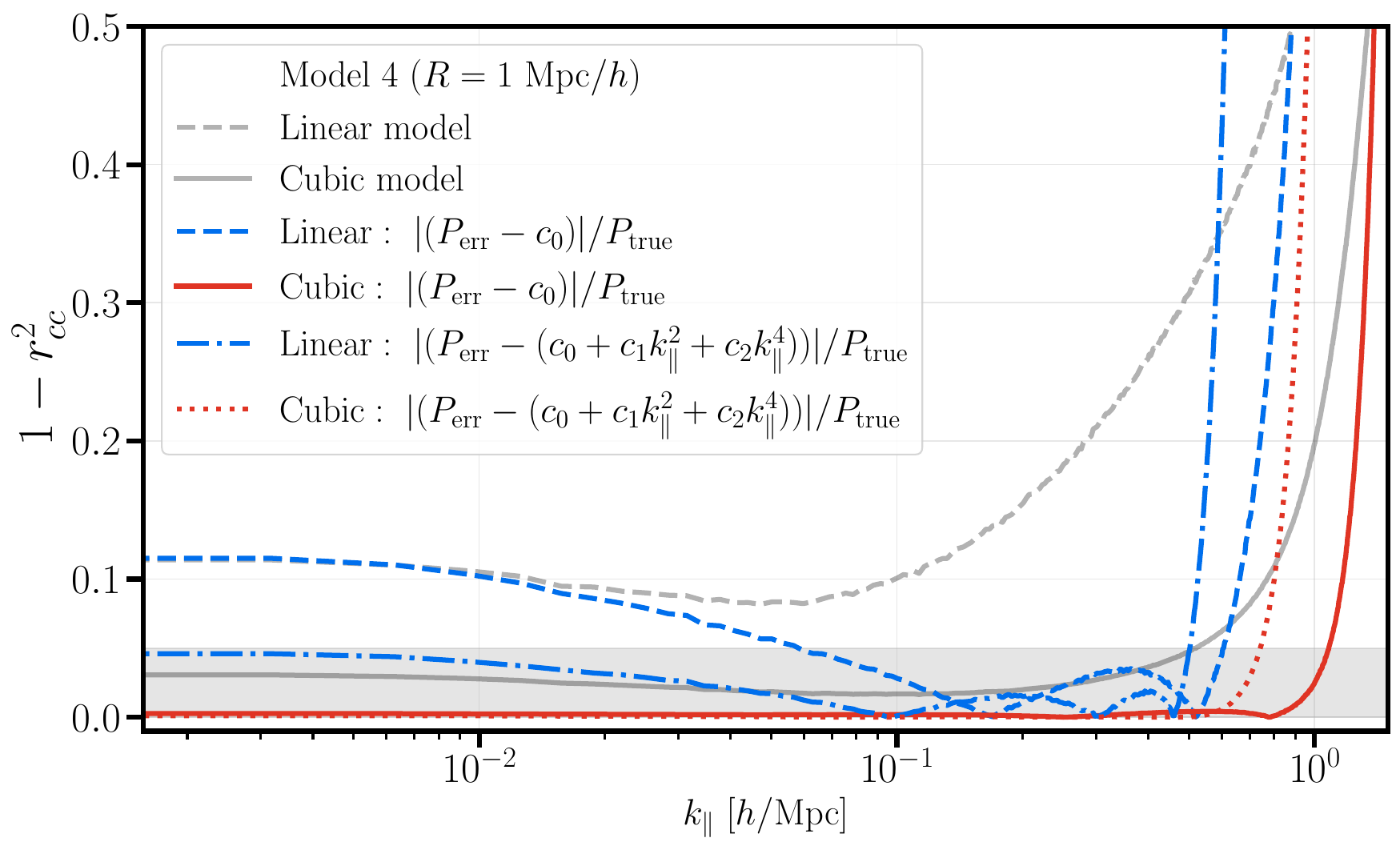}\hfill
    \vspace{-0.1in}
    \caption{\textbf{Model-dependence of Cross-Correlation Coefficient:}
    Same as Fig.~\ref{fig:abacus_pk_models} showing the cross-correlation coefficients for models two, three and four.
    }
    \label{fig:abacus_rcc_models}
\end{figure*}

\begin{table*}
\centering
\begin{tabular}{l|cccccccc}
\hline\hline
TF & $c_0$ & $c_{01}$ & $c_1$ & $c_{12}$ & $c_{14}$ & $c_4$ & $c_{22}$ & $c_{44}$ \\
\hline $\mathbf{Model\, II}$ &&&&&&\\
$\beta_1$ & $-0.134$ & $-0.142$ & $\phantom{-}0.073$ & $\phantom{-}0.253$ & $-0.052$ & $-0.008$ & $-0.156$ & $\phantom{-}0.055$ \\
$\beta_2$ & $\phantom{-}0.097$ & $\phantom{-}0.060$ & $-0.017$ & $\phantom{-}0.074$ & $-0.080$ & $-0.018$ & $-0.058$ & $-0.045$ \\
$\beta_{\mathcal{G}_2}$ & $-0.075$ & $-0.151$ & $\phantom{-}0.018$ & $\phantom{-}0.041$ & $\phantom{-}0.077$ & $\phantom{-}0.017$ & $\phantom{-}0.002$ & $\phantom{-}0.097$ \\
$\beta_3$ & $-0.006$ & $-0.011$ & $-0.016$ & $-0.108$ & $\phantom{-}0.087$ & $\phantom{-}0.016$ & $\phantom{-}0.092$ & $\phantom{-}0.008$ \\
$\beta_{KK_\parallel}$ & $-0.018$ & $-0.152$ & $-0.181$ & $\phantom{-}0.563$ & $-0.659$ & $-0.056$ & $-0.897$ & $\phantom{-}0.841$ \\
$\beta_{\eta}$ & $-0.199$ & $\phantom{-}0.066$ & $\phantom{-}0.121$ & $-0.241$ & $\phantom{-}0.101$ & $-0.011$ & $-0.043$ & $\phantom{-}0.178$ \\
$\beta_{\eta^2}$ & $\phantom{-}0.064$ & $-0.034$ & $-0.098$ & $\phantom{-}0.071$ & $\phantom{-}0.097$ & $\phantom{-}0.014$ & $\phantom{-}0.019$ & $\phantom{-}0.036$ \\
$\beta_{\delta\eta}$ & $-0.057$ & $\phantom{-}0.005$ & $-0.016$ & $-0.244$ & $-0.008$ & $-0.007$ & $\phantom{-}0.478$ & $-0.296$ \\
\hline $\mathbf{Model\, III}$ &&&&&&\\
$\beta_1$ & $-0.139$ & $-0.295$ & $\phantom{-}0.050$ & $\phantom{-}0.407$ & $-0.057$ & $-0.002$ & $-0.099$ & $-0.022$ \\
$\beta_2$ & $\phantom{-}0.060$ & $\phantom{-}0.105$ & $-0.022$ & $\phantom{-}0.244$ & $-0.160$ & $-0.008$ & $-0.129$ & $-0.082$ \\
$\beta_{\mathcal{G}_2}$ & $-0.072$ & $-0.300$ & $\phantom{-}0.041$ & $-0.110$ & $\phantom{-}0.205$ & $\phantom{-}0.009$ & $\phantom{-}0.149$ & $\phantom{-}0.100$ \\
$\beta_3$ & $-0.002$ & $-0.012$ & $-0.012$ & $-0.020$ & $\phantom{-}0.018$ & $\phantom{-}0.011$ & $\phantom{-}0.074$ & $-0.003$ \\
$\beta_{KK_\parallel}$ & $-0.168$ & $-0.515$ & $\phantom{-}0.106$ & $\phantom{-}0.461$ & $-0.137$ & $-0.093$ & $-1.051$ & $\phantom{-}0.777$ \\
$\beta_{\eta}$ & $-0.252$ & $-0.046$ & $\phantom{-}0.265$ & $-0.442$ & $\phantom{-}0.245$ & $-0.091$ & $\phantom{-}0.255$ & $-0.030$ \\
$\beta_{\eta^2}$ & $-0.032$ & $\phantom{-}0.010$ & $-0.163$ & $\phantom{-}0.415$ & $-0.076$ & $\phantom{-}0.052$ & $-0.008$ & $-0.073$ \\
$\beta_{\delta\eta}$ & $\phantom{-}0.045$ & $\phantom{-}0.054$ & $\phantom{-}0.027$ & $-0.211$ & $-0.290$ & $-0.032$ & $\phantom{-}0.223$ & $\phantom{-}0.030$ \\
\hline $\mathbf{Model\, IV}$ &&&&&&\\
$\beta_1$ & $-0.135$ & $-0.329$ & $\phantom{-}0.051$ & $\phantom{-}0.436$ & $-0.052$ & $-0.002$ & $-0.120$ & $-0.017$ \\
$\beta_2$ & $\phantom{-}0.064$ & $\phantom{-}0.119$ & $-0.025$ & $\phantom{-}0.257$ & $-0.164$ & $-0.008$ & $-0.142$ & $-0.088$ \\
$\beta_{\mathcal{G}_2}$ & $-0.072$ & $-0.331$ & $\phantom{-}0.040$ & $-0.100$ & $\phantom{-}0.221$ & $\phantom{-}0.010$ & $\phantom{-}0.147$ & $\phantom{-}0.111$ \\
$\beta_3$ & $\phantom{-}0.000$ & $-0.012$ & $-0.015$ & $-0.032$ & $\phantom{-}0.022$ & $\phantom{-}0.011$ & $\phantom{-}0.083$ & $\phantom{-}0.002$ \\
$\beta_{KK_\parallel}$ & $-0.152$ & $-0.594$ & $\phantom{-}0.078$ & $\phantom{-}0.727$ & $-0.223$ & $-0.078$ & $-1.337$ & $\phantom{-}0.893$ \\
$\beta_{\eta}$ & $-0.286$ & $-0.044$ & $\phantom{-}0.323$ & $-0.752$ & $\phantom{-}0.396$ & $-0.116$ & $\phantom{-}0.509$ & $-0.143$ \\
$\beta_{\eta^2}$ & $-0.011$ & $-0.001$ & $-0.164$ & $\phantom{-}0.476$ & $-0.113$ & $\phantom{-}0.049$ & $-0.048$ & $-0.045$ \\
$\beta_{\delta\eta}$ & $\phantom{-}0.043$ & $\phantom{-}0.068$ & $\phantom{-}0.039$ & $-0.273$ & $-0.364$ & $-0.037$ & $\phantom{-}0.275$ & $\phantom{-}0.037$ \\
\hline\hline
\end{tabular}
\caption{\textbf{Coefficients of Best-Fit Abacus Transfer Functions Models:} Same as Tab.~\ref{tab:abacus_tf_results} for models II-IV.} \label{tab:abacus_tf_results_2}
\end{table*}

\begin{figure}
    \centering
\includegraphics[width=0.329\linewidth]{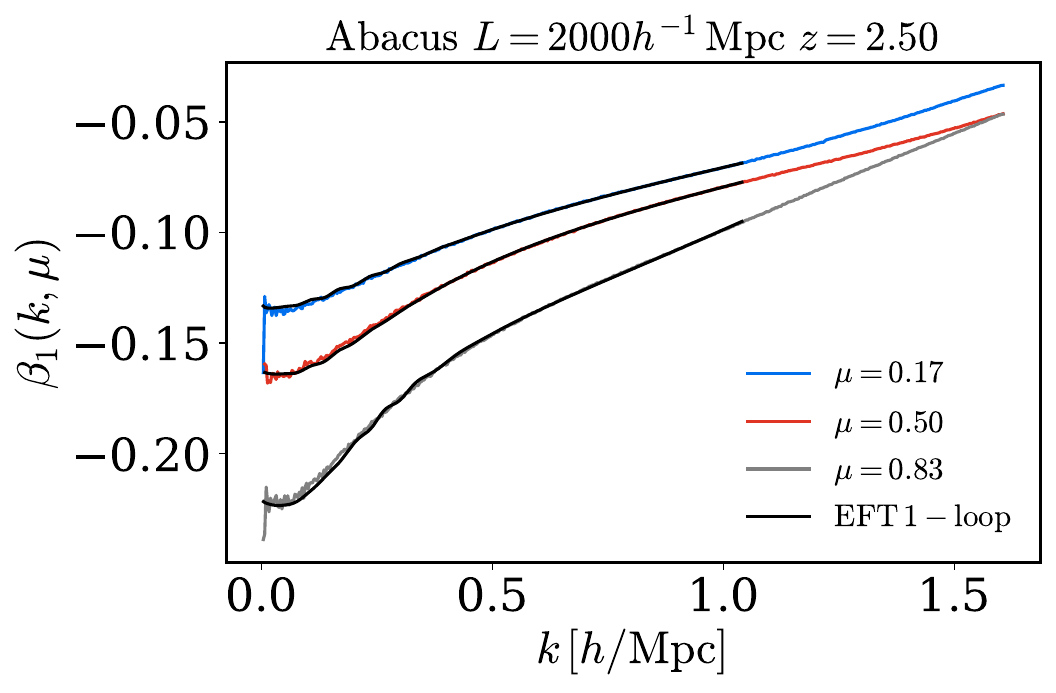}\hfill
\includegraphics[width=0.329\linewidth]{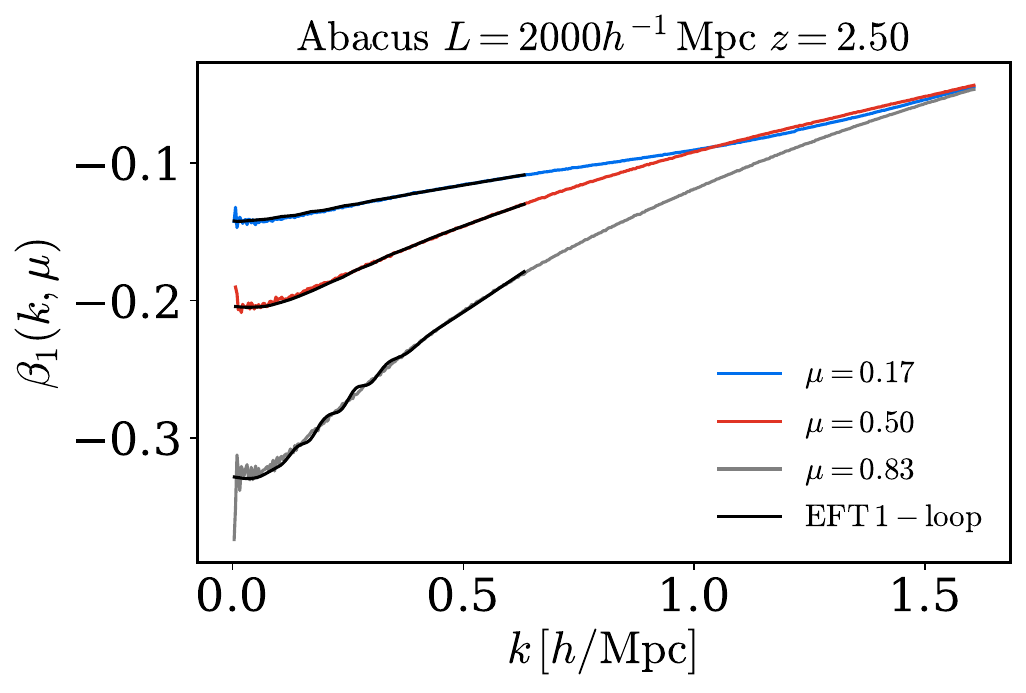}\hfill
\includegraphics[width=0.329\linewidth]{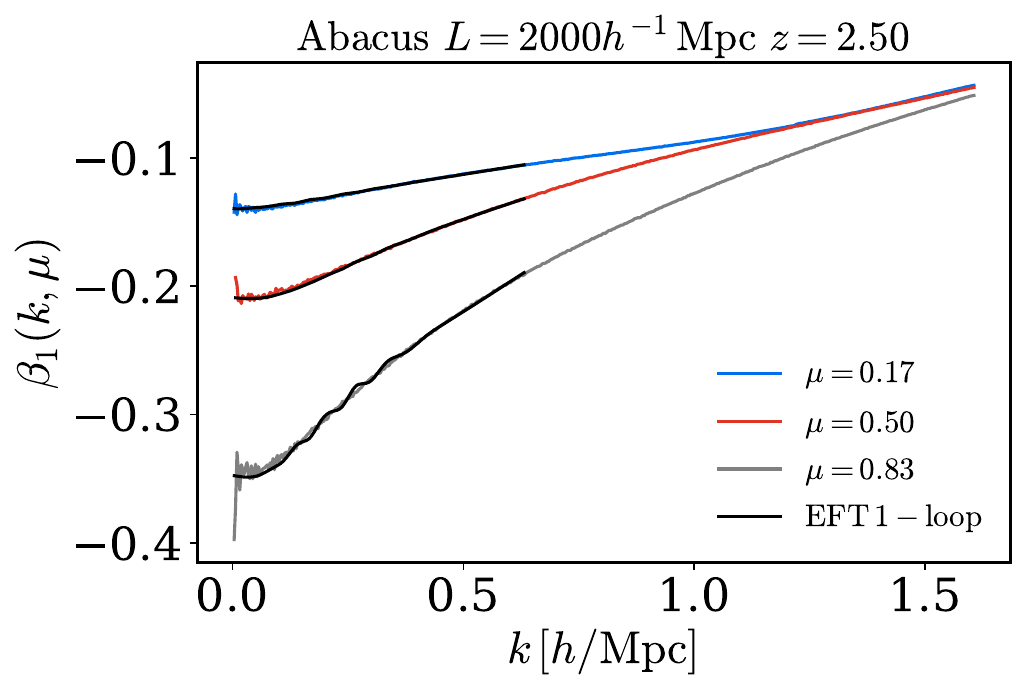}\hfill
\vspace{-0.1in}
    \caption{\textbf{Best-fit EFT TF Abacus:} Same as Fig.~\ref{fig:bf_tr_b1} for models II-IV for Abacus from left to right. Note that models III and IV use a lower $\kmax=0.75\hMpcinv$ for the fits. The perturbative transfer functions fit the data to high accuracy.}
    \label{fig:bf_tr_b1_abacus}
    \vspace{-0.1in}
\end{figure}

\bibliographystyle{JHEP}
\bibliography{apssamp, short.bib, references}
\end{document}